\documentclass[accepted,booklet]{itpthesis}

\usepackage[linktoc=all]{hyperref}

\usepackage[T1]{fontenc}
\usepackage[ascii]{inputenc}
\usepackage[USenglish]{babel}
\usepackage{aliascnt,doi,microtype,amsmath,amsthm,amsfonts,amssymb,bbm,graphicx,minibox,braket,wasysym,mathtools,tikz,tikz-3dplot,suffix,enumitem,makeidx,algpseudocode,booktabs,mathdots,rotating,multicol,caption,setspace,tabularx,ifthen}
\usepackage[vcentermath]{youngtab}
\usepackage[refpage,noprefix,noncompatible]{nomencl}

\allowdisplaybreaks[4]

\hypersetup{pdftitle={Reductions to IID in Device-independent Quantum Information Processing}, pdfauthor={Rotem Arnon-Friedman}}

\usepackage{tikz}
\usepackage{pgfplots}
\usepgfplotslibrary{fillbetween}
\usepackage{pgf}
\usetikzlibrary{patterns}
\usepackage{float}
\usepackage{graphicx}
\usepackage{mathrsfs}

\usepackage[chapter]{algorithm}
\floatname{algorithm}{Protocol}
\usepackage{algcompatible}

\usepackage{todonotes}

\usepackage{hhline}
\newcommand{\specialcell}[2][c]{
  \begin{tabular}[#1]{@{}l@{}}#2\end{tabular}}

\newcommand{\newjointcountertheorem}[3]{\newaliascnt{#1}{#2}\newtheorem{#1}[#1]{#3}\aliascntresetthe{#1}}
\newtheorem{thm}{Theorem}[chapter]
\newjointcountertheorem{lem}{thm}{Lemma}
\newjointcountertheorem{cor}{thm}{Corollary}
\newjointcountertheorem{prp}{thm}{Proposition}
\newjointcountertheorem{cnj}{thm}{Conjecture}
\newjointcountertheorem{que}{thm}{Question}
\newjointcountertheorem{pro}{thm}{Problem}
\newjointcountertheorem{fct}{thm}{Fact}
\newjointcountertheorem{obs}{thm}{Observation}
\newjointcountertheorem{alg}{thm}{Algorithm}
\newjointcountertheorem{asm}{thm}{Assumption}
\theoremstyle{definition}
\newjointcountertheorem{defn}{thm}{Definition}
\newjointcountertheorem{ntn}{thm}{Notation}
\theoremstyle{remark}
\newjointcountertheorem{rem}{thm}{Remark}
\newjointcountertheorem{nte}{thm}{Note}
\newjointcountertheorem{exl}{thm}{Example}

\AtBeginDocument{

}

\newtheorem{innercustomlem}{Lemma}
\newenvironment{customlemma}[1]
  {\renewcommand\theinnercustomlem{#1}\innercustomlem}
  {\endinnercustomlem}

\usepackage{fixmath}
\newcommand{\mr}[1]{\mathbold{#1}} 

\renewcommand{\P}{\mathrm{P}} 
\renewcommand{\O}{\mathrm{O}}
\newcommand{\Q}{\mathrm{Q}} 
\newcommand{\Tr}{\mathrm{Tr}} 
\newcommand{\idn}{\mathbb{I}} 
\newcommand{\G}{\mathrm{G}} 
\newcommand{\hilb}{\mathscr{H}} 

\renewcommand{\braket}[2]{\langle #1| #2 \rangle}

\newcommand{\passtest}{T} 
\newcommand{\agq}{aid} 
\newcommand{\inSigma}{in^{\Sigma}} 
\newcommand{\insigma}{in^{\sigma}}

\addtocontents{toc}{~\hfill\textbf{Page}\par}
\addtocontents{lof}{~\hfill\textbf{Page}\par}
\addtocontents{lot}{~\hfill\textbf{Page}\par}

\usepackage{subfiles}

\begin{document}

\title{\vspace{-5pt} \Huge Reductions to IID in \\ Device-independent \\ Quantum Information Processing \vspace{-5pt}}

\author{Rotem Arnon-Friedman}
\previousdegree{Master of Science \\ Tel-Aviv University}
\authorinfo{born January 12, 1986 \\ citizen of Israel}
\referees{Prof.~Dr.~Renato Renner, examiner\\  Prof.~Dr.~Nicolas Gisin, co-examiner\\ Prof.~Dr.~Ran Raz, co-examiner\\ Prof.~Dr.~Andreas Winter, co-examiner\\}
\degreeyear{\vspace{-5pt} 2018}

\ethdissnumber{25543}
\maketitle

\begin{dedication}
	To Neer, Avishai, and Omri
\end{dedication}

\begin{abstract}{Abstract}
%



	The field of \emph{device-independent} quantum information processing concerns itself with devising and analysing protocols, such as quantum key distribution and quantum tomography, without referring to the quality of the physical devices utilised to execute the protocols. 
	Instead, the analysis is based on the observed correlations that arise during a repeated interaction with the devices and, in particular, their ability to violate the so called Bell inequalities. 

	Since the analysis of device-independent protocols holds irrespectively of the underlying physical device, it implies that \emph{any} device can be used to execute the protocols: If the apparatus is of poor quality, the users of the protocol will detect it and abort; otherwise, they will accomplish their goal.  
	This strong statement comes at a price~--- the analysis of device-independent protocols is, a priori, extremely challenging. Having good techniques at hand is  thus crucial.

	The thesis presents an approach that can be taken to simplify the analysis of device-independent information processing protocols. 
	The idea is the following: Instead of analysing the most general  device leading to the observed correlations, one should first analyse a significantly simpler device that, in each interaction with the user, behaves in an identical way, independently of all other interactions. We call such a device an independently and identically distributed~(IID) device. 
	As the next step, special techniques are used to prove that, without loss of generality, the analysis of the IID device implies similar results for the most general device.
	Such techniques reduce the problem of analysing the general scenario to that of analysing an IID one and, hence, we term them \emph{reductions to IID}.

	We present two mathematical techniques that can be used as reductions to IID in the device-independent setting: \emph{de Finetti reductions for correlations} and the \emph{entropy accumulation theorem}. 
	Each technique is  accompanied by a showcase-application that exemplifies the reduction's usage and benefits. 
	Specifically, we use our de Finetti reduction to prove a \emph{non-signalling} (super-quantum) \emph{parallel repetition theorem}, belonging to a family of theorems discussed in theoretical computer science.
	The entropy accumulation theorem is used to prove the security of \emph{device-independent quantum cryptographic protocols}. 
	
	Performing the analysis via a reduction to IID instead of directly analysing the most general scenarios leads to simpler proofs and significant quantitive improvements, matching the tight results proven when analysing  IID devices. 
	In particular, our analysis of device-independent quantum key distribution protocols produces essentially optimal key rates and noise tolerance, crucial for all future experimental implementations of device-independent  cryptography.

\end{abstract}

\begin{abstract}{Zusammenfassung}

Die ger\"{a}teunabh\"{a}ngige Quanteninformationsverarbeitung besch\"{a}ftigt sich mit der Entwicklung und Analyse von Protokollen, wie z.B. dem Quantenschl\"{u}sselaustausch oder der Quantentomographie, welche unabh\"{a}ngig von der Qualit\"{a}t der eingesetzten physikalischen Ger\"{a}te ist. Stattdessen basiert die Analyse auf beobachteten Korrelationen, die durch wiederholte Wechselwirkung mit den Ger\"{a}ten entstehen, insbesondere ihrer F\"{a}higkeit Bellsche Ungleichungen zu verletzen.

Da die Analyse ger\"{a}teunabh\"{a}ngiger Protokolle unabh\"{a}ngig von den eingesetzten Ger\"{a}ten ist, k\"{o}nnen \emph{beliebige} Ger\"{a}te benutzt werden um die Protokolle auszuf\"{u}hren: Weist das genutzte Ger\"{a}t eine schlechte Qualit\"{a}t auf, detektiert das Protokoll dies und bricht ab. Ansonsten wird das Protokoll erfolgreich sein. Diese starke Aussage hat jedoch ihren Preis: die Analyse ger\"{a}teunabh\"{a}ngiger Protokolle ist extrem herausfordernd. Deswegen sind gute Methoden f\"{u}r deren Analyse essentiell.

Diese Dissertation stellt eine Herangehensweise zur Vereinfachung der Analyse ger\"{a}teunabh\"{a}ngiger Protokolle vor, basierend auf folgender Idee: Statt die allgemeinsten Ger\"{a}te zu analysieren, welche zu den beobachteten Korrelationen f\"{u}hren, wird zun\"{a}chst ein deutlich einfacheres Ger\"{a}t analysiert, das sich in jeder Wechselwirkung mit dem Benutzer identisch verh\"{a}lt, unabh\"{a}ngig von allen anderen Wechselwirkungen. Wir nennen solch ein Ger\"{a}t ein identisch und unabh\"{a}ngig verteiltes (IID) Ger\"{a}t. Dann werden spezielle mathematische Methoden benutzt um zu zeigen, dass diese einfachere Analyse
 \"{a}hnliche Ergebnisse wie die Analyse der allgemeinsten Ger\"{a}te liefert. Solche Methoden reduzieren die Analyse des allgemeinen Problems auf die des IID-Problems. Daher bezeichnen wir sie als \emph{Reduktionen auf das IID-Problem}.

Wir pr\"{a}sentieren zwei Reduktionen auf das IID-Problem: \emph{de Finetti Reduktionen f\"{u}r Korrelationen} sowie den \emph{Entropieanh\"{a}ufungssatz}. Beide Methoden werden durch Beispielanwendungen illustriert. Spezifisch benutzen wir die de Finetti Reduktion um einen sogenannten\emph{``non-signalling parallel repetition''}-Satz zu beweisen, welcher zu einer Familie von S\"{a}tzen geh\"{o}rt, die in der theoretischen Informatik diskutiert werden. Der Entropieanh\"{a}ufungssatz wird benutzt um die Sicherheit von \emph{ger\"{a}teunabh\"{a}ngigen Quantenkryptografieprotokollen} zu beweisen.

Indem man die Analyse durch Reduktionen auf den IID Fall durchf\"{u}hrt, anstelle einer Analyse der allgemeinsten Szenarien, erh\"{a}lt man einfachere Beweise und signifikante quantitative Verbesserungen, welche mit den strengen Resultaten f\"{u}r IID Ger\"{a}te \"{u}bereinstimmen. Insbesondere f\"{u}hrt unsere Analyse von ger\"{a}teunabh\"{a}ngigen Protokollen f\"{u}r Quantenschl\"{u}sselaustausch zu ann\"{a}hernd optimalen Schl\"{u}sselraten und Fehlertoleranzen, welche essentiell f\"{u}r alle zuk\"{u}nftigen experimentellen Implementierungen von ger\"{a}teunabh\"{a}ngiger Kryptographie sind.

\end{abstract}

\begin{acknowledgements}

	I am lucky to be surrounded by people who inspire me, believe in me, and allow me to grow. 
	There is no better way of spending our most precious resource, time, and thus to them I am grateful. 
	
	I would like to thank Renato Renner, my supervisor, who offered me an opportunity of a lifetime and opened the door to the never-ending quest of solving intriguing and challenging questions. 
	Renato was a great inspiration; I was constantly amazed by his contributions to science and his way of thinking, as well as how determined he is to come up with a proof even when it seems impossible 
	(these days, whenever I get stuck, I just tell myself that Renato would never give up!).
	Most of all, I am thankful to Renato for believing in me, moving mountains to allow me to focus on my research, and supporting me when I needed to focus on other things. 
	
	I appreciate and  thank  the co-examiners, Andreas Winter, Nicolas Gisin, and Ran Raz, for taking the time to read my thesis, as well as Ernest Tan, Fr\'{e}d\'{e}ric Dupuis, Jie Lin, Marco Tomamichel, and Thomas Vidick for their valuable comments on parts of the thesis. 
	
	I am thankful to 
	my collaborators. 
	Out of them, a special thanks goes to my office mate Christopher Portmann, who was always happy to discuss  the details of the details,
	and to Thomas Vidick for being an inspiring and motivating collaborator,  an invaluable mentor, and a great friend. 
	
	I was honored to be part of the QIT group at ETH, consisting of many talented people. Being part of this family allowed me to interact and learn from past and current members of the group to whom I thank.
	
	
	My great appreciation to Marko Gebbers for enlightening me, helping me separate the wheat from the chaff, and challenging me to become a better version of myself. I am forever grateful.

	Owing much more than that, I wish to thank my mother, father, and sister, as well as my best friends, for always being there for me, even from far away, and for having an unbelievable amount of patience.
	I cannot thank enough my mother and mother-in-law who were always happy to fly to Zurich to help with the kids when I wanted to travel for a conference and  my father who illustrated an endless number of boxes (and sheeps) for my papers, talks, and this thesis.

	Finally, no words can describe my gratitude and love to the  person who supported and believed in me from the first moment (pushing me into the building to meet Renato) to the last sentence of this thesis. The one who allows me to grow, accomplish myself, and win both worlds. The one who made it all possible~--- my husband.

\end{acknowledgements}

\cleardoublepage\pagestyle{empty}\phantomsection\pdfbookmark{\contentsname}{toc} %
\tableofcontents

\cleardoublepage\pagestyle{fancy}\startnumbering

\chapter{Introduction}\label{ch:intro}

	\section{Motivation}
	
		\subsection{Device-independent information processing}\label{sec:motivation_di}

			The study of quantum information unveils new possibilities for remarkable forms of computation, communication, and cryptography by investigating different ways of manipulating quantum states.
			Crucially, the analysis of quantum information processing tasks must be based, in one way or another, on the actual physical processes used to implement the considered task; the physical processes must be inherently quantum as otherwise no advantage can be gained compared to classical information processing. 
			In most applications, the starting point of the analysis is an explicit and exact characterisation of the quantum apparatus, or device, used to implement the task of interest.

			As an example, consider the task of quantum key distribution (QKD). In a QKD protocol, the goal of the honest parties, called Alice and Bob, is to create a shared key, unknown to everybody else but them. 
			The protocol is intrinsically quantum: To execute it Alice and Bob hold entangled quantum states in their laboratories and perform quantum operations, or measurements, on the quantum states. 
			Informally, proving the security of a QKD protocol amounts to showing that no adversary can hold (significant) information about the produced key. 
			To prove security one usually needs to have a complete description of the quantum devices, i.e., the quantum states and measurements, used by Alice and Bob. 
			For example, the security proof of the celebrated BB84 protocol~\cite{bennett1984proceedings} builds on the assumptions that Alice and Bob hold two-qubit states and are able to measure them in a specific way.
		 When these assumptions are dropped, the protocol is no longer secure~\cite{pironio2009device}.
			Thus, if Alice and Bob wish to use their quantum devices in order to implement a QKD protocol they need to first make sure that the device is performing the exact operations described by the protocol. 
			
			Unfortunately, in practice we are unable to fully characterise the physical devices used in quantum information processing tasks. 
			Even the most skilled experimentalist will recognise that a fully characterised, always stable, large-scale quantum device that implements a QKD protocol is extremely hard to build. 
			If the honest users' device is different from the device analysed in the accompanying security proof, security is no longer guaranteed and imperfections can be exploited to attack the protocol.

			Noise and imperfections cannot be completely avoided when implementing quantum information processing tasks. 
			Furthermore, imperfections being imperfections, one also cannot expect to perfectly characterise them. That is, we cannot say for sure what exactly is about to go wrong in the quantum devices: Maybe the measurements are not well-calibrated, perhaps some noise introduces correlations between particles which are intended to be independent, or interaction with the environment may possibly lead to decoherence.
			Even the advent of fault-tolerant computation, if achievable one day, cannot resolve all types of errors if no promise is given regarding the number of errors and their, possibly adversarial, nature. 			
			Once we come to terms with the above, a natural question arises: 

			\begin{center}
		 		  \bf{Can quantum information processing tasks be accomplished by utilising uncharacterised, perhaps even adversarial, physical devices?}
			\end{center}
			
			An adversarial, or malicious, device is one implemented by a hostile party interested in, e.g., breaking the cryptographic protocol being executed. 
			Clearly, this is an extreme scenario to consider. Note, however, that even if the manufacturer of the device is to be trusted, he may still be incompetent~--- the physical apparatus will be subject to uncharacterised imperfections  even though the manufacturer is honest and has good intentions.

%

			The field of device-independent information processing addresses the above question. 
			In the device-independent framework we treat the physical devices, on which a minimal set of constraints is enforced,\footnote{Clearly, one cannot perform any cryptographic task if the device includes a transmitter that just sends all the information to the adversary. Few minimal assumptions regarding the device will be needed; see Section~\ref{sec:untrusted_devices}. Depending on the considered task, some of the assumptions can be enforced in practice while others may require some minimal level of trust.}  as \emph{black boxes}~--- Alice and Bob hold a box and can interact with it classically (as explained below) to execute the considered protocol, but they cannot open it to assess its internal workings.\footnote{Notice that even if Alice and Bob did have some information about the physical apparatus, the device-independent framework does not allow them to take advantage of this information in the analysis. For example, Alice and Bob may be able to distinguish a device that uses the polarisation of a photon to encode a qubit from one based on superconducting qubits (even the author is able to do that). Yet, this information is not  to be used  when treating the device as a black box.}
			They have no knowledge regarding the physical apparatus and do not trust that it works as alleged by the manufacturer of the device.  
			
			What \emph{can} Alice and Bob do with the black box? They can interact with it by pushing buttons, each associated with some classical input (e.g., a bit)  and record the classical outputs produced by the box in response to pressing its buttons. 
			Thus, the only information available to Alice and Bob is the observed classical data created during their interaction with the black box. (Hence the name ``device-independent'').

			Since the device is not to be trusted, the classical information collected by Alice and Bob during the interaction with the box must allow them, somehow, to test the possibly faulty or malicious device and decide whether using it, e.g., to create their keys by executing a QKD protocol, poses any security risk. 
			A protocol or task is said to be device-independent if it guarantees that by interacting with the device according to the specified steps the parties will either abort, if they detect a fault, or accomplish the desired task (with high probability).
		
			The possibility of  device-independent information processing is quite surprising. Indeed, restricting ourselves to classical physics and classical information, it is impossible to derive device-independent statements.\footnote{Consider for example the case of device-independent QKD. Classical devices can always be pre-programmed by the adversary to output a fixed key of her choice.} 
			The most important ingredients for device-independent protocols are the existence of Bell inequalities and quantum ``non-local'' correlations that violate them~\cite{bell1964einstein}. These two facts are far from  trivial and play a fundamental role in quantum theory. 
			In the context of device-independent information processing, a Bell inequality acts as a ``test for quantumness'' that allows the users of the device to verify that their device is ``doing something quantum'' and cannot be simulated by classical means. This ``quantumness'', of a specific form discussed below, is what allows us to, e.g., prove security of a QKD protocol.
			
			A Bell inequality can be thought of as a multi-player game, also called a non-local game, played by the parties using the device they share.
			A non-local game goes as follows. A referee asks each of the (cooperating) parties a question chosen according to a given probability distribution. 	The parties need to supply answers which fulfil a pre-determined requirement according to which the referee accepts or rejects the answers. 
			In order to do so, they can agree on a strategy beforehand, but once the game begins communication between the parties is not allowed. If the referee accepts their answers the players win. 
			The goal of the parties is, naturally, to maximise their winning probability in the game.
			
			Different devices held by the parties implement different strategies for the game and may lead to different winning probabilities.
			In the device-independent setting we are interested in games that have a special ``feature''~--- there exists a quantum device which achieves a  winning probability in the game  that is greater than all classical, local, devices.
			If the honest parties learn, by interacting with the device, that their device can win the game with probability higher than that of all classical devices, they conclude it cannot be explained by classical physics alone.\footnote{We postpone the formal and more technical discussion to a later point; an enthusiastic reader may jump ahead to Section~\ref{sec:pre_bell_ineq}.}

			Crucially, the winning probability in the game does not merely indicate that the device is doing something quantum but how non-classical it is. 
			Relations are known between the probability of winning some non-local games and  various other quantities. Some examples for quantities of interest are the entropy produced by the device, the amount of entanglement consumed to play the game, or the distance (under an appropriate distance measure) of the device from a specific fully characterised quantum device. 
			Such relations lie at the heart of any analysis of device-independent information processing tasks.

			Although above we only mentioned device-independent QKD as an example for a device-independent task, the framework of device-independence does not only concern the more-than-average paranoid cryptographers. 
			The framework fits any scenario in which, a priori, we do not want to assume anything about the utilised devices and their underlying physical nature. To reassure the reader, we give three additional examples. 
			
			Bell inequalities were originally introduced  in the context of the foundations of quantum mechanics in order to resolve the EPR paradox~\cite{einstein1935can}. 
			When trying to test quantum theory against an alternative classical world that admits a ``local hidden variable model'' (or, in other words, falsify all classical explanations of a behaviour of a physical system), one cannot assume that quantum theory holds to begin with and must treat the device as a black box without assuming to know its internal workings. 
		
			A second example is that of blind tomography, also termed self-testing. Assume a quantum state is being produced in some experimental setting. Quantum tomography is the process of estimating which state is being created by performing measurements on copies of the state and collecting the statistics~\cite{tomography}. To get a meaningful estimation, a certain set of measurements needs to be used, depending on the dimension of the state. In other words, in order to estimate and characterise the quantum state, we must be able to first characterise the measurement devices.  
			\emph{Blind} quantum tomography refers to the process in which the measurements are also unknown. In such a case, nothing but the observed statistics can be used~\cite{mayers1998quantum,bancal2015physical}.
		
			Another interesting example is that of verification of computation~--- given a device claimed to be a quantum computer, how can human beings, who cannot perform quantum computations by themselves, verify that this is indeed the case? There are different ways of addressing this question, but in all cases we would like to make statements without presuming that the considered devices are performing any particular quantum operations (see, e.g.,~\cite{reichardt2013classical}).

			The device-independent framework  becomes relevant whenever one wishes to make concrete statements without referring to the underlying physical nature of the utilised devices and the types of imperfections or errors that may occur. 
			The derived statements are extremely strong.  Device-independent security, for example, is regarded as the gold standard for quantum cryptography, since attacks exploiting the mismatch between security proof and implementation are no longer an issue. 
			Making such strong statements comes at a price. The analysis of device-independent tasks is, a priori, extremely challenging: We treat the devices as black boxes and thus the proofs need to account for an almost arbitrary, even adversarial, behaviour of the devices. 
			Having good techniques for the analysis at hand is therefore crucial. This is further discussed in the following section.

		\subsection{Reductions to IID}\label{sec:motivation_reduction}
		
			In the device-independent setting one does not have a description of the specific device used in the considered task and, hence, must analyse the behaviour of arbitrary devices. For example, when proving security of cryptographic protocols we clearly need to consider \emph{any} possible device that the adversary may prepare. 
			Unfortunately, analysing the behaviour of arbitrary devices can be wearying at best and infeasible at worst. 
			Let us start by explaining why this is the case.
			
			As mentioned above, the ability to achieve device-independent information processing tasks is based on the existence of  non-local games and quantum strategies to play them that can beat any classical strategy. 
			To perform complex tasks, such as device-independent cryptography, employing the device to play a \emph{single} non-local game is clearly not enough; we cannot conclude any meaningful information regarding the device by asking it to produce outputs only for a single game. 
			To put quantum information to work we must consider protocols in which the device is used to play \emph{many} non-local games. This way, the parties executing the protocol can collect statistics and test their device. If the device does not pass the test the parties abort the protocol (see Protocol~\ref{pro:intro_qkd} below for an example). 
			
			The reason for the difficulty of the analysis lies in the fact that one needs to examine the overall behaviour of the device during the entire execution of the protocol, consisting of playing many games with the device, instead of its behaviour in a single game. 
			 As the device is uncharacterised its actions when playing one game may depend on other games.  
			 
			In general, there are two families of devices able to play many games that one can consider~--- parallel and sequential devices. A parallel device is one which can be used to play \emph{all} the games \emph{at once}. That is, the parties executing the protocol are instructed to give all the inputs, for all the games, to the device and only then the device produces the outputs for all the games. In such a case, the actions of the device in one game may depend on \emph{all} other games. 
			
			A sequential device, on the other hand, is used to play the games \emph{one after the other}, i.e., the parties give the device the first inputs and wait for its outputs and only then proceed to play the next game. In between the games, some communication may be allowed between the parties and the different components of the device. In the case of a sequential device, the behaviour  of the device in one game may depend on all \emph{previous} games as well as communication taking place during the time between the games.\footnote{The formal definitions of parallel and sequential devices are given in Chapter~\ref{ch:multi_box}.}
			In both cases, the input-output behaviour of the devices gets quite complicated. 
			
\begin{figure}
				\centering
				\includegraphics[width=\textwidth]{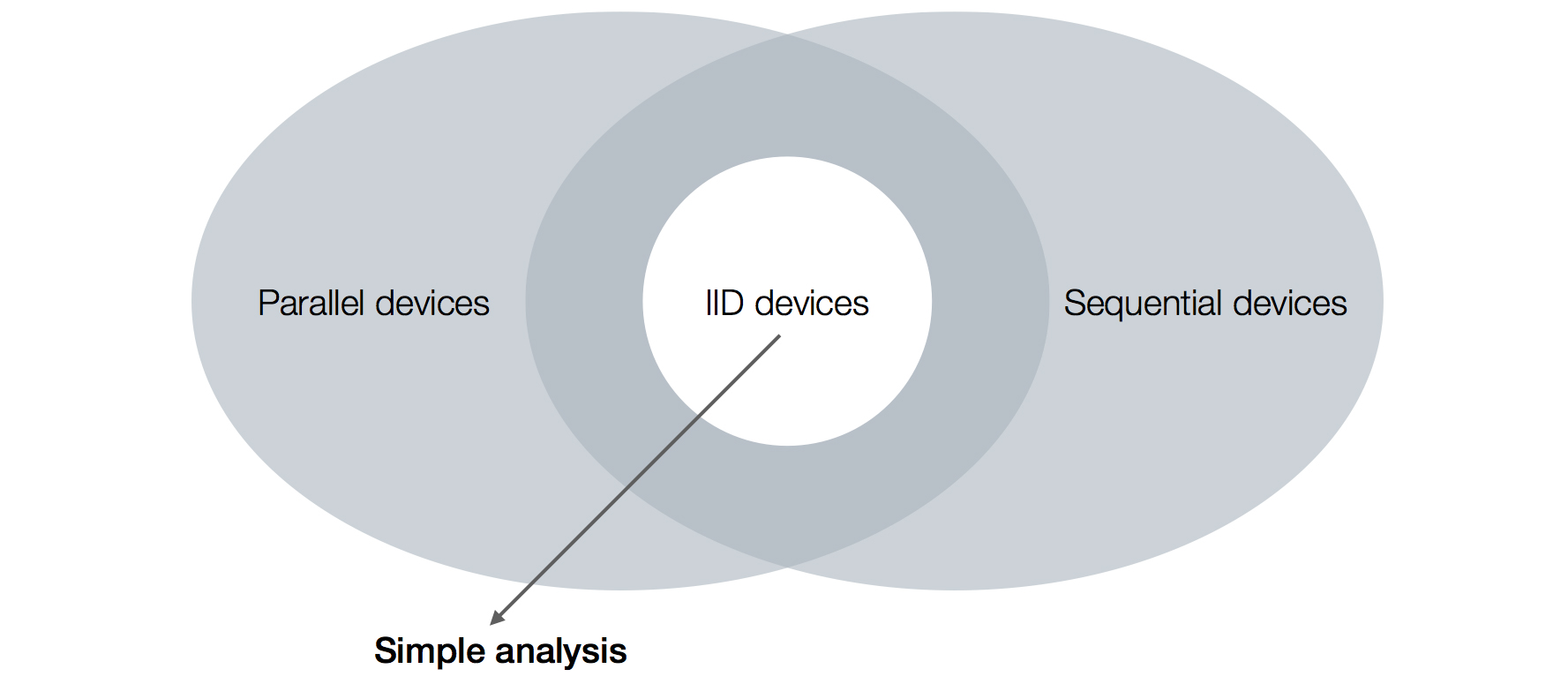}
				\caption{The relation between the different sets of devices. The intersection of the sets of sequential and parallel devices includes the set of IID devices. The analysis of IID devices, i.e., that done under the IID assumption, is rather simple.}
				\label{fig:boxes_sets_iid_intro}
			\end{figure}

			One common assumption introduced to simplify the analysis of device-independent information processing tasks is the so called ``independent and identically distributed''~(IID) assumption. 
			As the name suggests, a device is said to be an IID device if it plays each of the games independently of the others and utilises the same strategy for all games. An IID device is a special case of both parallel and sequential devices and, since it is highly structured, analysing its behaviour can be significantly simpler than analysing the more general devices; see Figure~\ref{fig:boxes_sets_iid_intro}.

			The IID assumption heavily restricts the structure of the device. It is therefore not clear at all that analysing device-independent information processing tasks under the IID assumption is sufficient. 
			Returning to the example of device-independent cryptography, an adversary who can prepare arbitrary devices (let it be sequential or parallel) may be strictly stronger, i.e., can get more information about the outputs of the honest parties, than an adversary restricted to IID devices.
			Thus, simplifying the analysis by using the IID assumption comes at the cost of weakening the final statement. 
			
			The main question addressed in this thesis is the following:
			\begin{center}
			   \bf{Can the analysis of device-independent information processing tasks be \emph{reduced} to that performed under the IID assumption?}
			\end{center}
			
			The term \emph{reduction} is widely used in theoretical computer science and is meant to describe the process of showing that one problem is as hard/easy as another. In our case, we ask whether analysing general devices is as easy as analysing IID devices or, in other words, does an analysis performed under the IID assumption imply results concerning general devices (i.e., statements which are not restricted to the IID case). 
			A priori, it is not at all obvious that this is the case; clearly, not all devices are IID devices. A positive answer to the above question means that \emph{even though} there exist devices that cannot be described as IID ones, it is sometimes possible to restrict the attention solely to IID devices and the rest will follow. 
			
			The idea of applying a reduction to IID as a proof technique was conceived\footnote{Perhaps surprisingly, as far as the author is aware the idea of a ``reduction to IID'' does not appear or used in classical information processing and cryptography.}  in~\cite{christandl2007one}, following which a concrete reduction relevant for applications was developed in~\cite{renner2008security} and used to reduce the security proof of QKD protocols to that done under the IID assumption.\footnote{In the context of QKD, security under the IID assumption is called security against collective attacks.}  As such,~\cite{renner2008security} acts as the first example for a proof using a reduction to IID.
			
			
			Analysing information processing tasks via a reduction to IID has several significant advantages.
			Analysing IID devices is relatively easy and almost always intuitive. Thus, having tools that allow us to extend the analysis to the general case greatly simplifies proofs.\footnote{The reductions themselves are not necessarily simple, but that is fine. They are technical tools that are only proved once and can then be used to simplify many other proofs. The researcher using the reduction does not need to reprove anything.}  The simplicity, in turn, allows for clear and modular statements as well as quantitively strong results.\footnote{This is in agreement with Occam's razor; while there is no notion of the ``right proof'' out of several possible proofs (assuming they are all mathematically correct), the simplest proof usually turns out to be the most useful and insightful one.}
			
			The importance of quantitively strong results is obvious, especially when discussing  quantum information processing tasks: If we wish to benefit from the new possibilities brought by the study of quantum information, we must be able to implement the protocols in practice. Without strong quantitive bounds on, e.g., key rates and tolerable noise levels, we cannot take the device-independent field from theory to practice. 
			Clarity and modularity should also not be dismissed. Science is not a ``one-man's job''; clarity and modularity  are crucial when advancing science as a community. Indeed, complex and fine-tuned proofs are hard to verify, adapt to other cases of interest, and quantitively improve.
			
			Another advantage of reducing a general analysis to  IID is that it allows us to separate the wheat from the chaff. 
			The essence of the arguments used in proofs of information processing tasks almost always enter the game in the analysis of the IID case. 
			Proofs that address the most general scenarios directly (i.e., not via a reduction to IID) are at risk of obscuring the ``physics'' by more technical mathematical steps. 
			When using a reduction to IID this is (mostly) not the case~--- the essence, or the interesting part, lies in the analysis of IID devices while the technicalities are pushed into the reduction itself.

			With the above advantages, the development and application of reductions to IID flourished in quantum information processing. Yet, the benefits did not reach the subfield of device-independent quantum information processing. The reason was clear~--- all the techniques used as reductions to IID had to make assumptions regarding the investigated system, which are too restrictive when studying uncharacterised devices.

			As we will show in the thesis, reductions to IID can also be developed and employed in device-independent quantum information processing.
			We present two techniques that can be used as reductions to IID, accompanied by  two showcase-applications that illustrate how the reductions can be used and their benefits in terms of the derived theorems.
			The following section presents the content of the thesis in more detail.

	\section{Content of the thesis}
	
		The goal of the thesis is to explain how reductions to IID can be performed in the context of device-independent information processing. 
		To this end, after explaining the different mathematical objects that one needs to consider and their relevance, we discuss the IID assumption and its implications in the device-independent setting. 
		We then present two techniques, or tools, that can be used as reductions to IID in the analysis of device-independent information processing tasks, one relevant for parallel devices and the other for sequential ones. 
		
		To better comprehend the topic and exemplify the usage of the two reductions, we consider two applications as showcases, namely, parallel repetition of non-local games and device-independent cryptography.
		These are studied in detail throughout the chapters of the thesis, while taking the perspective of reductions to IID.

		\subsection{Reductions}
		
\begin{figure}
				\centering
				\includegraphics[width=\textwidth]{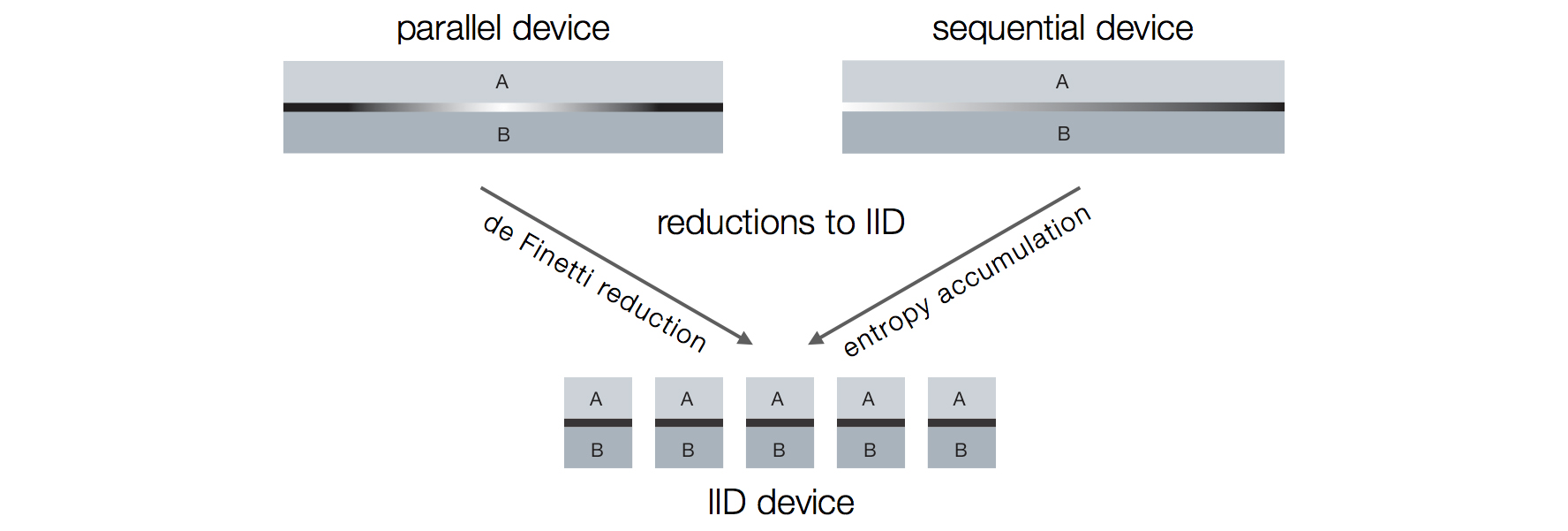}
				\caption{Reductions to IID in device-independent information processing. de Finetti reductions can be used to reduce the study of parallel devices to IID device (see Chapters~\ref{ch:reductions_par} and~\ref{ch:par_rep_showcase}), while the entropy accumulation theorem can be used  when dealing with sequential devices (Chapters~\ref{ch:reductions_seq} and~\ref{ch:crypto_showcase}).}
				\label{fig:reductions_big_pic_intro}
			\end{figure}

			Two types of reductions are presented. The reductions are applicable in different scenarios and give statements of different forms; see Figure~\ref{fig:reductions_big_pic_intro}.
			
			\subsubsection{de Finetti reduction for correlations}
			
				The first reduction, the topic of Chapter~\ref{ch:reductions_par}, is called ``de Finetti reduction for correlations'' and was developed in~\cite{arnon2013finetti}. The de Finetti reduction is relevant for the analysis of \emph{permutation invariant parallel devices}. 
				Permutation invariance is an inherent symmetry in many information processing tasks, device-independent tasks among them. Thus, analysing permutation invariant devices is of special interest. 
				
				In short, in our context, a de Finetti reduction is a theorem that relates any permutation invariant parallel device to a special type of device, termed de Finetti device, which behaves as a convex combination of IID devices (see Chapter~\ref{ch:reductions_par} for the formal definitions). 
				The given relation acts as a reduction to IID when considering tasks admitting a permutation invariance symmetry and in which a parallel device needs to be analysed. Our showcase of parallel repetition of non-local games fits this description and thus can benefit from our de Finetti reduction.
				
				\sloppy
				Various quantum de Finetti theorems were know prior to our work and were successfully used to substantially simplify the analysis of many quantum information tasks. However, they cannot be applied in the device-independent setting, since they make many assumptions regarding the permutation invariant quantum states being analysed and therefore cannot accommodate uncharacterised devices. 
				The unique property of the reduction presented in Chapter~\ref{ch:reductions_par} is that, apart from permutation invariance, it makes no assumptions whatsoever regarding the systems of interest and is therefore applicable in the analysis of device-independent information processing.  
			
				For pedagogical reasons, we choose to present in the thesis a de Finetti reduction which is relevant to the case of bipartite devices, i.e., devices which are shared between two parties, Alice and Bob. 
				The statements can be extended to any number of parties, as shown in~\cite{arnon2013finetti}; the proofs of the general case do not include fundamental insights on top of those used in the bipartite case but require somewhat heavy notation. 
				We therefore omit the more general theorems and proofs  (while supplying the full analysis of the bipartite case in Chapter~\ref{ch:reductions_par}), with the hope of making the content more inviting for readers unfamiliar with the topic.
				
				Apart from presenting the reduction and the possible ways of using it, Chapter~\ref{ch:reductions_par} also includes a discussion of  ways in which it may be possible to extend or modify the reduction (to be more specific, we mainly present impossibility results). This content does not appear in detail in other published papers and can be relevant for future studies of the topic.   
							
			\subsubsection{Entropy accumulation theorem}
			
				The second reduction to IID that can be used in the device-independent setting is the entropy accumulation theorem (EAT)~\cite{dupuis2016entropy} and is the topic of Chapter~\ref{ch:reductions_seq}. 
				The EAT can be seen as an extension of the entropic formulation of the asymptotic equipartition property (AEP)~\cite{holenstein2011randomness,tomamichel2009fully}, applicable only under the IID assumption, to more general sequential processes. 
				
				The AEP, presented in Chapter~\ref{ch:iid_assumption}, basically asserts that when considering IID random variables, the smooth min- and max-entropies of the random variables converge to their von Neumann (or Shannon, in the classical case) entropy, as the number of copies of the random variable increases. 
				The AEP is of great importance when analysing, both classical and quantum, information processing tasks under the IID assumption: It explains why the von Neumann entropy is so important in information theory~--- the smooth entropies, which describe operational tasks, converge to the von Neumann entropy when considering a large number of independent repetitions of the relevant task.\footnote{A commonly used example is that of ``data compression''. There, one would like to encode an $n$ bit string using less bits. If we allow for some small error when decoding the data, the smooth max-entropy roughly describes the number of bits needed. However, for a large enough number of independent repetitions, less bits suffice and the exact amount is governed by the Shannon entropy.}
								
				Moving on from the IID setting, the EAT considers a certain class of quantum sequential processes. That is, in our context, it is relevant when studying \emph{sequential devices}.\footnote{To be more precise, some requirements regarding the process, or protocol, in which the sequential device is to be used must hold. This is explained in details in Chapter~\ref{ch:reductions_seq}.}
				Similarly to the AEP, when applicable, the EAT allows one to bound the total amount of the smooth min- and max-entropies using the same bound on the von Neumann entropy calculated for the IID analysis, i.e., the one used when applying the AEP. 
				In this sense, the EAT can be seen as a reduction to IID~--- with the aid of the EAT the analysis done under the IID assumption using the AEP can be extended to the one relevant for sequential devices.
				
				The proof of the EAT is not presented in the thesis (and should not be attributed to the author). 
				We focus on motivating, presenting, and explaining the EAT in the form relevant for device-independent quantum information processing~\cite{arnon2018practical} (as well as quantum cryptography in general), so it can be later used in our showcase of device-independent cryptography. 
				The pedagogical presentation of the EAT given in Chapter~\ref{ch:reductions_seq} does not appear in full in any other published material and we hope that it will make the theorem more broadly accessible.
		
			Before presenting our showcases, let us remark that both of the reductions mentioned above are not ``black box'' reductions, in the sense that one cannot simply say that if a problem is solved under the IID assumption then it is solved in the general case. 
			In particular, one should be familiar with the exact statements of the reductions (though not with their proofs) as well as the analysis of the considered task  under the IID  assumption in order to apply the reductions (or even just check whether they are applicable or not).
			When discussing the reductions in Chapters~\ref{ch:reductions_par} and~\ref{ch:reductions_seq}, we explicitly explain in what sense the presented tools count as reductions to IID techniques. 
		
		\subsection{Showcases}
		
			We use two showcases throughout the thesis in order to exemplify the approach of reductions to IID and the more technical usage of the presented reductions.
			The showcase of parallel repetition of non-local games uses the de Finetti reduction technique while the showcase of device-independent cryptography builds on the EAT. 
			As mentioned in Section~\ref{sec:motivation_reduction} above, we believe that analysing device-independent tasks using a reduction to IID has its benefits. The derived theorems are, arguably, more intuitive and insightful and, in addition, give strong quantitive results.  			
			
			We shortly discuss below each of our showcases. We present informal theorems describing the results proven for the showcases. The informal theorems shed light on the fundamental nature and strength of the approach of reductions to IID.
			
			\subsubsection{Non-signalling parallel repetition}
				
				Our first showcase is that of non-signalling parallel repetition. Chapter~\ref{ch:par_rep_showcase} presents our formal statements and proofs, which previously appeared in~\cite{arnon2016non}.
				As before, we focus in the thesis on the bipartite case for pedagogical reasons;~\cite{arnon2016non} includes the general analysis, which is valid for any number of parties playing the game.

			 	Non-local games, as mentioned in Section~\ref{sec:motivation_di}, are games played by several cooperating parties, also called players. 
			 	A referee asks each of the players a question chosen according to a given probability distribution. 	The players need to supply answers which fulfil a pre-determined requirement according to which the referee accepts or rejects the answers. 
				In order to do so, they can agree on a strategy beforehand, but once the game begins communication between the parties is no longer allowed. If the referee accepts their answers the players win.
				 
				 In the language used so far, we can think of a device as implementing a strategy for the game. 
				Depending on the field of interest, one can consider classical, quantum, or non-signalling devices, the latter referring to devices on which the only restriction is that they do not allow the players to communicate.				
				We focus below on the case of non-signalling strategies, or devices. 
			
				One of the most interesting questions regarding non-local games is the question of parallel repetition. Given a non-local game with optimal winning probability~$1-\alpha$ using non-signalling strategies, we are interested in analysing the optimal winning probability of a non-signalling strategy in the repeated, or threshold, game. 
				A threshold game is a game in which the referee asks the players to play $n\in \mathbb{N}$ instances of the non-local game, all at once, and the players' goal is to win more than~$1-\alpha+\beta$ fraction of the games, for $\beta >0$ a parameter of the threshold game. 
				The parallel repetition question concerns itself with upper-bounding the optimal winning probability in the threshold game, as the number of games~$n$ increases.\footnote{This is actually a generalisation of the more commonly known parallel repetition question, in which one wishes to upper-bound the probability of winning \emph{all} the $n$ games.}
				
				One trivial strategy that the players can use in the threshold game is a strategy employing a non-signalling IID device. That is,  they simply answer each of the $n$ questions independently using the optimal non-signalling device used to play a single game. 
				Using an IID device, the fraction of successful answers is highly concentrated around $1-\alpha$ and the probability to win more than a $1-\alpha+\beta$ fraction of the games decreases exponentially fast with $n\beta^2$, as follows from the optimal formulation of the Chernoff bound. 
				
				However, since the players receive from the referee all the questions to the $n$ instances of the non-local game at once, an IID device is not the most general device that they can use. 
				Instead, they can use any non-signalling parallel device to implement their strategy. 
				As parallel devices are strictly more general than IID ones, using parallel devices in fact allows them to win the threshold game with higher probability than in the IID case.\footnote{When first encountering the question of parallel repetition it may seem surprising that the players can do better using a parallel device, but this is indeed the case; see Section~\ref{sec:par_rep_intro_thresh_thm} a concrete example.} 
				Still, one may ask how the winning probability behaves for a sufficiently large number of repetitions $n$ and, especially, whether it decreases in  a similar fashion as for IID strategies.
				
				To answer the above question, we wish to reduce the study of strategies employing parallel devices to those using IID devices. A crucial observation that allows us to do so is that the threshold game itself admits a permutation invariance symmetry (i.e., the order of questions-answers tuples does not matter; see Chapter~\ref{ch:par_rep_showcase} for the details) and, therefore, we can assume without loss of generality that the optimal strategy is also permutation invariant. Now that we can restrict our attention to permutation invariant parallel devices,  de Finetti reductions become handy and can be used as a tool for reduction to IID.
				
				In Chapter~\ref{ch:par_rep_showcase} we consider the case of non-signalling strategies for complete-support games. A complete-support game is one in which all possible combinations of questions being sent to the players have some non-zero probability of being asked by the referee. 
				We prove the following via a reduction to IID:
				\begin{thm}[Informal]\label{thm:intro_par_rep}
					Given a game with optimal non-signalling winning probability $1-\alpha$, for any $\beta>0$, the probability to win more than a fraction $1-\alpha+\beta$ of~$n$ games played in parallel using a non-signalling strategy is exponentially small in~$n\beta^2$, as in the IID case. 
				\end{thm}
				Perhaps surprisingly, while the parallel repetition question is a well-investigated one, an exponential decrease that matches the IID case, as far as we are aware, was not known prior to our work (also not for classical or quantum strategies). In the context of reductions to IID, however, achieving the same behaviour as in the IID case is not unexpected. 
			
				To prove Theorem~\ref{thm:intro_par_rep} we first prove another statement that has a ``reduction to IID flavour'' and is perhaps of more fundamental nature. To present it, however, we need to first set some notation.\footnote{We are jumping ahead now with the aim of being able to explain Theorem~\ref{thm:intro_ns_marginals} to readers who are already somewhat familiar with device-independent information processing and non-signalling systems. For a reader unfamiliar with these topics, the mathematical statements may seem puzzling without further explanations. We will get back to the discussed theorem in Chapter~\ref{ch:par_rep_showcase}, after giving all the preparatory information throughout the thesis. A reader unfamiliar with the used terminology can therefore skip the current discussion without the risk of missing out.}
				
				As mentioned above, we focus on two-player games, i.e.,  games played by Alice and Bob (and the referee). 
				A parallel device used for the threshold game can be described using a conditional probability distribution $\P_{\mr{A}\mr{B}|\mr{X}\mr{Y}}$, where $\mr{A}=A_1,\dots,A_n$ is the random variable describing Alice's answers in the threshold game ($A_i$ being her answer in the $i$'th game) and, similarly,  $\mr{B}=B_1,\dots,B_n$  describes Bob's answers, 	and~$\mr{X}=X_1,\dots,X_n$ and $\mr{Y}=Y_1,\dots,Y_n$ are Alice's and Bob's questions, respectively. 
				
				When we say that a parallel device is non-signalling, we mean that it cannot be used as means of communication \emph{between the parties}.
				The behaviour of the device in one game, however, may depend on the other games.\footnote{In other words, the local strategy of each player does require ``communication between the games'': In order to (locally) answer the $i$'th question received from the referee, the player needs to know his $j$'th question (with $i\neq j$).}
				Mathematically, this means that, while the marginals $\P_{\mr{A}|\mr{X}}$ and $\P_{\mr{B}|\mr{Y}}$ are proper conditional probability distributions, objects such as $\P_{A_1|X_1}$ are not well-defined.

				During the threshold game, the device used by the players produces the observed data in the $n$ games: $\mr{a}=a_1,\dots,a_n$,  $\mr{b}=b_1,\dots,b_n$,  $\mr{x}=x_1,\dots,x_n$,  and $\mr{y}=y_1,\dots,y_n$. These are distributed according to  $\Q_{XY}^{\otimes n}\P_{\mr{A}\mr{B}|\mr{X}\mr{Y}}$, where $\Q_{XY}$ denotes the distribution used by the referee to choose the questions in a single non-local game. $\Q_{XY}^{\otimes n}$ is then the IID distribution according to which the questions are chosen in the threshold game.
				
				 The observed data $\mr{a},\mr{b},\mr{x},\mr{y}$ can be used to calculate frequencies and define a ``frequencies' conditional probability distribution'', which we denote by $\O_{ABXY}^{\textup{freq}(\mr{a},\mr{b},\mr{x},\mr{y})}$,  as:
				\begin{equation*}
					\O_{ABXY}^{\textup{freq}(\mr{a},\mr{b},\mr{x},\mr{y})} (\tilde{a}\tilde{b}\tilde{x}\tilde{y}) = \frac{\Big|\left\{i : (a_i,b_i,x_i,y_i) = (\tilde{a},\tilde{b},\tilde{x},\tilde{y}) \right\}\Big|}{n} 
				\end{equation*}
				and define
				\begin{equation}\label{eq:freq_single_round_dF_intro}
					\O_{AB|XY}^{\textup{freq}(\mr{a},\mr{b},\mr{x},\mr{y})} = \frac{\O_{ABXY}^{\textup{freq}(\mr{a},\mr{b},\mr{x},\mr{y})}}{\Q_{XY}} \;.
				\end{equation}
				
				$\O_{AB|XY}^{\textup{freq}(\mr{a},\mr{b},\mr{x},\mr{y})} $ can be seen as a (not necessarily physical) device, or a strategy, for a single game. 
				Starting with IID devices, which can be written in the form of\footnote{An IID device is illustrated in the bottom of Figure~\ref{fig:reductions_big_pic_intro}. We can then think of each copy~$\O_{AB|XY}$ as describing a single copy of the smaller boxes in the figure, while $\P_{\mr{A}\mr{B}|\mr{X}\mr{Y}} = \O_{AB|XY}^{\otimes n}$ described the device including all the copies together.} $\P_{\mr{A}\mr{B}|\mr{X}\mr{Y}} = \O_{AB|XY}^{\otimes n}$, it holds that if the device $\O_{AB|XY}$ is non-signalling then~$\P_{\mr{A}\mr{B}|\mr{X}\mr{Y}}$ is non-signalling and vice versa. 
				This also implies that, for sufficiently large~$n$,~$\O_{AB|XY}^{\textup{freq}(\mr{a},\mr{b},\mr{x},\mr{y})}$ is non-signalling with high probability. 
				
				For a non-IID, but non-signalling, device~$\P_{\mr{A}\mr{B}|\mr{X}\mr{Y}}$, however, it is not clear at all that $\O_{AB|XY}^{\textup{freq}(\mr{a},\mr{b},\mr{x},\mr{y})}$ should be non-signalling as well. 
				Using a reduction to IID, the following theorem is proven:
				\begin{thm}[Informal]\label{thm:intro_ns_marginals}
					Let $\P_{\mr{A}\mr{B}|\mr{X}\mr{Y}}$ be a non-signalling  permutation invariant  parallel device and $\O_{AB|XY}^{\textup{freq}(\mr{a},\mr{b},\mr{x},\mr{y})}$ as in Equation~\eqref{eq:freq_single_round_dF_intro}. Then, for sufficiently large $n$, $\O_{AB|XY}^{\textup{freq}(\mr{a},\mr{b},\mr{x},\mr{y})}$ is close to a non-signalling device with high probability.
					In particular, this means that the observed data produced by a  non-signalling permutation invariant parallel device can be seen as if, with high probability, it was sampled using an IID device $\O_{AB|XY}^{\otimes n}$ in which every single device $\O_{AB|XY}$  is close to a non-signalling one.
				\end{thm}

				Theorem~\ref{thm:intro_par_rep} follows directly from Theorem~\ref{thm:intro_ns_marginals} by noting that the number of games won in a given use of the device can be directly read from $\O_{AB|XY}^{\textup{freq}(\mr{a},\mr{b},\mr{x},\mr{y})}$ and that if $\O_{AB|XY}^{\textup{freq}(\mr{a},\mr{b},\mr{x},\mr{y})}$ is close to a non-signalling device then its winning probability cannot be too far from the optimal non-signalling winning probability $1-\alpha$.

			\subsubsection{Device-independent quantum cryptography}
			
				Chapter~\ref{ch:crypto_showcase} is devoted to the analysis of our second showcase~--- device-independent cryptography. The chapter's content previously appeared in~\cite{arnon2016simple}.
				The most challenging cryptographic task in which device-independent security has been considered is device-independent QKD (DIQKD); we will use this task as our main example. 
				In DIQKD the goal of the honest parties, called Alice and Bob, is to create a shared key, unknown to everybody else but them. 
				To execute the protocol they hold a device consisting of two parts: Each part belongs to one of the parties and is kept in their laboratories. Ideally, the device performs measurements on some entangled quantum states it contains. 
							
				The basic structure of a DIQKD protocol is presented as Protocol~\ref{pro:intro_qkd}. The protocol consists of playing $n$ non-local games, one after the other, with the given untrusted device and calculating the average winning probability from the observed data (i.e., Alice and Bob's inputs and outputs). If the average winning probability is below the expected winning probability $\omega_{\mathrm{exp}}$ defined by the protocol, Alice and Bob conclude that something is wrong and \emph{abort} the protocol. 
				Otherwise, they apply  classical post-processing steps that allow them to create identical and uniformly distributed keys. (The full description of the considered DIQKD protocol is presented and discussed in the following chapters).
		
				\begin{algorithm}[t]
					\caption{Device-independent quantum key distribution protocol (simplified example)}\label{pro:intro_qkd}
					\begin{algorithmic}[1]
						\STATEx \textbf{Given:}  A device for Alice and Bob that can play the chosen Bell game repeatedly
							
						\STATEx
						
						\STATE For every round $i\in[n]$ do Steps~\ref{prostep:input_qkd_int}-\ref{prostep:use_device_qkd_int}:
							\STATE\hspace{\algorithmicindent} Alice and Bob choose $X_i,Y_i$ at random.  \label{prostep:input_qkd_int}
							\STATE\hspace{\algorithmicindent} They input $X_i,Y_i$ to the device and record the outputs  $A_i,B_i$. \label{prostep:use_device_qkd_int}
						\STATEx
						
						\STATE \textbf{Parameter estimation:} Alice and Bob estimate the average winning probability in the game from the observed data. If it is below the expected winning probability, $\omega_{\mathrm{exp}}$, they \textcolor{red}{\textbf{abort}}.
						
						\STATEx
						
						\STATE \textbf{Classical post processing:} Alice and Bob apply an error correction protocol and a privacy amplification protocol (both classical) on their raw data $\mr{A}$ and $\mr{B}$. 
					
					\end{algorithmic}
				\end{algorithm}
				
				The central task when proving security of DIQKD consists in bounding the information that an adversary, called Eve, may obtain about Alice's raw data~$\mr{A}=A_1,\dots,A_n$ used to create the final key (see Protocol~\ref{pro:intro_qkd}). 
				More concretely, one needs to establishing a lower bound on the smooth conditional min-entropy $H_{\min}^{\varepsilon}(\mr{A}|E)$, where $E$ is Eve's quantum system, which can be initially correlated to the device used by Alice and Bob in the protocol and $\varepsilon>0$ is one of the security parameters of the protocol (see Section~\ref{sec:pre_di_crypt}).
				The quantity $H_{\min}^{\varepsilon}(\mr{A}|E)$ determines the maximal length of the secret key that can be created by the protocol. Hence, proving security amounts to lower-bounding $H_{\min}^{\varepsilon}(\mr{A}|E)$.
				Evaluating the smooth min-entropy $H_{\min}^{\varepsilon}(\mr{A}|E)$ of a large system is often difficult, especially in the device-independent setting where Alice and Bob are using an uncharacterised device, which may also be manufactured by Eve. 
				
				The IID assumption is commonly used in order to simplify the calculation of~$H_{\min}^{\varepsilon}(\mr{A}|E)$. In the IID case we can assume that Alice and Bob use an IID device to execute the protocol and, hence, each $A_i$ is produced independently of all other outputs. 
				Furthermore, one can assume that Eve's quantum information also takes the IID form $E=E_1,\dots,E_n$, where each $E_i$ holds information only regarding $A_i$.
				Then, the AEP, briefly mentioned above, can be used to calculate an upper-bound on~$H_{\min}^{\varepsilon}(\mr{A}|E)$ and, by this, prove security. 
				
				The most general adversarial device to consider is, clearly, not an IID one. 
				Due to the sequential nature of the protocol, the relevant devices to consider are sequential devices. 
				As sequential devices are more complex than IID ones, security proofs for DIQKD that proved security by addressing the most general device directly, e.g.,~\cite{reichardt2013classical,vazirani2014fully}, had to use techniques which are far more complicated than the ones used for security proofs under the IID assumption, e.g., in~\cite{pironio2009device}.
				Consequently, the derived security statements were of limited relevance for practical experimental implementations; they are applicable only in an unrealistic regime of parameters, e.g., small amount of tolerable noise and large number of signals.

			We take the approach of reductions to IID in order to prove the security of our DIQKD protocol.
			In particular, we leverage the sequential nature of the protocol, as well as the specific way in which classical statistics are collected by Alice and Bob, to prove its security by reducing the analysis of sequential devices to that of IID devices using the EAT. 
			The resulting theorem can be informally stated as follows:
			\begin{thm}[Informal]
				Security of DIQKD in the most general case follows from security under  the IID assumption. 
				Moreover, the dependence of the key rate on the number of rounds of the protocol, $n$, is the same as the one in the IID case, up to terms that scale like $1/\sqrt{n}$.
			\end{thm}
			
			On the fundamental level, the theorem establishes the a priori surprising fact that general quantum adversaries are no stronger than an adversary restricted to preparing IID devices. 
			As mentioned in Section~\ref{sec:motivation_reduction}, this does not mean that the most general device that an adversary can prepare is an IID device. Instead, it means that the adversary (at least asymptotically) does not benefit form preparing more complex devices.

			On the quantitive level, taking the path of a reduction to IID results in a proof with several advantages. 
			In particular, it allows us to give simple and modular security proofs of DIQKD (as well as other device-independent protocols) and to extend tight results known for DIQKD under the IID assumption to the most general setting, thus deriving essentially optimal key rates and noise tolerance.
			This is crucial for experimental implementations of device-independent protocols. Our quantitive results have been applied to the analysis of the first experimental implementation of a protocol for randomness generation in the fully device-independent framework~\cite{liu2017high}.

	\section{How to read the thesis}
	
		We review the structure of the thesis.
		Depending on the reader's main interest and prior knowledge, different chapters of the thesis may or may not be relevant. 
				
		Chapters~\ref{ch:pre_general} and~\ref{ch:pre_di} give preliminary information.
		Chapter~\ref{ch:pre_general} presents general introductory information and notation. 
		We remark that in most parts of the thesis, general intuition is sufficient and the exact mathematical definitions are not that important in order to understand the \emph{essence}. 
		Therefore, even a reader unfamiliar with, e.g., the quantum formalism or the mathematical definitions of the various entropies, may skip  Chapter~\ref{ch:pre_general} in the first reading and get back to the relevant definitions appearing in it only when wishing to get a better understanding of the complete technical details. 
		
		Chapter~\ref{ch:pre_di} deals with basic information and terminology related to device-independent information processing. Readers who are unfamiliar with, e.g., non-locality, should first of all read this chapter. Readers already familiar with some device-independent tasks may skip the chapter and come back to it  if needed. 
	
		Chapter~\ref{ch:intro_showcases} acts as an \emph{introduction} to our showcases; no theorems or proofs are given there. Thus, readers who are familiar with the question of parallel repetition and the task of DIQKD may pass over this chapter. 
		
		Chapters~\ref{ch:single_round_box} and~\ref{ch:multi_box} concern themselves with the mathematical objects that we consider in the thesis~---  the ``black boxes'' that model the different types of devices. 
		Chapter~\ref{ch:single_round_box} defines what we call a ``single-round box'', which is, in a sense, a device that can be used to play only a single non-local game. The single-round box acts as an abstract object that allows us to study the fundamental aspects of non-locality, without needing to deal with complex protocols. 
		As we will see, it captures the ``physics'' of the problem at hand. Hence, studying single-round boxes is the first step in any analysis of device-information processing task. 
		In Chapter~\ref{ch:multi_box}, we formally define parallel and sequential boxes, which give the mathematical model for parallel and sequential devices, and discuss the relations between them. 
		
		After setting the stage, we are ready to start discussing the method of reductions to IID. 
		The first step in this direction is done in Chapter~\ref{ch:iid_assumption}, where we discuss the IID assumption and see how it can be used to simplify the analysis of device-independent tasks and, in particular, our showcases. 
		This chapter also presents the asymptotic equipartition property, which acts as a valuable mathematical tool when working under the IID assumption.
		
		The tools used as reductions, i.e., the de Finetti reduction and the entropy accumulation theorem, are the topics of Chapters~\ref{ch:reductions_par} and~\ref{ch:reductions_seq}, respectively. 
		Chapters~\ref{ch:par_rep_showcase} and~\ref{ch:crypto_showcase} are devoted to the analysis of our showcases via a reduction to IID. 
		
		Clearly, many open questions and directions for future works arise. We discuss open questions specific for our showcases within the relevant chapters. In addition, the thesis ends with an outlook in Chapter~\ref{ch:outlook} including questions that, in order to answer, require further development of the toolkit of reductions to IID.

		A reader interested in the topic of reductions to IID in general is recommended to read the thesis from the beginning to the end, following the order of the chapters. 
		On the other hand, a reader who is mainly interested in one of the showcases may focus only on the sections relevant for the showcase of interest. To assist such readers, we list in Table~\ref{tb:read_sugg} the relevant sections (in the order in which they should be read) for each of the showcases. 
		
		\begin{table}
		\begin{tabular}{ l | l }
		  Reader's interest & Recommended sections \\ \hline
		  Reductions to IID & All chapters \\
		  Parallel repetition & \ref{sec:pre_par_rep}, \ref{sec:sing_box_model}, \ref{sec:parallel_mr_boxes}, \ref{sec:iid_assump_desc}, \ref{sec:par_rep_under_iid}, \ref{ch:reductions_par}, \ref{ch:par_rep_showcase}  \\
		  Device-independent cryptography & \ref{sec:pre_di_crypt}, \ref{ch:single_round_box}, \ref{sec:seq_boxes}, \ref{sec:iid_assump_desc}, \ref{sec:aep_both}, \ref{sec:crypt_under_iid}, \ref{ch:reductions_seq}, \ref{ch:crypto_showcase}
		\end{tabular}
		\caption{Reading suggestion according to the reader's main interest.}\label{tb:read_sugg}
		\end{table}

\chapter{Preliminaries: basics and notation}\label{ch:pre_general}


	\section{General notation}

		The relevant notation for sets and vectors is summarised below. 
		\begin{itemize}
			\item $\mathbb{N}$, $\mathbb{R}$, and $\mathbb{C}$ are the sets of natural, real, and complex numbers, respectively. 
			\item $[a,b]$ denotes the closed set of real numbers $a\leq x \leq b$.
			\item $[n]$ denotes the set $\{1,2,\dots,n\}$.
			\item When an object $x_i$ is defined for all~$i\in[n]$, $\left\{x_i\right\}_{i\in[n]}$ denotes the set $\{x_1,x_2,\dots,x_n\}$. 
			\item Other sets are mostly denoted by calligraphic letters, e.g., $\mathcal{S}$.
			\item $\mathcal{S}\subseteq\mathcal{P}$ means that $\mathcal{S}$ is a subset  of $\mathcal{P}$.  $\mathcal{S}\subset\mathcal{P}$ means that $\mathcal{S}$ is a proper subset  of $\mathcal{P}$. 
			\item $\mathcal{S}\setminus\mathcal{P}=\{s: s\in\mathcal{S} \land s\notin\mathcal{P}\}$ stands for the difference between the two sets. 
			\item $\mathcal{S}\times\mathcal{P}=\{(s,p): s\in\mathcal{S} \land p\in\mathcal{P}\}$ is the multiplication of the sets. Furthermore, $\mathcal{S}\times \mathcal{S}$ is denoted by $\mathcal{S}^2$ and $\mathcal{S}^n$ is defined analogously for any $n$. 
			\item For sets $\mathcal{S},\mathcal{P}$ we denote by $\mathrm{Hom}(\mathcal{S},\mathcal{P})$ the set of all homomorphisms from $\mathcal{S}$ to~$\mathcal{P}$. The set of all endomorphisms is denoted by $\mathrm{End}(\mathcal{S})$, i.e., $\mathrm{End}(\mathcal{S})=\mathrm{Hom}(\mathcal{S},\mathcal{S})$.
			\item Vectors (of different objects) are marked in bold. For example, we use $\mr{x}=x_1,x_2,\dots,x_n$.
			\item 	Let $f: \mathcal{S} \rightarrow \mathbb{R}$ be a function over some set $\mathcal{S} \subset \mathbb{R}^{n}$. The infinity norm of the gradient of $f$ is defined as
				\begin{equation*}
					\|  \nabla f \|_\infty = \sup \left\{ \frac{\partial}{\partial s_{i}} f(\mr{s}) : \mr{s} \in \mathcal{S}, \, i \in [n] \right\} \;.
				\end{equation*}
		\end{itemize}

			We use the following general notation.
		\begin{itemize}
			\item $\land$, $\lor$, and $\neg$ denote the logical and, or, and negation, respectively.
			\item $\oplus$ denotes the XOR operation. 
			\item We denote by $\log$ the logarithm in base 2. 
			\item ${n \choose {k_1, \dots, k_m}}$ is the multinomial coefficient, i.e., ${n \choose {k_1, \dots, k_m}} = \frac{n!}{k_1! \dots , k_m!}$, where~$!$ is the factorial operation.
			\item A function $f: \mathbb{N}\rightarrow \mathbb{R}$ is called negligible if for every positive polynomial $p(\cdot)$, there exists an $n_0$ such that for all $n>n_0$, $f(n)< \frac{1}{p(n)}$. In the thesis, in all cases where the term negligible is used $f(n)$ decreases exponentially fast with~$n$.
		\end{itemize}

	\section{Probability distributions and random variables}
	
		We use both probability distributions and random variables (RV) and interchange the two when convenient. Specifically, 
		\begin{itemize}
			\item Capital letters, e.g., $X$, denote RV. When implicit, a RV $X$ takes values from the set denoted by the same letter, i.e., $\mathcal{X}$. 
			\item $\P_X$ denotes the probability distribution corresponding to the RV $X$. To distinguish different probability distributions we sometimes replace $\P$ by other letters, such as $\O$ and $\Q$.  
			\item $\P_X(x)$ is the probability that $X=x$. 
			\item When a probability distribution is used without a need of referring to the event space etc., we simply use $\{p_i\}_{i\in \mathcal{I}}$ for some $\mathcal{I}$ (usually clear from the context or irrelevant) while keeping in mind that $p_i\geq 0$ for all $i\in \mathcal{I}$ and $\sum_i p_i =1$. 
			\item When discussing more complex events $\Omega\subseteq\mathcal{X}$ over $\mathcal{X}$, we use $\Pr_{x\sim X}\left[\Omega\right]$ to denote the probability of the event $\Omega$ when sampling according to $\P_X$. When it is clear from the context according to which probability distribution the sampling is done we may drop the subscript and write only $\Pr\left[\Omega\right]$. For example, when applying Chernoff-type bounds, we use standard notation such as $\Pr\left[\sum_i X_i > t\right]$  instead of $\P_{X_1\dots X_n}\left[\sum a_i > t\right]$.
			\item The expectation value $\mathbb{E}[X]$ of $X$ is given by $\mathbb{E}[X]=\sum_{x\in\mathcal{X}} x \P_X(x)$.
		\end{itemize}

		When considering two RVs $X$ and $Y$, jointly distributed according to $\P_{XY}$, the \emph{marginal} $\P_X$ is defined via
		\[
			\P_X(x) = \sum_y \P_{XY}(x,y) \;.
		\]
		The \emph{conditional distribution} of $X$ given $Y=y$, given by
		\begin{equation}\label{eq:cond_prob_def}
			\forall x, \qquad \P_{X|Y=y}(x) = \frac{\P_{XY}(x,y)}{\P_Y(y)} \;.
		\end{equation}
		We mostly use  $\P_{X|Y}(x|y)$ to denote $\P_{X|Y=y}(x)$ and the shorthand notation $\P_{X|Y}=\P_{XY}/\P_Y$ instead that of Equation~\eqref{eq:cond_prob_def}. 
		
		Throughout the thesis, we use the following operations on probability distributions:
		\begin{itemize}
			\item For any $q\in[0,1]$, $\P_X$, and $\mathrm{R}_X$, the convex combination $\mathrm{S}_{X} = q \P_{X} + (1-q) \mathrm{R}_{X}$ is defined via
					\[
						\forall x, \qquad \mathrm{S}_{X} (x) = q  \P_{X}(x) + (1-q) \mathrm{R}_{X}(x) \;.
					\]
			\item For any $n\in\mathbb{N}$ and $\P_X$, $\P_X^{\otimes n}$ denoted the probability distribution over $\mathcal{X}^{n}$ defined via
				\[
					\forall \mr{x}, \qquad \P_X^{\otimes n}(\mr{x}) = \prod_i \P_X(x_i) \;,
				\]
				where $\mr{x}=x_1,x_2,\dots,x_n$.
		\end{itemize}

		\subsection{Independent and identical random variables}
		
			Consider two RV $X$ and $Y$ defined over $\mathcal{X}$ and $\mathcal{Y}$ respectively. We say that the two are independent if and only if for all $(x,y)\in\mathcal{X}\times\mathcal{Y}$, $\P_{XY}(x,y)=\P_X(x)\cdot\P_Y(y)$.
			For $\mathcal{X}=\mathcal{Y}$, we say that $X$ and $Y$ are identical if and only if $\P_X=\P_y$.

			A sequence of RVs  $\mr{X}=X_1,X_2,\dots,X_n$, each over $\mathcal{X}$ are said to be independently and identically distributed (IID) RVs if and only if they are all independent and identical to one another.

		\subsection{Concentration inequalities}
		
			When considering IID RVs, concentration inequalities are of special importance. Roughly speaking, concentration inequalities give bounds on how fast the observed frequencies converge to the expected value when sampling IID RVs. The formal statements relevant for the thesis are given below.
	
			\begin{lem}[Hoeffding's inequality]
				Consider a RV $X$ defined over $\mathcal{X}=\{0,1\}$ and let $X_1,X_2,\dots,X_n$ be a sequence of $n$ identical and independent copies of $X$. Then, 
				\[
					\Pr\left[  \sum_i X_i - n\mathbb{E}[X]  \geq tn \right] \leq \exp \left( - 2nt^2\right) \;
				\]
				and
				\[
					\Pr\left[  \Big| \sum_i X_i  - n\mathbb{E}[X]  \Big| \geq  tn \right] \leq 2 \exp \left( - 2nt^2\right) \;.
				\]
			\end{lem}

			Sanov's inequality can be seen as a concentration inequality for conditional probability distributions, in the following sense. 
			Let $\O_{A|X}$ be a conditional probability distribution over $\mathcal{A}$ and $\mathcal{X}$, $\Q_X$ be a probability distribution over $\mathcal{X}$ and denote  $\O_{AX}=\Q_X\O_{A|X}$. 
			Fix $n\in\mathbb{N}$.
			Consider a scenario in which we sample $\mr{a}=a_1,\dots,a_n$ and $\mr{x}=x_1,\dots,x_n$ using $\O_{AX}^{\otimes n}$ and estimate $\O_{A|X}$ from the sample by calculating $\O_{AX}^{\textup{freq}(\mr{a},\mr{x})}$ defined by
			\begin{equation*}
				\O_{AX}^{\textup{freq}(\mr{a},\mr{x})} (\tilde{a}\tilde{x}) = \frac{\Big|\left\{i : (a_i,x_i) = (\tilde{a},\tilde{x}) \right\}\Big|}{n} 
			\end{equation*}
			and define
			\begin{equation}\label{eq:sanov_freq_pre}
				\O_{A|X}^{\textup{freq}(\mr{a},\mr{x})} = \frac{\O_{AX}^{\textup{freq}(\mr{a},\mr{x})}}{\Q_{X}} \;.
			\end{equation}

			\begin{lem}[Sanov's inequality]\label{lem:sanovs_theorem}
				For every $\O_{AX}$ and $n$,
				\[
					\mathrm{Pr}_{\mr{a}\mr{x}\sim\O_{AX}^{\otimes n}}\left[ \big| \O_{A|X}^{\textup{freq}(\mr{a},\mr{x})} -  \O_{A|X} \big|_1 > \epsilon \right] \leq \delta(n,\epsilon) 
				\]
				where $\O_{A|X}^{\textup{freq}(\mr{a},\mr{x})}$ is as in Equation~\eqref{eq:sanov_freq_pre}, $\delta(n,\epsilon)=(n+1)^{|\mathcal{A}|\cdot|\mathcal{X}|-1}e^{-n\epsilon^2/2}$, and 
				\[
						\big|\O_{A|X}^{\textup{freq}(\mr{a},\mr{x})} -  \O_{A|X} \big|_1 =  \sum_{\tilde{x}} \Q_X(\tilde{x}) \sum_{\tilde{a}} \big| \O_{A|X}^{\textup{freq}(\mr{a},\mr{x})}(\tilde{a}|\tilde{x}) - \O_{A|X}(\tilde{a}|\tilde{x}) \big| \;.
					\]
			\end{lem}

	\section{Quantum formalism}\label{sec:pre_quantum}
	
		The basic notation used in the thesis related to the quantum formalism is listed below. We remark, however, that understanding   what is meant by a ``state'' and ``measurements'' on the intuitive level  will almost always suffice in order to understand the essence of the thesis. 
		The exact definitions below are given for the sake of completeness. Clearly, they do not cover all concepts and definitions employed in quantum physics and quantum information theory. Readers who are not familiar with the topics and would like to get a more comprehensive understanding are directed to~\cite{nielsen2002quantum}.

		We use the Dirac notation: $\ket{\psi}$ denotes a column vector while $\bra{\psi}$ is a row vector. 
		$\braket{\phi}{\psi}$ and $\ket{\phi}\bra{\psi}$ denote inner and outer products of the two vectors, respectively.
					
		\subsection{Operators}
		
			We use the following standard notation and definitions.			
			\begin{itemize}
				\item The identity matrix, or operator, of dimension $d$ is denoted by $\idn_d$. Alternatively, instead of indicating the dimension, we use, e.g., $\idn_X$ to denote the identity operator acting in a specific space associated to $X$ (see below). When the space or dimension is clear from the context we simply write $\idn$. 
				\item A \emph{Hermitian}, or self-adjoint, operator $A$ is an operator satisfying $A=A^{\dagger}$.
				\item A \emph{unitary} operator $A$ is an operator satisfying $AA^{\dagger}=A^{\dagger}A=\idn$.
				\item The trace of a square matrix $A$, i.e., the sum of the elements on the main diagonal of $A$, is denoted by $\Tr(A)$.
				\item $A \succeq 0$, for $A$ Hermitian, means that the eigenvalues of $A$ are non-negative. $A \succeq B$ stands for $A-B \succeq 0$.
				\item The 1-norm is defined as $\| A \|_1=\Tr|A| = \Tr\sqrt{A^{\dagger}A}$, where $A^{\dagger}$ denotes the conjugate transpose of $A$.
				\item For a diagonal matrix $A$ with eigenvalues $\{a_i\}_i$, $\log(A)$ is the diagonal matrix with eigenvalues $\{\log(a_i)\}_i$.			
			\end{itemize}

		\subsection{Hilbert spaces}

			The postulates of quantum mechanics tell us that all quantum states ``belong'' to a complex vector space called a \emph{Hilbert space}. All quantum states and operations will be defined with respect to the considered Hilbert spaces. We give the formal definitions below.
		
			\begin{defn}[Hilbert space]
				A Hilbert space $\hilb$ is a complex vector space, i.e., 
				\[
					\ket{\psi},\ket{\phi}\in \hilb\ \text{and}\ \lambda_1,\lambda_2\in \mathbb{C} \quad \rightarrow \quad
					\lambda_1 \ket{\psi}+\lambda_2\ket{\phi} \in \hilb
				\]
 				such that for all $\ket{\psi},\ket{\phi}\in \hilb$, there exists $\braket{\phi}{\psi}\in \mathbb{C}$  for which
 				\begin{enumerate}
					\item it is linear in $\ket{\psi}$: $\braket{\phi}{\lambda_1\psi_1+\lambda_2 \psi_2}= \lambda_1 \braket{\phi}{\psi_1}+\lambda_2 \braket{\phi}{\psi_2}$\ ,
					\item $\overline{\braket{\phi}{\psi}}=\braket{\psi}{\phi}$, where the bar  denotes the complex conjugate ,
					\item for all $\ket{\psi} \in \hilb$, $\braket{\psi}{\psi}\geq 0$ and $\braket{\psi}{\psi}= 0 \leftrightarrow \ket{\psi}=0$.
					\end{enumerate}	
					The norm of a vector $\ket{\psi}$ is defined as $\| \ket{ \psi}\|_1=\sqrt{\braket{\psi}{\psi}}$.
			\end{defn}

			\begin{defn}[Orthonormal basis]
				An orthonormal basis of $\hilb$ is a set of vectors $\{ \ket{\phi_i}\}_{i\in I}$ such that
				\begin{itemize}
					\item $\braket{\phi_i}{\phi_j}= \delta_{ij}$\ {for all}\ $i,j\in I$\ { and}
					\item $\braket{\psi}{\phi_i}=0$\ {for all}\ $i\in I$ $\rightarrow  \psi=0$\ .
				\end{itemize}
			\end{defn}
			We will usually consider Hilbert spaces of finite dimensions, meaning $I$ is a set with a finite amount of elements.

			\begin{defn}[Projector]
				Let $\hilb$ be a Hilbert space and $\hilb^{\prime}$ a subspace of $\hilb$ with $ \{\ket{\phi_i}\}_{i\in \mathcal{I}'}$ an orthonormal basis of $\hilb^{\prime}$. The projector of $\hilb$ onto $\hilb^{\prime}$ is the operator 
				\[
					P_{\hilb^{\prime}} = \sum_{i\in  \mathcal{I}'}\ket{\phi_i}\bra{\phi_i} \;.
				\]
			\end{defn}

				Given $\hilb_A$ and $\hilb_B$, the tensor product Hilbert space $\hilb_A\otimes\hilb_B$ is defined such that for $\ket{\psi}\in \hilb_A$ and $\ket{\phi}\in \hilb_B$, it associates a vector $\ket{\psi}\otimes \ket{\phi} \in \hilb_A\otimes\hilb_B$ with the property that
				\begin{enumerate}
					\item $c\cdot \left(\ket{\psi}\otimes \ket{\phi} \right) =(c\cdot \ket{\psi})\otimes \ket{\phi}  =\ket{\psi}\otimes (c\cdot  \ket{\phi}) $
					\item $(\ket{\psi_1}+\ket{\psi_2})\otimes \ket{\phi}	=\ket{\psi_1}\otimes \ket{\phi}+ \ket{\psi_2}\otimes \ket{\phi}$
					\item $\ket{\psi}\otimes (\ket{\phi_1}+\ket{\phi_2})=\ket{\psi}\otimes \ket{\phi_1}+\ket{\psi}\otimes\ket{\phi_2}$
				\end{enumerate}
					for all $c\in \mathbb{C}$, $\ket{\psi_1},\ket{\psi_2}\in \hilb_A$ and
					 $\ket{\phi_1},\ket{\phi_2}\in \hilb_B$. 

			\subsection{Quantum states}	
			
				\subsubsection{Pure and mixed states}
				
					There are two ``types'' of quantum states one can consider~--- pure and mixed states. 
					
					A \emph{pure quantum state} is associated with a vector belonging to an Hilbert space,~$\ket{\psi} \in \hilb$, with normalisation $\| \ket{ \psi}\|_1=1$.  
					
					Instead of working only with vectors, we can define quantum states as matrices, or operators.  
					\begin{defn}[Density operator]
						A density operator, or simply a quantum state,~$\rho \in \mathrm{End}(\hilb)$ is a Hermitian positive operator with trace 1. That is,
						\[
							\rho = \rho^{\dagger} \;; \quad \rho \succeq 0 \;; \quad \Tr(\rho)=1 \;.
						\]
						For a given Hilbert space $\hilb$, we denote by $\mathcal{S}(\hilb)$ the set of all density operators defined over $\hilb$.
					\end{defn}
			
					Any pure state $\ket{\psi} \in \hilb$ can be written as a density operator~$\rho=\ket{\psi}\bra{\psi}\in\mathcal{S}(\hilb)$. 
					Density operators can describe more general states, called \emph{mixed quantum states}, which can be thought of as a convex combination of pure states:
					\[
						\rho = \sum_i p_i \ket{\psi_i}\bra{\psi_i} \;.
					\]
					Note, however, that different convex combinations can result in the same mixed state $\rho$ and, thus, $\rho$ does not pin-down a specific decomposition to pure states.

				A \emph{qubit} is a quantum state belonging to $\mathcal{S}(\hilb)$ for a two-dimensional Hilbert space~$\hilb$. The basis states are denoted by $\ket{0}$ and $\ket{1}$.
				
			\subsubsection{Composite systems}

				One can consider quantum states over tensor products of Hilbert spaces. Such states are called multipartite states. For example, a bipartite state is a quantum state~$\rho_{AB} \in\mathcal{S}(\hilb_A\otimes\hilb_B)$ for some Hilbert spaces $\hilb_A$ and $\hilb_B$. The state $\rho_{AB}$ can describe a state shared between two parties, Alice and Bob.
				The most important thing to notice in the context of the thesis is that given a bipartite state $\rho_{AB}$, its marginals are also quantum states; these are called the reduced density operators. 
				\begin{defn}[Reduced density operators]
					Given $\rho_{AB} \in \mathcal{S}(\hilb_A\otimes\hilb_B)$, its reduced density operator over $\hilb_A$ is given by
					\begin{align*}
						\rho_A &=\Tr_B \left( \rho_{AB} \right) = \sum_{i}\left(\idn_A\otimes \bra{\phi_i}\right) {\rho_{AB}}\left(\idn_A\otimes \ket{\phi_i}\right) 
					\end{align*}
					where $\{\ket{\phi_i}\}_i$ is a basis of $\hilb_B$, and similarly for $\rho_B$.
				\end{defn}
				 Thinking of $\rho_{AB}$ as shared between Alice and Bob, Alice's local state is then $\rho_A$ while Bob's local state is $\rho_B$. 
				
				Given a state $\rho_A$ we can consider its \emph{purification}.
				\begin{defn}[Purification]
					The purification of a state $\rho_A \in \mathcal{S}(\hilb_A)$ is a \emph{pure} bipartite state $\rho_{AB} \in \mathcal{S}(\hilb_A\otimes\hilb_B)$ for which $\Tr_B(\rho_{AB})=\rho_A$. 
				\end{defn}
					Note that by applying a unitary on $B$ the state on $A$ is not being modified and the overall state remains pure. Thus, after the unitary operation, we are still holding a purification. In this sense, we usually say that all purifications are equivalent up to the application of a unitary on the purifying system $B$.
	
			\subsubsection{Classical systems}
	
				A classical system, defined by a RV $A$ with probability distribution $\P_A$, can be represented by the density operator
				\[
					\rho_A = \sum_{a\in \mathcal{A}} \P_A(a) \ket{a}\bra{a} \;,
				\]
				where $\{\ket{a}\}_a$ is an orthonormal basis of a Hilbert space $\hilb_A$.
				
				One example of a classical system that is of common use is the state associated with the uniform distribution $U_m$ over $\{0,1\}^m$. This distribution can be written as the state $\rho_{U_m} = \frac{1}{m} \idn_m$, called the \emph{completely mixed state} on $m$ qubits.

				A \emph{classical-quantum state} is a bipartite state in which one register is classical and the other is quantum. Formally,
				\begin{defn}[Classical-quantum state]
					A classical-quantum state $\rho_{AE}\in\mathcal{S}(\hilb_A\otimes\hilb_E)$, classical on $A$, is a state of the form
					\[
						\rho_{AE} = \sum_a \P_A(a) \ket{a}\bra{a}\otimes \rho_E^a \;, 
					\]
					where $\{\ket{a}\}_a$ is an orthonormal basis of the Hilbert space $\hilb_A$ and, for all $a\in \mathcal{A}$, $\rho_E^a\in\mathcal{S}(\hilb_E)$.
				\end{defn}
				
				Given a classical-quantum state $\rho_{AE}$ as above, we can consider the quantum state arising from \emph{conditioning} on an event defined over $\mathcal{A}$. For example, conditioning on the event $A=a$, the quantum state is~$\rho_E^a$. 
				Conditioning can also be done when considering more complicated events. For~$\Omega$ some event over $\mathcal{A}$, the state conditioned on $\Omega$ is 
				\[
					\rho_{AE|\Omega} = \frac{1}{\Pr[\Omega]} \sum_{a\in\Omega} \P_{A|\Omega}(a) \otimes \rho_E^a \;,
				\]
				where $\Pr[\Omega]=\sum_{a\in\Omega} \P_A(a)$ is the probability of $\Omega$ according to $\rho_{AE}$ and $\P_{A|\Omega}(a)=\Pr[A=a \land \Omega]/\Pr[\Omega]$ is the probability of $a$ given $\Omega$.

				\subsubsection{Entanglement}

					Given a bipartite state $\rho_{AB} \in \mathcal{S}(\hilb_A\otimes\hilb_B)$, shared between two parties, one can study the type of correlations that appear between the two parties. 
					A state is said to be \emph{separable} if it can be written as 
					\begin{equation}\label{eq:seperable_state}
						\rho_{AB} =  \sum_i p_i \; \rho_A^i \otimes \rho_B^i 
					\end{equation}
					for some probabilities $p_i$, $\rho_A^i\in\mathcal{S}(\hilb_A)$, and $\rho_B^i\in\mathcal{S}(\hilb_B)$.
					That is,  a separable state is a convex combination of tensor product states. 
					Using the above we notice that a pure state $\ket{\psi}_{AB}$ is separable if and only if it is a tensor product of two pure states~$\ket{\psi}_{AB}=\ket{\psi_A}\otimes\ket{\psi_B}$.
					
					Not all quantum states are separable. 
					A bipartite state  $\rho_{AB} \in \mathcal{S}(\hilb_A\otimes\hilb_B)$ is said to be \emph{entangled} if it cannot be written in the form of Equation~\eqref{eq:seperable_state}. 
					Such states exhibit correlations which cannot be explained by classical means.							
						
					Of specific interest to us are  maximally entangled states of two qubits, also  called Bell states, denoted by
					\begin{align*}
						 \ket{\Phi^+} = \frac{1}{\sqrt{2}}\left(\ket{00}+\ket{11}\right) \quad , \quad \ket{\Phi^-} = \frac{1}{\sqrt{2}}\left(\ket{00}-\ket{11}\right) \;, \\
						\ket{\Psi^+} = \frac{1}{\sqrt{2}}\left(\ket{01}+\ket{10}\right) \quad , \quad \ket{\Psi^-} = \frac{1}{\sqrt{2}}\left(\ket{01}-\ket{10}\right) \;.
					\end{align*}
					Here, $\ket{00}$ stands for $\ket{0}\otimes\ket{0}\in\hilb_A\otimes\hilb_B$, with $\hilb_A$ and $\hilb_B$ two-dimensional Hilbert spaces. $\ket{01}$, $\ket{10}$, and $\ket{11}$ are similarly defined.
					
		
		\subsection{Quantum operations}
		
			\subsubsection{Unitary evolution}
		
				The evolution of a closed, or isolated, quantum system is described by unitary operations. By ``a closed system'' we mean that the transformation of the system of interest is independent of the ``rest of the world'', or the environment. 
				We have:
				\begin{itemize}
					\item For any unitary $U$, $U$ evolves a pure state $\ket{\psi^1}\in\hilb$ to a pure state $\ket{\psi^2}\in\hilb$ according to $\ket{\psi^2}=U\ket{\psi^1}$.
					\item More generally, for mixed states, starting with $\rho^1\in\mathcal{S}(\hilb)$ we have $\rho^2 = U \rho^1 U^{\dagger}\in\mathcal{S}(\hilb)$.
					\item For a bipartite state $\rho^1_{AB}$, we can evolve each subsystem locally by $\rho^2_{AB}=(U_A \otimes U_B) \rho^1_{AB} (U^{\dagger}_A \otimes U^{\dagger}_B)$. 
					\item As unitary operations are reversible ($UU^{\dagger}=U^{\dagger}U=\idn$), the evolution of closed systems is always reversible. 
				\end{itemize}

			\subsubsection{Quantum measurements}
			
				To describe a quantum measurement one can use the so called \emph{Kraus operators}. 
				 \begin{defn}[Kraus operators]
					A set of Kraus operators $\{K_i\}_{i\in \mathcal{I}}$  is a set of operators such that $\sum_{i\in \mathcal{I}} K_i^{\dagger}K_i = \idn$.
				\end{defn}
				
				\begin{defn}[Quantum measurement: Kraus representation]
					Given a state $\rho$ and a set of Kraus operator $\{K_i\}_{i\in \mathcal{I}}$ describing a measurement, the outcome of the measurement on $\rho$ is a RV $I$, defined over the set $\mathcal{I}$, where each outcome $i\in\mathcal{I}$ is associated with the operator $K_i$. 
					The probability of observing the outcome $i$ when measuring $\rho$ with $\{K_i\}_i$ is given by 
					\[	
						\Pr(i) = \Tr( K_i \rho K_i^{\dagger}) \;.
					\]
					The post-measurement state is given by
					\[
						\rho_i = \frac{K_i \rho K_i^{\dagger}}{\Tr( K_i \rho K_i^{\dagger})} \;.
					\]
				\end{defn}
				
				We can further identify an operator $M_i=K_i^{\dagger}K_i$ and work with it, instead of the Kraus operators, to ease notation in some cases. These operators, called  \emph{positive operator valued measures} (POVMs), can then be used to describe the relevant measurements. 
				
				\begin{defn}[Positive operator valued measure]
					A positive operator valued measure (POVM) is a set of positive Hermitian operators $\{M_i\}_{i\in \mathcal{I}}$ such that $\sum_{i\in \mathcal{I}} M_i = \idn$.
				\end{defn}
		
				\begin{defn}[Quantum measurement: POVM representation]
					Given a state $\rho$ and a POVM $\{M_i\}_{i\in \mathcal{I}}$ describing a measurement, the outcome of the measurement on $\rho$ is a RV $I$, defined over the set $\mathcal{I}$, where each outcome $i\in\mathcal{I}$ is associated with the operator $M_i$. The probability of observing the outcome $i$ when measuring $\rho$ with $\{M_i\}_i$ is given by 
					\[	
						\Pr(i) = \Tr( M_i  \rho) \;.
					\]
				\end{defn}
				
				Given a POVM $\{M_i\}_{i\in \mathcal{I}}$ there are many different decomposition to Kraus operators. While the specific decomposition is not relevant for knowing the measurement statistics, they are needed in order to describe the post-measurement state. 
				
				In most of the scenarios considered in this thesis we will only be interested with the observed measurements statistics and therefore we will use POVMs to describe a measurement. When there will be a need to consider the post-measurement state we will switch to Kraus operators. Which form on quantum measurement is being used is usually clear from the context and hence we simply call all of them \emph{measurement operators}.

					The Pauli operators, denoted by $\sigma_x$, $\sigma_y,$ and $\sigma_z$, are an example for measurement operators for qubits:
					\begin{equation}~\label{eq:pauli_operators_def}
						\sigma_x= \begin{pmatrix} 0 & 1 \\  1 & 0  \end{pmatrix} 
						\quad \;; \quad
						\sigma_y= \begin{pmatrix} 0 & -i \\  i & 0  \end{pmatrix} 
						\quad \;; \quad
						\sigma_z= \begin{pmatrix} 1 & 0 \\  0 & 1  \end{pmatrix}  \;.
					\end{equation}

			
			\subsubsection{Quantum channels}
		
				Quantum channels, or maps, are functions describing the evolution of quantum states.  In order for a map  $\mathcal{E}$ to describe a real physical process, transferring one quantum state  $\rho\in \mathcal{S}(\hilb_A)$ to another  $\mathcal{E}(\rho) \in \mathcal{S}(\hilb_B)$,\footnote{Note that $\hilb_B$ may be different than $\hilb_A$. For a unitary evolution, discussed before, this was not the case.} it must  fulfil certain conditions. Specifically, it must be completely positive and trace preserving (CPTP).  
				\begin{defn}[Quantum channel]\label{def:cptp_map}
					A linear map $\mathcal{E} \in \mathrm{Hom}(\mathcal{S}(\hilb_A), \mathcal{S}(\hilb_B))$ is a quantum channel if it is:
					\begin{enumerate}
						\item Completely positive (CP): for any $\rho_{AR} \in \mathcal{S}(\hilb_A \otimes \hilb_R)$ with $\rho_{AR} \succeq 0$, 
							\[
								(\mathcal{E} \otimes \idn_{R}) (\rho_{AR}) \succeq 0 \;,
							\] 
							where $\hilb_{R}$ is any additional Hilbert space and $\idn_{R}$ is the identity map on that Hilbert space.
						\item Trace preserving (TP): for any $\rho \in \mathcal{S}(\hilb_A)$, $\Tr \left(\mathcal{E}(\rho) \right) = \Tr (\rho)$.
					\end{enumerate}
				\end{defn}

		\section{Distance measures}
		
			The trace distance of two states is given by $\Delta(\rho,\tau)=\frac{1}{2}\|\rho-\tau\|_1$. Operationally, the trace distance quantifies the \emph{distinguishing advantage} when trying to distinguish $\rho$ from $\tau$. Consider a situation in which either the state $\rho$ or the state $\tau$ are chosen uniformly at random and given to someone who has no information as to which state was chosen and needs to output a guess. The probability of succeeding in this task depends on how far $\rho$ and $\tau$ are from one another via 
			\[
				\Pr[\text{correct guess}] = \frac{1}{2}\left(1+ \frac{1}{2}\|\rho-\tau\|_1\right) = \frac{1}{2} + \Delta(\rho,\tau) \;.
			\]
			
			We will be interested below in the so called purified distance. The purified distance involves sub-normalised states, i.e., states with $\Tr(\rho)\leq 1$. For this, one first needs to extend the definition of the trace distance to describe also the distance between two sub-normalised states. 
			\begin{defn}[Generalised trace distance]
				The trace distance between two sub-normalised states $\rho$ and $\tau$ is given by
					\[
						\Delta(\rho,\tau)=\frac{1}{2}\|\rho-\tau\|_1 + \frac{1}{2} |\Tr(\rho-\tau)| \;. 
					\]
			\end{defn}

			Another important measure of distance (though not a metric) is the fidelity. The fidelity of two quantum states is given by $F(\rho,\tau) = \left(\Tr|\sqrt{\rho}\sqrt{\tau}|\right)^2$.
			The fidelity is related to the trace distance by
			\[	
				1 - \sqrt{F(\rho,\tau)} \leq \Delta(\rho,\tau) \leq \sqrt{1-F(\rho,\tau)}
			\]
			Here, again, we can define the fidelity between two sub-normalised states.
			\begin{defn}[Generalised fidelity]
				The	fidelity between two sub-normalised states $\rho$ and $\tau$ is given by
				\[
					F(\rho,\tau) = \left(\Tr|\sqrt{\rho}\sqrt{\tau}| + \sqrt{(1-\Tr(\rho))(1-\Tr(\tau)) } \right)^2 \;.
				\]
			\end{defn}

			The last distance measure that will be of importance for us is the purified distance~\cite{tomamichel2010entropyduality}. This measure will be used to define the smooth entropies below and will always be considered with  sub-normalised states.
			\begin{defn}[Purified distance]\label{def:pur_dist}
				The	purified distance between two sub-normalised states $\rho$ and $\tau$ is given by
				\[
					P(\rho,\tau) = \sqrt{1-F(\rho,\tau)} \;.
				\]
			\end{defn}

	\section{Entropies}
		
		
		\subsection{Shannon and von Neumann Entropy}
			
			\begin{defn}[Shannon entropy]
				Given RVs $A$ and $B$ defined over $\mathcal{A}$ and $\mathcal{B}$, respectively, the Shannon entropy of $A$ is given by 
				\[
					H(A) = - \sum_{a\in \mathcal{A}} \P_A(a)\log\left(\P_A(a)\right) \;.
				\]
				The conditional Shannon entropy of $A$ given $B$ is defined to be
				\[
					H(A|B) = H(AB) - H(B) = \sum_{b\in \mathcal{B}} \P_B(b) H(A|b) \;.
				\]
			\end{defn}
			
			In the case of a RV defined over $\{0,1\}$ with $\P_A(0)=p$ the Shannon entropy is reduced to the so called ``binary entropy'' $h(p)=-p\log (p) -(1-p)\log (1-p)$. 
			
			The von Neumann entropy is the extension of the Shannon entropy to quantum states.
			\begin{defn}[von Neumann entropy]
				Given a quantum state $\rho_{AB}\in \mathcal{S}(\hilb_A\otimes\hilb_B)$, the von Neumann entropy of $A$ is given by 
				\[
					H(A)_{\rho} = - \Tr\left(\rho \log \rho \right) \;.
				\]
				The conditional von Neumann entropy of $A$ given $B$ is defined to
				\[
					H(A|B)_{\rho} = H(AB)_{\rho} - H(B)_{\rho} \;.
				\]
			\end{defn}
			When the state on which the entropy is evaluated is clear from the context we drop the subscript and write, e.g., $H(A|B)$.


			\begin{defn}[Mutual information]
					For a quantum state $\rho_{ABC}$, the conditional mutual information between  $A$ and $B$ conditioned $C$ is given by
					\begin{align*}
						I(A:B|C)_{\rho} &= H(A|C)_{\rho} + H(B|C)_{\rho} - H(AB|C)_{\rho} \\
						&=H(A|C)_{\rho} - H(A|BC)_{\rho} \;.
					\end{align*}
			\end{defn}
			\sloppy
			There are other equivalent ways of defining Markov chains for quantum states~\cite{hayden2004structure}, but for our purposes this definition suffices.
			
			The conditional mutual information fulfils the following properties:
			\begin{enumerate}
				\item Strong subadditivity:  $I(A:B|C)_{\rho} \geq 0$ for any $\rho$.
				\item Data processing: for any quantum channels $\mathcal{E}:\mathcal{S}(\hilb_A)\rightarrow \mathcal{S}(\hilb_{A'})$  and $\mathcal{F}:\mathcal{S}(\hilb_B)\rightarrow \mathcal{S}(\hilb_{B'})$,
					\begin{equation*}
						I(A:B|C)_{\rho} \geq I(A':B'|C)_{\rho'} \;,
					\end{equation*}
					where $\rho'_{A'B'C} = \mathcal{E} \otimes \mathcal{F} \otimes \idn_C (\rho_{ABC}) $.
				\item $I(A:B|C) = 0$ if and only if  $A$ and $B$ are independent given $C$, i.e., $\P_{AB|C}=\P_{A|C}\cdot\P_{B|C}$.	
			\end{enumerate}
			
			\begin{defn}\label{eq:markov_chain_def}
				A tripartite quantum state $\rho_{ABC}$ is said to fulfil the Markov chain condition $A\leftrightarrow C \leftrightarrow B$	if $I(A:B|C) = 0$. 
			\end{defn}

		\subsection{Min- and max-entropies}
		
				We will work with the smooth min- and max-entropies, formally defined as follows.
				\begin{defn}[Smooth conditional entropies]
					For any $\varepsilon\in[0,1]$ the $\varepsilon$-smooth conditional min- and max-entropy of a state $\rho_{AB}$ are given by 
					\begin{align*}
						& H^{\varepsilon}_{\min}(A|B)_{\rho_{AB}} = \log \inf_{\sigma_{AB} \in \mathcal{B}^\varepsilon(\rho_{AB})} \inf_{\tau_B} \| \sigma_{AB}^{\frac{1}{2}}\tau_B^{-\frac{1}{2}}\|_{\infty}^2  \\
						& H^{\varepsilon}_{\max}(A|B)_{\rho_{AB}} = \log \inf_{\sigma_{AB} \in \mathcal{B}^\varepsilon(\rho_{AB})} \sup_{\tau_B} \| \sigma_{AB}^{\frac{1}{2}}\tau_B^{-\frac{1}{2}}\|_1^2 \;,
					\end{align*}
					for $\mathcal{B}^\varepsilon(\rho_{AB})$ the set of sub-normalised states $\sigma_{AB}$ with $P(\rho_{AB},\sigma_{AB}) \leq \varepsilon$, where~$P$ is the purified distance as in Definition~\ref{def:pur_dist}.
				\end{defn}
				
				In practice, we will not need the fully general definitions above (which are stated for completeness).  
				When considering the min-entropy, we will be interested in the case where the $A$ system is classical. This leads to more intuitive definitions.
				When~$A$ is classical and $B$ is trivial, one can simply write
				\[
					H_{\min}(A) = -  \log\left[ \max_{a}\P_A(a)\right] \;.
				\]

				 For quantum $B$, the state can be written as $\rho_{AB}=  \sum_a p_a \ket{a}\!\!\bra{a}\otimes \rho_B^a$. 
				 Then, the conditional min-entropy is the directly related to the guessing probability of $A$ given~$B$ via
					\[	
						H_{\text{min}}(A|B) = -\log p_{\text{guess}}(A|B)\;,
					\]
					where $p_{\text{guess}}(A|B)$ is the maximum probability of guessing $A$ given the quantum system $B$: 
					\[
						p_{\text{guess}}(A|B) = \max_{\{M^a_B\}_a} \Big| \sum_a p_a \Tr (M^a_B\rho^a_E) \Big| \;,
					\]
					and the maximisation is performed over all POVMs $\{M^a_B\}_a$ on $B$. 
					The smooth conditional min-entropy can be written by maximising the min-entropy over all close sub-normalised states, i.e., 
					 \[
						H_{\text{min}}^\varepsilon(A|B)_{\rho_{AB}} = \max_{\sigma_{AB} \in \mathcal{B}^\varepsilon(\rho_{AB})} H_{\text{min}}(A|B)_{\sigma_{AB}} \;. 
					\]

				Moving on to the max-entropy, we will mainly be interested in the case of classical registers. 
				In the classical case, the following holds for the max-entropy 		
				\[
					H_{\max}(A) \leq -  \log\left[ \min_{a | \P_A(a)\neq 0} \P_A(a)\right] \;. 
				\]

				Evaluating the smooth conditional max-entropy will be done by considering a closely related quantity, namely the classical smooth zero-entropy. 
				\begin{defn}[Classical zero-entropy]\label{defn:zero_entropy}
					For classical RVs $A$ and $B$ distributed according to $\mathrm{P}_{AB}$, 
					\[	
						H_{0}(A|B)= \max_{b} \log \left| \text{Supp}\left(\mathrm{P}_{A|B=b}\right)\right| \;,
					\]
					 where $\text{Supp}\left(\mathrm{P}_{A|B=b}\right)=\{ a : \mathrm{P}_{A|B=b}\left(a\right) > 0\}$. 
					 The smooth version of the zero-entropy is given by 
					\[
						H^{\varepsilon}_{0}(A|B)=\min_{\Omega} \max_{b} \log \left| \text{Supp}\left(\mathrm{P}_{A|\Omega,B=b}\right)\right| \;,
					\]
					where the minimum ranges over all events $\Omega$ with probability at least $1-\varepsilon$. 
				\end{defn}

				Finally, we remark that for any quantum state $\rho_{AB}$, 
				\[
					 H_{\text{max}}(A|B) \geq H(A|B) \geq H_{\text{min}}(A|B) \;.
				\]
				The same ordering does not necessarily hold for the smooth entropies.


\chapter{Preliminaries: device-independent concepts}\label{ch:pre_di}


	The goal of this chapter is to present the basic information needed while reading the thesis. It is by no means a comprehensive review of the topic of device-independent information processing. A reader completely unfamiliar with the concepts of non-locality and device-independent protocols is encouraged to read the survey~\cite{brunner2014bell} and lecture notes~\cite{scarani2013device}.

	As explained in the introduction, the device-independent framework allows one to examine certain properties of physical devices without referring to their internal workings. 
	Instead of describing a device using its hardware and actions we think of it as a \emph{box} with buttons, on which the user can press in order to give classical inputs to device, and a display, from which the user can read the classical outputs produced by the device. 
	Then, the only information available to the user of the box is the observed data, i.e., the input-output behaviour of the box. 
	
	The input-output behaviour of the box can be described mathematically using a \emph{conditional probability distribution} $\P_{O|I}$, where $I$ describes the possible inputs of the box and $O$ the possible outputs. For example, if the box has three buttons we can think of $I$ as being a random variable over $\{0,1,2\}$. If the box displays a bit as its output then $O$ is a random variable over $\{0,1\}$. $\P_{O|I}$ then describes the, possibly probabilistic, actions of the box. For example, a box with $\P_{O|I}(0|0) = 1/2$ and~$\P_{O|I}(1|0) = 1/2$ outputs $0$ or $1$, each with probability $1/2$, when the user presses the button associated with the input $0$.

	The following sections are devoted to explaining the types of  boxes that one can consider and their properties. In Section~\ref{sec:pre_black_box} we define three important classes of boxes according to their input-output behaviour. In Section~\ref{sec:pre_bell_ineq} we introduce the topic of Bell inequalities, which lies at the heart of all device-independent information processing tasks. In Section~\ref{sec:untrusted_devices} we formally discuss the concept of untrusted devices and, in particular, how a possibly malicious box is modelled.  
	
	\section{Black boxes}\label{sec:pre_black_box}
		
		In this thesis we  mainly consider bipartite boxes. We think of a bipartite box as a box with two components, each belonging to a different party -- one component for Alice and one for Bob. 
		Crucially later on, the components are separated in space so Alice and Bob may locate their parts of the box in different places. 
		Both of Alice's and Bob's component have buttons and a display.
		 Alice has the possibility of supplying an input to \emph{her component} and reading the output produced by \emph{her component}. Bob has no access to Alice's component. 
		 Similarly,  Bob has the possibility of supplying an input to \emph{his component} and reading the output produced by \emph{his component}, while Alice has no access to Bob's component.
		 
		Mathematically the  bipartite nature of the box presents itself by considering conditional probability distributions $\P_{AB|XY}$, where $X$ and $A$ denote Alice's inputs and outputs, respectively, while $Y$ and $B$ denote Bob's inputs and outputs, respectively. 
		$\P_{AB|XY}$ includes all the information about the input-output behaviour of the box and the correlations between Alice's and Bob's outputs.
		
		A priori, there are no restrictions on $\P_{AB|XY}$, i.e., it can be \emph{any} conditional probability distribution. One may restrict the type of boxes being considered by imposing certain constraints on $\P_{AB|XY}$ that depend on the physical theory being studied. 
		Specifically, we are interested in boxes that describe classical, quantum, and non-signalling devices (as explained below). A quantum box, for example, may exhibit correlations between Alice and Bob that cannot be created by classical means. 
		When considering the space of conditional probability distributions $\P_{AB|XY}$ the constraints imposed on the box define sets to which the different type of boxes belong. 
		The constraints defining the sets of interest are explained below.

		\subsection{Non-signalling boxes}\label{sec:pre_ns_corr}
		
			When considering general conditional probability distributions $\P_{AB|XY}$ any dependence between $A$, $B$, $X$, and $Y$ is allowed. 
			In particular, even though we think of Alice and Bob as holding two separated parts of the box, Alice's output $A$ may depend on both inputs $X$ and $Y$. 
			In practice this means that in order for Alice's box to produce an output, following Alice's choice of input $X$, the box first needs to get Bob's input $Y$  as well. That is, until a signal including Bob's input arrives to Alice's component, no actions will be taken by Alice's component of the box and Alice will need to wait. 
			
			In most cases, the above is not a desired behaviour; usually one expects the component of one user to produce an output as a response to pressing the button on \emph{that component} alone. 
			Mathematically this requirement is phrased using the so called ``non-signalling conditions'' that imply that the marginals $\P_{A|X}$ and $\P_{B|Y}$ are a well-defined conditional probability distribution. In other words, the behaviour of Alice's part of the box is described by $\P_{A|X}$, which is independent of Bob's input $Y$. Thus, Alice's box does not need to receive $Y$ before producing $A$. 
			A box fulfilling the non-signalling conditions between Alice and Bob is called a non-signalling box and is defined as follows. 
			\begin{defn}[Non-signalling box]\label{def:ns_box}
				A non-signalling box is a conditional probability distribution $\P_{AB|XY}$ for which the non-signalling conditions
				\begin{align}
					\sum_b \P_{AB|XY}(a,b|x,y) = \sum_b \P_{AB|XY}(a,b|x,y') \label{eq:pre_ns_bob} \\
					\sum_a \P_{AB|XY}(a,b|x,y) = \sum_a \P_{AB|XY}(a,b|x',y) \label{eq:pre_ns_alice}
				\end{align}
				hold for all $a\in\mathcal{A}$, $ b\in\mathcal{B}$, $ x,x'\in\mathcal{X}$ and $y,y'\in\mathcal{Y}$.
			\end{defn}
			
			Denote by $\P^y_{A|X}(a|x,y)= \sum_b \P_{AB|XY}(a,b|x,y)$ the behaviour of Alice's part of the box, which may a priori depend on Bob's choice of input $y$. 
			Equation~\eqref{eq:pre_ns_bob} states that $\P^y_{A|X}=\P^{y'}_{A|X}$ and, hence, the  conditional probability distribution describing Alice's part of the box is independent of Bob's input, i.e., whether Bob inputs $y$ or $y'$. We can therefore drop the superscript and simply consider $\P_{A|X}$~--- a well defined marginal. 
			Similarly, Equation~\eqref{eq:pre_ns_alice} implies that $\P_{B|Y}$ is independent of Alice's input and faithfully describes Bob's part of the box. 
						
			On the more fundamental level, the non-signalling conditions describe the assumption that the box cannot be used to send instantaneous signals between Alice and Bob. 
			Alice and Bob may locate their components arbitrarily far away from one another. If we require the two components to produce outputs right away, then signals including information about the inputs used by the other party have no time to get from one part of the box to the other and influence its actions. 
			In such a case, using the above notation, if $\P^y_{A|X}\neq\P^{y'}_{A|X}$, then Alice may conclude from her observed statistics whether Bob used $y$ or $y'$, even though this information did not have enough time to travel from Bob to Alice. 
			It follows that a non-signalling box is a device that cannot be used as means of communication between Alice and Bob. 			

		
			A closely related definition that will be of use later is that of  a non-signalling extension.
			Given Alice's component, one can consider an extension of it to an additional party Bob. Specifically, we will be interested in what we call a non-signalling extension of a box, defined below.
			\begin{defn}[Non-signalling extension]\label{def:ns_extension}
					A non-signalling extension of a box $\P_{A|X}$ is a non-signalling box $\P_{AB|XY}$ such that for all $a\in\mathcal{A}$, $x\in\mathcal{X}$, and $y\in\mathcal{Y}$, $\sum_{b} \P_{AB|XY}(a,b|x,y) = \P_{A|X}(a|x)$.
			\end{defn}
			In words, given $\P_{A|X}$, $\P_{AB|XY}$ is a non-signalling box with the ``correct marginal'' on Alice's side (while Bob's marginal $\P_{B|Y}$ can be arbitrary).

		Before moving on we point to the simplicity of the non-signalling conditions in Equations~\eqref{eq:pre_ns_bob} and~\eqref{eq:pre_ns_alice}. The non-signalling conditions are linear. As a result,  the set of non-signalling boxes is a polytope. The faces of the polytope are defined by the various non-signalling conditions as well as the positivity and normalisation constrains fulfilled by any conditional probability distribution.

		\subsection{Quantum boxes}
		
			One can further restrict the modelled device by considering quantum boxes, i.e., boxes that exhibit quantum correlations. Such boxes are relevant when considering device-independent processing tasks in which all the resources are quantum. 
			
			Quantum correlations are correlations that can be explained within the formalism of quantum physics. To put it differently, we think of a box as a device that ``holds'' some bipartite quantum state $\rho_{Q_AQ_B}$, shared between Alice and Bob.\footnote{We distinguish the quantum state from the correlations throughout the thesis: $Q_A$ and $Q_B$ denote quantum registers belonging to Alice and Bob while $A$ and $B$ denote their classical outputs.} Alice's component of the device performs some local quantum measurements on her marginal state $\rho_{Q_A}$ and similarly for Bob.  Formally:
			\begin{defn}[Quantum box]\label{def:quant_box}
				A quantum box is a conditional probability distribution $\P_{AB|XY}$ such that there exist a bipartite state $\rho_{Q_AQ_B}$ and sets of POVMs for Alice and Bob $\{M^{x}_a\}_{a\in\mathcal{A}}$ for all $x\in\mathcal{X}$ and  $\{M^{y}_b\}_{b\in\mathcal{B}}$ for all $y\in\mathcal{Y}$, respectively, for which
				\begin{equation}\label{eq:quant_box_stat}
					\P_{AB|XY}(a,b|x,y) = \Tr\left(M^{x}_a \otimes M^{y}_b \; \rho_{Q_AQ_B} \right) \quad \forall  a,b,x,y\;.
				\end{equation}
			\end{defn}
			
			We make the following remarks regarding Definition~\ref{def:quant_box}.
			First, while we assume that the bipartite box is quantum, we do not assume anything regarding its internal workings. In particular, we only assume here that the state $\rho_{Q_AQ_B}$ is defined over a bipartite Hilbert space\footnote{The definition of a quantum box over a bipartite Hilbert space $\hilb_{Q_A} \otimes \hilb_{Q_B}$ is the standard one in the context of non-relativistic quantum mechanics. When studying relativist quantum mechanics one considers a single Hilbert space $\hilb$ and two commuting measurements acting on it (instead of tensor product measurements). The two definitions coincide when restricting the attention to finite dimensional Hilbert spaces but otherwise different in general~\cite{slofstra2017set}.} $\hilb_{Q_A} \otimes \hilb_{Q_B}$ (since we consider bipartite boxes) but we do not restrict the dimensions of $\hilb_{Q_A}$ and~$\hilb_{Q_B}$.
			
			Second, the non-signalling assumption is ``encoded'' in the bipartite structure of Alice and Bob's state $\rho_{Q_AQ_B}$ together with tensor product structure of their measurements as in Equation~\eqref{eq:quant_box_stat}. That is, the conditional probability distribution~$\P_{AB|XY}$ is by definition non-signalling. Hence, the set of quantum boxes is a subset of the set of non-signalling boxes.

		\subsection{Classical boxes}
		
			A classical box is described by a conditional probability distribution that can be explained in terms of shared randomness alone. That is, we think of Alice's and Bob's component of the box as holding a shared random string (in contrast to a shared quantum state). Each component decides on its output depending on its input and the shared string. Formally:
			
			\begin{defn}[Classical box]\label{def:classical_box}
					A classical box is a conditional probability distribution $\P_{AB|XY}$ that can be written in the form 
					\begin{equation}\label{eq:local_correlations}
						\P_{AB|XY}(a,b|x,y) = \int_{\Lambda} \mathrm{d}\lambda \Pr\left[\Lambda = \lambda\right] \P_{A|X\Lambda}(a|x\lambda) \cdot \P_{B|Y\Lambda}(b|y\lambda) \;,
					\end{equation}
					where $\Lambda$ is the random variable describing the randomness shared by the two components of the box.
			\end{defn}
		
			One can assume without loss of generality that $\P_{A|X\Lambda}$ and $\P_{B|Y\Lambda}$ are deterministic. That is, for all $\lambda$, $x$, and $a$, either $\P_{A|X\Lambda}(a|x,\lambda) = 0$ or $\P_{A|X\Lambda}(a|x,\lambda) = 1$, and similarly for Bob. This holds since we can always ``push'' the non-deterministic behaviour of the components to the shared randomness $\lambda$ itself. 
			As the number of \emph{deterministic} assignments of $a$ to each $x$ is finite (assuming $\mathcal{A}$ and $\mathcal{X}$ are finite), it follows that one can also express all classical boxes as 
			\[ 
				\P_{AB|XY}(a,b|x,y) = \sum_{\lambda} \Pr\left[\Lambda = \lambda\right] \P_{A|X\Lambda}(a|x\lambda) \cdot \P_{B|Y\Lambda}(b|y\lambda) \;,
			\]
			for $\lambda$ belonging to a finite set and deterministic $\P_{A|X\Lambda}$ and $\P_{B|Y\Lambda}$. 
		
			In the context of Bell inequalities, discussed below, $\Lambda$ is called the ``hidden variable'' that explains the correlations between Alice's and Bob's parts of the box. Conditioned on the value of $\Lambda$, the two components, $\P_{A|X\Lambda}$ and $\P_{B|Y\Lambda}$, are independent of one another, as seen in Equation~\eqref{eq:local_correlations}.
			 Classical boxes, or correlations, are also termed ``local correlations''.\footnote{Though common, this is a rather confusing and unjustified terminology. As clear from Equation~\eqref{eq:quant_box_stat}, quantum correlations are also local, in the sense that each component performs a local operation on its part of the state.}
			 
			 It is easy to see that when considering scenarios with a \emph{single} party, i.e., boxes~$\P_{A|X}$, all conditional probability distributions can be written in  the form of Equation~\eqref{eq:local_correlations}. Thus, all single-party boxes are classical boxes. This is not an interesting scenario and, in particular, no device-independent information processing task can be performed in such a case. Thus, boxes of two parties or more are always considered. 
			
		\subsection{Correlations' space}

\begin{figure}
				\centering
				\includegraphics[width=\textwidth]{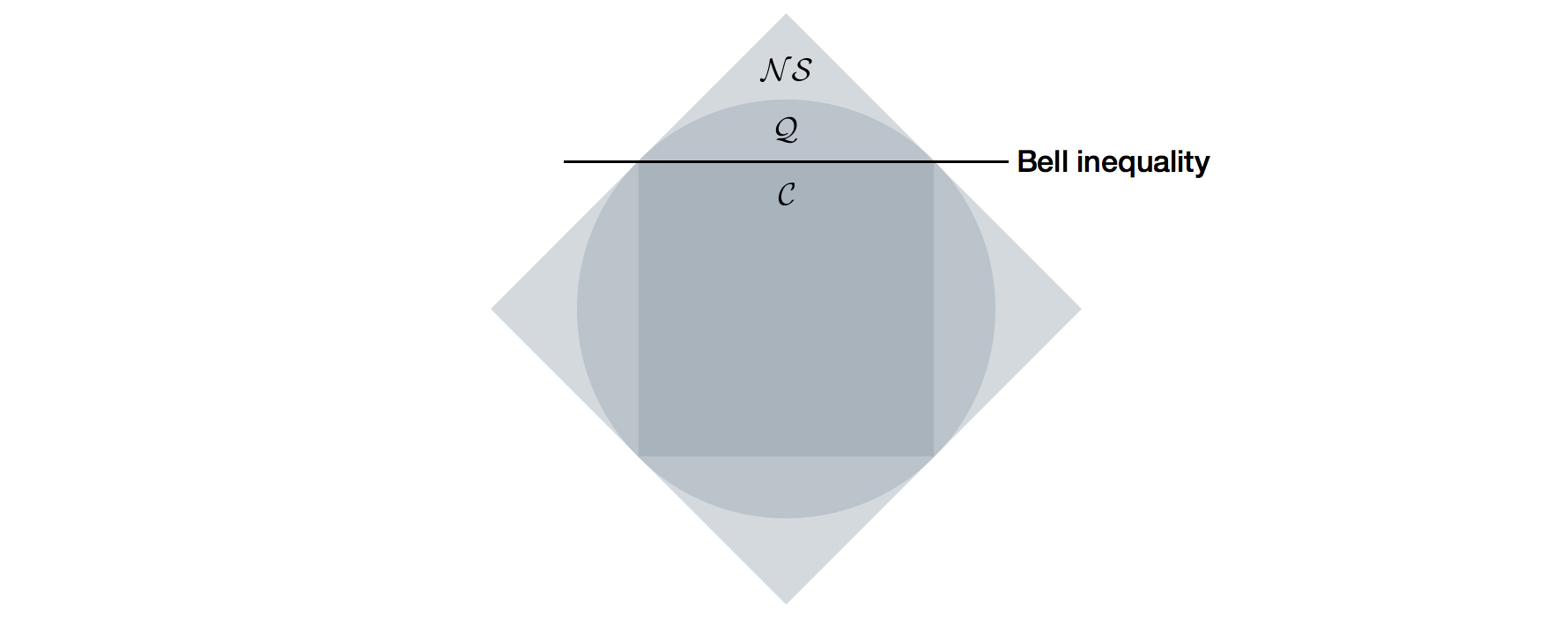}
				\caption{Illustration of the sets of boxes. $\mathcal{C}$, $\mathcal{Q}$, and $\mathcal{NS}$ denote the sets of classical, quantum, and non-signalling boxes, respectively. All sets are convex,  $\mathcal{C}$ and $\mathcal{NS}$  being polytopes, and the relation $\mathcal{C} \subsetneq \mathcal{Q} \subsetneq \mathcal{NS}$ holds. Bell inequalities can be used to separate classical boxes from quantum ones.}
				\label{fig:polytopes}
			\end{figure}

			
			Let $\mathcal{C}$, $\mathcal{Q}$, and $\mathcal{NS}$ denote the sets of classical, quantum, and non-signalling boxes, respectively. 
			It is easy to see that all of these sets are convex: given two classical boxes $\P^1_{AB|XY}$ and $\P^2_{AB|XY}$, the box $\P_{AB|XY} = p \P^1_{AB|XY} + (1-p) P^2_{AB|XY}$ is also classical, and similarly for quantum and non-signalling boxes. 
			The convex sets of classical and non-signalling boxes can be described  as the convex combination of a finite number of extremal point and hence $\mathcal{C}$ and $\mathcal{NS}$ are polytopes. This is not the case for the quantum set  $\mathcal{Q}$. See Figure~\ref{fig:polytopes} for an illustration.
			
			As clear from the definition of the various types of boxes, any classical box is also a quantum box and any quantum box is also a non-signalling box. 
			Furthermore, there are examples for quantum boxes that \emph{cannot} be written in the form of Equation~\eqref{eq:local_correlations} and for non-signalling boxes that  \emph{cannot} be written in the form of Equation~\eqref{eq:quant_box_stat}. 
			It follows that the sets fulfil the relation
			\[
				\mathcal{C} \subsetneq \mathcal{Q} \subsetneq \mathcal{NS} \;,
			\]
			as in Figure~\ref{fig:polytopes}.
			
			Bell inequalities, discussed in the next section, give us a way of separating classical boxes from quantum ones\footnote{Separating quantum boxes from non-signalling ones is a far more complicated task; see, e.g.,~\cite{navascues2008convergent}.}~--- an essential tool in any device-independent information processing task.
			
	\section{Bell inequalities}\label{sec:pre_bell_ineq}
		
		From now on our discussion is restricted to boxes which fulfil, at the least, the non-signalling conditions. That is, we are considering only boxes which cannot be used as means of communication between two seperated parties, Alice and Bob.
		
		The polytope of classical boxes $\mathcal{C}$ is a strict subset of the set of quantum and non-signalling boxes. 
		As such, some of the affine hyperplanes defining $\mathcal{C}$ separate  $\mathcal{C}$ from $\mathcal{Q}$.\footnote{Other hyperplanes represent the trivial conditions of positivity of normalisation of the conditional probability distributions which are relevant for all sets.} 
		Informally, when we say that a hyperplane separates  $\mathcal{C}$ from $\mathcal{Q}$ we mean that all classical boxes are on one side of the hyperplane, while the other side can only include quantum and non-signalling boxes. 
		
		The condition of being on ``one side of the hyperplane'' is written in the form of an inequality
		\begin{equation}\label{eq:bell_ineq}
			\forall \P_{AB|XY}\in \mathcal{C}, \qquad  \sum_{a,b,x,y} s(a,b,x,y) \P_{AB|XY}(a,b|x,y) \leq S \;,
		\end{equation}
		for some given constants $S$ and $s(a,b,x,y)$, for all $a,b,x,y$. 
				
		Given a box $\P_{AB|XY}$, the simple form of Equation~\eqref{eq:bell_ineq} allows us to test, by calculating $\sum_{a,b,x,y} s(a,b,x,y) \P_{AB|XY}(a,b|x,y)$, if the box \emph{cannot} be a classical one.
		In other words, if the inequality is violated, i.e., 
		\[
			\sum_{a,b,x,y} s(a,b,x,y) \P_{AB|XY}(a,b|x,y) > S \;,
		\]
		 then $ \P_{AB|XY}$ cannot be written in the form of  Equation~\eqref{eq:local_correlations}.

		As first noticed by~\cite{bell1964einstein}, some quantum boxes, arising from measurements performed on entangled states, are capable of violating inequalities as in Equation~\eqref{eq:bell_ineq}.\footnote{Notice that the statement that  some quantum states violate Bell inequalities is independent from the statement that classical boxes cannot violate the inequality; it could have been the case that no box is able to violated such inequalities. This would have implied that all quantum correlations can be written in the form of  Equation~\eqref{eq:local_correlations} and, hence, can be described as arising from some shared randomness, or an ``hidden variable'', $\lambda$.}
		Bell suggested to use such states in an experiment, proposed to test the EPR paradox~\cite{einstein1935can}, that will allow us to check whether there is some classical piece of information, that we are just unaware of or cannot observe, that can explain the apparent ``non-local'' correlations exhibit by certain quantum states.
		Such experiments, called today ``loophole-free Bell tests''~\cite{hensen2015loophole,shalm2015strong,giustina2015significant}, have verified the violation of Bell inequalities and by this refuted the possibility of classical explanations of the behaviour of some entangled quantum states.

		The inequalities which are fulfilled by any classical box while being violated by \emph{some} quantum boxes are called \emph{Bell inequalities}; see Figure~\ref{fig:polytopes}.
		All of the above implies that a Bell  inequality acts as a test for ``quantumness'' or, more precisely, ``non-classicalness'' and its violation acts as a certificate for passing the test.
		 As such, it is crucial for any device-independent information processing task in which we need to rule out the possibility of executing the considered task with a classical device.

		\subsection{Non-local games}\label{sec:pre_games}
		
			Bell inequalities, as in Equation~\eqref{eq:bell_ineq}, can also be phrased as special types of games, called non-local, or Bell, games. 
			In a game,  a referee asks Alice and Bob, the players of the game, a question each, chosen according to a given probability distribution; each player only sees her/his question. The players then need to supply answers which fulfil a pre-determined requirement according to which the referee accepts or rejects the answers. 
			To win the game the players can agree on a strategy beforehand but, once the game begins, communication between the players is not allowed. If the referee accepts their answers the players win. 
			
			Formally, a game $\G=(\mathcal{X},\mathcal{Y},\mathcal{A},\mathcal{B},\Q_{XY},w)$ is defined by sets of possible questions $\mathcal{X}$ and answers $\mathcal{A}$ for Alice,  sets of possible questions $\mathcal{Y}$ and answers $\mathcal{B}$ for Bob, a probability distribution $\Q_{XY}$ over the questions, according to which the referee choses the questions, and a winning condition $w:\mathcal{A}\times\mathcal{B}\times\mathcal{X}\times\mathcal{Y}\rightarrow\{0,1\}$, where~$w(a,b,x,y)=0$ means that the referee rejects $(a,b,x,y)$, i.e., the players lose, and~$w(a,b,x,y)=1$ means that the players win with  $(a,b,x,y)$.

			A strategy for the game is naturally described by a box $\P_{AB|XY}$ held by the players~--- the referee chooses questions $(x,y)\in\mathcal{X}\times\mathcal{Y}$ and the players need to reply with answers $(a,b)\in\mathcal{A}\times\mathcal{B}$. The fact that the players are not allowed to communicate during the game means that, at the least, the box $\P_{AB|XY}$ is constrained by non-signalling conditions between Alice and Bob, i.e., $\P_{AB|XY}$ belongs to the non-signalling polytope $\mathcal{NS}$.\footnote{Depending on the context, one can further restrict the allowed strategies by considering classical or quantum boxes.}	
			The winning probability of a box  $\P_{AB|XY}$ in the game $\G$ is given by
			\begin{equation*}
				\omega\left(\P_{AB|XY}\right) = \sum_{a,b,x,y} \Q_{XY}(x,y) \P_{AB|XY}(a,b|x,y) w(a,b,x,y) \;.
			\end{equation*}
			When the discussed box $\P_{AB|XY}$ is clear from the context we simple write $\omega$ to denote its winning probability.\footnote{Notice the notation: $w$ denotes a winning condition (function) while $\omega$ is the winning probability (a number). $W$ will be used to denote the random variable describing whether a game is won or lost. In any case, the difference between these three objects is always clear from the text.}

			In the context of device-independent information processing we interpret a Bell inequality as a special type of game. 
			What makes the game special is that it is designed so that any classical box used by the players leads to a winning probability of at most $\omega_c<1$, while there exists a quantum box that can be used by the players to achieve a greater winning probability, $\omega_q > \omega_c$. 
			Instead of a Bell inequality as in Equation~\eqref{eq:bell_ineq} we have
			\begin{equation}\label{eq:winning_ineq}
				\forall \P_{AB|XY}\in \mathcal{C}, \qquad \omega\left(\P_{AB|XY}\right) \leq \omega_c \;.
			\end{equation}
			Violating a Bell inequality then translates to violating Equation~\eqref{eq:winning_ineq} by winning the respective game with probability greater than $\omega_c$. 
			In both cases, the conclusion is the same; if $\P_{AB|XY}$  violates Equation~\eqref{eq:winning_ineq} then $\P_{AB|XY}\notin\mathcal{C}$.

			Before discussing an explicit example of a non-local game, one remark is in order.
			Above, we thought of Alice and Bob as the ones preparing the box, according to their strategy in the game, and the referee was asking them questions to test their winning probability. 
			Alternatively, we can think of Alice and Bob as holding an uncharacterised box and they are the ones testing the box, by choosing the questions themselves. In that case, Alice and Bob basically take the role of the referee (while the box takes the role of Alice and Bob). 
			(In the showcase of non-signalling parallel repetition, in Chapter~\ref{ch:par_rep_showcase}, we use the first terminology, while the showcase of device-independent quantum cryptography, in Chapter~\ref{ch:crypto_showcase},  the second is terminology is the more appropriate one).

		\subsection{The CHSH game}\label{sec:chsh_game}
			
			We now present an explicit non-local game that will be of use in the thesis.
			The Clauser-Horne-Shimony-Holt (CHSH) game~\cite{clauser1969proposed} is probably the most famous non-local game. 
			In the game, Alice's and Bob's inputs and outputs are bits, $a,b,x,y\in\{0,1\}$ and the inputs are distributed uniformly at random, i.e., $\Q_{XY}(x,y)=1/4$ for all $x$ and $y$. 
			The winning conditions is given by:
			\begin{equation}\label{eq:chsh_win_cond}
				w_{\text{CHSH}} = \begin{cases}
					1 & \text{ and } a\oplus b = x \cdot y \\
					0 & \text{otherwise.}
				\end{cases}
			\end{equation}

			The optimal classical box, or strategy, achieves a winning probability of $0.75$. An example for such a strategy is one in which the outputs are always $(a,b)=(0,0)$. 
			
			An optimal quantum strategy consists in measuring the maximally entangled state $\ket{\Phi^+} = \left(\ket{00}+\ket{11}\right)/\sqrt{2}$ with the following measurements: Alice's measurements $x=0$ and $x=1$ correspond to the Pauli operators $\sigma_z$ and $\sigma_x$ respectively and Bob's measurements $y=0$ and $y=1$ to $\left(\sigma_z + \sigma_x\right)/\sqrt{2}$ and $\left(\sigma_z - \sigma_x\right)/\sqrt{2}$ respectively.
			A box implementing the above achieves winning probability $\omega = \frac{2+\sqrt{2}}{4}\approx 0.85$. 
			Perhaps surprisingly, any box that achieves the optimal quantum winning probability (or close to it) must be implementing a strategy identical to the above up to local isometries (or close to such a strategy)~\cite{popescu1992states,mayers2003self,mckague2012robust}. 
			
			The CHSH game can also phrased in the form of a Bell \emph{inequality}. The most common way of writing the CHSH inequality is as follows. 
			Given $\P_{AB|XY}$, for any pair of inputs $(x,y)$, let
			\begin{equation*}
				\begin{split}
					E_{xy} = &  \P_{AB|XY}(0,0|x,y) + \P_{AB|XY}(1,1|x,y) \\
					& \qquad - \P_{AB|XY}(0,1|x,y) - \P_{AB|XY}(1,0|x,y) 
				\end{split}
			\end{equation*}
			and denote the CHSH \emph{value} by
			\[
				\beta \left( \P_{AB|XY} \right) = E_{00} + E_{01} + E_{10} - E_{11} \;.
			\]
			The CHSH inequality reads
			\[
				\forall \P_{AB|XY}\in \mathcal{C}, \qquad  \beta \left( \P_{AB|XY} \right)  \leq 2 \;.
			\]
			The interesting regime is  $\beta\in [2,2\sqrt{2}]$, where $\beta=2$ is the optimal classical violation while $\beta=2\sqrt{2}$ is the quantum one. The relation between the winning probability in the CHSH game and the CHSH value is given by $\omega=1/2+\beta/8$ and we have~$\omega \in \left[ \frac{3}{4}, \frac{2+\sqrt{2}}{4}\right]$.
			
			\sloppy
			When discussing device-independent quantum cryptography in Chapter~\ref{ch:crypto_showcase}, we use a variant of the CHSH game previously used in~\cite{pironio2009device,vazirani2014fully}. In this game Alice has two inputs $\mathcal{X}=\{0,1\}$ while Bob has three possible inputs $\mathcal{Y}=\{0,1,2\}$. 
			The output sets are $\mathcal{A}=\mathcal{B}=\{0,1\}$. 
			The winning condition is the following:\footnote{For the inputs $(x,y)=(1,2)$ one can set either $w_{\text{CHSH}}=1$ or $0$ (it is not relevant later on); for completeness we choose $w_{\text{CHSH}}=1$ in this case, following previous works.}
			\begin{equation}\label{eq:chsh_win_cond_qkd}
				w_{\text{CHSH}} = \begin{cases}
					1 & x,y\in\{0,1\} \text{ and } a\oplus b = x \cdot y \\
					1 & (x,y)=(0,2) \text{ and } a=b \\
					1 & (x,y)=(1,2) \\
					0 & \text{otherwise.}
				\end{cases}
			\end{equation}
			The optimal quantum strategy for this game is the same as in the standard CHSH game, except that if Bob's input is a $y=2$ he applies the same measurement as Alice's measurement for input~$0$. Since the underlying state is maximally entangled this ensures that their outputs always match when~$(x,y)=(0,2)$. 
			
			Conditioned on Bob's input not being $2$, the game played is the CHSH game and the optimal classical and quantum winning probabilities are as above. 			
			
	\section{Untrusted devices}\label{sec:untrusted_devices}
		
		Formally defining an \emph{untrusted device}, or untrusted box, is essential  when analysing device-independent information processing tasks. The current section is devoted to explaining what is meant by this term and what are the assumptions regarding  an untrusted device.
		To understand the definition of an untrusted device, it is perhaps best to consider a cryptographic scenario in which the device may be manufactured by the malicious party, the adversary, and hence is not to be trusted. 
		The same definition of a device is used also when we do not have an explicit adversary in mind; the device itself is still uncharacterised (in the sense explained below) but we are free to ignore the additional subsystem given to the adversary in what follows. 
		We therefore employ below the terminology used in the cryptographic setting.
				
		 As before, we consider the case of two honest parties.
		A device $D$ is modelled by a bipartite box $\P_{AB|XY}$, shared between the honest parties,  Alice and Bob, who try to accomplish a certain task. We think of the box as being prepared by the adversary Eve and hence we call it untrusted. 
		Since Eve is the one manufacturing the device it allows her, in particular, to keep an extension of Alice and Bob's device. Formally, we consider a non-signalling extension of $\P_{AB|XY}$ to a tripartite box $\P_{ABC|XYZ}$  (recall Definition~\ref{def:ns_extension}): we have
		\[
			\P_{AB|XY}(a,b|x,y) = \sum_c \P_{ABC|XYZ}(abc|xyz) \quad \forall a,b,x,y,z  
		\]
		and Eve ``holds'' the marginal $\P_{C|Z}$. 
		Eve can use her component $\P_{C|Z}$ as she wishes. For example, in a cryptographic protocol, Eve can eavesdrop on all the classical communication between the honest parties during the run of the protocol and only later choose to use her box with input $z$ that depends on all other information available to her. 
		
		When considering non-signalling (super-quantum) boxes, the only constraint on the extension $\P_{ABC|XYZ}$ is that it is non-signalling between the three parties and that the marginal of Alice and Bob is equal to the box $\P_{AB|XY}$. 
		In the quantum case, $\P_{AB|XY}$ describes both the state shared between Alice and Bob $\rho_{Q_AQ_B}$ and measurements devices used to measure $\rho_{Q_AQ_B}$. Eve  then holds a purification\footnote{A purification $\rho_{Q_AQ_BE}$ is the most general extension of a quantum state $\rho_{Q_AQ_B}$, in the sense that it gives Eve the maximal amount of information regarding Alice and Bob's marginal state. Hence, in the cryptographic setting we always say that Eve holds the purifying system $E$, without loss of generality~--- any adversary holding a system $E'$ which is not the purifying system $E$ can only be weaker than that holding $E$.} of Alice and Bob's quantum state in a quantum register in her possession. The tripartite box $\P_{ABC|XYZ}$ describes the pure state $\rho_{Q_AQ_BE}$ together with the measurements of Alice and Bob as well as the measurements that can be used by Eve to measure her marginal~$\rho_E$.\footnote{We emphasise again that Eve is not required to measure her quantum state at any particular point.}

		Although the device is untrusted, we always assume that the following requirements hold. 
		
		\paragraph*{Alice and Bob can interact with the device as expected.} 
		
		In any considered scenario, the type of interaction with the device $D$ is defined explicitly. In particular, every protocol clearly states how the users should interact with the device utilised to run the protocol; for example, a protocol may require the users to play $n$ games with the device (by pressing buttons and recording the outputs) \emph{one after the other}. 
		The different types of interactions and the resulting conditions on the untrusted device are discussed in Chapter~\ref{ch:multi_box}.
		Note that this requirement can be verified -- if the honest parties try to use the device in the specified way and the device does not react as expected (e.g., it does not produce outputs or produces outputs from a different alphabet) then it is clear that something is wrong. In this sense, the requirement that it is possible to interact with the device as expected is not really an assumption regarding the device, but rather a formality that allows us to be explicit when talking about untrusted devices.

		\paragraph*{Communication (signalling) between the components of the device.} 
		The communication between Alice, Bob, and Eve's components of the device is restricted in the following way:
		\begin{enumerate}
			\item Alice and Bob's components of $D$ cannot signal to Eve's component. \label{it:com_eve}
			\item Alice and Bob can decide when to allow communication (if any) between their components.\label{it:com_alice_bob}
			\item Alice and Bob can decide when to receive communication (if any) from Eve's component.\label{it:from_eve} 
		\end{enumerate}
		
		The requirement given in Item~\ref{it:com_eve} is necessary for device-independent \emph{cryptography}; without it the device could directly send to Eve all the raw data it generated.
		
		Item~\ref{it:com_alice_bob} implies that Alice and Bob's component must be (at least) bipartite. This is necessary to assure that the violation of the considered Bell inequality is meaningful. In the quantum case, this requirement is identified with the ``assumption'' that we can write Alice and Bob's quantum state as a bipartite state $\rho_{Q_AQ_B}$ and that the measurements made in Alice's and Bob's components of the device are in tensor product with one another. 
		
		Items~\ref{it:com_alice_bob} and~\ref{it:from_eve} give Alice, Bob, and Eve's components the \emph{possibility} to communicate in certain stages of the protocol (see Section~\ref{sec:qkd_model_of_dev} for an explicit example). This is not a restrictive nor necessary assumption. 
		This possibility to communicate is added since it is advantageous to actual implementations of certain protocols. For instance, allowing the different components of the device to communicate in certain stages of some protocols opens the possibility of distributing resources, such as entanglement,  ``on the fly'' for each round of the protocol, instead of maintaining large quantum memories.

		\paragraph*{Other assumptions.} 
		Apart from the above description of the untrusted device, the following list includes the standard assumptions used in device-independent information processing tasks (in particular, device-independent cryptography):
		\begin{enumerate}
			\item The honest parties have a trusted random number generator (that can be used to choose the inputs for playing the games, for example).
			\item The honest parties have a trusted classical post-processing units to make the necessary (classical) calculations during the considered task.
			\item There is a public, but authenticated, classical channel connecting the honest parties (if the considered task requires that the parties communicate classically with one another).
			\item In cryptographic scenarios~--- the honest parties' physical locations are secure and can be isolated if needed (unwanted information cannot leak outside to Eve or between their devices).
			\item Depending on the considered scenario~--- the actions of the device can be described within the non-signalling or quantum formalism.
		\end{enumerate}
		
		
		In contrast to an \emph{untrusted device}, we sometimes use the terminology \emph{honest device} or \emph{honest implementation}. A device is said to be honest if it implements the considered protocol by using a certain pre-specified strategy. In that case, the actions of the device are known and fixed. See Section~\ref{sec:honest_qkd_imp} for an example.


\chapter{Introduction to the showcases}\label{ch:intro_showcases}

	\section{Introduction to non-signalling parallel repetition}\label{sec:pre_par_rep}

	Non-local games, introduced in Section~\ref{sec:pre_games}, are relevant in many areas of both theoretical physics and theoretical computer science. 
	In the context of parallel repetition, we think of a game $\G$ as follows. 
	A referee asks each of the cooperating parties, also called players, a question chosen according to a given probability distribution. 
	The players then need to supply answers which fulfil a pre-determined requirement according to which the referee accepts or rejects the answers. 
	In order to do so, they can agree on a strategy beforehand, but once the game begins communication between the players is not allowed. If the referee accepts their answers the players win. 
	The goal of the players is, naturally, to maximise their winning probability in the game.  
	
	According to the field of interest, one can analyse any non-local game under different restrictions on the players (in addition to not being allowed to communicate). 
	In classical computer science the players are usually assumed to have only classical resources, or strategies. That is, they can use only local operations and shared randomness. 
	In contrast, one can also consider quantum strategies: before the game starts the players create a multipartite quantum state that can be shared among them. When the game begins each player locally measures their own part of the state and bases the answer on their measurement result.   
	Another, more general, type of strategies are those where the players can use any type of correlations that do not allow them to communicate, i.e., non-signalling strategies. 
	
	One of the most interesting questions regarding non-local games is the question of parallel repetition. Given a game $\G$ with optimal winning probability $1-\alpha$ (using either classical, quantum, or non-signalling strategies), we are interested in analysing the winning probability in the repeated game, denoted by $\G^n$. 
	In $\G^n$ the referee gives the players $n$ independent tuples of questions at once, to which the players should reply. 
	The players win $\G^n$ if they win all of the $n$ games. 
	Another, more general and natural, winning criterion is that the players answer a certain fraction $1-\alpha+\beta$ of the $n$ game instances correctly.
	One can then ask what is the probability that the players succeed in the repeated game, as the number of repetitions $n$ increases and whether, in particular, this probability decreases exponentially fast with $n$, similarly to what happens when playing each of the games independently. 
	While the question of parallel repetition is easy to phrase, its answer is far from trivial (and, in fact, up to date there is no general answer that holds for all games). 
	
	The device-independent framework fits perfectly to the study of the parallel repetition question: 
	We can think of a box, i.e., a conditional probability distribution, as describing a strategy of the players. The requirement that the players are not allowed to communicate easily translates to non-signalling conditions between the parties holding the box. 
	Furthermore, the claims that we wish to make regarding the probability of winning the repeated game are oblivious of the exact description of the strategy and hence treating the strategy as a \emph{black box} makes sense. In particular, studying the behaviour of the strategy without having an explicit description of it is necessary in order to be able to use parallel repetition results to, e.g., analyse experiments that aim at ruling-out local realism while performing several Bell violation experiments in parallel or for hardness amplification~\cite{haastad2001some} in complexity theory and cryptography.

	We define and explain the question of parallel repetition below. Our showcase, presented in Chapter~\ref{ch:par_rep_showcase}, focuses on the case of non-signalling parallel repetition. 
	Note, however, that all statements made in the following two sections are general and applicable to any type of strategies. (One only needs to interchange the words non-signalling and quantum or classical).

	\subsection{Parallel repeated games}\label{sec:par_rep_def}
	
		For simplicity and as in the rest of the thesis, we consider two player non-local games. All of the statements below can be extended to an arbitrary number of players.
		
		We define a two-player game, with the players named Alice and Bob, similarly to a non-local game.\footnote{Multi-player games and non-local games are one and the same; we define a two-player game here just to set the terminology used when discussing the showcase of parallel repetition.}
		\begin{defn} [Two-player game]\label{def:pr_game}
			A two-player game $\G=(\mathcal{X},\mathcal{Y},\mathcal{A},\mathcal{B},\Q_{XY},R)$ is defined by:
			\begin{enumerate}
				\item A set of possible questions for each player: $\mathcal{X}$ for Alice and $\mathcal{Y}$ for Bob.
				\item A probability distribution $\Q_{XY}$ over the questions, according to which the referee choses the questions.
				\item A set of possible answers for each player: $\mathcal{A}$ for Alice and $\mathcal{B}$ for Bob.
				\item A winning condition $R:\mathcal{A}\times\mathcal{B}\times\mathcal{X}\times\mathcal{Y}\rightarrow\{0,1\}$.
			\end{enumerate}
		\end{defn}
		
		In the repeated game, denoted by $\G^n$, the referee asks Alice and Bob $n$ questions, all at once. The questions are chosen independently for each game and the answers are checked independently. 
		In most works dealing with parallel repetition, the winning condition of the repeated game is defined such that Alice and Bob win $\G^n$ if and only if they win \emph{all} $n$ repetitions of $\G$ (hence the name).
		We will use a more general winning condition in which only a certain fraction of the games needs to be won. We call such games \emph{threshold games}. 
		\begin{defn} [Threshold game]\label{def:threshold_game}
			Any two-player game $\G=(\mathcal{X},\mathcal{Y},\mathcal{A},\mathcal{B},\Q_{XY},R)$ induces a two-player threshold game $\G_{1-\gamma}^n=(\mathcal{X}^n,\mathcal{Y}^n,\mathcal{A}^n,\mathcal{B}^n,\Q_{XY}^{\otimes n},R_{1-\gamma}^n)$, for $0\leq\gamma\leq 1$, where the winning criterion $R_{1-\gamma}^n$ is defined by:
			\begin{equation*}
				R_{1-\gamma}^n(\mr{a},\mr{b},\mr{x},\mr{y}) = 1 \Leftrightarrow \frac{1}{n}\sum_{i\in[n]} R(a_i,b_i,x_i,y_i) \geq 1-\gamma \;.
			\end{equation*} 
		\end{defn}

		A strategy for a game is simply a box, i.e., a conditional probability distribution, defining the input-output behaviour of the players. 
		Throughout the thesis, a strategy for a single game $\G$ is denoted by $\O_{AB|XY}$. The winning probability of a strategy $\O_{AB|XY}$ in game $\G$ is given by 
		\begin{equation}\label{eq:game_win_prob}
			w \left( \O_{AB|XY} \right) =\sum_{a,b,x,y} \Q_{XY}(x,y)  \O_{AB|XY}(a,b|x,y) R(a,b,x,y) 
		\end{equation}
		When we say that the optimal non-signalling winning probability in a game $\G$ is~$1-\alpha$ we mean that
		\[
			\max_{\O_{AB|XY}} w \left( \O_{AB|XY} \right) = 1 - \alpha \;,
		\] 
		where the maximisation is over all non-signalling strategies $\O_{AB|XY}$. 		
		
		 A strategy for the threshold game $\G_{1-\gamma}^n$ is denoted by $\P_{\mr{A}\mr{B}|\mr{X}\mr{Y}}$.
		$\P_{\mr{A}\mr{B}|\mr{X}\mr{Y}}$'s winning probability in the threshold game is given by
		\begin{equation*}
			w \left( \P_{\mr{A}\mr{B}|\mr{X}\mr{Y}} \right) = \sum_{\mr{a},\mr{b},\mr{x},\mr{y}} \Q_{XY}^{\otimes n}(\mr{x},\mr{y})  \P_{\mr{A}\mr{B}|\mr{X}\mr{Y}}(\mr{a},\mr{b}|\mr{x},\mr{y}) R_{1-\gamma}^n(\mr{a},\mr{b},\mr{x},\mr{y}) \;.
		\end{equation*}

	\subsection{Threshold theorems}\label{sec:par_rep_intro_thresh_thm}
	
		The parallel repetition question is the following. 
		Let $\G$ be a non-local game whose optimal non-signalling winning probability is $1-\alpha$ and let $\G_{1-\gamma}^n$ be the threshold game defined by $\G$ (Definition~\ref{def:threshold_game}). 
		We then ask~--- \emph{what is the probability that a non-signalling strategy $\P_{\mr{A}\mr{B}|\mr{X}\mr{Y}}$ wins $\G_{1-\gamma}^n$?}
		The more ``standard'' parallel repetition question is retrieved by setting $\gamma =0$. 
		The interesting scenario to consider is the one in which $1-\gamma = 1 - \alpha + \beta$ for~$\beta>0$. 
		
		The players can always use the trivial independent and identically distributed~(IID) strategy for $\G_{1-\alpha + \beta}^n$: they simply answer each of the $n$ questions independently according to the optimal non-signalling strategy for $\G$. 
		In this case, the fraction of successful answers is highly concentrated around $1-\alpha$ and the probability to win all games simultaneously is $\left(1-\alpha\right)^n$.
		  Thus, for any $\beta>0$, the winning probability in $\G_{1-\alpha + \beta}^n$ decreases exponentially fast with $n$. 
		
		Can the players do better when using a correlated, i.e., a \emph{non-IID}, strategy?
		There are many examples showing that the answer is yes for certain games. 
		One of the most interesting examples to a person studying non-locality is that of two repetitions of the CHSH game. As mentioned in Section~\ref{sec:chsh_game}, the optimal \emph{classical} strategy in the CHSH game achieves a winning probability of $3/4$. If the players play two CHSH games in parallel and use the optimal classical strategy of a single game twice, the probability that they win both games is $9/16$. 
		However, there exists a better classical strategy~\cite{barrett2002quantum}: 
		\begin{equation*}
		\begin{split}
			\text{Alice's actions:} \quad 
				&\begin{cases}
						(a_1,a_2) = (1,1) & (x_1,x_2)\in\{ (0,0), (1,0), (1,1) \} \\
						(a_1,a_2) = (1,0) & (x_1,x_2)=(0,1) \;,
				\end{cases} \\
			\text{Bob's actions:} \quad 
				&\begin{cases}
						(b_1,b_2) = (1,1) & (y_1,y_2)\in\{ (0,0), (0,1), (1,1) \} \\
						(b_1,b_2) = (0,1) & (y_1,y_2)=(1,0) \;.
				\end{cases}
		\end{split}
		\end{equation*}
		One can easily check that this strategy wins both games with probability $10/16$ and, hence, is better than playing the two games independently. 
		More examples are known for classical games, e.g.,~\cite{feige1991success, raz2011counterexample}, as well as for quantum and non-signalling games~\cite{kempe2010no}. 
		
		Still, one may ask whether the players can achieve a \emph{significantly} higher winning probability compared to the IID strategy as $n$ increases. 
		In the IID case the probability of winning more than a fraction $1-\alpha + \beta$ of the games decreases exponentially fast with $n$ and $\beta^2$; see Section~\ref{sec:par_rep_under_iid} for the simple analysis.
		Does this type of decrease also hold when considering strategies $\P_{\mr{A}\mr{B}|\mr{X}\mr{Y}}$ that may correlate the different rounds? 
		If correlated strategies for $\G_{1-\alpha + \beta}^n$ are not substantially better than independent ones, even in an asymptotic manner, we learn that ``one cannot fight independence with correlations''. As long as the questions are asked, and the answers are verified, in an independent way, creating correlations between the different answers using a correlated strategy cannot help much. 
		
		The first  exponential parallel repetition theorem was derived for classical two-player games and appeared in~\cite{raz1998parallel}: it was shown that if the classical optimal winning probability in a game $\G$ is smaller than 1, then the probability to win all the games in the repeated game, using a classical strategy, decreases exponentially with the number of repetitions $n$. 
		This was improved and adapted to the non-signalling case in~\cite{holenstein2007parallel}. Another improvement was made in~\cite{rao2011parallel}, where a threshold theorem for the classical two-player case was proven: the probability to win more than a fraction~$1-\alpha+\beta$ of the games for any $\beta>0$ is exponentially small in $n$. 

		\sloppy
		Following the same proof technique as~\cite{raz1998parallel,holenstein2007parallel,rao2011parallel},  ~\cite{buhrman2013parallel} gave a threshold theorem for multi-player non-signalling complete-support games. Their threshold theorem was the first result where more than two players were considered. 
		In~\cite{arnon2016non,lancien2016parallel} a completely different proof technique, based on de Finetti reductions, was used to derive similar (improved in some respects) results as~\cite{holenstein2007parallel,buhrman2013parallel}.\footnote{The de Finetti reduction used in these proofs is the topic of Chapter~\ref{ch:reductions_par}; the non-signalling threshold theorem of~\cite{arnon2016non} acts as one of our showcases and is discussed in Chapter~\ref{ch:par_rep_showcase}.}
		~\cite{holmgren2017counterexample} gave a counter example to a general non-signalling parallel repetition~--- they show that for a certain three-player game without complete-support the probability of winning$~n$ instances of the game played in parallel remains \emph{constant}. This implies that the results of~\cite{buhrman2013parallel,arnon2016non,lancien2016parallel} cannot be extended to games  \emph{without complete-support} (as they hold for any number of parties). 

	\sloppy
		The question of parallel repetition in the quantum case is less well understood than its classical and non-signalling versions. The only results applicable to all two-player games is that of~\cite{yuen2016parallel}, which states that the probability of winning all~$n$ instances of the game decreases inverse polynomially with~$n$. Exponential decrease is known for different classes of two-player games~\cite{cleve2008perfect, kempe2010unique,dinur2015parallel,chung2015parallel} or modifications thereof~\cite{kempe2011parallel,bavarian2015anchoring,bavarian2016parallel}.

	\section{Introduction to device-independent quantum cryptography}\label{sec:pre_di_crypt}

		Classical cryptography relies on computational assumptions, such as the hardness of factoring, to deliver a wide range of functionalities. The advent of quantum information brought forward a completely different possibility: security based only on the fundamental laws of physics. For example, the quantum key distribution (QKD) protocols by Bennett and Brassard~\cite{bennett1984proceedings} and  Ekert~\cite{ekert1991quantum} allow mutually trustful users connected only by an authenticated classical channel, and an arbitrary quantum channel, to establish a private key whose security is guaranteed by the laws of quantum mechanics. With their private key, the users can communicate with perfect security using, e.g., a one-time pad.

		The security of cryptographic protocols such as QKD relies on certain assumptions regarding the physical implementation, such as the quantum states and measurements used in the apparatus implementing the protocol. 
		In real life, however, the manufacturer of the device can have limited technological abilities (and hence cannot guarantee that the device's actions are exact and non-faulty) or even be malicious. Furthermore, the quantum device may be far too complex for the honest parties running the protocol to open and assess whether it works as alleged.
		In the cryptographic setting, imperfections in the physical apparatus are of a real concern, even when the manufacture himself is honest and has good intentions. Indeed, when trying to implement quantum devices we find that creating perfect states and measurements is practically impossible. In the presence of an adversary, imperfections and noise in the implementation can and are being exploited to gain information on the outputs of the cryptographic protocols~\cite{fung2007phase,lydersen2010hacking,weier2011quantum,gerhardt2011full}. 
		This means that if one does not trust that the quantum devices are exactly as  supposed to be, due to a potentially incompetent or malicious manufacture, then the security of the protocols no longer holds. 
		
		To solve this issue the quantum cryptography community took one step further. 
		In contrast to ``standard'' quantum cryptographic protocols that are proven to be secure only for specific implementations of the used devices, device-independent quantum cryptographic protocols achieve an unprecedented level of security with guarantees that hold (almost) irrespective of the quality, or trustworthiness, of the physical devices used to implement them and hence count as the ``gold standard'' of quantum cryptography~\cite{ekert2014ultimate}.
		In device-independent cryptography we let the adversary, called Eve, prepare the quantum devices used in the protocol.
		The honest parties, Alice and Bob, must therefore treat the possibly faulty or malicious device as an untrusted device (as defined in Section~\ref{sec:untrusted_devices}) with which they can only interact according to the protocol.
		The protocol must allow them to test the untrusted device and decide whether using it to run the considered cryptographic protocol poses any security risk. The protocol guarantees that by interacting with the device according to the specified steps the honest parties will either abort, if they detect a fault, or produce secure outputs (with high probability). 
		Clearly, the security proof cannot rely on the inner-workings of the device as it may be malicious. Hence, if we are able to prove that the produced outcomes are secure to use, then the statement is inherently independent of the implementation of the physical device (hence the name ``device-independent security'').
 
 		At first sight, it seems impossible to prove that the outputs of a cryptographic protocol are secure to use when the adversary is the one to manufacture the device. As known for quite some time now, the solution is to base device-independent protocols on the violation of Bell inequalities~\cite{ekert1991quantum,mayers1998quantum,barrett2005no}. 	
		As explained in Section~\ref{sec:pre_games}, a Bell inequality can be thought of as a game played by the honest parties using the device they hold. Different devices lead to different winning probabilities when playing the game.
		The game has a special property -- there exists a quantum device which achieves a  winning probability $\omega$ greater than all classical, local, devices.
		Hence, if the honest parties observe that their device wins the game with probability $\omega$ they conclude it must be non-local.\footnote{A recent sequence of breakthrough experiments have verified the quantum advantage in non-local games in a loophole-free way~\cite{hensen2015loophole,shalm2015strong,giustina2015significant}. In the context of device-independent cryptography, the fact that the experiments are ``loophole-free'' means that the experiments were executed without making assumptions that could otherwise be exploited by Eve to compromise the security of a cryptographic protocol.} A non-local game therefore acts as a ``test for quantumness''.
		The idea of basing the security of cryptographic protocols (QKD especially) on the violation of Bell inequalities originates in the celebrated work of Ekert~\cite{ekert1991quantum}. Later, Mayers and Yao~\cite{mayers1998quantum} recognised that devices that maximally violate a certain Bell inequality could be fully characterised, up to local degrees of freedom, and thus need not be trusted a priori.

		Device-independent security relies on the following deep but well-established facts. High winning probability in a non-local game not only implies that the measured system is non-local but, more importantly, that the kind of non-local correlations it exhibits are ``private'' -- the higher the winning probability, the less information any adversary can have about the devices' outcomes. 
		The amount of entropy, or secrecy, generated in a single round of the protocol can therefore be calculated from the winning probability in the game.
		Let us gain some intuition regarding the relation between the winning probability in a non-local game and the knowledge of the adversary about, e.g., Alice's output in the game by considering two extreme cases -- the optimal classical and quantum strategies for the CHSH game.
			
		If the device is classical its local strategy used to win the game can be written as (recall Definition~\ref{def:classical_box}):
		\[
			\P_{AB|XY}(ab|xy) = \int_{\Lambda} \mathrm{d}\lambda \Pr\left[\Lambda = \lambda\right] \P_{A|X\Lambda}(a|x\lambda) \cdot \P_{B|Y\Lambda}(b|y\lambda) \;,
		\]
		where $\lambda$ describes the ``hidden variable'' (or shared randomness). $\Pr\left[\Lambda = \lambda\right]$ as well as $\P_{A|X\Lambda}$ and $\P_{B|Y\Lambda}$ are chosen by the adversary and, in particular,   $\P_{A|X\Lambda}$ and $\P_{B|Y\Lambda}$ may be deterministic. 
		It is easy to see that in such a case Eve may simply keep a copy of $\lambda$ for herself by extending $\P_{AB|XY}$ to include her system $E$ in the following way:
		\[
			\P_{ABE|XY}(abe|xy) = \int_{\Lambda} \mathrm{d}\lambda \Pr\left[\Lambda = \lambda\right] \P_{A|X\Lambda}(a|x\lambda) \cdot \P_{B|Y\Lambda}(b|y\lambda) \cdot \P_{E|\Lambda}(e|\lambda) \;,
		\]
		where $\P_{E|\Lambda}(e|\lambda)=1$ if $e=\lambda$ and $\P_{E|\Lambda}(e|\lambda)=0$ otherwise. 
		Since $\P_{A|X\Lambda}$ is deterministic, $\lambda$ (and $x$, which is considered to be known to the adversary in most cryptographic protocols) reveals all the information about Alice's outcome $a$. Hence, Eve has full information about Alice's outcome.

		However, if the device is implementing the \emph{optimal} quantum strategy then the underlying quantum state and measurements are fully characterised (recall Section~\ref{sec:chsh_game}). 
		In particular, the state shared between Alice and Bob must be the maximally entangled  state. As such, any quantum state held by Eve must be completely uncorrelated with Alice and Bob's state, i.e., it is of the form 
		\[
			\rho_{Q_AQ_BE} = \ket{\Phi^+}\bra{\Phi^+}_{Q_AQ_B}\otimes\rho_E \;,
		\]
		and hence is uncorrelated with Alice's measurement outcome. Furthermore, measuring the maximally entangled state using the optimal measurements  employed by the device results in a uniformly distributed bit on Alice's side. In total we get that Alice's output is completely random from Eve's perspective. 
		In Section~\ref{sec:randomness_single_round} we will see a quantitive relation between the knowledge of any adversary and the winning probability in the CHSH game which goes beyond the above two extreme cases.
		
		We use the task of device-independent QKD (DIQKD) as one of the showcases considered in this thesis.
		In DIQKD the goal of the honest parties, called Alice and Bob, is to create a shared key, unknown to everybody else but them. 
		To execute the protocol they hold a device consisting of two parts: each part belongs to one of the parties and is kept in their laboratories. Ideally, the device performs measurements on some entangled quantum states it contains. 			
		The basic structure of a DIQKD protocol was presented as Protocol~\ref{pro:intro_qkd}. 
		The protocol consists of playing $n$ non-local games with the given untrusted device and calculating the average winning probability from the observed data (i.e., Alice and Bob's inputs and outputs). If the average winning probability is below the expected winning probability $\omega_{\mathrm{exp}}$ defined by the protocol, Alice and Bob conclude that something is wrong and \emph{abort} the protocol. 
		Otherwise, they apply  classical post-processing steps that allow them to create identical and uniformly distributed keys. (The full description of the DIQKD protocol considered in the analysis performed in the following chapters is given in Section~\ref{sec:diqkd-protocol} below).
		
%
%
%
%
%
%

		\sloppy
		\cite{barrett2005no} were the first to derive a ``proof of concept''\footnote{The protocol of~\cite{barrett2005no} could not tolerate any amount of noise and produced just one secret bit when using the device many times (i.e. the key rate is zero); we therefore consider it to be a ``proof of concept'' showing that device-independent security is possible to achieve.} of the security of DIQKD. 
		Following that an extended line of research has explored the application of the device-independence paradigm to multiple cryptographic tasks. A partial list includes QKD~\cite{barrett2005no,pironio2009device,vazirani2014fully,reichardt2013classical,miller2014robust}, randomness expansion~\cite{Colbeck09,pironio2010random,vazirani2012certifiable,coudron2013infinite,miller2014robust} and amplification~\cite{colbeck2012free,gallego2013full,chung2014physical,brandao2016realistic,kessler2017device}, verified quantum computation~\cite{reichardt2013classical,gheorghiu2015robustness,hajduvsek2015device,coladangelo2017verifier}, bit commitment~\cite{aharon2015device} and weak string erasure~\cite{kaniewski2016device}.

		The following sections present the preliminary knowledge needed when considering our showcase of device-independent quantum cryptography in the upcoming chapters. Specifically, in Section~\ref{sec:security_defs} we explain what is meant when talking about the \emph{security} of DIQKD and present the formal security definitions. Sections~\ref{sec:diqkd-protocol} and~\ref{sec:qkd_main_task} describe our DIQKD protocol and explain what is the main challenge in any security proof. 
		Section~\ref{sec:honest_qkd_imp} includes a possible implementation of the protocol in the honest (i.e., non-adversarial) case while Section~\ref{sec:qkd_model_of_dev} describes the assumptions made regarding a potentially malicious device.
		The security analysis itself is presented as a showcase in later chapters. In particular, the full security proof, which previously appeared in~\cite{arnon2016simple}, is given in Chapter~\ref{ch:crypto_showcase}.

		
		\subsection{DIQKD security definitions}\label{sec:security_defs}
		
			 A DIQKD protocol consists of an interaction between two trusted parties, Alice and Bob, and an untrusted device as defined in Section~\ref{sec:untrusted_devices}. 
			At the end of the protocol each party outputs a key of length $\ell$, $\tilde{K}_A$ for Alice and $\tilde{K}_B$ for Bob.  
			The goal of the adversary, Eve, is to gain as much information as possible about Alice and Bob's keys without being detected (i.e., in the case where the protocol is not being aborted). 

			
			Correctness, secrecy, and overall security of a DIQKD protocol are defined as follows (see also~\cite{portmann2014cryptographic,beaudry2015assumptions}):
			
			\begin{defn} [Correctness]\label{def:qkd_correctness}
				A DIQKD protocol is said to be $\varepsilon_{corr}$-correct, when implemented using a device $D$, if Alice and Bob's keys,  $\tilde{K}_A$ and $\tilde{K}_B$ respectively, are identical with probability at least $1-\varepsilon_{corr}$. That is, $\Pr ( \tilde{K}_A\neq \tilde{K}_B ) \leq \varepsilon_{corr}$.\footnote{We use the convention that when the protocol aborts, $\tilde{K}_A= \tilde{K}_B=\perp$.}
			\end{defn}
		
			\begin{defn}[Secrecy]\label{defn:qkd_secrecy}
				A DIQKD protocol is said to be $\varepsilon_{sec}$-secret, when implemented using a device~$D$, if for a key of length $\ell$,\footnote{$\ell$ can be thought of as a parameter of the protocol. In what follows, we set $\ell$ in terms of the other parameters of the protocol, such that secrecy holds for the protocol.} 
				\[
					\left(1-\Pr[\text{abort}]\right) \| \rho_{\tilde{K}_A E} - \rho_{U_{\ell}} \otimes \rho_{E} \|_1 \leq \varepsilon_{sec}\;,
				\]
				 where $\Pr[\text{abort}]$ is the probability that the protocol aborts when running using device $D$ and $\rho_{\tilde{K}_A E}$ is Alice and Eve's quantum state in the end of the protocol, conditioned on not aborting, with $E$ a quantum register holding Eve's state that may initially be correlated with $D$.
			\end{defn}
			$\varepsilon_{sec}$ in the above definition can be understood as the probability that some non-trivial information leaks to the adversary~\cite{portmann2014cryptographic}. 
			
			If a protocol is $\varepsilon_{corr}$-correct and $\varepsilon_{sec}$-secret (for a given $D$), then it is $\varepsilon_{\mathrm{QKD}}^s$-correct-and-secret for any $\varepsilon_{\mathrm{QKD}}^s\geq \varepsilon_{corr}+\varepsilon_{sec}$.
			
			\begin{defn}[Security]\label{def:security_QKD}
				A DIQKD protocol is said to be $(\varepsilon_{\mathrm{QKD}}^s,\varepsilon_{\mathrm{QKD}}^c)$-secure if:
				\begin{enumerate}
					\item (Completeness) There exists an honest implementation of the device $D$ such that the protocol does not abort with probability greater than $1-\varepsilon_{\mathrm{QKD}}^c$. 
					\item (Soundness) For \emph{any} implementation of the device $D$, the protocol is $\varepsilon_{\mathrm{QKD}}^s$-correct-and-secret.
				\end{enumerate}
			\end{defn}
			
			The protocols that we consider below take into account possible noise in the honest implementation. That is, even when there is no adversary at all, the actual implementation of the devices might not be perfect. 
			This should be taken into account when proving the \emph{completeness} of the protocol -- completeness must be proven for noisy but honest devices (as otherwise the protocol is of no real use). By doing so we get that the completeness of the protocol implies its \emph{robustness} to the desired amount of noise.
			
			\sloppy
			Lastly, a remark regarding the \emph{composability} of this security definition is in order.  A security definition is said to be composable~\cite{canetti2001universally,ben2004general,portmann2014cryptographic} if it implies that the protocol can be used arbitrarily and composed with other protocols (proven secure by themselves), without compromising security. 
			Obviously, if Alice and Bob wish to use the keys they produced in a DIQKD protocol in some other cryptographic protocol (i.e., they compose the two protocols), it is necessary for them to use protocols which were proven to have composable security. 
			
			For the case of (device-\emph{dependent}) QKD, Definition~\ref{def:security_QKD} was rigorously proven to be composable~\cite{portmann2014cryptographic}. This suggests that the same security definition should also be the relevant one in the device-independent context and, indeed, as far as we are aware, it is the sole definition used in works on DI cryptography. 
			Nevertheless, the claim that Definition~\ref{def:security_QKD} is composable for device-independent protocols as well has never been rigorously proven. 
			Even worse, there is some evidence indicating that the definition is \emph{not} composable when the same devices are being reused in the composition. Let us briefly explain that. 
			
			\cite{barrett2013memory} highlighted a simple fact: A malicious device may store the raw data used to create the key in a first execution of the DIQKD protocol and then, when reusing the device to execute the protocol \emph{for the second time} (or any other protocol for that matter), leak the raw data from the first run.\footnote{This should not be confused with ``reusing'' the device in a given execution of the protocol, i.e., playing many non-local games with the same physical device.}
			Our security definition, Definition~\ref{def:security_QKD}, deals only with a single execution of the protocol and, hence, does not address this type of attack. 
			In other words, even when proving that the considered protocol is secure according to Definition~\ref{def:security_QKD}, the above attack can still be performed by a malicious device when composing two protocols that utilise the same device. This implies that, as is, the security definition is not composable. 
			Note that the same issue does \emph{not} arise when considering device-\emph{dependent} protocols; there, by assumption, the devices do not keep any information in their memory after the end of the execution of the protocol. 
			
			Even given the above, Definition~\ref{def:security_QKD} seems like the most promising security definition to date. We therefore stick to it here.
		 This implies that, as in all other works, \emph{after the end of the protocol} the device cannot be used again in an arbitrary way.

		
		\subsection{DIQKD protocol}\label{sec:diqkd-protocol}

			\begin{algorithm}[t]
				\caption{CHSH-based DIQKD protocol}
				\label{pro:diqkd_chsh_intro}
				\begin{algorithmic}[1]
					\STATEx \textbf{Arguments:} 
						\STATEx\hspace{\algorithmicindent} $D$ -- untrusted device of two components that can play CHSH repeatedly
						\STATEx\hspace{\algorithmicindent} $n \in \mathbb{N}_+$ -- number of rounds
						\STATEx\hspace{\algorithmicindent} $\gamma \in (0,1]$ -- expected fraction of test rounds 
				
						\STATEx\hspace{\algorithmicindent} $\omega_{\mathrm{exp}}$ -- expected winning probability in an honest implementation    
						\STATEx\hspace{\algorithmicindent} $\delta_{\mathrm{est}} \in (0,1)$ -- width of the confidence interval for parameter estimation
						
						\STATEx\hspace{\algorithmicindent} $\mathrm{EC}$ -- error correction protocol 
						 \STATEx\hspace{\algorithmicindent} $\mathrm{PA}$ -- privacy amplification protocol 
						
					\STATEx
					
					\STATE For every round $i\in[n]$ do Steps~\ref{prostep:choosing_est_test}-\ref{prostep:use_device_qkd}:
						\STATE\hspace{\algorithmicindent} Alice and Bob choose a random $T_i\in\{0,1\}$ such that $\Pr(T_i=1)=\gamma$.  \label{prostep:choosing_est_test}
						\STATE\hspace{\algorithmicindent} If $T_i=0$, Alice and Bob choose $(X_i,Y_i)=(0,2)$ and otherwise $X_i,Y_i\in \{0,1\}$ uniformly at random.  
						\STATE\hspace{\algorithmicindent} Alice and Bob use $D$ with $X_i,Y_i$ and record their outputs as $A_i$ and $\tilde{B}_i$ respectively.\label{prostep:use_device_qkd}
					
					\STATEx
					
					\STATE \textbf{Error correction:} Alice and Bob apply the error correction protocol $\mathrm{EC}$. If $\mathrm{EC}$ aborts they abort the protocol. Otherwise, they obtain raw keys denoted by $K_A$ and $K_B$. \label{prostep:ec}
					\STATE \textbf{Parameter estimation:} Using $\tilde{\mr{B}}$ and $K_B$, Bob sets $W_i = w_{\text{CHSH}}\left({K_B}_i,\tilde{B}_i,X_i,Y_i\right)$ for the test rounds and $W_i = \perp$ otherwise. He aborts if $\sum_{j:T_j=1} W_j < \left(\omega_{\mathrm{exp}}\gamma - \delta_{\mathrm{est}}\right) \cdot n;$. \label{prostep:abort_chsh_qkd}
					\STATE \textbf{Privacy amplification:} Alice and Bob apply the privacy amplification protocol $\mathrm{PA}$ on $K_A$ and $K_B$ to create their final keys $\tilde{K}_A$ and $\tilde{K}_B$ of length $\ell$. \label{prostep:pa}	
				\end{algorithmic}
				\end{algorithm}

				Our protocol for DIQKD is described as Protocol~\ref{pro:diqkd_chsh_intro}. 
				An honest implementation of a device that can be used to run the protocol is described in Section~\ref{sec:honest_qkd_imp}. 
				
				In the first part of the protocol Alice and Bob use their devices to produce the raw data by playing $n$ CHSH games one after the other.
				Specifically, in each round Alice and Bob randomly choose whether the round is going to be a test round or a generation round ($T_i=1$ or $T_i=0$, respectively, in Protocol~\ref{pro:diqkd_chsh_intro}). This can be done using classical communication or shared public randomness. In both cases, this information becomes available to Eve during the execution of the protocol. (Crucially, she does not know in advance, i.e., before supplying the devices to Alice and Bob, which rounds are going to be test rounds). 
				The inputs used by Alice and Bob in each round depend on whether it is a test or generation round; see Protocol~\ref{pro:diqkd_chsh_intro}.
				
				In the second part of the protocol Alice and Bob apply classical post-processing steps to produce their final keys. We choose classical post-processing steps that optimise the key rate but may not be optimal in other aspects, e.g., computation time. The protocol and the analysis presented in Chapter~\ref{ch:crypto_showcase} can easily be adapted for other choices of classical post-processing. 
				
				We now describe the three post-processing steps, error correction, parameter estimation, and privacy amplification in detail.\footnote{In many QKD protocols there is an additional step called ``sifting''; in the sifting step Alice and Bob announce their choice of measurements in the different rounds so that they can ignore the rounds that do not contribute to parameter estimation or the generation of the key (for example, in protocols like BB84~\cite{bennett1984proceedings} Alice and Bob ignore the rounds in which they chose non-identical measurements). 
				Sifting is not necessary in our case since in Step~\ref{prostep:choosing_est_test} of Protocol~\ref{pro:diqkd_chsh_intro} Alice and Bob choose $T_i$ together (or exchange its value between them) in every round of the protocol and choose their inputs accordingly. This is in contrast to choosing Alice and Bob's inputs from a product distribution and then adding a sifting step. It follows from our proof technique that making $T_i$ public as we do does not compromise the security of the protocol.}

				\subsubsection{Error correction} 
					
					An essential property of any QKD protocol is its correctness -- Alice and Bob should hold identical keys in the end of the protocol (see Definition~\ref{def:qkd_correctness}). Since the raw data of the two parties may differ in parts, Alice and Bob need to run an error correction protocol (also termed an ``information reconciliation protocol'' in the literature).  
					An error correction protocol\footnote{Note that we are discussing \emph{classical} error correction protocols, not to be confused with the task of quantum error correction~\cite{gottesman2010introduction}.} starts by the exchange of classical information between Alice and Bob that should help the parties agree on the final key. When the communication is only from one party to the other, the protocol is said to be a ``one-way error correction protocol''.  
					By sending classical information about the raw data over a public classical channel the uncertainty of the adversary regarding the key decreases. A good error correction protocol therefore needs to minimise the amount of communication, or leakage, while still allowing to correct the errors with high probability.

					In the considered DIQKD protocol, Alice and Bob use an error correction protocol $\mathrm{EC}$ to obtain identical raw keys $K_A$ and $K_B$ from their raw data $\mr{A},\tilde{\mr{B}}$.\footnote{It will become clear in Section~\ref{sec:diqkd_proof} why we use here $\tilde{B}_i$ rather than $B_i$. Although it is not relevant at the moment, we keep it like this for the sake of consistency.}  We use a one-way error correction protocol, based on universal hashing, which minimises the amount of leakage to the adversary~\cite{brassard1993secret,renner2005simple}  (see also~\cite[Section~3.3.2]{beaudry2015assumptions} for more details). 
					To implement $\mathrm{EC}$ Alice chooses an hash function and sends the chosen function and the hashed value of her bits to Bob. We denote this classical communication by $O$ and the number of bits of $O$ by $\mathrm{leak_{EC}}$. Bob uses $O$, together with all his prior knowledge $\tilde{\mr{B}}\mr{X}\mr{Y}\mr{T}$, to compute a guess $\hat{\mr{A}}$ for Alice's bits $\mr{A}$.\footnote{The idea is basically the following -- given the output of the hash function, there is a small set of possible strings (from the domain of the function) compatible with it; Bob then chooses the one which is most compatible to his prior knowledge about Alice's key~\cite[Section~4]{brassard1993secret}.}
					If $\mathrm{EC}$ fails to produce a good guess the protocol aborts; in an \emph{honest} implementation this happens with probability at most $\varepsilon_{\mathrm{EC}}^c$. 
					The probability of Alice and Bob not aborting and while holding non-identical keys is at most $\varepsilon_{\mathrm{EC}}$.
					
					\sloppy The following guarantee holds for the described protocol~\cite{renner2005simple,renner2008security}:
					\begin{equation}\label{eq:ec_leakage}
						\mathrm{leak_{EC}} \leq H_{0}^{\varepsilon'_{\mathrm{EC}}}\left(\mr{A}|\tilde{\mr{B}}\mr{X}\mr{Y}\mr{T}\right)_{\rho^{\text{honest}}} + \log\left(\frac{1}{\varepsilon_{\mathrm{EC}}}\right)  \;,
					\end{equation}
					for any $\varepsilon_{\mathrm{EC}}^c, \; \varepsilon'_{\mathrm{EC}}, \; \varepsilon_{\mathrm{EC}} \in [0,1]$ such that $ \varepsilon'_{\mathrm{EC}} = \varepsilon_{\mathrm{EC}}^c - \varepsilon_{\mathrm{EC}}$ and where $H_{0}^{\varepsilon'_{\mathrm{EC}}}(\mr{A}|\tilde{\mr{B}}\mr{X}\mr{Y}\mr{T})_{\rho^{\text{honest}}}$ is the smooth zero-entropy (Definition~\ref{defn:zero_entropy}) evaluated on the state $\rho^{\text{honest}}$ used in an \emph{honest} implementation of the protocol.\footnote{For quantum channels with an IID noise model $H_{0}^{\varepsilon'_{\mathrm{EC}}}\left(\mr{A}|\tilde{\mr{B}}\mr{X}\mr{Y}\mr{T}\right)_{\rho^{\text{honest}}}$ can be bounded by above using the asymptotic equipartition property, discussed in Section~\ref{sec:quant_aep}. The explicit calculation is done in Section~\ref{sec:key_rates}.}
					Equation~\eqref{eq:ec_leakage} presents the tradeoff between the probability of having non-identical keys after the end of the protocol ($\varepsilon_{\mathrm{EC}}$), the probability of the protocol not succeeding in the honest case ($\varepsilon_{\mathrm{EC}}^c$), and the number of bits leaked to the adversary in the process ($\mathrm{leak_{EC}}$). 
					The amount of communication during the error correction protocol is chosen, before running the DIQKD protocol, such that  Equation~\eqref{eq:ec_leakage} holds. 
					If more errors than expected in the honest implementation occur when running the DIQKD protocol (due to the use of adversarial or too noisy devices), then Bob may not have a sufficient amount of information to obtain a good guess of Alice's bits and hence will not be able to correct the errors. If so, this will be detected with probability at least $1 - \varepsilon_{\mathrm{EC}}$ and the protocol will abort.

				\subsubsection{Parameter estimation}
				
					The goal of the parameter estimation step is to check whether the device $D$, used to run the protocol, is sufficiently good in order to produce a \emph{secret} key. In the case of device-independent protocols the quantity to be considered is the number of games won during the run of the protocol. If the number of games won is not large enough, the honest parties conclude that the device cannot be used to produce a secure key (an adversary may be present). 
					Specifically, we require that the number of games won, $\sum_{j:T_j=1} W_j$, fulfils 
					\begin{equation}\label{eq:diqkd_not_abort_cond}
						\sum_{j:T_j=1} W_j \geq \left(\omega_{\mathrm{exp}}\gamma - \delta_{\mathrm{est}}\right) \cdot n \;,
					\end{equation}
					where $\omega_{\mathrm{exp}}$, $\gamma$, and $\delta_{\mathrm{est}}$ are parameters of the protocol. 
					$\gamma$ is the probability of a test round while $\omega_{\mathrm{exp}}$ is the expected winning probability (of an honest device). Thus, the multiplication $\omega_{\mathrm{exp}}\gamma$ gives the expected fraction of games won out of \emph{all} rounds of the protocol.  $\delta_{\mathrm{est}}$ describes the desired confidence interval (which cannot be zero since we consider a finite number of rounds $n$).
				
					After the error correction step described above, Bob has all of the relevant information to perform parameter estimation from his data alone, without any further communication with Alice.\footnote{In many QKD protocols error correction is performed \emph{after} the parameter estimation step. In such cases, Alice and Bob reveal the data collected in the test rounds and use it for parameter estimation. Further information is then communicated during the error correction step.} 
					Using his raw data $\tilde{\mr{B}}$ and his guess of Alice's key $K_B$, Bob sets 
					\[
						W_i = \begin{cases}
							\perp & T_i =0 \\
							w_{\text{CHSH}}\left(\hat{A}_i,\tilde{B}_i,X_i,Y_i\right)=w_{\text{CHSH}}\left({K_B}_i,\tilde{B}_i,X_i,Y_i\right) & T_i=1 \;, 
						\end{cases}
					\]
					where $w_{\text{CHSH}}$ is the CHSH winning condition given in Equation~\eqref{eq:chsh_win_cond_qkd}.
					Bob aborts if the fraction of successful game rounds is too low, that is, if Equation~\eqref{eq:diqkd_not_abort_cond} is \emph{not} fulfilled. 
					
					As Bob does the estimation using his guess of Alice's bits, the probability of aborting in this step in an honest implementation, $\varepsilon_{\mathrm{PE}}^c$, is bounded by
					\begin{equation}\label{eq:pe_completeness}
						\begin{split}
								\varepsilon_{\mathrm{PE}}^c&\leq \Pr\Big( \sum_{j:T_j=1} W_j < \left(\omega_{\mathrm{exp}} \gamma - \delta_{\mathrm{est}}\right) \cdot n \Big| K_A=K_B \Big) \\
						&\quad + \Pr \big( K_A \neq K_B \text{ and $\mathrm{EC}$ does not abort}\big)  \;.
						\end{split}
					\end{equation}
								
				\subsubsection{Privacy amplification} 
					
					The final classical post-processing step is that of privacy amplification. The goal of privacy amplification is to take Alice's raw key\footnote{Note that Alice's and Bob's raw keys, $\mr{A}$ and $\hat{\mr{A}}$ respectively, are identical with high probability, due to the error correction step. As we now explain, in the privacy amplification step Alice and Bob can perform the exact same actions so that they end with identical final keys (assuming that the error correction step was successful). Thus, we describe here only Alice's actions, while keeping in mind that Bob is going to perform the same steps on his raw key.} $\mr{A}$, on which the adversary may have partial information, and transform it to a secret final key, as required by the secrecy definition of the protocol (Definition~\ref{defn:qkd_secrecy}).
					To this end, Alice applies a quantum-proof randomness extractor, defined as follows.
					\begin{defn}[quantum-proof strong extractor]\label{defn:quant_proof_extr}
						A function $\mathrm{Ext}:\{0,1\}^{n}\times \{0,1\}^{d}\rightarrow\{0,1\}^{\ell}$ that takes as an input a string $\mr{A}\in \{0,1\}^{n}$ together with a seed $S\in\{0,1\}^{d}$ and outputs a string $\tilde{K}_A\in\{0,1\}^{\ell}$ (for $\ell\leq n$) is called a \emph{quantum-proof $(m,\varepsilon_{\mathrm{PA}})$-strong extractor} if for any $\rho_{\mr{A}E}$ with $H_{\min}(\mr{A}|E)_{\rho}\geq m$ and uniformly distributed seed $S$ we have
						\begin{equation}\label{eq:extractor_defn}
							 \| \rho_{\mathrm{Ext}(\mr{A},S)SE} - \rho_{U_{\ell}} \otimes \rho_{SE} \|_1 \leq \varepsilon_{\mathrm{PA}} \;.
						\end{equation}
					\end{defn}
					Several constructions of extractors have been shown to be fulfil the above definition, among them~\cite{renner2005universally, konig2008bounded, fehr2008randomness, de2012trevisan}. Different constructions are used in different scenarios; for example, some constructions minimise the length $d$ of the seed while others maximise the output length $\ell$ or the computation time needed to apply the extractor. 
					
					Before continuing, an important (though somewhat technical) remark is in order. An extractor, as above, is defined with respect to the min-entropy. However, it is the \emph{smooth} min-entropy $H_{\min}^{\varepsilon_{\mathrm{s}}}(\mr{A}|E)_{\rho}$, rather than the min-entropy, that is known to give a tight bound on the maximum amount of uniform randomness that can be extracted from $\mr{A}$ while being independent from $E$~\cite{konig2009operational}. 
					If one is interested in using an extractor when starting with a lower bound on the smooth min-entropy, then some parameters should be adapted. In particular,~$\varepsilon_{\mathrm{PA}}$ appearing in Equation~\eqref{eq:extractor_defn} is the error probability of the extractor when it is applied on a normalised state satisfying the relevant min-entropy condition. For universal hashing~\cite{renner2005universally} for example, when only a bound on the smooth min-entropy is supplied the smoothing parameter $\varepsilon_{\mathrm{s}}$ should be added to the error $\varepsilon_{\mathrm{PA}}$ (as done below). When working with other extractors one should adapt the parameters accordingly; see~\cite[Section 4.3]{arnon2015quantum}.

					For simplicity we use universal hashing~\cite{renner2005universally,tomamichel2010leftover} as our privacy amplification protocol $\mathrm{PA}$.\footnote{Any other quantum-proof strong extractor, e.g., Trevisan's extractor~\cite{de2012trevisan}, can be used for this task and the analysis done in Chapter~\ref{ch:crypto_showcase} can be easily adapted.}
					The secrecy of the final key $\tilde{K}_A=\mathrm{Ext}(\mr{A},S)$ depends only on the privacy amplification protocol used and the value of $H^{\varepsilon_{\mathrm{s}}}_{\min} ( \mr{A} | \mr{X}\mr{Y}\mr{T} O E )$, evaluated on the state at the end of the protocol, conditioned on not aborting. For universal hashing, for any $\varepsilon_{\mathrm{PA}},\varepsilon_{\mathrm{s}}\in(0,1)$ a secure key of maximal length~\cite{tomamichel2010leftover} 
						\begin{equation}\label{eq:universal_hashing_length_intro}
							\ell = H^{\varepsilon_{\mathrm{s}}}_{\min} ( \mr{A} | \mr{X} \mr{Y} \mr{T} O E ) -2\log\frac{1}{\varepsilon_{\mathrm{PA}}} 
						\end{equation}
						is produced with probability at least $1-\varepsilon_{\mathrm{PA}} - \varepsilon_{\mathrm{s}}$.

			\subsection{Main task of a security proof}\label{sec:qkd_main_task}	
			
				After presenting the DIQKD protocol and the relevant security definition we are equipped with the necessary information needed to explain what the main task is when proving security of a DIQKD protocol.  
				First, note that in order to prove security one needs to prove both the correctness (Definition~\ref{def:qkd_correctness}) and  the secrecy (Definition~\ref{defn:qkd_secrecy}) of the protocol. Correctness follows almost directly from the error correction step performed in the protocol. We therefore focus below on the secrecy of the protocol. 
				
				Returning to the secrecy requirement of a DIQKD protocol given in Definition~\ref{defn:qkd_secrecy} and the definition of a quantum-proof extractor as in Definition~\ref{defn:quant_proof_extr}, we see that by applying the extractor we assure that  the output of the extractor $\tilde{K}_A=\mathrm{Ext}(\mr{A},S)$ is~$\varepsilon_{\mathrm{PA}}$-close to an ideal key, i.e., a uniform key of $m$ bits that is completely independent of the overall side-information $SE$\footnote{We include the seed $S$ as part of the side-information and ask that the output of the extractor is close to uniform even conditioned on the seed $S$. Extractors that fulfil this requirement are called ``strong extractors'' (while those that fulfil the weaker condition $ \| \rho_{\mathrm{Ext}(\mr{A},S)E} - \rho_{U_{\ell}} \otimes \rho_{E} \|_1 \leq \varepsilon_{\mathrm{PA}}$ are termed ``weak extractors''). When considering QKD protocols, one needs to use a strong extractor since the seed $S$ is to be communicated between Alice and Bob and hence should be considered as information which leaks to the adversary.} and hence the protocol is secret. 
				
				For the extractor to work, the raw data $\mr{A}$ must exhibit a sufficient amount of min-entropy (by definition). 
				Relations for specific extractor, such as the one given in Equation~\eqref{eq:universal_hashing_length_intro}, determine the length of the key that can be extracted for a given amount of (smooth) min-entropy. 
				Therefore, the main task of any security proof of a protocol applying an extractor boils down to computing a lower bound on the (smooth) min-entropy. 		
				 Indeed, the security proofs presented in Section~\ref{sec:crypt_under_iid} and Chapter~\ref{ch:crypto_showcase} are focused on deriving such bounds.

		\subsection{The honest implementation}\label{sec:honest_qkd_imp}

			The honest implementation of the device $D$ describes the way the device acts when an adversary is not present. In other words, this is the device Alice and Bob expect to share when the manufacture of the device is not malicious and ``everything goes according to the plan''. 
			In the analysis of DIQKD the description of the honest implementation is used in two places. Firstly, the completeness of the protocol (recall Definition~\ref{def:security_QKD}) is proven with respect to the chosen honest implementation.
			Secondly, it is used to set the amount of communication between Alice and Bob during the error correction step, according to the relation presented in Equation~\eqref{eq:ec_leakage}.
			We remark that these are the only two places in the proof where the choice of honest implementation is taken into account and both are used solely for \emph{choosing} the parameters  of the protocol. Critically, the soundness proof does not depend in any way on the choice of honest implementation. 
		
			The chosen honest implementation may also be noisy. In fact, in an experiment, the mathematical description of the honest device, or honest boxes, should be chosen to fit the behaviour of the physical systems as accurately as possible. An accurate description allows us carefully choose the parameters of the DIQKD protocol (e.g.,~$\omega_{\mathrm{exp}}$) such that  the produced key rate is maximised while keeping the probability of the protocol aborting, when utilising the honest device, small. That is, an accurate description allows us to construct a protocol which is useful in practice. 
			
			Most commonly, one chooses the honest implementation to be an IID one.  That is, that device $D$ acts in an IID manner: in \emph{every round~$i\in[n]$} of the protocol $D$ performs the measurements $\mathcal{M}_{x_i}^{a_i} \otimes \mathcal{M}_{y_i}^{b_i}$ on Alice and Bob's state $\sigma_{Q_AQ_B}$. That is, the device is initialised with an IID  bipartite state, $\sigma_{Q_AQ_B}^{\otimes n}$, on which the device makes IID measurements. 
			The state $\sigma_{Q_AQ_B}$ and measurements are such that the winning probability achieved in the CHSH game in a single round is $\omega_{\mathrm{exp}}$.\footnote{Note that in our notation, the noise that affects the winning probability in the CHSH game is already included in $\omega_{\mathrm{exp}}$.}

			As a concrete example, one possible realisation of such an implementation is the following. Alice and Bob share the two-qubit Werner state 
			\[
				\sigma_{Q_AQ_B} = (1-\nu) \ket{\phi^+}\bra{\phi^+} + \nu\frac{\mathbb{I}}{4}
			\]
			 for $\ket{\Phi^+} = 1/\sqrt{2}\left(\ket{00}+\ket{11}\right)$ and $ \nu \in [0,1]$. 
			 The state $\sigma_{Q_AQ_B}$ arises, e.g., from the state $\ket{\Phi^+}$ after going through a depolarisation channel. 
			 We can therefore think of the over all state $\sigma_{Q_AQ_B}^{\otimes n}$ as resulting from the transmission of $\ket{\Phi^+}^{\otimes n}$ using an IID noisy channel. 
			For every $i\in[n]$, Alice's measurements $X_i=0$ and $X_i=1$ correspond to the Pauli operators\footnote{Even though both are denoted by $\sigma$, do not confuse our bipartite state $\sigma_{Q_AQ_B}$ describing the honest state with the Pauli operators $\sigma_x$ and $\sigma_z$ defined in Equation~\eqref{eq:pauli_operators_def}.} $\sigma_z$ and $\sigma_x$ respectively and Bob's measurements $Y_i=0$, $Y_i=1$, and $Y_i=2$ to the Pauli operators $\frac{\sigma_z + \sigma_x}{\sqrt{2}}$, $\frac{\sigma_z - \sigma_x}{\sqrt{2}}$ and $\sigma_z$ respectively. 
			The winning probability in the CHSH game (restricted to $X_i,Y_i\in\{0,1\}$) using these measurements on $\sigma_{Q_AQ_B}$ is 
			\[
				\omega_{\mathrm{exp}}=\frac{2+\sqrt{2}(1-\nu)}{4} 
			\]
			and the quantum bit error rate is given by 
			\[
				Q = \Pr[A_i\neq B_i|(X_i,Y_i)=(0,2)] = \frac{\nu}{2} \;.
			\]

		\subsection{Model of an arbitrary device}\label{sec:qkd_model_of_dev}
			
			As previously mentioned, Alice and Bob's device is considered to be an untrusted device, as defined in Section~\ref{sec:untrusted_devices}. 
				
			On top of the general statements made in Section~\ref{sec:untrusted_devices}, we can further describe the untrusted device in the case of DIQKD as follows.
		 	Alice and Bob  interact with $D$ according to Protocol~\ref{pro:diqkd_chsh_intro}. Alice and Bob's components of $D$ implement the protocol by making sequential measurements on quantum states. In each round of the protocol, we say that the device is implementing some strategy for the CHSH game. 
			 The device may have memory, and thus apply a different strategy each time the game is played, depending on the previous rounds. Therefore, the measurement operators may change in each round, and the state on which the measurements are performed may be the post-measurement state from the previous round, a new state, or any combination of these two. 
		
			To be specific, we consider the following scenario. \emph{In-between} different rounds of the protocol, Alice and Bob's components of the device are allowed to communicate freely. During the execution of a single round, however, no communication is allowed. In particular, when the game is being played, there is no communication between the components once the honest parties' inputs are chosen and until the outputs are supplied by the device. That is, communication is allowed in every round $i$ right after Step~\ref{prostep:use_device_qkd} is done, and until the beginning of round $i+1$, i.e., before $T_{i+1}$ is chosen in Step~\ref{prostep:choosing_est_test}.
			Furthermore, in-between rounds Eve may send information to the device, but not receive any from it.
			In actual implementations this implies that entanglement can be distributed ``on the fly'' for each round of the protocol, instead of maintaining large quantum memories.
			
			Section~\ref{sec:untrusted_devices} includes a list of standard assumptions made when working with device-independent protocols. The following list includes the assumptions that are made when proving the security of DIQKD:
			\begin{enumerate}
				\item Alice and Bob have a trusted random number generator.
				\item Alice and Bob have trusted classical post-processing units.
				\item There is a public, but authenticated, classical channel connecting the honest parties.
				\item Alice's and Bob's physical locations are secure (unwanted information cannot leak outside to Eve.
				\item Quantum physics is correct.
			\end{enumerate}

\chapter{Single-round box}\label{ch:single_round_box}

In the device-independent framework we use ``boxes'' to describe the physical devices, or resources, of interest. 
	A box, formally modelled as a conditional probability distribution (recall Section~\ref{sec:pre_black_box}), is always defined with respect to a \emph{specific} task or protocol. More specifically, note the following:
	\begin{enumerate}
		\item To define a box $\P_{AB|XY}$ we need to fix the sets of the inputs $\mathcal{X},\mathcal{Y}$ and the outputs $\mathcal{A},\mathcal{B}$ of the box. These sets are chosen according to the task in which the box is being used. For example, if a box is used to play a single CHSH game then the sets are all chosen to be $\{0,1\}$. The box's action is undefined when it is used with, e.g., the input $x=2$. \label{it:box_dep_meas_set} 
		
		\item The location of the used devices in space (or space-time) also sets the conditions that the box describing the devices must fulfil. For example, if a protocol demands two devices, separated in space,  that cannot communicate during the execution of the protocol then the defined box should fulfil certain non-signalling conditions.\footnote{Interestingly, if one considers protocols with more than two parties in which the devices can only be used in specific space-time coordinates and merely assumes that the box modelling the devices respects relativistic causality (in the sense that it cannot lead to casual loops) then the conditions defining the box are different than the non-signalling ones~\cite{horodecki2016relativistic}. This acts as another example for how the specific use of the devices effects the mathematical model of the box.}
		
		\item When considering boxes that are used to execute a complex protocol, in which many games are being played with the box (as done in the succeeding chapters), we also need to take into account the type of interaction when defining the box. 
		For example, some protocols require boxes with which we can interact sequentially -- in each round of the protocol we give one input to the box, wait for the output, and only then give the next input. 
		Other protocols involve boxes which  accepts all the inputs and only then produces all the outputs. If we only give one input to such a box we do not expect it to output anything and its action is undefined. Thus, these differences in the behaviour of the boxes depend on the way we intend to use it in the task of interest and effect the mathematical model of the considered boxes.

	\end{enumerate}

	To grasp the dependence of the box on the considered task, as described above, one can contrast it with the standard formalism used to define quantum states and measurements. 
	For example, the definition of a quantum state in terms of a density operator is completely independent of the way we might want to measure it. Consider, for example, a quantum state used to play the CHSH game with the measurements $\sigma_x$ and $\sigma_z$ for one of the parties. Even though we only intend to perform these measurements, the formalism also tells us what will happen if we choose to measure $\sigma_y$ instead. This stands in contrast to Item~\ref{it:box_dep_meas_set} above.\footnote{One can rightfully say that this property of boxes, among several other properties, renders them an ``unphysical description'' of real systems and resources. With this respect, the formalism of the so called ``generalised probabilistic theories''~\cite{barrett2007information,chiribella2010probabilistic} is a more appropriate mathematical setting to discuss physical theories which extend, or abstract, quantum physics. In contrast, boxes are merely a simplified mathematical model sufficient for certain analyses.}
	
	The current chapter as well as Chapter~\ref{ch:multi_box} are devoted to the way one models the different boxes used in device-independent information processing, depending on the considered setting and interaction with the boxes. 
	In Chapter~\ref{ch:multi_box} we will be interested in boxes, or devices, which can be used to implement certain protocols. Before we explain how such boxes can be described let us focus on a simpler object -- the ``single-round box''.

	We think of a single-round box as a small device that can be used to play a \emph{single} round of a Bell game. That is, in the case of the CHSH game, for example, Alice and Bob can input their bits $x,y\in\{0,1\}$ to the box and receive the outcomes $a,b\in\{0,1\}$. After that the box can no longer be used (i.e., Alice and Bob cannot play another game with it). Mathematically, such a box can be described by a non-signalling conditional probability distribution  $\P_{AB|XY}$ as explained in Section~\ref{sec:pre_black_box}. Physically, an example of a single-round box is a single EPR pair together with a set of possible measurements for each party.
	
	A single-round box is \emph{not} a useful resource in the \emph{operational sense}. Since our starting point in the device-independent setting is that we do not know how the device operates, we must interact with it to test it. However, since a single-round box allows us to play just a single game we can hardly conclude anything regarding its inner-working. One can imagine Alice and Bob playing the CHSH game with their box and observing $(a,b,x,y)=(0,0,0,0)$. Then what? It can always be the case that they are sharing a classical device that always outputs $(a,b)=(0,0)$ for the inputs $(x,y)=(0,0)$.  Thus, Alice and Bob cannot learn anything regarding, e.g., the randomness of their outputs, from this single game. As the information collected in a single game is not sufficient to test the box we start, instead, with an \emph{assumption} regarding the box, e.g., that it can be used to win the CHSH game with winning probability~$\omega$. As will be shown below, various fundamental properties can be concluded by starting with such an assumption.

	Although a single-round box is not a valuable resource in practice, it is useful as a simple abstract object that allows us to study the fundamental implications of violating a Bell inequality (while putting aside many technical details that arise when considering the complex devices used in protocols). Furthermore, it is the goal of this thesis to explain how ``single-round box statements'' can be lifted to operational statements regarding more complex scenarios such as the analysis of device-independent protocols.

	\begin{figure}
		\centering
		\includegraphics[width=\textwidth]{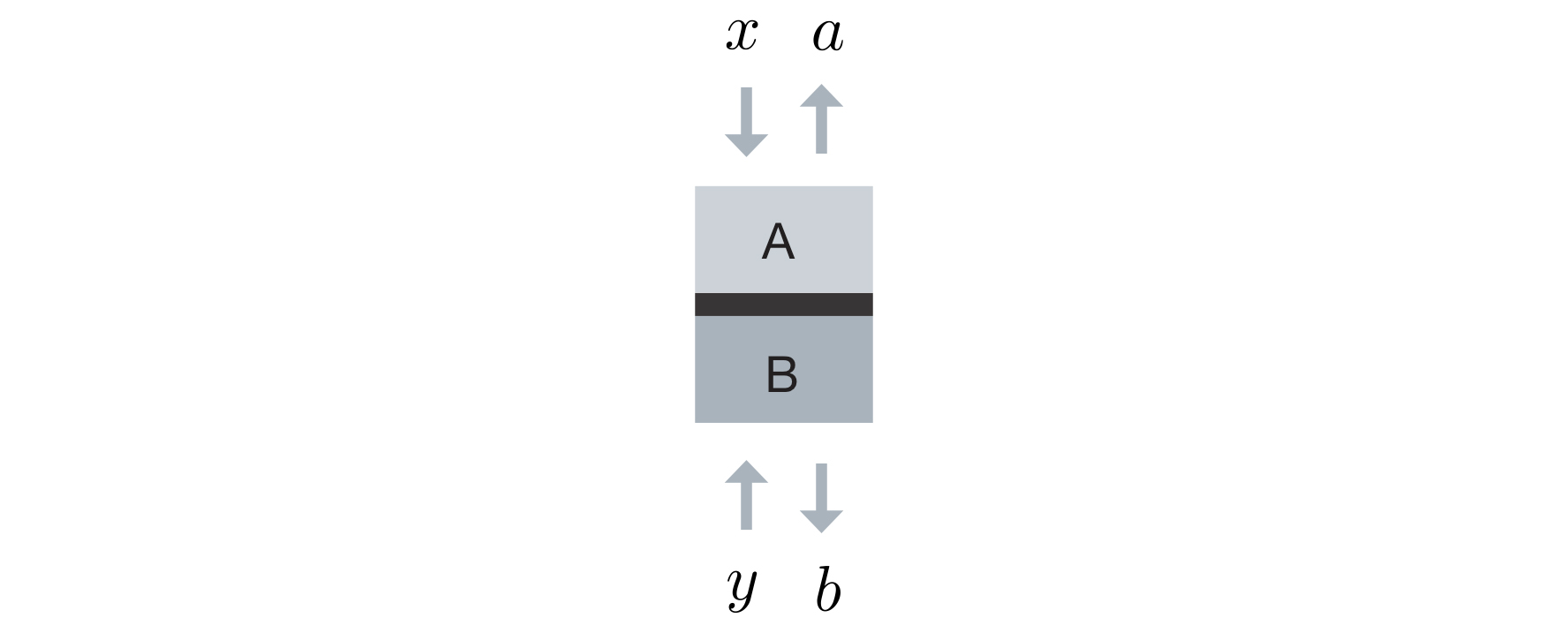}
		\caption{A single-round box. We think of a single-round box as a small device, shared between Alice and Bob, which can be used to play a single round of a Bell game, such as the CHSH game. It is described by a conditional probability distribution $\P_{AB|XY}$.}
		\label{fig:single_round_box}
	\end{figure}
	
	\section{The model}\label{sec:sing_box_model}
		
		Mathematically, we model a single-round black box by a non-signalling conditional probability distribution  $\P_{AB|XY}$ that can be used to play a single Bell game $\G$ defined over the sets of inputs $\mathcal{X},\mathcal{Y}$ and outputs $\mathcal{A},\mathcal{B}$ for Alice and Bob (see Section~\ref{sec:pre_games} for complete definitions). $\P_{AB|XY}$ is also sometimes referred to as a strategy for~$\G$.  
		
		As mentioned above, when considering single-round boxes one usually assumes that the box $\P_{AB|XY}$ can be used to win the game with a certain winning probability~$\omega$. That is, $\P_{AB|XY}$ is such that
		\begin{equation}\label{eq:win_prob_assump}
			\mathbb{E}_{x,y} \sum_{\substack{a,b | \\ w(a,b,x,y)=1}} \P_{AB|XY}(ab|xy) = \omega \;,
		\end{equation}
		where the expectation $\mathbb{E}_{x,y}$ is defined with respect to the input distribution of the considered game and $w:\mathcal{A}\times\mathcal{B}\times \mathcal{X}\times\mathcal{Y}\rightarrow\{0,1\}$ is the winning function of the game.

		Depending on the context, one can consider quantum single-round boxes or non-signalling ones. 
	\subsubsection{Quantum single-round boxes}
		
			When we say that a single-round box is quantum we mean that its inner-working can be described within the quantum formalism. Specifically:
			
			\begin{defn}[Quantum single-round box]\label{def:quant_single_box}
				Given a Bell game $\G$, a quantum single-round box is a quantum box $\P_{AB|XY}$, as in Definition~\ref{def:quant_box}, defined for the inputs and outputs of the game $\G$ -- $\mathcal{X},\mathcal{Y},\mathcal{A},\mathcal{B}$. 
				That is, there exist a bipartite state $\rho_{Q_AQ_B}$ and measurements $\{M^{x}_a\}$ and $\{M^{y}_b\}$ such that
				\begin{equation}\label{eq:quant_single_box_stat}
					\P_{AB|XY}(ab|xy) = \Tr\left(M^{x}_a \otimes M^{y}_b \; \rho_{Q_AQ_B} \right) \quad \forall  a,b,x,y\;.
				\end{equation}
				The quantum single-round box is said to win $\G$ with winning probability $\omega$ when the state and measurements are such that Equation~\eqref{eq:win_prob_assump} holds.
			\end{defn}
			
			Note that mathematically a quantum single-round box is merely a quantum box (Definition~\ref{def:quant_box}). What makes it \emph{single-round} is that $\P_{AB|XY}$ is defined for the inputs and outputs of a single game $\G$.

			When considering cryptographic applications where a quantum adversary is present we extend the box to the adversary. That is, we let $\rho_{Q_AQ_BE}$ be the purification of $\rho_{Q_AQ_B}$ where $E$ is a quantum register belonging the the adversary and $\rho_{Q_AQ_B}=\Tr_E\left(\rho_{Q_AQ_BE}\right) $ is Alice and Bob's marginal satisfying Equations~\eqref{eq:win_prob_assump} and~\eqref{eq:quant_single_box_stat}. 	
		
		\subsubsection{Non-signalling single-round boxes}	
		
			Instead of restricting our attention to quantum boxes we can also consider non-signalling single-round boxes. These are defined in a similar way to their quantum counterparts.
			\begin{defn}[Non-signalling single-round box]\label{def:ns_single_box}
				Given a Bell game $\G$, a non-signalling single-round box is a non-signalling box $\P_{AB|XY}$, as in Definition~\ref{def:ns_box}, defined for the inputs and outputs of the game $\G$ -- $\mathcal{X},\mathcal{Y},\mathcal{A},\mathcal{B}$. That is,  for all $a\in\mathcal{A}$, $ b\in\mathcal{B}$, $ x,x'\in\mathcal{X}$ and $y,y'\in\mathcal{Y}$,
				\begin{align*}
					&\sum_b \P_{AB|XY}(a,b|x,y) = \sum_b \P_{AB|XY}(a,b|x,y') \\
					&\sum_a \P_{AB|XY}(a,b|x,y) = \sum_a \P_{AB|XY}(a,b|x',y) \;.
				\end{align*}
				The non-signalling single-round box is said to win $\G$ with winning probability $\omega$ when $\P_{AB|XY}$ is such that Equation~\eqref{eq:win_prob_assump} holds.
			\end{defn}
	
			Here as well one can consider an extension of the single-round box to an additional party describing a non-signalling (super-quantum) adversary. This will not be needed in this thesis so we do not explain how this is done. The interested reader is referred to~\cite[Section 3.2]{hanggi2010device_thesis}.
		

		\section{Showcase: device-independent quantum cryptography}\label{sec:randomness_single_round}
			
			
			As mentioned above, a single-round box is useful as a simple abstract object that allows us to study the fundamental implications of violating a Bell inequality. More specifically, certain properties of the box can be concluded if we assume to know the probability of winning a Bell game using a single-round box described by~$\P_{AB|XY}$. We consider out showcase of device-independent cryptography as an example. 
			
			The most crucial observation when considering device-independent cryptographic protocols is the fact that high winning probability in a Bell game not only implies that the measured system is non-local, but more importantly that the kind of non-locality it exhibits cannot be shared: the higher the winning probability, the less information any eavesdropper can have about the outcomes produced by the box. 
			
			There are different ways of making such a statement quantitive. One possible way (that will also be of relevance later on) is to consider the conditional von Neumann entropy $H(A|XYE)$ where $A$ is the random variable describing Alice's outcome bit, $X$ and $Y$ are the random variables describing the inputs of Alice and Bob and~$E$ is a quantum register holding the quantum side information belonging to the adversary. If the adversary is completely oblivious to the value of a bit $A$ even given $X$, $Y\;$ and $E$ then takes its maximal value $H(A|XYE)=1$.
			
			A tight trade-off between the winning probability of a single-round box~$\omega$ and the entropy~$H(A|XYE)$ generated by the box was derived in~\cite{pironio2010random,acin2012randomness} and is stated in the following lemma.
			\begin{lem}[\cite{pironio2010random,acin2012randomness}\footnote{Lemma~\ref{lem:single_round_secrecy} is stated in the form appearing in~\cite{arnon2016simple}. To see how the original results of~\cite{pironio2010random} can be used to derive the lemma as we state it, follow the proof given in Appendix~\ref{appsec:crypto_single_proof}.}]\label{lem:single_round_secrecy}
				For any quantum single-round box $\P_{AB|XY}$ with winning probability $\omega\in\left[\frac{3}{4},\frac{2+\sqrt{2}}{4}\right]$ in the \textup{CHSH} game,
				\begin{equation}\label{eq:one_box_entropy}
	 				H(A|XYE) \geq 1 - h\left( \frac{1}{2} + \frac{1}{2}\sqrt{16\omega \left(\omega-1\right) +3}  \right) \;,
				\end{equation}
				where $E$ denotes the quantum side-information belonging to the adversary and $h(\cdot)$ is the binary entropy function.
			\end{lem}

\begin{figure}
				\centering
				\begin{tikzpicture}[scale=0.75]
				\begin{axis}[
					height=7cm,
					width=11cm,
					xlabel=$\omega$,
					xmin=0.75,
					xmax=0.853553,
					ymax=1,
					ymin=0,
				     xtick={0.76,0.78,0.80,0.82,0.84},
			          ytick={0,0.2,0.4,0.6,0.8,1},
					legend style={at={(0.20,0.95)},anchor=north,legend cell align=left,font=\footnotesize} 
				]
				
			
				\addplot[blue,thick,smooth] coordinates {
				(0.75, 0.) (0.752071, 0.0120347) (0.754142, 0.0242374) (0.756213, 0.0366113) (0.758284, 0.0491598) (0.760355, 0.0618861) (0.762426, 0.074794) (0.764497, 0.0878872) (0.766569, 0.10117) (0.76864, 0.114646) (0.770711, 0.128319) (0.772782, 0.142196) (0.774853, 0.156279) (0.776924, 0.170575) (0.778995, 0.185089) (0.781066, 0.199826) (0.783137, 0.214793) (0.785208, 0.229996) (0.787279, 0.245441) (0.78935, 0.261137) (0.791421, 0.277091) (0.793492, 0.293311) (0.795563, 0.309806) (0.797635, 0.326586) (0.799706, 0.343661) (0.801777, 0.361042) (0.803848, 0.378741) (0.805919, 0.396771) (0.80799, 0.415147) (0.810061, 0.433884) (0.812132, 0.452998) (0.814203, 0.47251) (0.816274, 0.49244) (0.818345, 0.51281) (0.820416, 0.533648) (0.822487, 0.554982) (0.824558, 0.576846) (0.82663, 0.599279) (0.828701, 0.622324) (0.830772, 0.646033) (0.832843, 0.670469) (0.834914, 0.695705) (0.836985, 0.721832) (0.839056, 0.748965) (0.841127, 0.777251) (0.843198, 0.806888) (0.845269, 0.838156) (0.84734, 0.871481) (0.849411, 0.907587) (0.851482, 0.948007) (0.853553, 1.)
				};
				\addlegendentry{$H(A|XYE)$}
				
				\addplot[red,thick,smooth,dotted] coordinates {
				(0.75, 0.) (0.752071, 0.00603838) (0.754142, 0.0122045) (0.756213, 0.0185027) (0.758284, 0.0249374) (0.760355, 0.0315132) (0.762426, 0.0382352) (0.764497, 0.0451088) (0.766569, 0.0521395) (0.76864, 0.0593332) (0.770711, 0.0666965) (0.772782, 0.0742361) (0.774853, 0.0819592) (0.776924, 0.0898735) (0.778995, 0.0979875) (0.781066, 0.10631) (0.783137, 0.114851) (0.785208, 0.12362) (0.787279, 0.132628) (0.78935, 0.141889) (0.791421, 0.151414) (0.793492, 0.161218) (0.795563, 0.171316) (0.797635, 0.181726) (0.799706, 0.192467) (0.801777, 0.203558) (0.803848, 0.215022) (0.805919, 0.226885) (0.80799, 0.239174) (0.810061, 0.251921) (0.812132, 0.265161) (0.814203, 0.278934) (0.816274, 0.293285) (0.818345, 0.308265) (0.820416, 0.323936) (0.822487, 0.340367) (0.824558, 0.357638) (0.82663, 0.375848) (0.828701, 0.395113) (0.830772, 0.415575) (0.832843, 0.437409) (0.834914, 0.460839) (0.836985, 0.486151) (0.839056, 0.513728) (0.841127, 0.544096) (0.843198, 0.578018) (0.845269, 0.616671) (0.84734, 0.662045) (0.849411, 0.718043) (0.851482, 0.794796) (0.853553, 1.)
				};
				\addlegendentry{$H_{\min}(A|XYE)$}
						
				\end{axis}  
				\end{tikzpicture}

				\caption{Secrecy vs.\@ winning probability $\omega$ in the CHSH game for a \emph{single-round box}. Two lower-bounds are shown: one for the conditional von Neumann entropy $H(A|XYE)$~\cite{pironio2010random} and the other for the conditional min-entropy $H_{\min}(A|XYE)$~\cite{masanes2011secure}; both bounds are tight. As soon as the winning probability is above the classical threshold of 75\% some secret randomness is produced.}
				\label{fig:entropy_one_box}
			\end{figure}
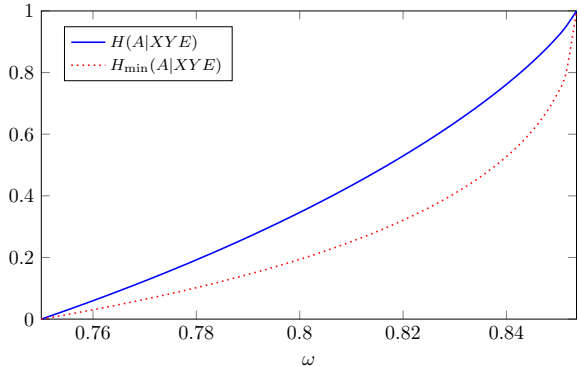

			The relation stated in Equation~\eqref{eq:one_box_entropy} is plotted in Figure~\ref{fig:entropy_one_box}. One can see that the entropy increases as the winning probability $\omega$ increases. That is, the amount of secret randomness in Alice's outcome is directly related to the winning probability of the single-round box. In particular, we observe that $H(A|XYE)=0$ (i.e., the adversary knows the value of $A$) for the optimal classical winning probability and $H(A|XYE)=1$ (i.e., $A$ looks completely random to the adversary) for the optimal quantum winning probability.
			\footnote{These two extreme cases are easy to understand. When the box employs a classical strategy the adversary can simply hold a copy of $A$. When the box employs the optimal quantum strategy the used state is the maximally entangled state. Then, due to monogamy of entanglement, the adversary is completely decoupled from the Alice and Bob's state. For more details see Section~\ref{sec:pre_di_crypt}.}
			Note that there can be many different boxes $\P_{AB|XY}$ (and hence extensions to the adversary) with the same winning probability~$\omega$.  That is, the assumption regarding the winning probability of the box does not pin down the full probability distribution. The bound given in Equation~\eqref{eq:one_box_entropy} is thus very strong -- it says that for \emph{any} single-round box with winning probability $\omega$ and \emph{any} purification to the adversary the stated lower bound holds.

			Instead of considering the von Neumann entropy as above, one can also study lower-bounds on the conditional min-entropy $H_{\min}(A|XYE)$ as a function of the winning probability of a single-round box -- as was done in~\cite{masanes2011secure}. We plot the resulting bound in Figure~\ref{fig:entropy_one_box}. As can be seen in the figure, for non-optimal Bell violation the min-entropy can be significantly lower than the von Neumann entropy. Indeed, the min-entropy is always upper-bounded by the von Neumann entropy (hence the name). Still, in some cases a bound on the min-entropy, rather than the von Neumann entropy, is needed or, at the least, is easier to derive. In particular, lower-bounds on the min-entropy for \emph{single-round boxes} can be found using general techniques based on the semidefinite programming hierarchies of~\cite{navascues2008convergent} while, up to date, there is no general technique to derive (or even estimate) such bounds on the von Neumann entropy.

			Similar bounds were derived also for other Bell inequalities. For example, lower-bounds on the min-entropy produced by a single-round box were found as a function of the violation of the Mermin inequality~\cite[Equation~(6)]{gallego6514full} and the tilted-CHSH inequality~\cite[Lemma~2]{bamps2017device}. Another result in the same spirit is that of~\cite[Section~5]{nieto2016device}, where a bound on the min-entropy is derived as a function of several Bell inequalities all at once.\footnote{That is, instead of assuming that we know just the winning probability of the single-round box in a specific game, we assume we know its winning probabilities in several different games. In the context of single-round boxes this is a stronger assumption regarding the device. However, in actual application this is not an issue, as will be mentioned later on.}
			Lower-bounds on the von Neumann entropy were derived as a function of the violation of the MDL inequalities~\cite[Section~3]{kessler2017device} and the MABK inequality~\cite[Lemma~S5]{ribeiro2017fully}.

			Before continuing to the next chapter, we emphasise once again that single-round statements as mentioned above should not be understood as operational statements. If we are given a single-round box but we do not assume to know its winning probability~$\omega$ then we cannot conclude anything about its properties (e.g., the entropy of the outputs). 
			When considering, for example, device-independent cryptographic protocols one must test the device in order to estimate whether it can violate a Bell inequality or not. This is done by playing several games with the device and collecting statistic regarding its input-output behaviour. For this purpose we need to consider \emph{multi-rounds} boxes, as done in the following sections.


\chapter{Multi-round box}\label{ch:multi_box}

		In the previous chapter we discussed the \emph{single-round box}, which can be seen as a simple abstract object that allows us to study the fundamental aspects of non-locality. 
		When studying actual device-independent information processing tasks, however, one must consider more complex objects that describe the behaviour of the devices while performing the task of interest. 
		More concretely, in actual applications we usually interact with a device by playing \emph{many} games. 
		Even in the simplest setting where one would like to merely verify the violation of a Bell inequality, as in experiments performing loophole-free Bell tests,  a Bell game is played many times so that sufficient amount of data can be collected to estimate the violation in a satisfactory statistical manner. Playing just a single game is clearly not enough. 
		Another example is device-independent protocols, such as quantum key distribution. All protocols include a phase in which the users (or honest parties) are playing many games with their device in order to decide whether it can be used for the considered task.
		Hence, considering boxes that can be used to play just a single game is not enough. Instead, we need to work with \emph{multi-round boxes}. 
		
		Multi-round boxes can be described using a conditional probability distribution~$\P_{\mr{A}\mr{B}|\mr{X}\mr{Y}}$ over the inputs and outputs of many rounds of a game. 
		That is, for $n$  the number of games which one would like to play with the box (e.g., the number of rounds of a protocol), $\mr{A}=A_1A_2\dots A_n$ is a random variable over $\mathcal{A}^{n}$ and $\mr{B},\mr{X}$, and $\mr{Y}$ are similarly defined. 
	
		\sloppy
		As explained in the beginning of Chapter~\ref{ch:single_round_box}, the way we model a box, and in particular a multi-round box, depends on the type of interaction that we would like to perform with it. 
		We consider two different forms of interactions: parallel and sequential interactions. 
		Different tasks require different types of boxes. 
		Parallel boxes are used, for example, in self-testing~\cite{natarajan2017quantum}, parallel quantum key distribution~\cite{jain2017parallel}, and certification of entanglement~\cite{arnon2017noise}. 
		Some examples for settings in which sequential boxes are considered are delegated computation~\cite{reichardt2013classical} and randomness amplification~\cite{kessler2017device}. 
		In the scope of this thesis, Chapters~\ref{ch:reductions_par} and~\ref{ch:par_rep_showcase} deal with parallel boxes while Chapters~\ref{ch:reductions_seq} and~\ref{ch:crypto_showcase} focus on sequential boxes.
	
	\section{Parallel interaction}\label{sec:parallel_mr_boxes}
	
\begin{figure}
			\centering
			\includegraphics[width=\textwidth]{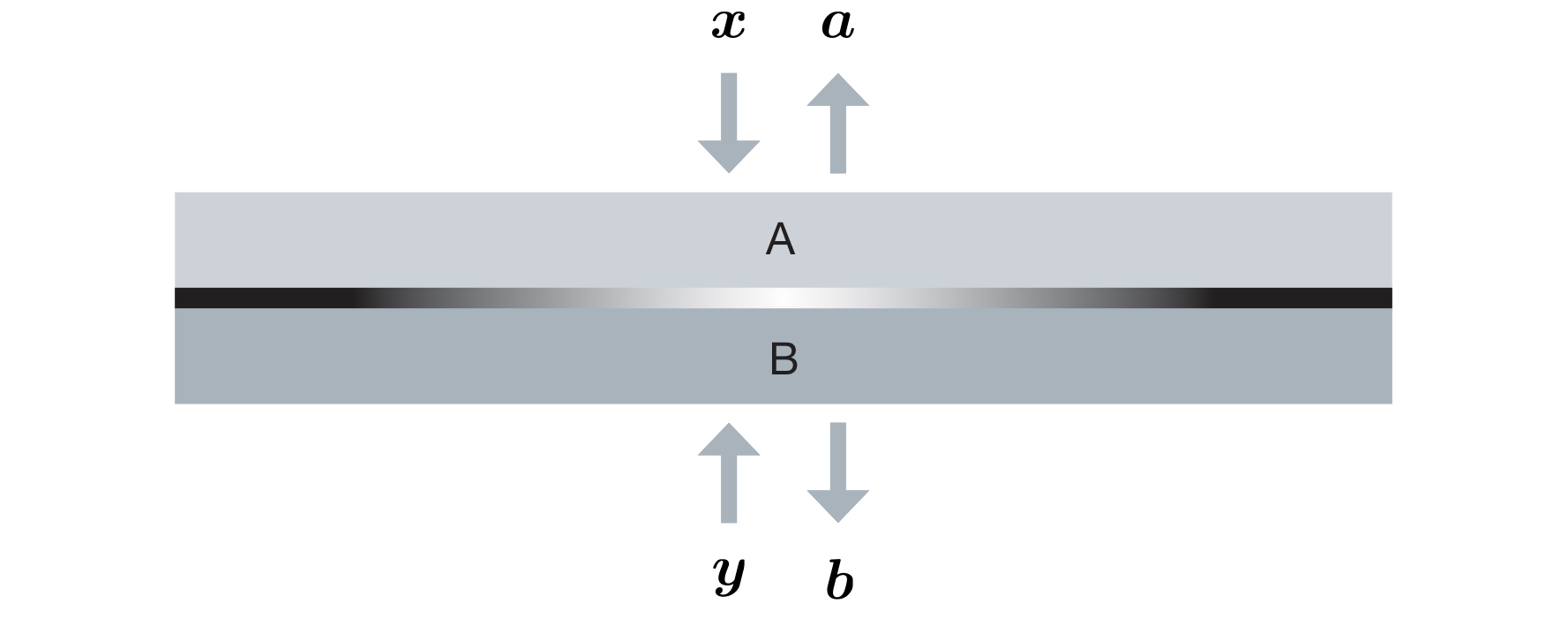}
			\caption{Parallel multi-round box. We think of a parallel multi-round box as a large device, shared between Alice and Bob, which can be used to play many rounds of a Bell game, all at once. Such a box is expecting to get the inputs for all rounds,~$\mr{x}$ and~$\mr{y}$, at the same time, and it will then produce all the outputs,~$\mr{a}$ and~$\mr{b}$ for Alice and Bob.}
			\label{fig:parallel_box}
		\end{figure}

		The simplest  to describe form of interaction is the ``parallel interaction''. In such an interaction the box is ``expecting'' to get the $n$ inputs of all the rounds, $\mr{x}$ and $\mr{y}$, at the same time and is expected to give all the outputs, $\mr{a}$ and $\mr{b}$, together; see Figure~\ref{fig:parallel_box}.
		If the box is only given inputs of a single game, e.g., $x_1,y_1$, it is not expected to return any output. This behaviour of the box will present itself in the mathematical model of the box, as we explain below.

			For a given a game $\G$, a parallel multi-round box is a device with which Alice and Bob can play $n$ instances of $\G$ in parallel (i.e., at the same time). Mathematically this translates to a conditional probability distribution $\P_{\mr{A}\mr{B}|\mr{X}\mr{Y}}$, non-signalling between Alice and Bob, defined over the inputs and outputs of $n$ games. For example, when considering the CHSH game, $\mr{A}, \mr{B}, \mr{X},$ and $\mr{Y}$ are all random variables over~$\{0,1\}^n$.

			As explained in Section~\ref{sec:pre_ns_corr}, the non-signalling conditions between Alice and Bob imply that Alice and Bob's marginals, $\P_{\mr{A}|\mr{X}}$ and $\P_{\mr{B}|\mr{Y}}$ respectively, are well-defined. The fact that we are talking about a \emph{parallel} multi-round box means that no further structure can be assumed. In particular, other marginals, e.g., $\P_{A_1|X_1}$ or $\P_{A_2B_2|X_2B_2}$, are not necessarily well-defined. Intuitively this stands for the fact that the box is expecting to get all the inputs together and only then it produces the outputs; the output for $A_1$ can therefore depend, for example, on the value of $X_5$ and not on just that of~$X_1$. Hence the conditional probability distribution $\P_{A_1|X_1}$ is not properly defined.

			\subsubsection{Non-signalling parallel boxes}
			
				One can  consider a parallel multi-round box which is only restricted by the non-signalling conditions. We then get the following definition. 
				
				\begin{defn}[Non-signalling parallel multi-round box]\label{def:ns_parallel_box}
					Given a Bell game $\G$, a non-signalling parallel multi-round box is a non-signalling box $\P_{\mr{A}\mr{B}|\mr{X}\mr{Y}}$, as in Definition~\ref{def:ns_box}, defined for the inputs and outputs of $n$ rounds of the game $\G$ -- $\mathcal{X}^n,\mathcal{Y}^n,\mathcal{A}^n,\mathcal{B}^n$. 
					That is,  for all $\mr{a}\in\mathcal{A}^n$, $ \mr{b}\in\mathcal{B}^n$, $ \mr{x},\mr{x}'\in\mathcal{X}^n$ and $\mr{y},\mr{y}'\in\mathcal{Y}^n$,
					\begin{equation}\label{eq:multi_round_ns}
					\begin{split}
						&\sum_\mr{b} \P_{\mr{A}\mr{B}|\mr{X}\mr{Y}}(\mr{a}\mr{b}|\mr{x}\mr{y}) = \sum_\mr{b} \P_{\mr{A}\mr{B}|\mr{X}\mr{Y}}(\mr{a}\mr{b}|\mr{x}\mr{y}') \\
						&\sum_\mr{a} \P_{\mr{A}\mr{B}|\mr{X}\mr{Y}}(\mr{a}\mr{b}|\mr{x}\mr{y}) = \sum_\mr{a} \P_{\mr{A}\mr{B}|\mr{X}\mr{Y}}(\mr{a}\mr{b}|\mr{x}'\mr{y}) \;.
					\end{split}
					\end{equation}
				\end{defn}
				
				As mentioned above, the only non-signalling conditions restricting the parallel box, are those between Alice and Bob appearing in Definition~\ref{def:ns_parallel_box}; we do not set any other assumptions regarding the box apart from that.

			\subsubsection{Quantum parallel boxes}
			
				Similarly to a quantum single-round box, as in Definition~\ref{def:quant_single_box}, a quantum parallel multi-round box is just a  quantum box (Definition~\ref{def:quant_box}) defined for the inputs and outputs of $n$ rounds of $\G$.
				
				\begin{defn}[Quantum parallel multi-round box]\label{def:quant_parallel_box}
					Given a Bell game $\G$, a quantum parallel multi-round box is a quantum box $\P_{\mr{A}\mr{B}|\mr{X}\mr{Y}}$, as in Definition~\ref{def:quant_box}, defined for the inputs and outputs of $n$ rounds of the game $\G$ -- $\mathcal{X}^n,\mathcal{Y}^n,\mathcal{A}^n,\mathcal{B}^n$.
					That is, there exist a bipartite state $\rho_{Q_AQ_B}$ and measurements $\{M^{\mr{x}}_{\mr{a}}\}$ and $\{M^{\mr{y}}_{\mr{b}}\}$ such that
					\begin{equation}\label{eq:quant_parallel_box_stat}
						\P_{\mr{A}\mr{B}|\mr{X}\mr{Y}}(\mr{a}\mr{b}|\mr{x}\mr{y}) = \Tr\left(M^{\mr{x}}_{\mr{a}} \otimes M^{\mr{y}}_{\mr{b}} \; \rho_{Q_AQ_B} \right) \quad \forall  \mr{a},\mr{b},\mr{x},\mr{y}\;.
					\end{equation}
				\end{defn}
		
				The non-signalling conditions in Equation~\eqref{eq:multi_round_ns} are automatically fulfilled by quantum parallel boxes defined above.
				We remark again that there are no further assumptions regarding the structure of the state and measurements apart from what appears in Equation~\eqref{eq:quant_parallel_box_stat}. Specifically,~$\rho_{Q_A}$ and $\rho_{Q_B}$ are not assumed to have some further subsystem structure and the measurements need not have a tensor product form such as $M^{x_1}_{a_1} \otimes \dots \otimes M^{x_n}_{a_n}$.

		
	\section{Sequential interaction}\label{sec:seq_boxes}
		
		In the previous section we discussed \emph{parallel} multi-rounds boxes. These are boxes that allow (and ``expect'') to be interacted with in a parallel way, i.e., by giving all the inputs to the box at the same time. As the parallel multi-round box receives all the inputs at once, the output for, e.g., the first game, $A_1$, can depend on the inputs for all games $X_1, X_2, \dots, X_n$.
		
		In this section we consider a different type of multi-round boxes --  \emph{sequential multi-round boxes}. 
		Such boxes are, in some sense, more structured than parallel multi-round boxes and accurately model the devices used in many device-independent scenarios. As such, sequential multi-round boxes are of relevance for applications. 
		Furthermore, the additional structure of sequential multi-round boxes will allow us to derive stronger results than those derived for their parallel counterparts. 
		
\begin{figure}
			\centering
			\includegraphics[width=\textwidth]{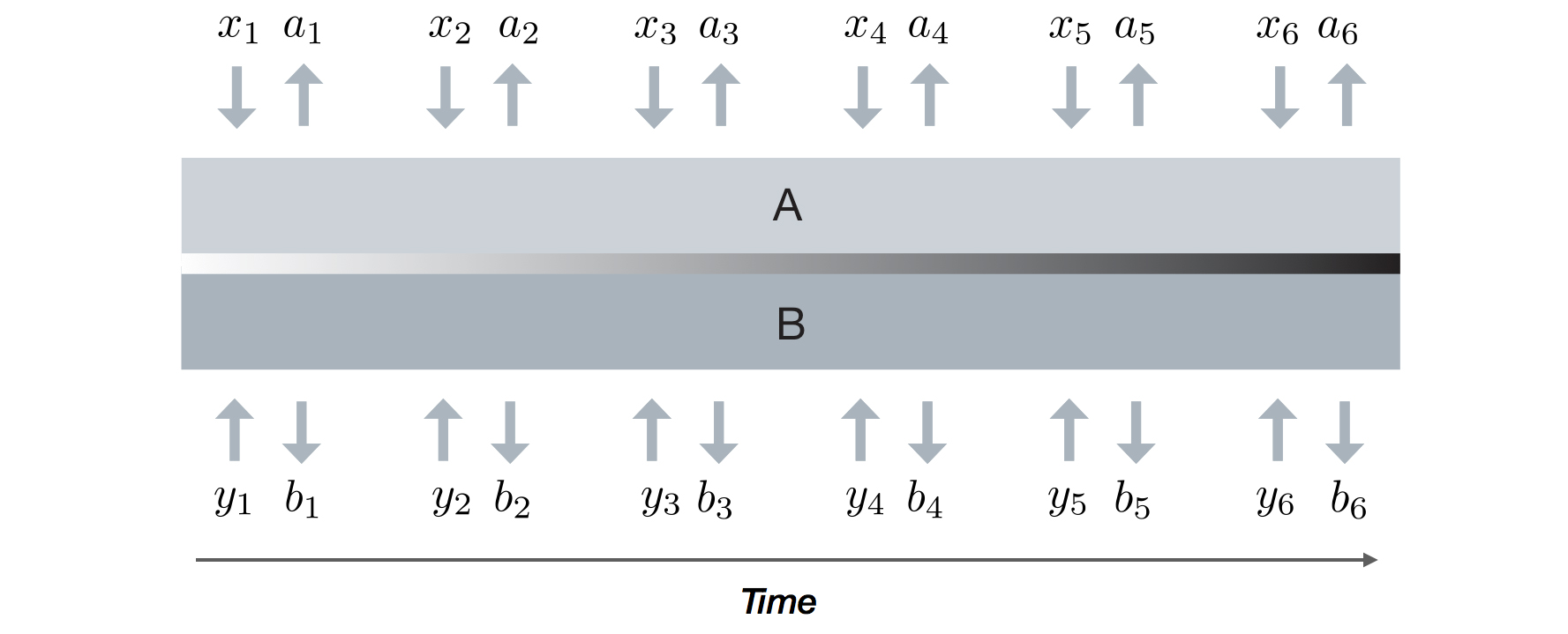}
			\caption{Sequential interaction with a multi-round box. Alice and Bob start by playing the first game with the box and only once they receive the outputs from the box they proceed to the second game, and so on.}
			\label{fig:seq_box}
		\end{figure}

		As mentioned above, the way we model a multi-round box depends on how we would like to interact with it. Most device-independent protocols proceed in rounds which are performed one after the other: Alice and Bob use their box in the first round of the protocol and only once they receive the outputs from the box they proceed to the second round, and so on; See Protocol~\ref{pro:intro_qkd} for an example.
		We call such an interaction with the box ``sequential interaction''. This is illustrated in Figure~\ref{fig:seq_box} (the reader may compare Figure~\ref{fig:seq_box} to the single-round box in Figure~\ref{fig:single_round_box} and the parallel multi-round box in Figure~\ref{fig:parallel_box}). 
		
		The chronological order which is implied by the sequential interaction enforces certain constraints on the behaviour of the box. In particular, while past events can influence future ones, the future cannot change the past. For example, the first output $A_1$ can depend on the first input $X_1$ but not on the inputs of the next rounds~$X_2, \dots, X_n$. The second output $A_2$ can depend both on $X_2$ and past events, such as the values assigned to $A_1$ and $X_1$, but not on the following inputs~$X_3, \dots, X_n$.
		
\begin{figure}
			\centering
			\includegraphics[width=\textwidth]{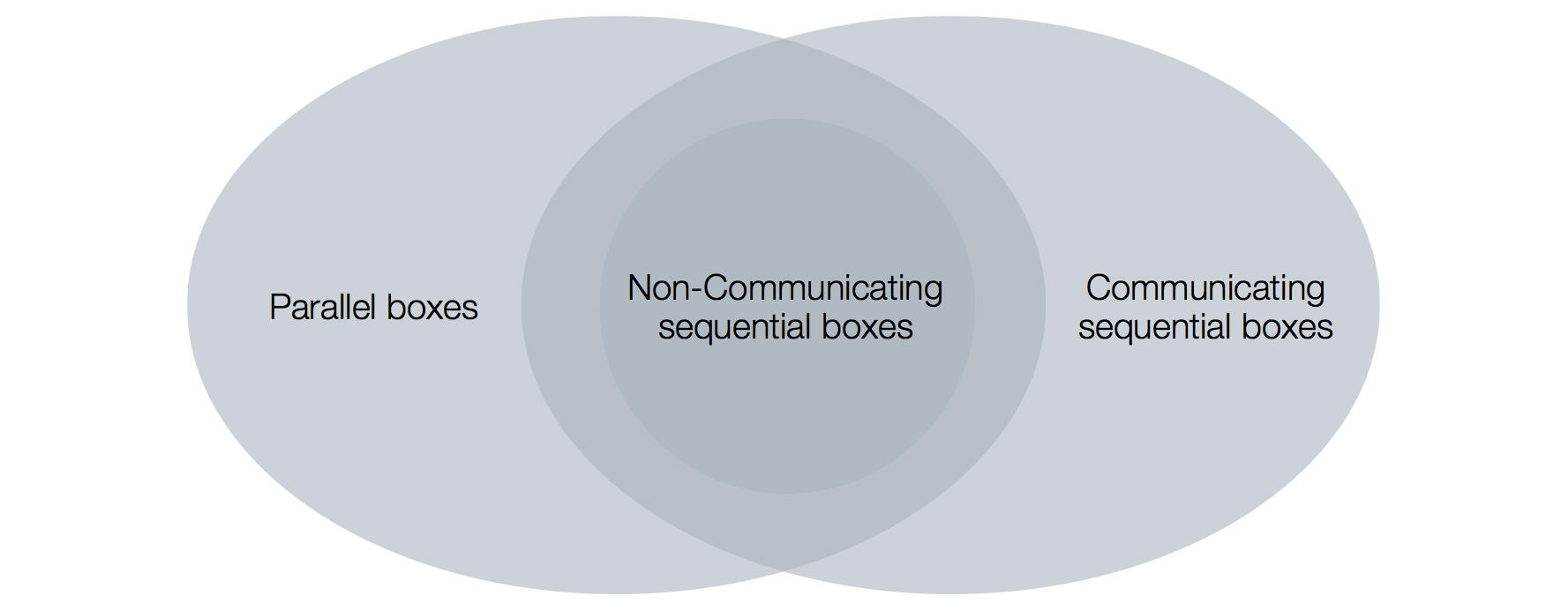}
			\caption{The relation between the different multi-round boxes.}
			\label{fig:boxes_sets}
		\end{figure}

		We define two different types of sequential boxes -- one which allows for communication between the rounds of interactions and one which does not. 
		A box that allows for communication between the rounds is a box in which Alice and Bob's devices can exchange classical or quantum information after finishing playing a game and before starting the next one. Such boxes should be considered when entanglement is to be distributed ``on the fly'', e.g., in protocols where Alice is expected to send half of an entangled state to Bob in each round, or when the devices are located far enough so they cannot communicate during a \emph{single} game but too close to make sure signals from one round cannot arrive to the other device until the end of \emph{all} games. 
		A box that does not allow for communication can be considered, e.g., in cryptographic settings in which any communication between the devices implies that \emph{all} information can leak to the adversary.
		We remark that parallel boxes and sequential boxes that allow for communications are incomparable to one another, while both are more general than sequential boxes without communication; see Figure~\ref{fig:boxes_sets}.  This is explained in more detail after formally defining the two types of sequential boxes.
		
		\subsection{Without communication between the rounds}\label{sec:seq_box_wo_comm}
			
			As in the case of a parallel multi-round box, a sequential multi-round box is described by a conditional probability distribution  $\P_{\mr{A}\mr{B}|\mr{X}\mr{Y}}$ defined over the inputs and outputs of $n$ rounds of the game $\G$ -- $\mathcal{X}^n,\mathcal{Y}^n,\mathcal{A}^n,\mathcal{B}^n$. The special thing about a sequential box is that the marginals describing the individual rounds of the game are well-defined and non-signalling between Alice and Bob.  That is, they are boxes by themselves.
			
			In this section we consider a model of sequential boxes in which Alice's and Bob's components are not allowed to communicate between the rounds of the game. For short, we call such boxes \emph{non-communicating sequential boxes}.
			Formally, to define a non-communicating sequential box we consider the marginals of $\P_{\mr{A}\mr{B}|\mr{X}\mr{Y}}$ describing a round~$i\in[n]$. The relevant marginals are 
			\begin{equation}\label{eq:seq_marg_wo_comm}
				\P_{A_iB_i|X_iY_iH^{i,\text{Alice}}H^{i,\text{Bob}}}
			\end{equation}
			where $H^{i,\text{Alice}}=\mr{X}_{1,\dots,i-1}\mr{A}_{1,\dots,i-1}$ and $H^{i,\text{Bob}}=\mr{Y}_{1,\dots,i-1}\mr{B}_{1,\dots,i-1}$ denote the ``histories'' of Alice and Bob's boxes in round $i$. These histories basically describe all the information that can be kept by the boxes from the previous rounds (we can think of such boxes as devices which record past events in their memory). 
			The history may include more information\footnote{For example, in device-independent quantum key distribution protocols the parties randomly choose in each round whether the round is used for testing the device or for generating key bits. This information can also be included in the history $H^i$.}  than past inputs and outputs; for simplicity we stick to the above choice. 
			
			A first requirement on a sequential box is that the marginals~\eqref{eq:seq_marg_wo_comm} are well-defined. This can be mathematically described by a set of non-signalling conditions.
				Explicitly, for every $i\in[n]$, we denote:
				\begin{enumerate}
					\item $\mathcal{P} = [i-1]$,  $\mr{a}_{\mathcal{P}} = a_1,\dots,a_{i-1}$, and similarly for $\mr{b}_{\mathcal{P}}$, $\mr{x}_{\mathcal{P}}$, and $\mr{y}_{\mathcal{P}}$.
					\item $\mathcal{F} = \{i+1,\dots,n\}$,  $\mr{a}_{\mathcal{F}} = a_{i+1},\dots,a_n$, and similarly for $\mr{b}_{\mathcal{F}}$, $\mr{x}_{\mathcal{F}}$, and $\mr{y}_{\mathcal{F}}$.
					\item For any $\mr{x}_{\mathcal{P}}$,  $\mr{y}_{\mathcal{P}}$, $x_i$, $y_i$, $\mr{x}_{\mathcal{F}}$, $\mr{y}_{\mathcal{F}}$, $\mr{x}'_{\mathcal{F}}$, and $\mr{y}'_{\mathcal{F}}$, 
					\begin{enumerate}
						\item $\mr{x}=\mr{x}_{\mathcal{P}},x_i,\mr{x}_{\mathcal{F}}$
						\item $\mr{x}'=\mr{x}_{\mathcal{P}},x_i,\mr{x}'_{\mathcal{F}}$
					\end{enumerate} 
					and similarly for $\mr{y}$ and $\mr{y}'$. 
				\end{enumerate}
				Then, we require that the following non-signalling conditions hold for all $\mr{a}_{\mathcal{P}}$, $\mr{b}_{\mathcal{P}}$, $\mr{x}_{\mathcal{P}}$,  $\mr{y}_{\mathcal{P}}$, $a_i$, $b_i$, $x_i$, $y_i$, $\mr{x}_{\mathcal{F}}$, $\mr{x}'_{\mathcal{F}}$, $\mr{y}_{\mathcal{F}}$, and $\mr{y}'_{\mathcal{F}}$,
				\begin{equation}\label{eq:time_order_marginals}
					\begin{split}	
					&	\sum_{\mr{a}_{\mathcal{F}},\mr{b}_{\mathcal{F}}} \P_{A_iB_i\mr{A}_{\mathcal{F}}\mr{B}_{\mathcal{F}}|\mr{A}_{\mathcal{P}}\mr{B}_{\mathcal{P}}\mr{X}\mr{Y}} \left( a_i,b_i, \mr{a}_{\mathcal{F}},\mr{b}_{\mathcal{F}} | \mr{a}_{\mathcal{P}},\mr{b}_{\mathcal{P}}, \mr{x}, \mr{y} \right) = \\
					& \qquad	\sum_{\mr{a}_{\mathcal{F}},\mr{b}_{\mathcal{F}}} \P_{A_iB_i\mr{A}_{\mathcal{F}}\mr{B}_{\mathcal{F}}|\mr{A}_{\mathcal{P}}\mr{B}_{\mathcal{P}}\mr{X}\mr{Y}} \left( a_i,b_i, \mr{a}_{\mathcal{F}},\mr{b}_{\mathcal{F}} | \mr{a}_{\mathcal{P}},\mr{b}_{\mathcal{P}}, \mr{x}', \mr{y}' \right)  \;.
					\end{split}
				\end{equation}

			Now that the marginals $\P_{A_iB_i|X_iY_iH^{i,\text{Alice}}H^{i,\text{Bob}}}$ are well-defined for all $i\in[n]$, we further ask that they are non-signalling between Alice and Bob, when each party holds only its own history. That is, $\P_{A_i|X_iH^{i,\text{Alice}}}$ and $\P_{B_i|Y_iH^{i,\text{Bob}}}$ need to be well-defined as well. 
			Explicitly, for each round $i\in[n]$, for all $a\in\mathcal{A}$, $ b\in\mathcal{B}$, $ x,x'\in\mathcal{X}, \;y,y'\in\mathcal{Y}$ and histories $h^{i,\text{Alice}},h^{i,\text{Alice}'}\in\mathcal{X}^{i-1}\times\mathcal{A}^{i-1}$ and $h^{i,\text{Bob}},h^{i,\text{Bob}'}\in\mathcal{Y}^{i-1}\times\mathcal{B}^{i-1}$,
			\begin{equation}\label{eq:seq_wo_comm_ns_cond}
			\begin{split}
				&\sum_b \P_{A_iB_i|X_iY_iH^{i,\text{Alice}}H^{i,\text{Bob}}}(a,b|x,y,h^{i,\text{Alice}},h^{i,\text{Bob}}) = \\ 
				& \hspace{30pt} \sum_b \P_{A_iB_i|X_iY_iH^{i,\text{Alice}}H^{i,\text{Bob}}}(a,b|x,y',h^{i,\text{Alice}},h^{i,\text{Bob}'}) \\
				&\sum_a \P_{A_iB_i|X_iY_iH^{i,\text{Alice}}H^{i,\text{Bob}}}(a,b|x,y,h^{i,\text{Alice}},h^{i,\text{Bob}}) = \\ 
				& \hspace{30pt} \sum_a \P_{A_iB_i|X_iY_iH^{i,\text{Alice}}H^{i,\text{Bob}}}(a,b|x',y,h^{i,\text{Alice}'},h^{i,\text{Bob}}) \;.
			\end{split}
			\end{equation}

			The fact that the boxes cannot communicate between the rounds presents itself by having two different histories, one for Alice and one for Bob. The above equations then imply that the actions of Alice's box in round $i$ depend only on Alice's history, i.e., on what happened in the previous rounds on Alice's side (while she is oblivious to Bob's history), and similarly for Bob.\footnote{This should be compared to the next section, where we will have just a single history $H^i$ for Alice and Bob together.}
			
			Note that we only ask the marginals $\P_{A_i|X_iH^{i,\text{Alice}}}$ and $\P_{B_i|Y_iH^{i,\text{Bob}}}$ to be well-defined. $\P_{A_i|X_i}$, on the other hand, are not necessarily valid boxes. 
			
			\subsubsection{Non-signalling non-communicating sequential boxes}
			
			\sloppy
				A non-signalling non-communicating sequential multi-round box is simply a box~$\P_{\mr{A}\mr{B}|\mr{X}\mr{Y}}$ fulfilling the above non-signalling constraints; there are no further requirements. 
				\begin{defn}[Non-signalling non-communicating sequential multi-round box]\label{def:ns_seq_box}
					Given a Bell game $\G$, a non-signalling non-communicating sequential multi-round box is a conditional probability distribution $\P_{\mr{A}\mr{B}|\mr{X}\mr{Y}}$ defined for the inputs and outputs of $n$ rounds of the game $\G$ -- $\mathcal{X}^n,\mathcal{Y}^n,\mathcal{A}^n,\mathcal{B}^n$ fulfilling the non-signalling conditions given in Equations~\eqref{eq:time_order_marginals} and~\eqref{eq:seq_wo_comm_ns_cond}.
				\end{defn}

			\subsubsection{Quantum non-communicating sequential boxes}
			
				The simplest way of defining a quantum non-communicating sequential box is to consider the initial state shared by Alice and Bob and the sequence of measurements that they perform. 
				
				More specifically, in each round Alice and Bob's boxes can perform a measurement on the post-measurement state of the previous round. We denote the state in the beginning of round $i\in[n]$ (i.e., before performing the measurements of the $i$'th round) by $\rho^{i,h^{i,\text{Alice}},h^{i,\text{Bob}}}_{Q_AQ_B}$. As clear from the notation, this state depends on the histories $h^{i,\text{Alice}},h^{i,\text{Bob}}$. We identify $\rho^{1}_{Q_AQ_B} = \rho_{Q_AQ_B} $ as the initial state of the box. 
				
				Furthermore, we denote the (Kraus) measurements performed in each round by~$\{K^{x}_a\}$ and~$\{K^{y}_b\}$.\footnote{Note that in contrast to the previous definitions, the measurement operators $K$ are now written as Kraus operators and not POVMs, since we are interested in the post-measurement state. See Section~\ref{sec:pre_quantum} for more details.} 
				One can think of the measurements $\{K^{x}_a\}$ as depending on the history~$h^{i,\text{Alice}}$ and similarly for Bob. 
				Alternatively, we can imagine that the history is already kept in some classical registers within the quantum state~$\rho^{i,h^{i,\text{Alice}},h^{i,\text{Bob}}}_{Q_AQ_B}$, i.e.,~$\rho_{Q_A}$ includes also the information $h^{i,\text{Alice}}$ and similarly for Bob. The measurements can thus be defined as first reading the history and then applying the relevant measurement depending on the history. This allows us to use the shorter notation in which the operators do not depend on the histories explicitly.   
				
				\sloppy
				Using the above notation, the relation between the state in round $i$ to that of round $i-1$ is simply (up to normalisation of the state)
				\begin{equation}\label{eq:seq_states}
				\begin{split}
					&\rho^{i,h^{i,\text{Alice}},h^{i,\text{Bob}}}_{Q_AQ_B} \propto \\
					& \quad \left( K^{x_{i-1}}_{a_{i-1}} \otimes K^{y_{i-1}}_{b_{i-1}} \right) \rho^{i-1,h^{i-1,\text{Alice}},h^{i-1,\text{Bob}}}_{Q_AQ_B} 
					\left( \left(K^{x_{i-1}}_{a_{i-1}}\right)^{\dagger} \otimes \left(K^{y_{i-1}}_{b_{i-1}} \right)^{\dagger}\right) \;,
				\end{split}
				\end{equation}
				where $h^{i,\text{Alice}}$ and $h^{i,\text{Bob}}$ uniquely determine $x_{i-1},a_{i-1},h^{i-1,\text{Alice}}$ and $y_{i-1},b_{i-1},h^{i-1,\text{Bob}}$, respectively (i.e., the values on the righthand-side of Equation~\eqref{eq:seq_states} should be consistent with the histories on the lefthand-side). 
				The conditions stated in Equation~\eqref{eq:seq_wo_comm_ns_cond} follow directly. 
				
				\begin{defn}[Quantum  non-communicating sequential multi-round box]\label{def:quant_seq_box_wo_comm}
					Given a Bell game $\G$, a quantum non-communicating sequential multi-round box is a conditional probability distribution $\P_{\mr{A}\mr{B}|\mr{X}\mr{Y}}$ defined for the inputs and outputs of $n$ rounds of the game $\G$, $\mathcal{X}^n,\mathcal{Y}^n,\mathcal{A}^n,\mathcal{B}^n$, such that
					there exist a bipartite state~$\rho_{Q_AQ_B}$ and measurements $\{K^{x}_a\}$ and $\{K^{y}_b\}$ defining a sequence of bipartite states for $i\in[n]$ as in Equation~\eqref{eq:seq_states}.
				\end{defn}
				
				As mentioned before, a  non-communicating sequential box is also a parallel one. Indeed, it is easy to see that a parallel box can always simulate the behaviour of a non-communicating sequential box.
				

	\subsection{With communication between the rounds}\label{sec:mr_seq_with_comm}
			
			In the previous section we considered sequential boxes in which Alice's and Bob's components are not allowed to communicate between the rounds. This implies that Alice's and Bob's components evolve separately in time and each of them has their own ``history'': $h^{i,\text{Alice}}$ for Alice and $h^{i,\text{Bob}}$ for Bob. 
			Now, we consider a scenario in which Alice's and Bob's components are allowed to communicate between the different games, i.e., after the outputs of round $i-1$ were supplied by the box and before the $i$'th inputs are given.\footnote{In Protocol~\ref{pro:intro_qkd}, for example, ``between the different games'' refers to the time \emph{after} Step~3 of round $i-1$ and \emph{before} Step~2 of round $i$, for all $i\in[n]$.}
			Considering boxes that are allowed to communicate is, in particular, relevant when considering realistic application of, e.g., device-independent cryptography. There, one would like to allow the experimentalists to distribute entanglement ``on the fly'' during the protocol. To send a new quantum state in each round the communication channels need to be open and an adversarial box may use this opportunity to communicate.

			Mathematically this setting can be formalised by allowing Alice and Bob to keep a common history register that includes the classical information of all past events \emph{on both sides}.
			More specifically, the marginal describing the $i$'th round of the game, for $i\in[n]$, is given by $\P_{A_iB_i|X_iY_iH^i}$, where $H^i$ denotes the history defined by the previous rounds.~$H^i$ includes $\mr{X}_{1,\dots,i-1}\mr{Y}_{1,\dots,i-1}\mr{A}_{1,\dots,i-1}\mr{B}_{1,\dots,i-1}$ as well as any other information available to Alice's and Bob's component. For simplicity we assume that $H^i=\mr{X}_{1,\dots,i-1}\mr{Y}_{1,\dots,i-1}\mr{A}_{1,\dots,i-1}\mr{B}_{1,\dots,i-1}$ similarly to what was done before.
			The only non-trivial communication to consider is one which depends on the history, since any other  information could have been included as part of the box to begin with. Therefore, we can assume without loss of generality that the communicated information is simply the entire history.

			As before, we first require that $\P_{A_iB_i|X_iY_iH^i}$  are well-defined, i.e., Equation~\eqref{eq:time_order_marginals} is fulfilled.
			In addition, $\P_{A_iB_i|X_iY_iH^i}$ needs to be non-signalling between Alice and Bob, when they both hold their common history. That is, $\P_{A_i|X_iH^i}$ and $\P_{B_i|Y_iH^i}$ are well-defined. 
			Formally: for each round~$i\in[n]$, for all $a\in\mathcal{A}$, $ b\in\mathcal{B}$, $ x,x'\in\mathcal{X}, \;y,y'\in\mathcal{Y}$ and $h^i\in\mathcal{A}^{i-1}\times\mathcal{B}^{i-1}\times\mathcal{X}^{i-1}\times\mathcal{Y}^{i-1}$,
			\begin{equation}\label{eq:ns_a_b_w_comm}
				\begin{split}
				&\sum_b \P_{A_iB_i|X_iY_iH^i}(a,b|x,y,h^i) = \sum_b \P_{A_iB_i|X_iY_iH^i}(a,b|x,y',h^i) \\
				&\sum_a \P_{A_iB_i|X_iY_iH^i}(a,b|x,y,h^i) = \sum_a \P_{A_iB_i|X_iY_iH^i}(a,b|x',y,h^i) \;.
			\end{split}
			\end{equation}
			 In contrast to Equation~\eqref{eq:seq_wo_comm_ns_cond}, in the above equations the behaviour of Alice's component in the $i$'th round may depend also on past events on Bob's side, as $H^i$ includes also~$\mr{Y}_{1,\dots,i-1}\mr{B}_{1,\dots,i-1}$, and similarly for Bob's part of the box.

			\subsubsection{Non-signalling communicating sequential boxes}		

				A non-signalling communicating sequential multi-round box is a box~$\P_{\mr{A}\mr{B}|\mr{X}\mr{Y}}$ fulfilling the above non-signalling constraints.
				\begin{defn}[Non-signalling communicating sequential multi-round box]\label{def:ns_seq_box}
					Given a Bell game $\G$, a non-signalling communicating sequential multi-round box is a conditional probability distribution $\P_{\mr{A}\mr{B}|\mr{X}\mr{Y}}$ defined for the inputs and outputs of $n$ rounds of the game $\G$ -- $\mathcal{X}^n,\mathcal{Y}^n,\mathcal{A}^n,\mathcal{B}^n$ fulfilling the non-signalling conditions given in Equations~\eqref{eq:time_order_marginals} and~\eqref{eq:ns_a_b_w_comm}.
				\end{defn}
				
				It is perhaps instructive to note that $\P_{\mr{A}\mr{B}|\mr{X}\mr{Y}}$ itself is \emph{not} a non-signalling box.; communication (i.e., signalling) between the rounds may be \emph{necessary} in order to implement the box. We give a trivial example in the end of the section.

			\subsubsection{Quantum communicating sequential boxes}
			
				When we say that a communicating sequential multi-round box is quantum we mean that in each round the behaviour of the box can be described within the formalism of quantum physics. 
			
				\begin{defn}[Quantum communicating  sequential multi-round box]\label{def:quant_seq_box_w_comm}
					Given a Bell game $\G$, a quantum sequential multi-round box is a conditional probability distribution $\P_{\mr{A}\mr{B}|\mr{X}\mr{Y}}$ defined for the inputs and outputs of $n$ rounds of the game~$\G$, $\mathcal{X}^n,\mathcal{Y}^n,\mathcal{A}^n,\mathcal{B}^n$, such that for all $i\in[n]$ the marginal $\P_{A_iB_i|X_iY_iH^i}$, for $H^i=\mr{X}_{1,\dots,i-1}\mr{Y}_{1,\dots,i-1}\mr{A}_{1,\dots,i-1}\mr{B}_{1,\dots,i-1}$, is a quantum box as in Definition~\ref{def:quant_box}. 
					That is, there exist a bipartite state $\rho^{h^i}_{Q_AQ_B}$ and measurements $\{M^{h^i,x}_a\}$ and $\{M^{h^i,y}_b\}$ such that
					\begin{equation}\label{eq:quant_seq_box_stat}
						\P_{A_iB_i|X_iY_iH^i}(ab|xyh^i) = \Tr\left(M^{h^i,x}_a \otimes M^{h^i,y}_b \; \rho^{h^i}_{Q_AQ_B} \right) \quad \forall  a,b,x,y,h^i\;.
					\end{equation}
				\end{defn}

				The box in Equation~\eqref{eq:quant_seq_box_stat} is written as $\P_{A_iB_i|X_iY_iH^i}$ so it is mathematically clear which marginals of $\P_{\mr{A}\mr{B}|\mr{X}\mr{Y}}$ are being discussed. On the level of the state and measurements one thinks of $\rho^{h^i}_{Q_AQ_B}$, $\{M^{h^i,x}_a\}$, and $\{M^{h^i,y}_b\}$ as depending on the history $h^i$, which allows the actions in each round to depend on the past. As in Section~\ref{sec:seq_box_wo_comm}, we may also consider a state $\rho^{h^i}_{Q_AQ_B}$ that keeps $h^i$ in one of its registers and measurements that first read the history and then apply the relevant operations; in such a case we may think of $\{M^x_a\}$, and $\{M^y_b\}$ independent of the history.
				
				It may seem from Definition~\ref{def:quant_seq_box_w_comm} that only the individual rounds are considered. The sequential nature of the box is concealed in the relations between the different rounds. It becomes apparent when noting that all the marginals describing the individual rounds should be consistent with the same overall box~$\P_{\mr{A}\mr{B}|\mr{X}\mr{Y}}$. 
				Alternatively, one can consider an equivalent definition of a quantum communicating sequential multi-round box that is perhaps more intuitive (but mathematically more complex): 
				Similarly to the evolution described in Equation~\eqref{eq:seq_states}, we start with some initial quantum state and make sequential measurements. In contrast to  Equation~\eqref{eq:seq_states}, however, we  allow for an additional general operation, which may depend on the history, to be performed on the post-measurement state of each round.
				The general operation between the rounds is what models the communication between the two parts of the box. 				
				
			Before concluding this section, let us mention the relations between the different types of multi-round boxes.
			The relations are shown in Figure~\ref{fig:boxes_sets}.
			It is obvious to see that communicating sequential boxes are more general than non-communicating sequential boxes.
			In contrast to  non-communicating sequential boxes, parallel boxes cannot simulate a general communicating sequential box.
			A trivial example is a communicating sequential box that always outputs $b_2=x_1$. Clearly, since a parallel box must, in particular, fulfil Equation~\eqref{eq:multi_round_ns}, it cannot simulate such a box. 
			On the other hand, communicating sequential boxes cannot simulate a general parallel box. For example, a communicating sequential box cannot simulate a parallel box for which $a_1=x_2$. 
			Thus, the two types of boxes are incomparable.

\chapter{Working under the IID assumption}\label{ch:iid_assumption}

	In this thesis, we are interested in analysing the behaviour of multi-round boxes when such boxes are used to play many non-local games, e.g., while running a cryptographic protocol. 
	In the previous chapter we discussed the different models of multi-round boxes (the parallel and sequential ones). As we saw, their behaviour can be quite complex. As a consequence, the analysis of protocols which use such boxes is (a priori) tedious in the good case and infeasible in the worst.
	
	In this chapter we discuss an assumption that can make the analysis of the scenarios of interest much simpler -- the so called ``independent and identically distributed'' (IID) assumption. The assumption states that the boxes behave independently and identically when playing the $n$ games. The IID assumption is commonly made in the literature as it significantly simplifies the behaviour of the considered boxes and allows us to gain better intuition and understanding of the problem at hand. As we explain below, there is no reason to believe that the IID assumption can be enforced in the device-independent setting; we use it just as a first stage before moving on to the general analysis. In Chapters~\ref{ch:reductions_par}-\ref{ch:reductions_seq} we will see that, in certain scenarios, some techniques can be used to reduce the general analysis to the one made under the IID assumption.

	We start by explaining the assumption itself.  
	Following that, we present a mathematical tool, namely the ``quantum asymptotic equipartition property'', which is of great use when considering IID random variables and quantum systems. 
	Finally, we discuss the analysis of our showcases under the IID assumption.
	
	\section{The IID assumption}\label{sec:iid_assump_desc}

		\begin{figure}
			\centering
			\includegraphics[width=\textwidth]{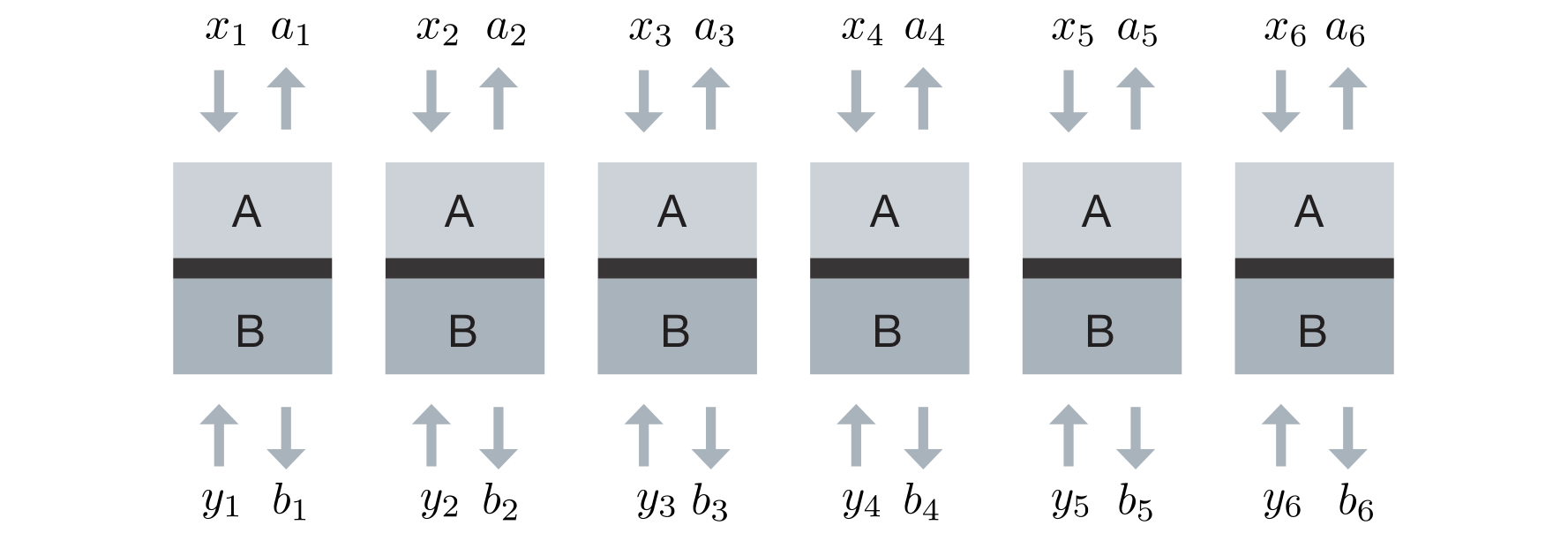}
			\caption{IID box. The box $\P_{\mr{A}\mr{B}|\mr{X}\mr{Y}}$ can be described as $n$ identical and independent copies of a single-round box. Each game is played with a different copy of the box. We can see each copy as a well defined subsystem.}
			\label{fig:iid_box}
		\end{figure}
	
		As in the previous chapter, we consider multi-round boxes. 
		An IID box, as the name suggests, is a multi-round box which behaves identically and independently in each game played with it. 
		Pictorially, we can think of an IID box as $n$ identical and independent copies of a single-round box, as shown in Figure~\ref{fig:iid_box}.  Comparing this to Figures~\ref{fig:parallel_box} and~\ref{fig:seq_box}, one sees that an IID box has more structure than the other, more general, multi-round boxes. In particular, in the case of an IID box we can talk about a subsystem structure of the box. In the quantum case, for example, if $\sigma$ denotes the state of a \emph{single-round box} in Figure~\ref{fig:iid_box} then the overall state of the IID box is $\sigma^{\otimes n}$. A similar tensor product structure also holds for the measurements describing the box. 
		Mathematically, an IID box is defined as follows.
		
		\begin{defn}[IID box]\label{def:iid_box}
			Given a non-local game $\G$, an IID box is a conditional probability distribution $\P_{\mr{A}\mr{B}|\mr{X}\mr{Y}}$ defined for the inputs and outputs of $n$ rounds of the game $\G$, $\mathcal{X}^n,\mathcal{Y}^n,\mathcal{A}^n,\mathcal{B}^n$, such that
			\begin{equation}\label{eq:iid_box}
				\P_{\mr{A}\mr{B}|\mr{X}\mr{Y}}(\mr{a},\mr{b}|\mr{x},\mr{y}) = \prod_{i\in[n]} \P_{AB|XY}(a_i,b_i|x_i,y_i)
			\end{equation}
			for some \emph{single-round box} $\P_{AB|XY}$. 
			An IID box is said to be quantum or non-signalling if the single round box $\P_{AB|XY}$ is quantum (Definition~\ref{def:quant_single_box}) or non-signalling (Definition~\ref{def:quant_single_box}), respectively.
		\end{defn}
		
\begin{figure}
			\centering
			\includegraphics[width=\textwidth]{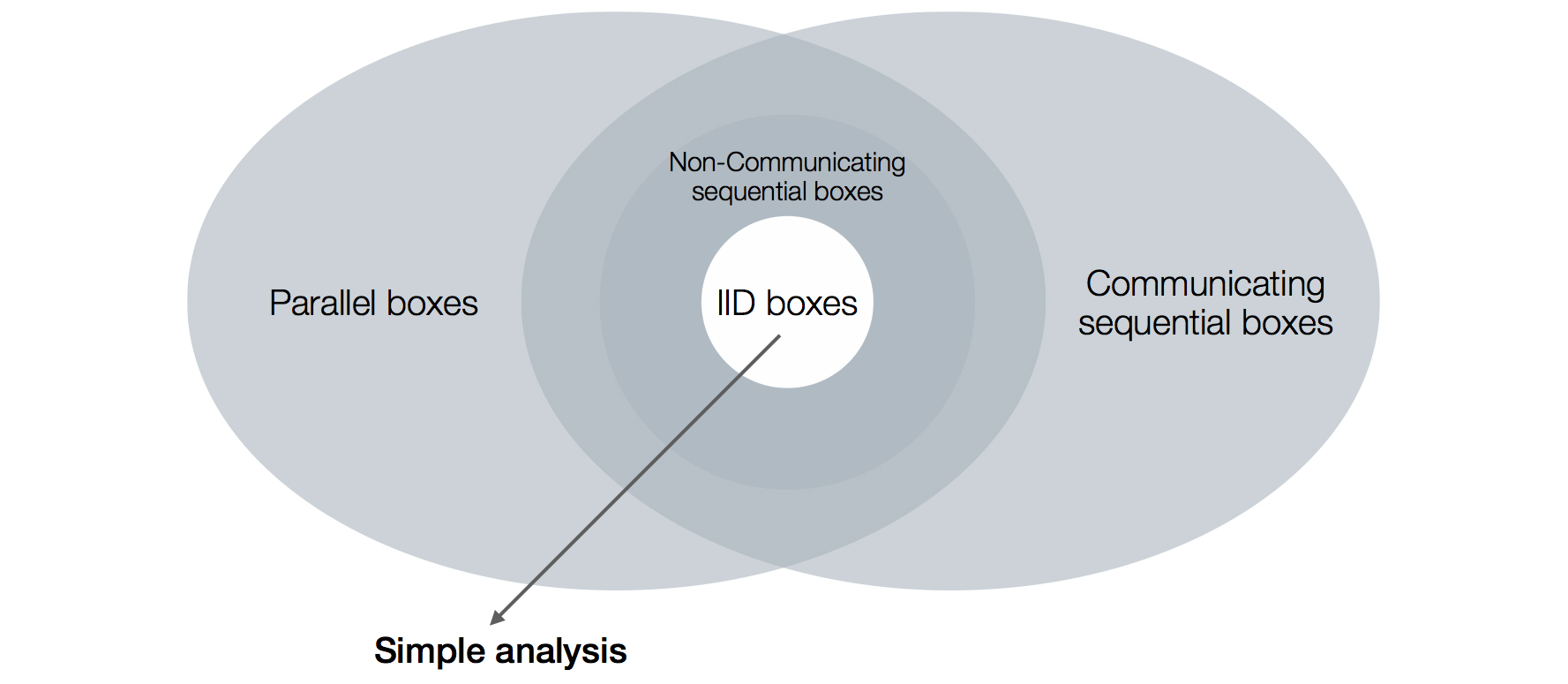}
			\caption{The relation between the sets of multi-round boxes. The intersection of the sets of sequential and parallel boxes includes the set of IID boxes. The analysis of IID boxes is rather simple.}
			\label{fig:boex_sets_iid}
		\end{figure}

		Note that the single-round box $\P_{AB|XY}$ in Equation~\eqref{eq:iid_box}  is the same for every round $i\in[n]$. This means that the behaviour of the box is identical in each round and independent of all other rounds.\footnote{As always, a box is a \emph{conditional} probability distribution and its definition is therefore independent of the distribution of the inputs, $\mr{x}$ and $\mr{y}$, which can be arbitrary (depending on how the box is being used). It is perhaps helpful to note that the idea here is that, while the inputs of the different rounds may be correlated in general (i.e., not IID), the box itself does not ``create'' further correlations between the rounds (in contrast to parallel and sequential boxes). In any case, in most scenarios the inputs are usually taken to be IID random variables as well.} 
		Hence, the behaviour of an IID box $\P_{\mr{A}\mr{B}|\mr{X}\mr{Y}}$ is solely characterised by the single-round box $\P_{AB|XY}$ and, thus, the substance of any analysis done for the IID box is the study of the single-round box. 
		This also implies that the box behaves exactly the same whether we give it all the inputs at once (parallel interaction) or one after the other (sequential interaction). An IID box is therefore both a parallel multi-round box and a sequential multi-round box; see Figure~\ref{fig:boex_sets_iid}.
		Given all of the above, it indeed makes sense that any analysis done solely for IID boxes can be much simpler than the general analysis in which one needs to deal with parallel or sequential multi-round boxes.
		
		When considering device-independent protocols one usually encounters IID boxes in two different contexts -- the so called ``completeness'' and ``soundness'' of the protocols (recall, e.g., Section~\ref{sec:pre_di_crypt}).  
		When proving the completeness of a protocol one shows that an ``honest implementation'' of the box does not cause the protocol to abort (with high probability). The honest implementation is the implementation of the box that one would like to have if the manufacture of the device is to be trusted and, most commonly, it is described as an IID box. Thus, investigating the behaviour of the protocol when an IID box is being used allows us to see what happens in the honest scenario when ``everything goes according to the plan''.
		
		The second context to discuss IID boxes is that of the soundness proof. There, one ought to show that the protocol acts as required for \emph{any} box, i.e., even for adversarial ones.\footnote{Recall that in the device-independent setting we assume that the adversary is the one constructing the box. Device-independent protocols are expected to abort, with high probability, when an adversarial device is detected.}
		Clearly, not all boxes are IID boxes and hence analysing the situation only for IID boxes is not sufficient. That is, by \emph{assuming} that all boxes behave in an IID manner we weaken the final statement. Still, working under the IID assumption allows us to gain better understanding of the full question at hand. 	
		
		It is important to remark that, even though quite convenient for the soundness analysis, the IID assumption cannot be justified a priori.
		Assuming that the box behaves in an IID way goes against the spirit of device-independence by imposing severe restrictions on the implementation of the box. In particular, the assumption implies that the multi-round box does not include any, classical or quantum, internal memory (i.e., its actions when playing one game cannot depend on the other games) and cannot display time-dependent behaviour. We therefore emphasise that working under the IID assumption is only a first step in the process of proving full soundness (as will be shown in the proceeding chapters).

	
	\section{Asymptotic equipartition property}\label{sec:aep_both}
		
		When analysing IID processes a useful mathematical tool is the so called ``asymptotic equipartition property'' (AEP). 
		The entropic formulation of the AEP used in this thesis basically asserts that when considering IID RV $\mr{A}=A_1,A_2,\dotsc,A_n$, all identical copies of the RV $A$, the smooth min- and max-entropies rates, $H^{\varepsilon}_{\min} (\mr{A})/n$ and $H^{\varepsilon}_{\max} (\mr{A})/n$,  converge to $H(A)$~\cite{holenstein2011randomness}. 
		Similarly, the quantum version of the AEP asserts that the same is true for IID quantum states $\left(\sigma_{A}\right)^{\otimes n}$ and, even more, it holds also when considering conditional entropies~\cite{renner2008security, tomamichel2009fully, tomamichel2012framework}.

		In many information theoretic tasks one needs to bound the smooth min- and max-entropies, as they describe operational quantities. In particular, this will be the case in one of the showcases investigated in the thesis. 
		When considering IID processes, as done in this chapter, the AEP allows us to reduce the analysis of the smooth entropies for \emph{IID boxes} to the analysis of the von Neumann entropy for a \emph{single-round box}.\footnote{An example of the analysis of the von Neumann entropy for single-round boxes was presented in Section~\ref{sec:randomness_single_round}. The AEP motivates the analysis done in that section when working under the IID assumption.} This explains why the AEP is a useful tool when working under the IID assumption.
				
		To comprehend the statement of the AEP and its significance we start by presenting and explaining the classical AEP. The quantum variant is then presented as an extension of the classical one.   
	
		\subsection{Classical asymptotic equipartition property}
		
			The (classical) AEP can be seen as the ``information theoretic version'' of the law of large numbers. Given IID RV $A_1,A_2,\dotsc,A_n$ the law of large numbers states that for large enough number of samples $n$, the average is close to the expected value in probability. Formally this can be written as
			\begin{equation}\label{eq:law_of_large_num}
				\forall \mu>0 \quad \lim_{n\rightarrow\infty}\Pr\left[\Big|\frac{1}{n}\sum_i A_i-\mathbb{E} \left[A\right]\Big| >\mu\right]=0 \;,
			\end{equation}
			for $A=A_1=\dots=A_n$ a single copy of the RV. 
			Similarly, the AEP, which is a direct consequence of the law of large numbers,\footnote{To see this, one can define a new RV, $\tilde{A_i}$, which, for all $a\in\mathcal{A}$ takes the value $\log\left(\Pr[a] \right)$ with probability $\Pr[a]$. Applying Equation~\eqref{eq:law_of_large_num} for the new IID RV $\tilde{A_1},\dots,\tilde{A_n}$, Equation~\eqref{eq:clas_aep_basic} follows.} states that for IID RV
			\begin{equation}\label{eq:clas_aep_basic}
				\forall \mu>0 \quad \lim_{n\rightarrow\infty} \Pr \left[ \Big| -\frac{1}{n} \log \left(\P_{\mr{A}}[\mr{a}]\right) - H(A) \Big| >\mu\right]=0 \;,
			\end{equation} 
			where we denoted $\mr{A}=A_1A_2\dots A_n$.

			Assume that we sample a sequence $\mr{a}$. What can we say about its probability $\P_{\mr{A}}[\mr{a}]$? 
			We learn from Equation~\eqref{eq:clas_aep_basic} that, for large enough~$n$,  
			\begin{equation}\label{eq:prob_typ_seq}
				2^{-n(H(A) + \mu)}  < \P_{\mr{A}}[\mr{a}] < 2^{-n(H(A) - \mu)}
			\end{equation}
			with high probability. This allows us to talk about ``typical sequences'' and ``typical sets''. A typical sequence is a sequence $\mr{A}$ for which Equation~\eqref{eq:prob_typ_seq} holds and the typical set includes all typical sequences. 
			Denote by $1-\varepsilon$ the probability that Equation~\eqref{eq:prob_typ_seq} holds or, in other words, the probability of the typical set. 
			In the limit $n\rightarrow\infty$, $\varepsilon\rightarrow 0$, the typical set has probability approximately~1, all elements of it appear with approximately $2^{-nH(A)}$ probability, and, hence, it includes approximately $2^{nH(A)}$ elements. (For formal proofs see~\cite[Chapter~3]{cover2012elements}). Thus, the AEP implies that when analysing probabilistic statements regarding a sequence of IID RV, one can focus on the typical events (and ignore the non-typical ones) without introducing much of an error. 
			
			Equation~\eqref{eq:prob_typ_seq} can be used to state the AEP in terms of the smooth min- and max-entropies; this form of the AEP is the one used in this thesis. 
			\begin{thm}[AEP\footnote{Note that this theorem is actually a \emph{non}-asymptotic version of the AEP, as it describes also the convergence rate for finite $n$ (i.e., it includes also the second order term). The limit, stated as Equation~\eqref{eq:class_aep_limit} below, follows trivially from the presented theorem.} (direct part)]\label{thm:class_aep}
				Let $\mr{A}=A_1A_2\dots A_n$ be a sequence of IID RV. Then, for any $\varepsilon\in(0,1)$ and $n$ large enough,
				\begin{align*}
					&\frac{1}{n} H^{\varepsilon}_{\min} (\mr{A})  \geq H(A) - \frac{\delta}{\sqrt{n}} \\
					&\frac{1}{n} H^{\varepsilon}_{\max} (\mr{A})  \leq H(A) + \frac{\delta}{\sqrt{n}} \;,
				\end{align*}
				where $\delta$ depends on $\varepsilon$ and $A$.\footnote{For the time being we are not interested in the explicit form of $\delta$; this will be discussed when relevant.}
			\end{thm}
			
			\sloppy
			To gain some intuition of how the smooth entropies enter to the above theorem we sketch the main arguments here in a somewhat hand-waving way.
			For the more accurate analysis we refer the interested reader to~\cite{renner2008security, holenstein2011randomness,tomamichel2012framework}.
			Recall that
			\begin{equation*}
				\begin{split}
					&H_{\min}(\mr{A}) = \min_{\mr{a}} -\log\left[\P_\mr{A}[\mr{a}]\right] \\
					&H_{\max}(\mr{A}) \leq \max_{\mr{a} | \P_{\mr{A}}[\mr{a}]\neq 0} -\log\left[\P_{\mr{A}}[\mr{a}]\right] \;.
				\end{split}
			\end{equation*}
			Thus, when considering only typical events, it follows from  Equation~\eqref{eq:prob_typ_seq} that
			\begin{align*}
				&\frac{1}{n} H_{\min}(\mr{A}) > H(A) - \mu \\
				&\frac{1}{n} H_{\max}(\mr{A}) < H(A) - \mu \;.
			\end{align*} 
			To account for non-typical events we need to incorporate their probability~$\varepsilon$. We  do so by switching to the smooth versions of the entropies while using~$\varepsilon$ as the smoothing parameter.\footnote{The above only (roughly) explains why the \emph{smooth} entropies are considered, without addressing the second order term of the AEP. The second order term  does not come from the law of large numbers but its refinement~--- the central limit theorem.}

			Theorem~\ref{thm:class_aep}, in combination with a converse bound\footnote{The converse bound roughly follows from the monotonicity of the so called $\alpha$-entropies. For details see~\cite[Section 6.4]{tomamichel2015quantum}.}, implies that when~$n$ goes to infinity both smooth entropies converge to the Shannon entropy:
			\begin{equation}\label{eq:class_aep_limit}
				\lim_{n\rightarrow \infty } \frac{1}{n} H^{\varepsilon}_{\min} (\mr{A}) = \lim_{n\rightarrow \infty } \frac{1}{n} H^{\varepsilon}_{\max} (\mr{A}) = H(A) \;.
			\end{equation}
			This explains why the Shannon entropy is so important in information theory~--- the smooth entropies, which describe operational tasks (recall, for example, Section~\ref{sec:pre_di_crypt}), converge to the Shannon entropy when considering a large number of independent repetitions of the relevant task. 
			A commonly used example is that of ``data compression''. There, one would like to encode an $n$ bit string using less bits. If we allow for some small error when decoding the data, roughly $H^{\varepsilon}_{\max} (\mr{A})$ bits are needed~\cite{renner2004smooth}. For a large enough IID sequence $A_1,A_2,\dotsc,A_n$, however, $nH(A)$ bits suffices~\cite{shannon2001mathematical}. 
						
			A final important comment about the entropic formulation of the AEP is with regards to the so called ``chain rules''. The Shannon entropy respects the chain rule
			$H(\mr{A}) = \sum_{i} H(A_i|\mr{A}_{<i})$, where $\mr{A}_{<i}$ denotes the sequence of all RV $A_j$ with $j<i$. In the case of IID RV this is reduced to $H(\mr{A}) = nH(A)$. 
			That is, \emph{the total amount of entropy of $\mr{A}$ is $n$ times the entropy of a single copy of $A$}. Thus, in order to calculate $H(\mr{A})$ we only need to know $H(A)$.  
			In contrast to the Shannon entropy, the smooth min- and max-entropies do not fulfil a similar chain rule. Theorem~\ref{thm:class_aep} tells us that, to first order in $n$, $H^{\varepsilon}_{\min} (\mr{A})=H^{\varepsilon}_{\max} (\mr{A}) = nH(A)$. Therefore, for sufficiently large~$n$, \emph{the total amount of the smooth min- and max- entropies of $\mr{A}$ are $n$ times the Shannon entropy of a single copy of $A$} and, here as well, we only need to know $H(A)$ to calculate $H^{\varepsilon}_{\min} (\mr{A})$ and $H^{\varepsilon}_{\max} (\mr{A})$.
			 					
		\subsection{Quantum asymptotic equipartition property}\label{sec:quant_aep}
		
			As the name suggests, the quantum AEP is an extended version of the AEP that applies to IID quantum states $\rho=\left(\sigma_{AB}\right)^{\otimes n}$ (the classical variant is then a special case of the quantum one). 
			The following theorem, developed in~\cite[Result 5]{tomamichel2012framework} (see also~\cite[Theorem 9]{tomamichel2009fully}), acts as the generalisation of Theorem~\ref{thm:class_aep} above; it extends the theorem to quantum states and, at the same time, incorporates conditioning on quantum systems.\footnote{The classical AEP, given as Theorem~\ref{thm:class_aep}, can be easily written also in terms of conditional entropies if the conditioning is done on \emph{classical} systems (then one can directly define the probability distribution of $A$ as the conditional one). This is not the case when conditioning on quantum systems. That is to say that the statement of the theorem which includes conditional entropies does not follow directly from a ``non-conditional'' variant.}

			\begin{thm}[Quantum AEP (direct part)~\cite{tomamichel2012framework}]\label{thm:quant_aep}
				Let $\rho=\left(\sigma_{AB}\right)^{\otimes n}$ be an IID quantum state. Then, for any $\varepsilon\in(0,1)$ and $n$ large enough,
				\begin{align}
					&\frac{1}{n} H^{\varepsilon}_{\min} (\mr{A}|\mr{B})_{\rho}  \geq H(A|B)_{\sigma} - \frac{\delta(\varepsilon,\nu)}{\sqrt{n}} \label{eq:quan_eap_min}\\
					&\frac{1}{n} H^{\varepsilon}_{\max} (\mr{A}|\mr{B})_{\rho}  \leq H(A|B)_{\sigma} + \frac{\delta(\varepsilon,\nu)}{\sqrt{n}} \;,\label{eq:quan_eap_max}
				\end{align}
				where $\delta(\varepsilon,\nu)=4\log\nu\sqrt{\log(2/\varepsilon^2)}$ for $\nu=2\sqrt{2^{H_{\max}(A|B)}}+1$.
			\end{thm}

			In combination with a converse bound \cite[Corollary~6.3]{tomamichel2015quantum}, we get the asymptotic equality of the conditional entropies:
			\begin{equation*}\label{eq:quant_aep_limit}
				\lim_{n\rightarrow \infty } \frac{1}{n} H^{\varepsilon}_{\min} (\mr{A}|\mr{B})_{\rho} = \lim_{n\rightarrow \infty } \frac{1}{n} H^{\varepsilon}_{\max} (\mr{A}|\mr{B})_{\rho} = H(A|B)_{\sigma} \;.
			\end{equation*}
			For the proofs of the quantum AEP the reader is directed to~\cite[Section 6.4]{tomamichel2015quantum}.
	
			The quantum AEP reveals the same important facts as its classical counterpart when considering IID quantum states -- it justifies the use of the von Neumann entropy in quantum information processing and tell us that, for sufficiently large~$n$, \emph{the total amount of the conditional smooth  entropies, $H^{\varepsilon}_{\min} (\mr{A}|\mr{B})_{\rho}$ and $H^{\varepsilon}_{\max} (\mr{A}|\mr{B})_{\rho}$,  are $n$ times the von Neumann entropy $H(A|B)_{\sigma}$ of a single copy of $\sigma$}. That is, instead of calculating the entropies of the full state $\rho$ one only needs to analyse the von Neumann entropy for a single copy of $\sigma$.	
			
			When considering applications in which the analysis should be done for a finite number of repetitions $n$, it is not sufficient to know that the smooth entropies converge to the von Neumann entropy; we also need to know how fast they converge. 
			The second order terms appearing in Equations~\eqref{eq:quan_eap_min} and~\eqref{eq:quan_eap_max}, i.e., the terms that scale with $1/\sqrt{n}$, account for the ``finite-size effects''. 
			While the $1/\sqrt{n}$ dependency is optimal, the constant $\delta$ is not tight.\footnote{To see that the $1/\sqrt{n}$ dependency is optimal follow, e.g., the proof of~\cite[Theorem~3.3.3]{renner2008security}. Second order terms with constants better than $\delta$ can be derived from~\cite{tomamichel2013hierarchy}.}

	\section{Using the IID assumption}\label{sec:using_IID_assump}
	
		In this section we discuss the analysis of our showcases when working under the IID assumption. 
		The analysis of the parallel repetition question, presented in Section~\ref{sec:par_rep_under_iid}, is somewhat trivial. 
		Our showcase of device-independent cryptography, considered in Section~\ref{sec:crypt_under_iid}, demonstrates the use of the quantum AEP in device-independent information processing tasks.

		\subsection{Showcase: non-signalling parallel repetition}\label{sec:par_rep_under_iid}

				In our terminology, parallel repetition results aim to upper-bound the probability that a parallel multi-round box can simultaneously win all the $n$ games played with it; recall Section~\ref{sec:pre_par_rep}. 
				The discussion below holds for classical, quantum, and non-signalling strategies. The word ``optimal'' then refers to the considered type of players.
				
				One simple strategy for the parallel repeated game is the IID strategy. This strategy takes the form of an IID box, which plays each of the~$n$ games independently and identically, as in Definition~\ref{def:iid_box}. That is, the box does not take advantage of the fact that it gets all the inputs at the same time. 
				For an \emph{optimal IID strategy}, i.e., the strategy which achieves the maximal probability of winning all games out of all IID strategies, the single-round box $\P_{AB|XY}$ appearing in Equation~\eqref{eq:iid_box} is the \emph{optimal single-game strategy}, that is, the one achieving winning probability of $1-\alpha$.
				
				It is easy to see that the probability that an IID box wins all the $n$ games simultaneously decreases exponentially fast with $n$. 
				Specifically, consider an IID box (or strategy) $\P_{\mr{A}\mr{B}|\mr{X}\mr{Y}}$ with
				\[
					\P_{\mr{A}\mr{B}|\mr{X}\mr{Y}}(\mr{a},\mr{b}|\mr{x},\mr{y}) = \prod_{i\in[n]} \P_{AB|XY}(a_i,b_i|x_i,y_i) 
				\]
				for some single-round box $\P_{AB|XY}$.
				Let $W_i$ denote the RV describing whether the $i$'th game is won ($W_i=1$) or not ($W_i=0$) and denote by $1-\alpha = \Pr\left[W_i =1\right]$ the winning probability of the \emph{single-round box} $\P_{AB|XY}(a_ib_i|x_iy_i)$ in a single game (as in Equation~\eqref{eq:win_prob_assump}).
				Due to the IID assumption, all the RV $W_i$ are independent and identically distributed. Thus, the probability that all the $n$ games are won is given by
				\[
					\Pr\left[\sum_i W_i = n\right] = \prod_i \Pr\left[W_i =1\right] = (1-\alpha)^n \;,
				\]			
				Clearly, for any $1-\alpha<1$, $(1-\alpha)^n$ decreases exponentially fast in $n$. 
				
				It is easy to show that also a concentration bound holds for any IID box: for any $0\leq\beta\leq\alpha$,  Hoeffding's inequality tells us that\footnote{Hoeffding's inequality tells us even more; it says that when using the optimal IID strategy the probability of winning \emph{less} than $1-\alpha-\beta$ fraction of the games is also decreasing exponentially fast.}
				\[
					\Pr\left[ \sum_i W_i \geq (1-\alpha + \beta)n\right] \leq \exp \left( - 2n\beta^2\right) \;,
				\]
				which decreases exponentially fast in $n$ as well.
				The answer to the parallel repetition question, \emph{under the IID assumption}, is therefore almost trivial.

	 	\subsection{Showcase: device-independent quantum cryptography}\label{sec:crypt_under_iid}
	 	
	 	\sloppy
	 		Following the first proof of concept of the security of device-independent quantum key distribution derived in~\cite{barrett2005no},  a long line of  works~\cite{acin2006bell,acin2006efficient,scarani2006secrecy,acin2007device,masanes2009universally,pironio2009device,hanggi2010efficient,hanggi2010device,masanes2011secure,masanes2014full}  considered the security of device-independent quantum and non-signalling cryptography under the IID assumption. 
		 	In this section we explain how the IID assumption is used when analysing device-independent quantum cryptographic protocols. Specifically, we consider here the task of device-independent randomness certification in the presence of a quantum adversary, which acts as the main building block of many device-independent cryptographic protocols, e.g., device-independent quantum key distribution. 
		 	We focus only on the parts of the security proof in which the IID assumption plays a crucial role and present them in a slightly simplified form. In particular, we consider large enough number of rounds $n$ and neglect finite-size effects for the moment. In Chapter~\ref{ch:crypto_showcase} we give full security proofs  (which do not rely on the IID assumption) and contrast the relevant parts with the analysis done here.

		 	When dealing with device-independent cryptography we first need to model the box used by the honest parties, Alice and Bob, and the adversary's knowledge about it. 
		 	Under the IID assumption, the state of Alice and Bob has an IID structure~$\rho_{\mr{Q_A}\mr{Q_B}}=\left(\sigma_{Q_AQ_B}\right)^{\otimes n}$ where each copy of $\sigma_{Q_AQ_B}$ is a bipartite state shared between Alice and Bob. 
		 	Moreover, we assume that the measurements performed in each round of the protocol are all identical and independent of one another, i.e., for all $\mr{a},\mr{x}$, $M_{\mr{a}}^{\mr{x}} = \left(M_{a}^{x}\right)^{\otimes n}$ and similarly for Bob's measurements. 
		 	The most general quantum adversary holds a purification of Alice and Bob's state. As all purifications are equivalent up to local unitaries on Eve's state, we can assume without loss of generality that the overall state of Alice, Bob, and Eve takes the IID form\footnote{It is the equivalence of all purifications that allows us to go from an IID assumption regarding Alice and Bob's state $\rho_{\mr{Q_A}\mr{Q_B}}$ to an IID assumption regarding the state $\rho_{\mr{Q_A}\mr{Q_B}\mr{E}}$, which also includes Eve. Interestingly, the same thing cannot be done when considering non-signalling boxes and adversaries. It follows from~\cite{arnon2012limits} that the extension of a non-signalling IID box to the adversary does not necessarily have an IID structure as well. (See also~\cite{masanes2014full}, where the box itself is assumed to have a subsystem structure similar to that of an IID box while the structure of the adversary's system is unrestricted).}
		 		\begin{equation}\label{eq:tripartite_iid_state}
		 			\rho_{\mr{Q_A}\mr{Q_B}\mr{E}}=\left(\sigma_{Q_AQ_BE}\right)^{\otimes n} \;.
		 		\end{equation}
		 		We remark that while we assume that $\rho$ has the above IID structure, the state~$\sigma_{Q_AQ_BE}$ is unknown. 
		 		
		 	Equation~\eqref{eq:tripartite_iid_state}, together with the IID form of the quantum measurements describing the device, indeed leads to an IID box $P_{\mr{A}\mr{B}|\mr{X}\mr{Y}}$ as in Equation~\eqref{eq:iid_box}.  In particular, this implies that $A_1,A_2,\dots,A_n$ are IID RV. Furthermore, it follows from Equation~\eqref{eq:tripartite_iid_state} that, for all $i\in[n]$, the quantum system $E_i$ holds information \emph{only} regarding the output $A_i$ of the same round (that is, $A_1$ and $E_2$, for example, are independent of one another).  
		 		
		 	Recall from Section~\ref{sec:qkd_main_task} that the central task when proving security of quantum cryptographic protocols is to bound the amount of information that Eve may obtain about certain values generated by the protocol, which are supposed to be unknown to her. 
		 	In the case of randomness certification the main technical step of all soundness proofs is to lower-bound the smooth min-entropy of Alice's outputs~$\mr{A}=A_1,A_2,\dots,A_n$ (see, e.g., Protocol~\ref{pro:intro_qkd}).
		 	Our goal is therefore to lower-bound~$H^{\varepsilon}_{\min}(\mr{A}|\mr{E})$, where $\mr{E}=E_1,E_2,\dots,E_n$ are Eve's IID quantum systems appearing in Equation~\eqref{eq:tripartite_iid_state}.

	 		\begin{figure}
	 			\centering
	 			\includegraphics[width=\textwidth]{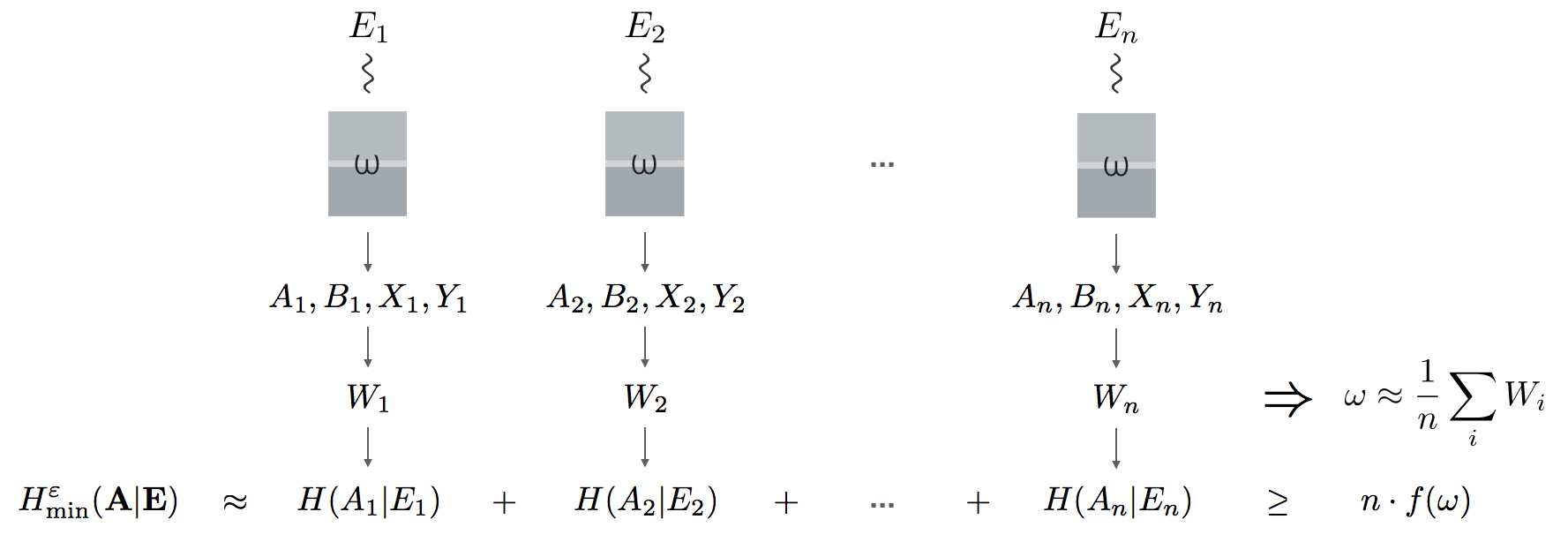}
	 			\caption{Sketch of a security proof under the IID assumption and for large enough~$n$. The honest parties hold an IID box. Each quantum system $E_i$, belonging to the adversary, can be entangled only to the $i$'th box. 
	 			The non-local game is being played with each of the independent and identical boxes. The statistics are then collected and used to estimate the winning probability $\omega$ of the single-round boxes. According to the quantum AEP, for large enough $n$, the total amount of smooth min-entropy is the sum of the von Neumann entropy of each round, which can be bounded as a function of the estimated winning probability $\omega$.}
	 			\label{fig:rand_iid_proof}
	 		\end{figure}

	 		The rough idea behind a security proof under the IID assumption  is illustrated in Figure~\ref{fig:rand_iid_proof} and is rather simple.  
	 		The first step is the estimation of the winning probability $\omega$ of the single-round box defining the IID box, i.e., the unknown state~$\sigma_{Q_AQ_B}$.
			Alice and Bob play the $n$ games with each of their independent quantum boxes and collect the statistics. Denoting by $W_i$ the RV describing whether the~$i$'th game is won or not, the IID assumption implies that $W_1,W_2,\dots,W_n$ are, as well, IID RV. Thus, it follows from Chernoff's bound that the average $\frac{1}{n}\sum_i W_i$ is close to the expected winning probability $\mathbb{E}[W]$, which is no other than the winning probability $\omega$ of a single copy of the state (see Equation~\eqref{eq:win_prob_assump}). That is, 
				\[
				   \omega \approx \frac{1}{n} \sum_i W_i \;.
				\]
	 				
	 		The second step is to lower-bound the conditional smooth min-entropy $H^{\varepsilon}_{\min}(\mr{A}|\mr{E})$ as a function of $\omega$. Due to the IID assumption, we can do so using the quantum AEP presented as Theorem~\ref{thm:quant_aep} above. Specifically, for large enough $n$ we have
	 		\begin{equation} \label{eq:rc_ent_rate_iid}
	 		\begin{split}
	 			H^{\varepsilon}_{\min} (\mr{A}|\mr{E})  & \approx n H(A|E) \\
	 			& \geq n f(\omega)  \;,
	 		\end{split}
	 		\end{equation}
	 		where $f(\omega)$ is some function of $\omega$ that lower-bounds  the conditional von Neumann entropy $H(A|E)_{\sigma}$ for any state $\sigma$ with the estimated winning probability $\omega$.\footnote{We previously wrote $\sigma$ as the tripartite state $\sigma_{Q_AQ_BE}$ while here we are referring also to the classical register $A$. What is meant by this notation is that $\sigma$ is a state which can lead to winning probability $\omega$ when measured with some given measurements $\left\{M_a^x\right\}$ and $\left\{M_b^y\right\}$. The result of measuring $Q_A$ with $\left\{M_a^x\right\}$ defines the RV $A$.}  For the CHSH game, such a function $f(\omega)$ was given in Lemma~\ref{lem:single_round_secrecy} as part of the discussion of single-round boxes.

	 		In certain protocols one would like to use different copies of the boxes in different ways. For example, in device-independent quantum key distribution the protocol includes ``test rounds'' and ``generation rounds''. Alice's usage of the box in a test round may be different than her usage in a generation round. The winning probability~$\omega$ is estimated from the statistics collected in the test rounds, as discussed above.
	 		Using the IID assumption we can conclude that the other boxes, utilised in the generation rounds, could have also been used to win the game with probability $\omega$, even though Alice and Bob do not test these boxes. 
	 		
	 		Clearly, the IID assumption plays a crucial role in the above proof sketch; it allows us to talk about a \emph{single-round box}, estimate its winning probability in a meaningful way, and, furthermore, to bound the total amount of smooth min-entropy of the outputs as the number of games played times the von Neumann entropy of the output of a single game. In total, the IID assumption allows us to reduce the analysis of the multi-round box to that of a single-round box -- the ``physics'' enters the analysis only in the single-round statement (e.g., Lemma~\ref{lem:single_round_secrecy}) while the rest is done using standard mathematical tools such as Chernoff's bound and the AEP.

	 		\subsubsection{Quantum key distribution key rates}

 				The main fundamental difference between device-independent randomness certification and device-independent quantum key distribution is that in the latter Alice and Bob should share identical secret keys in the end of the protocol. To this end, they need to apply an additional classical post-processing step, namely, error correction. The goal of the error correction step is to reconcile the differences between Alice's and Bob's keys so they share the same final key with high probability. 
 				
 				In classical error correction protocols utilising one-way communication, Alice sends some classical information about her key to Bob. This information, together with all of Bob's prior information, helps Bob conclude which key Alice is most likely to hold. If the information sent is not sufficient in order for Bob to derive a conclusion, the parties abort the protocol. 
 				Since Alice sends the additional information to Bob over a public (but authenticated) classical channel, this information also leaks to the adversary and hence increases her knowledge about Alice's key. In other words, the leakage reduces the conditional smooth min-entropy -- we now need to consider $H^{\varepsilon}_{\min} (\mr{A}|\mr{E}O)$, where $O$ denotes the leaked information, instead of~$H^{\varepsilon}_{\min} (\mr{A}|\mr{E})$ appearing in Equation~\eqref{eq:rc_ent_rate_iid}. 
 				
 				Notice the resulting tradeoff. To get good key rates, we wish to leak as little information as possible to the adversary (so we do not reduce the min-entropy by too much). On the other hand, we want the error correction step to succeed when Alice and Bob use the \emph{honest} box\footnote{If the box is malicious or simply noisier than we wished for, we anyhow expect the protocol to abort. Thus, we only ask that the error correction protocol does not abort with high probability when the honest implementation of the devices is used since, otherwise, it will affect the \emph{completeness} of the protocol.} and, thus, Alice needs to send a sufficient amount of information that will allow Bob to correct the errors. 
 				We therefore wish to minimise the amount of leakage needed for successful error correction. As explained in Section~\ref{sec:pre_di_crypt}, this turns out to be quantified by the conditional smooth zero-entropy~$H^{\varepsilon}_{0} (\mr{A}|\mr{B})$~\cite{renner2005simple}, which is closely related to the conditional smooth max-entropy $H^{\varepsilon}_{\max} (\mr{A}|\mr{B})$. 
 				
 				In many cases the honest box, which also incorporates the considered honest noise model, is chosen to be an IID box. (For example, a common choice is a box describing $n$ independent pairs of maximally entangled states which are being distributed over an IID noisy quantum channel.)
 				Hence, one can use the AEP to get
 				\begin{equation}\label{eq:ec_using_aep}
 					H^{\varepsilon}_{\max} (\mr{A}|\mr{B}) \approx n H(A|B) 
 				\end{equation}
 				and by this upper-bound the amount of leakage due to error correction. 
 				All and all, under the IID assumption and for sufficiently large $n$, the key rate is governed by 
 				\begin{align}
 					r & \gtrapprox \frac{1}{n} \left(H^{\varepsilon}_{\min} (\mr{A}|\mr{E}) - H^{\varepsilon}_{\max} (\mr{A}|\mr{B}) \right)\\
 						& \gtrapprox H(A|E)-H(A|B) \;.\label{eq:iid_asym_keyrate}
 				\end{align}
 							
 				\sloppy
 				Equation~\eqref{eq:iid_asym_keyrate} is usually referred to as the DW-formula since it first appeared in~\cite[Theorem 2.1]{devetak2005distillation}. 
 				In~\cite{renner2005simple}, the smooth entropies where used to describe the optimal key rates without employing the IID assumption and, by the use of the AEP, the results of~\cite{renner2005simple} imply  Equation~\eqref{eq:iid_asym_keyrate} (as we sketched above). 
 				Interestingly, \cite[Theorem 2.8]{devetak2005distillation} states that, up to some possible classical post-processing, Equation~\eqref{eq:iid_asym_keyrate} is tight for any protocol utilising error correction with one-way communication.  
 				
 				Two last remarks are in order. 
 				Firstly, we would like to emphasise that Equation~\eqref{eq:ec_using_aep} does not rely on the IID \emph{assumption} that we are making in order to simplify the soundness analysis. Here we are allowed to use the AEP since we \emph{choose} to consider an IID box as our \emph{honest} box. Other choices can also be made (if one, for example, wishes to analyse the protocol under a different honest noise model) and then the AEP might no longer be relevant. In Chapter~\ref{ch:crypto_showcase} we will drop the IID assumption used for the soundness analysis but will still choose an IID honest implementation for the completeness analysis.
 				
 				Secondly, since an adversary limited to preparing IID boxes is weaker than one that can make general multi-round boxes, tight key rates achieved under the IID assumption act as upper-bounds on the achievable key rates in the general setting. Thus, by calculating key rates using Equation~\eqref{eq:iid_asym_keyrate} we usually already get a feeling of what is the best we can hope for when performing the general analysis. Indeed,~\cite{pironio2009device} used Equation~\eqref{eq:iid_asym_keyrate} to derive tight key rates for device-independent quantum key distribution under the IID assumption  for $n\rightarrow\infty$. These will act as an upper-bound when considering the most powerful quantum adversary in Chapter~\ref{ch:crypto_showcase}.

	\section{Beyond IID}
	
		In this chapter we studied the behaviour of IID boxes and saw how the main ingredient in an analysis of IID boxes is the analysis of a single-round box. 
		In a way, one can say that once we understand the behaviour of a single-round box, we understand the ``physics'', or the essence, of the problem at hand. 
		Unfortunately, IID boxes are far from being the most general ones and so we are enforced to go beyond the IID analysis and consider more complicated objects, namely, the different types of multi-round boxes that we encountered in Chapter~\ref{ch:multi_box}.

\begin{figure}
			\centering
			\includegraphics[width=\textwidth]{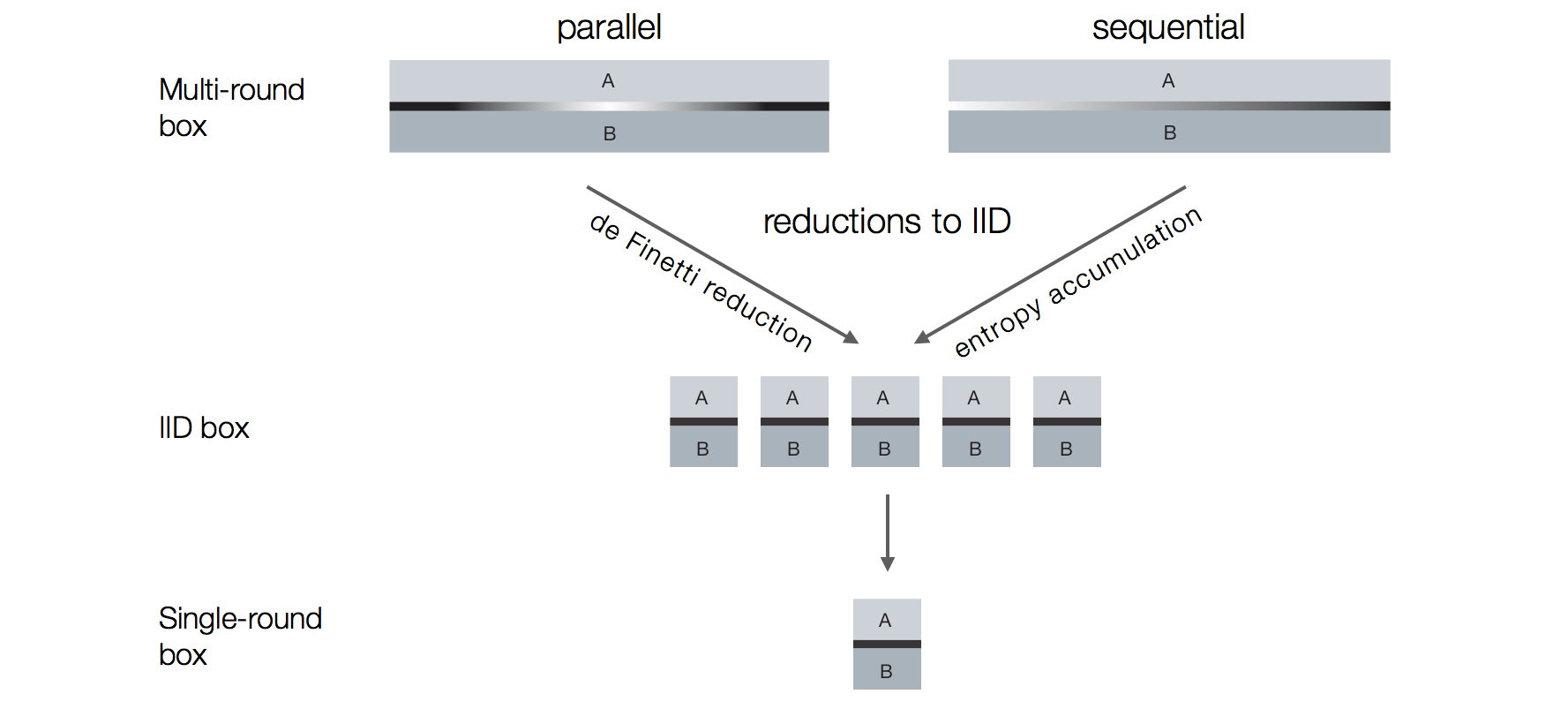}
			\caption{The big picture. The single-round box, at the bottom of the figure, is the simplest object to consider (Chapter~\ref{ch:single_round_box}). The IID box consists of many copies of the single-round box and, thus, can be easily analysed once we understand the behaviour of the single-round box (Chapter~\ref{ch:iid_assumption}). 
			The multi-round boxes are the most complex objects (Chapter~\ref{ch:multi_box}).
			``Reductions to IID'' techniques can be used to simplify the analysis of multi-round boxes by \emph{reducing} it to that of IID boxes. The ``de Finetti reduction'' technique (Chapter~\ref{ch:reductions_par}) is used when dealing with parallel boxes while the ``entropy accumulation theorem'' (Chapter~\ref{ch:reductions_seq}) is relevant for sequential boxes.
			In total, with the help of the different reductions, the main thing to study when considering device-independent information processing tasks is the behaviour of single-round boxes.}
			\label{fig:iid_reduc_big}
		\end{figure}

		As explained in Chapter~\ref{ch:intro}, it is the goal of this thesis to show that the analysis of IID boxes can be almost directly extended, at least in some cases, to the analysis of multi-round boxes via a \emph{reduction to IID}.
		In the following chapters we  present two techniques that can be used to reduce the analysis of parallel and sequential boxes to that of IID boxes; see Figure~\ref{fig:iid_reduc_big}.
		Specifically, Chapter~\ref{ch:reductions_par} deals with a technique called ``de Finetti reduction'' that relates permutation invariant parallel boxes to IID boxes~\cite{arnon2013finetti}. 
		Chapter~\ref{ch:reductions_seq} presents the so called ``entropy accumulation theorem'' that relates sequential boxes, fulfilling certain Markov-chain conditions, to IID ones~\cite{dupuis2016entropy}. 
		
		With the help of those techniques one can show that, in certain scenarios, the analysis of IID boxes is sufficient without loss of generality. 
		This is not to say that all multi-round boxes are IID boxes; clearly this is not the case. Instead, we claim that \emph{even though} there exist multi-round boxes that can not be described as IID boxes, one can sometimes restrict the attention solely to IID boxes and the rest will follow. 
		This will be clarified with the aid of our showcases in Chapters~\ref{ch:par_rep_showcase} and~\ref{ch:crypto_showcase}.



\chapter{Reductions to IID: parallel interaction}\label{ch:reductions_par}

	Multi-round parallel boxes, discussed in Section~\ref{sec:parallel_mr_boxes}, can display an almost arbitrary behaviour and hence are complicated to analyse. 
	However, some additional structure of the boxes can be assumed when certain types of symmetries are present in the considered information processing task.  
	In this chapter we focus on the analysis of parallel boxes that are \emph{permutation invariant}. Permutation invariance is an inherent symmetry in many information processing tasks, device-independent tasks among them. Thus, analysing permutation invariant boxes (as defined below) is of special interest. 
	
	A well known family of tools used to study permutation invariant systems\footnote{Depending on the context, the term system may refer to a probability distribution, a quantum state, or a box.} is the family of ``de Finetti-type theorems''. A de Finetti-type theorem is any theorem that relates (in one way or another) \emph{permutation invariant systems} to a more structured system, having the form of a \emph{convex combination of IID systems}, called a de Finetti system (or state). 
	The relation given by the theorem can be used, in certain cases, to argue that instead of analysing permutation invariant systems one can restrict the attention to the simpler to analyse (convex combination of) IID systems.
	A de Finetti theorem therefore acts as a reduction to IID. 
	
	\sloppy
	The first de Finetti theorem~\cite{deFinetti69} established that the collection of infinitely exchangeable sequences, i.e., distributions on infinite strings that are invariant under all permutations, exactly coincides with the collection of all convex combinations of IID distributions. 
	Subsequent results gave quantitative bounds of different forms~\cite{DiaconisF80finite, raggio1989quantum,caves2002unknown,renner2007symmetry,christandl2007one,christandl2009postselection,christandl2009finite,brandao2013quantum}. 
	de Finetti-type theorems had proven to be useful in various proofs. The quantum de Finetti theorems, for example, enable a substantially simplified analysis of many quantum information tasks such as quantum cryptography \cite{christandl2009postselection,leverrier2014composable}, tomography~\cite{christandl2012reliable}, channel capacities~\cite{berta2011reverse} and complexity~\cite{brandao2013quantum}.


	The de Finetti theorems listed above cannot be used in the device-independent setting for various reasons.\footnote{The mentioned theorems rely on some initial subsystem structure and/or a bound on the dimension of the subsystems. In the device-independent setting one cannot start with such assumptions regarding the considered boxes in general.}
	In this chapter we present a de Finetti-type theorem, which was introduced in~\cite{arnon2013finetti}, that is applicable when working with parallel boxes. 
	Our de Finetti theorem, termed ``de Finetti reduction for correlations'', is then used in the analysis of one of our showcases, namely, non-signalling parallel repetition, in Chapter~\ref{ch:par_rep_showcase}.
	
	The chapter is arranged as follows. We start by explaining the notion of permutation invariance in the device-independent context in Section~\ref{sec:perm_inva}. 
	The de Finetti reduction is presented and proven in Section~\ref{sec:dF_reductions}. Section~\ref{sec:using_dF_red} exemplifies how the reductions can be used in two different general ways (while Chapter~\ref{ch:par_rep_showcase} deals with a specific application). The theorems proven in Section~\ref{sec:using_dF_red} clarify in what sense we think of a de Finetti reduction as a reduction to IID in the device-independent setting.

	In accordance with the rest of the thesis, the chapter focuses only on the case of two parties. All the statements can be extended to any number of parties, as can be seen in~\cite{arnon2013finetti}.

	\section{Permutation invariance}\label{sec:perm_inva}

		As mentioned above, we are interested in considering permutation invariant parallel multi-round boxes. 
		Let $n$ be the number of games that can be played with the parallel box of interest  $\P_{\mr{A}\mr{B}|\mr{X}\mr{Y}}$. A~permutation $\pi$ is a bijective function $\pi: [n] \rightarrow [n]$. 
		We denote $\pi(\mr{x})=x_{\pi^{-1}(1)},x_{\pi^{-1}(2)},\dots,x_{\pi^{-1}(n)}$  and similarly for $\pi(\mr{y})$, $\pi(\mr{a})$, and $\pi(\mr{b})$.
		A~permutation invariant box\footnote{The definition and the derived theorem are independent of the nature of the box, i.e., if it is classical, quantum, non-signalling, or even signalling. This will be addressed in Section~\ref{sec:dF_reductions}.} is defined as follows. 
		\begin{defn}[Permutation invariant box]\label{def:permutation}
			Given a parallel multi-round box $\P_{\mr{A}\mr{B}|\mr{X}\mr{Y}}$, defined over $\mathcal{X}^n,\mathcal{Y}^n,\mathcal{A}^n,\mathcal{B}^n$, and a permutation $\pi:[n]\rightarrow[n]$ we denote by $\P_{\mr{A}\mr{B}|\mr{X}\mr{Y}}\circ\pi$ the box defined by 
			\begin{equation}\label{eq:perm_def}
				\forall \mr{a},\mr{b},\mr{x},\mr{y} \quad \left(\P_{\mr{A}\mr{B}|\mr{X}\mr{Y}}\circ\pi \right) ( \mr{a},\mr{b}|\mr{x},\mr{y})=\P_{\mr{A}\mr{B}|\mr{X}\mr{Y}}(\pi(\mr{a}),\pi(\mr{b})|\pi(\mr{x}),\pi(\mr{y})) \;.
			\end{equation}
			A parallel multi-round box $\P_{\mr{A}\mr{B}|\mr{X}\mr{Y}}$ is said to be permutation invariant if and only~if 
			\[
			\forall \pi \quad P_{\mr{A}\mr{B}|\mr{X}\mr{Y}} = \P_{\mr{A}\mr{B}|\mr{X}\mr{Y}}\circ\pi \;.
			\]
	\end{defn}

		\begin{figure}
		\centering
		\includegraphics[width=0.8\textwidth]{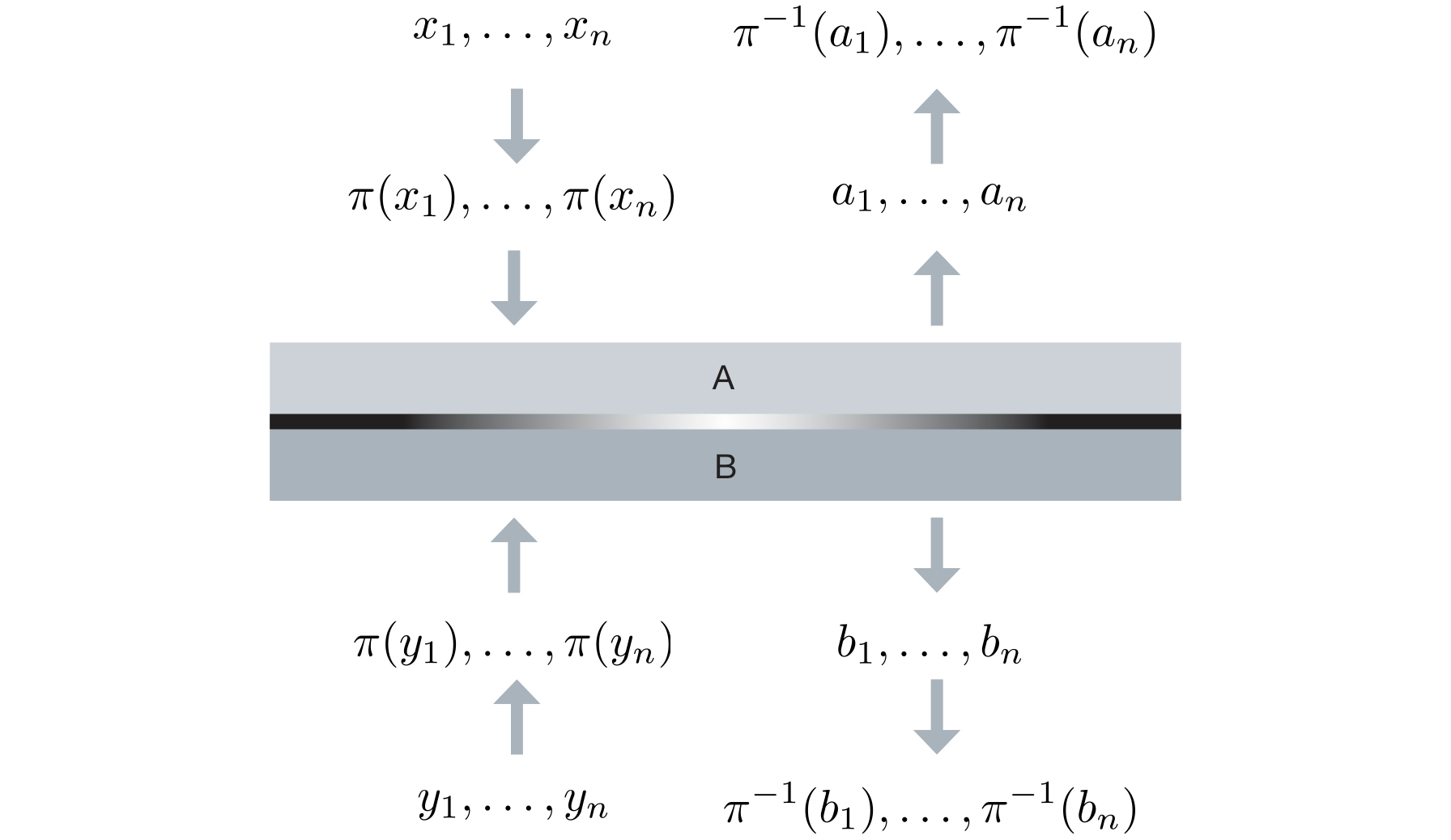}
		\caption{Permutation of a box $\P_{\mr{A}\mr{B}|\mr{X}\mr{Y}}$.  The permuted box, $\P_{\mr{A}\mr{B}|\mr{X}\mr{Y}}\circ\pi$ acts by first applying the permutation $\pi$ on the inputs, then producing the outputs using the initial box $\P_{\mr{A}\mr{B}|\mr{X}\mr{Y}}$, and lastly applying the inverse permutation on the outputs.
			The input output distribution of the box is then defined according to Equation~\eqref{eq:perm_def}. A box is said to be permutation invariant if for all $\pi$, $\P_{\mr{A}\mr{B}|\mr{X}\mr{Y}} = \P_{\mr{A}\mr{B}|\mr{X}\mr{Y}}\circ\pi$.}
		\label{fig:perm_box}
	\end{figure}

	Figure~\ref{fig:perm_box} illustrate the action of permuting a parallel box. The action of the permuted box can be understood as follows: First, the box applies the permutation~$\pi$ on the inputs. Second, it uses the initial box $\P_{\mr{A}\mr{B}|\mr{X}\mr{Y}}$ to produce the intermediate outputs. Lastly, it applies the inverse permutation $\pi^{-1}$ on the intermediate outputs and returns these final strings as the ultimate outputs.
	Note that only the inputs and the outputs of the box are being permuted, all using the same permutation $\pi$. In particular, we do not permute the parties, that is, Alice and Bob do not swap their inputs and outputs with one another.

	As we are merely permuting the classical inputs and outputs, the box itself need not to have a subsystem structure. That is, we do not require, e.g., $\P_{A_1|X_1}$ to be a valid system (i.e., a conditional probability distribution). This is in contrast to, e.g., quantum de Finetti-type theorems such as~\cite{renner2007symmetry,christandl2009postselection}, where the permutation is applied on the quantum states themselves.\footnote{In a quantum de Finetti statement a permutation takes a state $\ket{\phi_1}\otimes\dots\ket{\phi_n} $ to  $\ket{\phi_{\pi^{-1}(1)}}\otimes\dots\ket{\phi_{\pi^{-1}(n)}}$. That is, the quantum states themselves are being permuted. 
	}
	This distinction is relevant when wishing to discuss general parallel boxes (recall Section~\ref{sec:parallel_mr_boxes}).
		
	In some applications (e.g., the showcase considered in Chapter~\ref{ch:par_rep_showcase}) one can easily show that it is sufficient to consider permutation invariant boxes without loss of generality. 
	If this is not the case, it is also possible to \emph{enforce} permutation invariance. 
	A protocol, for example, can be modified to enforce the symmetry by adding a step in which a random permutation is applied\footnote{Depending on the considered scenario, the application of the permutation may be a purely theoretical step or needs to be done in practice.} on the box and by this make it permutation invariant. Precisely: given \emph{any} parallel box $\P_{\mr{A}\mr{B}|\mr{X}\mr{Y}}$, let
	\begin{equation*}
		\tilde{\P}_{\mr{A}\mr{B}|\mr{X}\mr{Y}} = \frac{1}{n!} \sum_\pi \P_{\mr{A}\mr{B}|\mr{X}\mr{Y}} \circ \pi 
	\end{equation*}
	be the result of applying a permutation $\pi$, chosen uniformly at random out of all permutations, on the original box.  
	It can be easily verified that $\tilde{\P}_{\mr{A}\mr{B}|\mr{X}\mr{Y}}$ is indeed a permutation invariant box. 
	
	\section{de Finetti reductions for correlations}\label{sec:dF_reductions}

		A de Finetti-type theorem is any theorem that relates a permutation invariant system to a much more structured system called a de Finetti system. 
		In our context, we consider permutation invariant and de Finetti boxes. 
		A \emph{de Finetti box} is defined as follows. 
		\begin{defn}[de Finetti box\footnote{As previously mentioned, we focus on the case of two parties. The definition extends to any number of parties trivially.}]\label{defn:dF_box}
			A de Finetti box is any box of the form of a convex combination of IID boxes. That is, it is a box $\tau_{\mr{A}\mr{B}|\mr{X}\mr{Y}}$, defined over $\mathcal{X}^n,\mathcal{Y}^n,\mathcal{A}^n,\mathcal{B}^n$, such that
			\[
				\tau_{\mr{A}\mr{B}|\mr{X}\mr{Y}} = \int \O_{AB|XY}^{\otimes n} \mathrm{d}\O_{AB|XY} \;,
			\]
			 where $\mathrm{d}\O_{AB|XY}$ is some measure on the space of bipartite boxes over $\mathcal{A}$, $\mathcal{B}$, $\mathcal{X}$,  and~$\mathcal{Y}$ and $\O_{AB|XY}^{\otimes n}$ is the IID box  defined by $\O_{AB|XY}$, i.e.,
			\[
				\O_{AB|XY}^{\otimes n}(\mr{a},\mr{b}|\mr{x},\mr{y}) = \prod_{i\in[n]} \O_{AB|XY} (a_i,b_i|x_i,y_i)\;.
			\]
		\end{defn}
	
		As seen from the above definition, by choosing different measures $\mathrm{d}\O_{AB|XY}$ we define different de Finetti boxes. Depending on the measure, $\tau_{\mr{A}\mr{B}|\mr{X}\mr{Y}}$ may be classical, quantum, non-signalling, or even signalling between the two parties. 
		If the measure $\mathrm{d}\O_{AB|XY}$ assigns weight only to, e.g., non-signalling boxes $\O_{AB|XY}$, then the de Finetti box $\tau_{\mr{A}\mr{B}|\mr{X}\mr{Y}}$ is non-signalling as well. The other direction does not necessarily hold -- there are convex combinations of signalling boxes that result in over-all non-signalling boxes.
		
		A de Finetti reduction is a de Finetti-type theorem of a specific form: it sets an \emph{inequality relation} between any permutation invariant box to a certain de Finetti box. 
		Specifically, the following theorem is a de Finetti reduction for any permutation invariant conditional probability distribution~\cite{arnon2013finetti}.\footnote{In~\cite{arnon2013finetti}, a more general version of Theorem~\ref{thm:dF_conditional} was proven, in which further symmetries of $\P_{\mr{A}\mr{B}|\mr{X}\mr{Y}}$ (on top of permutation invariance) can be exploited to construct more structured de Finetti boxes and prove de Finetti reductions with improved parameters. Theorem~\ref{thm:dF_conditional} was then derived as a corollary. To keep things (relatively) concise, we present in this thesis a direct proof of Theorem~\ref{thm:dF_conditional}.}
		\begin{thm}[de Finetti reduction for conditional probability distributions]\label{thm:dF_conditional}
			For any $\mathcal{X}$, $\mathcal{Y}$, $\mathcal{A}$, $\mathcal{B}$, and $n$ there exists a de Finetti box $\tau_{\mr{A}\mr{B}|\mr{X}\mr{Y}}$, defined over $\mathcal{X}^n,\mathcal{Y}^n,\mathcal{A}^n,\mathcal{B}^n$, 
			such that for every permutation invariant box $\P_{\mr{A}\mr{B}|\mr{X}\mr{Y}}$  
			\begin{equation}\label{eq:general_dF_red_ineq}
				\forall \mr{a},\mr{b},\mr{x},\mr{y} \quad \P_{\mr{A}\mr{B}|\mr{X}\mr{Y}}( \mr{a},\mr{b}|\mr{x},\mr{y}) \leq (n+1)^{|\mathcal{X}||\mathcal{Y}|\left(|\mathcal{A}||\mathcal{B}|-1\right)} \; \tau_{\mr{A}\mr{B}|\mr{X}\mr{Y}} (\mr{a},\mr{b}|\mr{x},\mr{y})  \;.
			\end{equation}
		\end{thm}

		To see why Theorem~\ref{thm:dF_conditional} is not trivial and what needs to be done to prove it, let us first consider a ``bad choice'' of a de Finetti box, $\tau^{\text{bad}}_{\mr{A}\mr{B}|\mr{X}\mr{Y}}$. Imagine that we choose our de Finetti box to be the uniform distribution over $\mathcal{A}^n\times\mathcal{B}^n$ for all $\mr{x}$ and~$\mr{y}$. With this chocie, $\tau^{\text{bad}}_{\mr{A}\mr{B}|\mr{X}\mr{Y}} (\mr{a},\mr{b}|\mr{x},\mr{y})=\left(|\mathcal{A}||\mathcal{B}|\right)^{-n}$ for all $\mr{a}$, $\mr{b}$, $\mr{x}$, and $\mr{y}$. Then, the only inequality relation that holds is
		\[
				\forall \mr{a},\mr{b},\mr{x},\mr{y} \quad \P_{\mr{A}\mr{B}|\mr{X}\mr{Y}}( \mr{a},\mr{b}|\mr{x},\mr{y}) \leq \left(|\mathcal{A}||\mathcal{B}|\right)^{n} \; \tau^{\text{bad}}_{\mr{A}\mr{B}|\mr{X}\mr{Y}} (\mr{a},\mr{b}|\mr{x},\mr{y})  \;,
		\]
		i.e., a relation with a pre-factor exponential in $n$. 
		By choosing a ``good'' de Finetti box, we are able to get a pre-factor polynomial in $n$ instead; this is crucial for applications of de Finetti reductions. 
		In Section~\ref{sec:using_dF_red} we show how Theorem~\ref{thm:dF_conditional} can be utilised as a reduction to IID in certain scenarios.\footnote{A curious reader may already take a glimpse of Theorems~\ref{thm:dF_test_bound} and~\ref{thm:dF_diamond_norm_bound}.}
		
		The proof of the theorem proceeds in two steps. In the first, an explicit de Finetti box $\tau_{\mr{A}\mr{B}|\mr{X}\mr{Y}}$ is constructed and a lower-bound on its entries is calculated. 
		In the second step the permutation invariance of $\P_{\mr{A}\mr{B}|\mr{X}\mr{Y}}$ is used to upper-bound its entries. The theorem follows by combining the two bounds. 
		
		In the proofs below we use the following notation. 
		\begin{enumerate}
			\item $|\mathcal{X}||\mathcal{Y}| = l$ and we identify each pair $(x,y)\in \mathcal{X}\times \mathcal{Y}$ with a label $j\in[l]$ by writing $(x,y)=j$.
			\item $|\mathcal{A}||\mathcal{B}| = m$ and we identify each pair $(a,b)\in \mathcal{A}\times \mathcal{B}$ with a label $k\in[m]$ by writing $(a,b)=k$.
			\item For all $j\in[l]$  and $k\in[m]$, $p^j_k\in[0,1]$ such that $\sum_k p^j_k =1$. 
			\item For all $j\in[l]$  and $k\in[m]$, $c^j_k=1-\sum_{t<k} p^j_t$.
			\item  For all $\mr{x}$, $\mr{y}$, and $j\in[l]$, $n^j=|\left\{i : (x_i,y_i)=j\right\}|$, i.e., $n^j$ denotes the number of indices  of $(\mr{x},\mr{y})$ in which the type of inputs is $(x,y)=j$.
			\item For all $\mr{x}$, $\mr{y}$, $\mr{a}$, $\mr{b}$, $j\in[l]$, and $k\in[m]$, $n^j_k=|\left\{i : (x_i,y_i)=j \land (a_i,b_i)=k \right\}|$, i.e., $n^j_k$ denotes the number of indices of $(\mr{x},\mr{y},\mr{a},\mr{b})$ in which the type of inputs  is $(x,y)=j$ and the type of outputs is $(a,b)=k$.
		\end{enumerate}	
		Note that by definition:
		\begin{enumerate}
			\item For all  $j\in[l]$  and $k\in[m-1]$, $p^j_k\in[0,c^j_k]$ and $p^j_m = c^j_m$.
			\item For all  $j\in[l]$, $n^j_m = n^j - \sum_{k=1}^{m-1} n^j_k$.
		\end{enumerate}
		
		According to Definition~\ref{defn:dF_box}, a de Finetti box is defined via the choice of measure~$\mathrm{d}\O_{AB|XY}$. 
		We think of a bipartite box $\O_{AB|XY}$ as a set of probabilities $p^j_k$, with the identification $\O_{AB|XY}(a,b|x,y) = p^j_k$ for $(x,y)=j$ and $(a,b)=k$.
		Thus, we can define a measure over $\O_{AB|XY}$ by a measure over the probabilities $p^j_k$. 
		Our chosen measure is
		\[
			\mathrm{d}\O_{AB|XY} = \prod_{j=1}^l 
			\frac{\mathrm{d}p^j_1}{c^j_1} \dots \frac{\mathrm{d}p^j_{m-1}}{c^j_{m-1}}\;,
		\]
		where $\mathrm{d}p^j_k$ is the uniform measure over $[0,c^j_k]$ for $c^j_k$ defined above.
		The resulting de Finetti box is given by
		\begin{equation}\label{eq:dF_general_explicit}
		\begin{split}
			&\tau_{\mr{A}\mr{B}|\mr{X}\mr{Y}} (\mr{a},\mr{b}|\mr{x},\mr{y}) = \int \O_{AB|XY}^{\otimes n} \mathrm{d}\O_{AB|XY} \\
			&\qquad  = \prod_{j=1}^l 
			\left[ \int_0^{c^j_1} \frac{\mathrm{d}p^j_1}{c^j_1} \left( p^j_1 \right)^{n^j_1} \right]
			\dots
			\left[ \int_0^{c^j_{m-1}} \frac{\mathrm{d}p^j_{m-1}}{c^j_{m-1}} \left( p^j_{m-1} \right)^{n^j_{m-1}} \right] \\
			& \qquad \qquad \quad  \cdot \left( p^j_m \right)^{n^j - \sum_{k=1}^{m-1} n^j_k} \;.
		\end{split}
			\end{equation}
		
		The measure $\mathrm{d}\O_{AB|XY}$ assigns some weight to \emph{all} conditional probability distributions $\O_{AB|XY}$. As a result, the de Finetti box in Equation~\eqref{eq:dF_general_explicit} is \emph{signalling}. 
		This is discussed in Section~\ref{sec:dF_imp_res} below.

		The following lower-bound on the entries of the above de Finetti box is proven in Appendix~\ref{appsec:proof_dF_red}:
		\begin{lem}\label{lem:dF_up_bound}
			For all $\mr{a}$, $\mr{b}$, $\mr{x}$, and $\mr{y}$, 
			\[
				\tau_{\mr{A}\mr{B}|\mr{X}\mr{Y}} (\mr{a},\mr{b}|\mr{x},\mr{y}) \geq \prod_{j=1}^l {n^j \choose {n^j_1, \dots, n^j_m}}^{-1} \frac{1}{(n^j + 1)^{m-1}} \;,
			\]
			where $\tau_{\mr{A}\mr{B}|\mr{X}\mr{Y}}$ is as in Equation~\eqref{eq:dF_general_explicit}.
		\end{lem}
		
		Next, we exploit the permutation invariance of $\P_{\mr{A}\mr{B}|\mr{X}\mr{Y}}$ to prove the following upper-bound on it:
		\begin{lem}\label{lem:perm_inv_low_bound}
			For every permutation invariant box $\P_{\mr{A}\mr{B}|\mr{X}\mr{Y}}$, as in Definition~\ref{def:permutation}, and for all $\mr{a}$, $\mr{b}$, $\mr{x}$, and $\mr{y}$,
			\[
				\P_{\mr{A}\mr{B}|\mr{X}\mr{Y}} (\mr{a},\mr{b}|\mr{x},\mr{y}) \leq \prod_{j=1}^l {n^j \choose {n^j_1, \dots, n^j_m}}^{-1} \;.
			\]
		\end{lem}
		\begin{proof}
			To prove the lemma we bound the value of a specific entry $\P_{\mr{A}\mr{B}|\mr{X}\mr{Y}} (\mr{a},\mr{b}|\mr{x},\mr{y})$ by counting how many entries $\P_{\mr{A}\mr{B}|\mr{X}\mr{Y}} (\tilde{\mr{a}},\tilde{\mr{b}}|\mr{x},\mr{y})$ must have the same value as $\P_{\mr{A}\mr{B}|\mr{X}\mr{Y}} (\mr{a},\mr{b}|\mr{x},\mr{y})$ due to permutation invariance. 
			The normalisation of $\P_{\mr{A}\mr{B}|\mr{X}\mr{Y}}$ then implies a bound on the value of the entries.
			
			Denote 
			\[
				\mathcal{N}(\mr{a},\mr{b},\mr{x},\mr{y}) = \Big| \left\{ (\tilde{\mr{a}},\tilde{\mr{b}})\in \mathcal{A}\times\mathcal{B} : \P_{\mr{A}\mr{B}|\mr{X}\mr{Y}} (\tilde{\mr{a}},\tilde{\mr{b}}|\mr{x},\mr{y}) = \P_{\mr{A}\mr{B}|\mr{X}\mr{Y}} (\mr{a},\mr{b}|\mr{x},\mr{y}) \right\} \Big| \;.
			\]
			The permutation invariance of $\P_{\mr{A}\mr{B}|\mr{X}\mr{Y}}$ implies that $\mathcal{N}(\mr{a},\mr{b},\mr{x},\mr{y})$ is lower-bounded by the number permutations $\pi$ for which $\pi(\mr{x})=\mr{x}$, $\pi(\mr{y})=\mr{y}$.
			To keep $\pi(\mr{x})=\mr{x}$ and~$\pi(\mr{y})=\mr{y}$, the relevant permutations $\pi$ are only allowed to permute indices with the same input type $(x,y)$. The number of such permutations is exactly $\prod_{j=1}^l {n^j \choose {n^j_1, \dots, n^j_m}}$. 
			Thus, 
			\[
				\mathcal{N}(\mr{a},\mr{b},\mr{x},\mr{y}) \geq \prod_{j=1}^l {n^j \choose {n^j_1, \dots, n^j_m}}
			\]
			 and 
			\[
				\P_{\mr{A}\mr{B}|\mr{X}\mr{Y}} (\mr{a}\mr{b}|\mr{x}\mr{y}) \leq \frac{1}{\mathcal{N}(\mr{a},\mr{b},\mr{x},\mr{y})} \leq \prod_{j=1}^l {n^j \choose {n^j_1, \dots, n^j_m}}^{-1} \;. \qedhere
			\]
			
		\end{proof}

		\begin{proof}[Proof of Theorem~\ref{thm:dF_conditional}.]
				Using Lemmas~\ref{lem:dF_up_bound} and~\ref{lem:perm_inv_low_bound} one can easily prove Theorem~\ref{thm:dF_conditional}. For all  $\mr{a}$, $\mr{b}$, $\mr{x}$, and $\mr{y}$, 
				\begin{align*}
					\frac{\P_{\mr{A}\mr{B}|\mr{X}\mr{Y}} (\mr{a},\mr{b}|\mr{x},\mr{y})}{\tau_{\mr{A}\mr{B}|\mr{X}\mr{Y}} (\mr{a},\mr{b}|\mr{x},\mr{y})} &\leq \frac{\prod_{j=1}^l {n^j \choose {n^j_1, \dots, n^j_m}}^{-1}}{\prod_{j=1}^l {n^j \choose {n^j_1, \dots, n^j_m}}^{-1} (n^j + 1)^{-(m-1)}} \\
					&\leq \prod_{j=1}^l (n^j +1)^{m-1} \\
					&\leq (n+1)^{l(m-1)} \;. \qedhere
				\end{align*}
		\end{proof}

		To end this section let us give a last remark regarding 	Theorem~\ref{thm:dF_conditional}. Notice the order of the quantifiers; \emph{there exists one} de Finetti box for which Equation~\eqref{eq:general_dF_red_ineq} holds for \emph{all} permutation invariant box. For the purpose of applications, one could also imagine a different statement in which for each permutation invariant box a de Finetti box is constructed (i.e., different permutation invariant boxes may be related to different de Finetti boxes). Such a statement has the potential of improving the obtained parameters and simplifying the use of the reduction in applications (see also~\cite{arnon2013finetti} for examples of such statements).

	\section{Ways of using the reductions}\label{sec:using_dF_red}
		
		The main motivation for considering de Finetti reductions as in Theorem~\ref{thm:dF_conditional} is to allow us to simplify the analysis of device-independent information processing tasks. 
		However, it is a priori not clear how one can bring an inequality as that in Equation~\eqref{eq:general_dF_red_ineq} into work.
		The aim of this section is to exemplify the usage of the inequality in a mathematical way by considering two types of abstract applications.  
		Chapter~\ref{ch:par_rep_showcase} discusses a more concrete application of the reduction to prove a non-signalling parallel repetition theorem.
		
		To derive the results presented in this section we use an alternative, but equivalent, version of the de Finetti reduction; this is the topic of Section~\ref{sec:dF_post_selection} below. Sections~\ref{sec:dF_fail_prob_test} and~\ref{sec:dF_diam_norm} present two ways of using the de Finetti reduction via the alternative formulation. 
		
		\subsection{Post-selecting permutation invariant boxes}\label{sec:dF_post_selection}

			\begin{lem}\label{lem:dF_post_selection}
				There exists a de Finetti box $\tau_{\mr{A}\mr{B}|\mr{X}\mr{Y}}$  and a non-signalling extension\footnote{Note that $\tau_{\mr{A}\mr{B}|\mr{X}\mr{Y}}$ may be signalling, as in our previous statements. The fact that we are considering non-signalling extensions only means that the marginals $\tau_{\mr{A}\mr{B}|\mr{X}\mr{Y}}$ and $\tau_{C|Z}$ of $\tau_{\mr{A}\mr{B}C|\mr{X}\mr{Y}Z}$  are well defined.} of it (Definition~\ref{def:ns_extension}) to a larger box $\tau_{\mr{A}\mr{B}C|\mr{X}\mr{Y}Z}$ such that for every permutation invariant box $\P_{\mr{A}\mr{B}|\mr{X}\mr{Y}}$ there exists an input $z$ and an output of this input $c_z$ for which 
				\[
					\forall \mr{a},\mr{b},\mr{x},\mr{y} \quad \tau_{\mr{A}\mr{B}C|\mr{X}\mr{Y}Z} (\mr{a},\mr{b},c_z|\mr{x},\mr{y},z) = \frac{1}{(n+1)^{l(m-1)}}\P_{\mr{A}\mr{B}|\mr{X}\mr{Y}} (\mr{a},\mr{b}|\mr{x},\mr{y}) \;,
				\]
				where $l=|\mathcal{X}||\mathcal{Y}|$ and $m=|\mathcal{A}||\mathcal{B}|$.
			\end{lem}
			
\begin{figure}
				\begin{center}
				\begin{tikzpicture}
						
					\node at (1.5,0.5) {$\tau_{\mr{A}\mr{B}C|\mr{X}\mr{Y}Z}$};
					\node at (1,1.7) {$\mr{x},\mr{y}$};
					\draw[->] (1,1.5) -- (1,1);
					\node[red] at (2,1.7) {z};
					\draw[->,red] (2,1.5) -- (2,1);
					\draw  (0,0) rectangle (3,1);
					\node at (1,-0.7) {$\mr{a},\mr{b}$};
					\draw[->] (1,0) -- (1,-0.5);
					\node[red] at (2,-0.7) {$c_z$};
					\draw[->,red] (2,0) -- (2,-0.5);
					
					\node at (3.5,0.5) {=};
					
					\node at (5,0.5) {$\P_{\mr{A}\mr{B}|\mr{X}\mr{Y}}$};
					\node at (5,1.7) {$\mr{x},\mr{y}$};
					\draw[->] (5,1.5) -- (5,1);
					\draw (4,0) rectangle (6,1);
					\node at (5,-0.7) {$\mr{a},\mr{b}$};
					\draw[->] (5,0) -- (5,-0.5);
					  	
				\end{tikzpicture}
				\end{center}
			\caption{post-selecting a box $\P_{\mr{A}\mr{B}|\mr{X}\mr{Y}}$ from an extension of $\tau_{\mr{A}\mr{B}|\mr{X}\mr{Y}}$. Conditioned on the output $c_z$, the resulting box is $\P_{\mr{A}\mr{B}|\mr{X}\mr{Y}}$.}
			\label{fig:postselectionl}
			\end{figure}
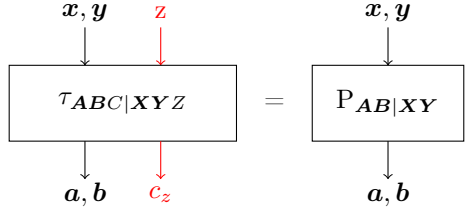

			This lemma states that there exists a de Finetti box $\tau_{\mr{A}\mr{B}|\mr{X}\mr{Y}}$ and a non-signalling extension of it $\tau_{\mr{A}\mr{B}C|\mr{X}\mr{Y}Z}$ such that any permutation invariant box $\P_{\mr{A}\mr{B}|\mr{X}\mr{Y}}$ can be \emph{post-selected} from it with probability greater or equal to $\frac{1}{(n+1)^{l(m-1)}}$. 
			When we say that  $\P_{\mr{A}\mr{B}|\mr{X}\mr{Y}}$ can be post-selected we mean that there exists an input $z$ to $\tau_{\mr{A}\mr{B}C|\mr{X}\mr{Y}Z}$ and an output $c_z$ of this input such that with probability $\tau_{C|Z}(c_z|z)\geq\frac{1}{(n+1)^{l(m-1)}}$ the resulting box (the ``post-measurement box'', using terminology borrowed from quantum physics) is $\P_{\mr{A}\mr{B}|\mr{X}\mr{Y}}$ (see Figure \ref{fig:postselectionl}). 
			Note that we consider a single extension $\tau_{\mr{A}\mr{B}C|\mr{X}\mr{Y}Z}$ of the box $\tau_{\mr{A}\mr{B}|\mr{X}\mr{Y}}$, and by choosing different inputs $z$ we can post-select different boxes $\P_{\mr{A}\mr{B}|\mr{X}\mr{Y}}$.
			
			It is easy to see how to derive Lemma \ref{lem:dF_post_selection} from Theorem~\ref{thm:dF_conditional} by using the formalism introduced in \cite{hanggi2009quantum,hanggi2010device} of partitions of a conditional probability distribution. We repeat here the relevant statements. 
			\begin{defn}
				A partition of a box $\Q_{\mr{A}\mr{B}|\mr{X}\mr{Y}}$ is a family of pairs $\left\{ \left( q_c,\Q^c_{\mr{A}\mr{B}|\mr{X}\mr{Y}} \right) \right\}_c$ where $q_c\geq 0$, $\sum_c q_c = 1$, and the boxes $\Q^c_{\mr{A}\mr{B}|\mr{X}\mr{Y}}$  are such that
				\[
					\Q_{\mr{A}\mr{B}|\mr{X}\mr{Y}} = \sum_c q_c \cdot \Q^c_{\mr{A}\mr{B}|\mr{X}\mr{Y}} \;.
				\]
			\end{defn}
			\begin{lem} [Lemma 9 in \cite{hanggi2009quantum}] \label{lem:extension_condition}
				Given a box $\Q_{\mr{A}\mr{B}|\mr{X}\mr{Y}}$, there exists a partition with element $\left( q_c,\Q^c_{\mr{A}\mr{B}|\mr{X}\mr{Y}} \right)$ if and only if 
				\[
					\forall \mr{a},\mr{b},\mr{x},\mr{y} \quad q_c\cdot \Q^c_{\mr{A}\mr{B}|\mr{X}\mr{Y}}(\mr{a},\mr{b}|\mr{x},\mr{y}) \leq \Q_{\mr{A}\mr{B}|\mr{X}\mr{Y}} (\mr{a},\mr{b}|\mr{x},\mr{y}) \;.
				\]
			\end{lem}
			
			\begin{lem}[Lemma 3.2 in \cite{hanggi2010device}] \label{lem:extension}
				Given a box $\Q_{\mr{A}\mr{B}|\mr{X}\mr{Y}}$, let $Z$ be the set of all partitions $\left\{ \left( q_{c_z}, \Q^{c_z}_{\mr{A}\mr{B}|\mr{X}\mr{Y}} \right) \right\}_{c_z}$ of $\Q_{\mr{A}\mr{B}|\mr{X}\mr{Y}}$. 
				Then, there exist a non-signalling extension $\Q_{\mr{A}\mr{B}C|\mr{X}\mr{Y}Z}$  of $\Q_{\mr{A}\mr{B}|\mr{X}\mr{Y}}$, an input $z$, and an output $c_z$ such that  
				\[
					\forall \mr{a},\mr{b},\mr{x},\mr{y} \quad \Q_{\mr{A}\mr{B}C|\mr{X}\mr{Y}Z}(\mr{a},\mr{b},c_z|\mr{x},\mr{y},z)=q_{c_z} \cdot \Q^{c_z}_{\mr{A}\mr{B}|\mr{X}\mr{Y}}(\mr{a},\mr{b}|\mr{x},\mr{y}) \;.
				\]
			\end{lem}
			
			Using the lemmas above and Theorem~\ref{thm:dF_conditional} we can now prove Lemma \ref{lem:dF_post_selection}. 
			\begin{proof}[Proof of  Lemma \ref{lem:dF_post_selection}.]
				The above lemmas together with Theorem~\ref{thm:dF_conditional} imply that for any permutation invariant box $\P_{\mr{A}\mr{B}|\mr{X}\mr{Y}}$, $\left( \frac{1}{(n+1)^{l(m-1)}},\P_{\mr{A}\mr{B}|\mr{X}\mr{Y}} \right) $ is an element of a partition of $\tau_{\mr{A}\mr{B}|\mr{X}\mr{Y}}$. Moreover, there exists a box $\tau_{\mr{A}\mr{B}C|\mr{X}\mr{Y}Z}$ and an input $z$ such that with probability $\frac{1}{(n+1)^{l(m-1)}}$ the resulting box is $\P_{\mr{A}\mr{B}|\mr{X}\mr{Y}}$:
				\[
						\forall \mr{a},\mr{b},\mr{x},\mr{y} \quad \tau_{\mr{A}\mr{B}C|\mr{X}\mr{Y}Z}(\mr{a},\mr{b},c_z|\mr{x},\mr{y},z) = \frac{1}{(n+1)^{l(m-1)}}\P_{\mr{A}\mr{B}|\mr{X}\mr{Y}} \;. \qedhere
				\]
			\end{proof}

			Lemma \ref{lem:dF_post_selection} is used in the following sections to illustrate two ways in which de Finetti reductions can be used in applications.
			
		\subsection{Failure probability of a test}\label{sec:dF_fail_prob_test}
		
			We start by considering the following abstract application. Let $\mathcal{T}$ be a test which interacts with a box $\P_{\mr{A}\mr{B}|\mr{X}\mr{Y}}$ and outputs ``success'' or ``fail'' with some probabilities. One can think about this test, which can be chosen according to the application being considered, as a way to quantify the success probability of a protocol when the box $\P_{\mr{A}\mr{B}|\mr{X}\mr{Y}}$ is given as input. For example, if one considers an estimation, or a tomography, protocol a test can be chosen to output ``success'' when the estimated box is close to the actual box \cite{christandl2009postselection}. 	Another type of test will be considered explicitly in Section~\ref{sec:approx_ns_marginals}.
			
				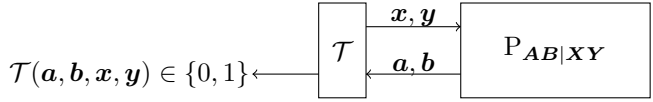
\begin{figure}
				\begin{center}
					\begin{tikzpicture}[scale=1.25]
					\begin{scope}[shift={(-0.5,-0.5)}]
						\node at (1,0.5) {$\P_{\mr{A}\mr{B}|\mr{X}\mr{Y}}$};
						\draw  (0,0) rectangle (2,1);
						
						\node at (-0.5,0.85) {$\mr{x},\mr{y}$};
						\draw[->] (-1,0.75) -- (0,0.75);
						\node at (-0.5,0.35) {$\mr{a},\mr{b}$};
						\draw[<-] (-1,0.25) -- (0,0.25);
						
						\draw (-1.5,0) rectangle (-1,1);
						\node at (-1.25,0.5) {$\mathcal{T}$};
						\draw[->] (-1.5,0.25) -- (-2.2,0.25);
						\node at (-3.5,0.25) {$\mathcal{T}(\mr{a},\mr{b},\mr{x},\mr{y})\in\{0,1\}$};
					\end{scope}
								  	
					\end{tikzpicture}
				\end{center}
			\caption{The test $\mathcal{T}$ interacts with  $\P_{\mr{A}\mr{B}|\mr{X}\mr{Y}}$ by supplying it with inputs $\mr{x},\mr{y}$ and collecting its outputs $\mr{a},\mr{b}$. The test then decides whether to output $0$ or $1$ depending on $\mr{x}$, $\mr{y}$, $\mr{a}$, and $\mr{b}$. If the output is $0$ then we say that the test failed.}
			\label{fig:dF_test_interaction}
			\end{figure}
		
			A test $\mathcal{T}$ interacts with  $\P_{\mr{A}\mr{B}|\mr{X}\mr{Y}}$ by supplying it with inputs $\mr{x},\mr{y}$, according to some probability distribution $\mathrm{Pr}_{\mathcal{T}}(\mr{x},\mr{y})$  over $\mathcal{X}^n\times\mathcal{Y}^n$, and collecting its outputs~$\mr{a},\mr{b}$. This is illustrated in Figure~\ref{fig:dF_test_interaction}. The test then decides whether to output~$0$ or~$1$ depending on~$\mr{x}$, $\mr{y}$, $\mr{a}$, and $\mr{b}$.
			Given a test $\mathcal{T}$, we denote by $\mathrm{Pr}_{\textup{fail}}(\P_{\mr{A}\mr{B}|\mr{X}\mr{Y}})$ the failure probability of the test, i.e., the probability that $\mathcal{T}$ outputs $0$ after interacting with~$\P_{\mr{A}\mr{B}|\mr{X}\mr{Y}}$:
			\begin{equation*}
				\mathrm{Pr}_{\textup{fail}}(\P_{\mr{A}\mr{B}|\mr{X}\mr{Y}})  = \sum_{\mr{x},\mr{y}} \mathrm{Pr}_{\mathcal{T}}(\mr{x},\mr{y}) \sum_{\mr{a},\mr{b} : \mathcal{T}(\mr{a},\mr{b},\mr{x},\mr{y})=0} \P_{\mr{A}\mr{B}|\mr{X}\mr{Y}}(\mr{a},\mr{b}|\mr{x},\mr{y}) \;.
			\end{equation*}
			The event of failing the test can therefore be defined as an event over~$\mathcal{X}^n\times\mathcal{Y}^n\times\mathcal{A}^n\times\mathcal{B}^n$.
			
			We consider permutation invariant tests, defined as follows. 
			\begin{defn}\label{def:permutation-invariant-test}
				A test $\mathcal{T}$ is permutation invariant if and only if for all boxes $\P_{\mr{A}\mr{B}|\mr{X}\mr{Y}}$ and all permutations $\pi$ we have
				\[
				 	\mathrm{Pr}_{\textup{fail}}(\P_{\mr{A}\mr{B}|\mr{X}\mr{Y}}) = \mathrm{Pr}_{\textup{fail}}(\P_{\mr{A}\mr{B}|\mr{X}\mr{Y}}\circ\pi) \;.
				\]
			\end{defn}
			
			Using the de Finetti reduction in Theorem~\ref{thm:dF_conditional} we can prove upper bounds of the following type: 
			
			\begin{thm}\label{thm:dF_test_bound}
				Let $\mathcal{T}$ be a permutation invariant test. Then for every box $\P_{\mr{A}\mr{B}|\mr{X}\mr{Y}}$ 
				\[
						\mathrm{Pr}_{\textup{fail}}(\P_{\mr{A}\mr{B}|\mr{X}\mr{Y}}) \leq (n+1)^{l(m-1)} \mathrm{Pr}_{\textup{fail}}(\tau_{\mr{A}\mr{B}|\mr{X}\mr{Y}}) \;,
				\]
				where $\tau_{\mr{A}\mr{B}|\mr{X}\mr{Y}}$ is the de Finetti box given in Equation~\eqref{eq:dF_general_explicit}.
			\end{thm}
			
			The importance of de Finetti reductions is already obvious from Theorem~\ref{thm:dF_test_bound}~--- if one wishes to prove an upper bound on the failure probability of the test $\mathcal{T}$, then instead of proving it for all boxes $\P_{\mr{A}\mr{B}|\mr{X}\mr{Y}}$, it is sufficient to prove it for the de Finetti box~$\tau_{\mr{A}\mr{B}|\mr{X}\mr{Y}}$ and ``pay'' for it with the additional polynomial pre-factor of~$(n+1)^{l(m-1)}$. 
			Since the de Finetti box can be written as a convex combination of IID boxes, this can highly simplify the calculations of the bound. In this sense the de Finetti reduction acts as a \emph{reduction to IID}.

			In many cases one finds that the bound on $\mathrm{Pr}_{\textup{fail}}(\tau_{\mr{A}\mr{B}|\mr{X}\mr{Y}})$ is exponentially small in $n$. For an estimation protocol, the failure probability of the test, when interacting with an IID box, can be shown to be exponentially small in the number of boxes~$n$ used for the estimation, using Chernoff bounds. This is also the case when dealing with security proofs -- the failure probability of a protocol, when a de Finetti box is given as input, is usually exponentially small in the number of boxes~$n$ used in the protocol. If this is indeed the case then the polynomial pre-factor of  $(n+1)^{l(m-1)}$ becomes irrelevant in the asymptotic limit of large $n$. In other words, an exponentially small bound on $\mathrm{Pr}_{\textup{fail}}(\tau_{\mr{A}\mr{B}|\mr{X}\mr{Y}})$ implies an exponentially small bound on $\mathrm{Pr}_{\textup{fail}}(\P_{\mr{A}\mr{B}|\mr{X}\mr{Y}})$.

			Let us prove Theorem~\ref{thm:dF_test_bound} using the de Finetti reduction given as Theorem~\ref{thm:dF_conditional}.
			\begin{proof}[Proof of Theorem~\ref{thm:dF_test_bound}.]
			We follow here a similar proof given in \cite{renner2010simplifying} for the quantum post-selection theorem \cite{christandl2009postselection}.
			 First, since the test $\mathcal{T}$ is permutation invariant it is sufficient to consider only permutation invariant boxes. To see this recall that for any box  $\P_{\mr{A}\mr{B}|\mr{X}\mr{Y}}$  and permutation $\pi$ we have $\mathrm{Pr}_{\textup{fail}}(\P_{\mr{A}\mr{B}|\mr{X}\mr{Y}}) = \mathrm{Pr}_{\textup{fail}}(\P_{\mr{A}\mr{B}|\mr{X}\mr{Y}}\circ\pi)$ according to Definition  \ref{def:permutation-invariant-test}. Therefore we also have by linearity\footnote{Linearity refers here to the linearity of the test in the box $\P_{\mr{A}\mr{B}|\mr{X}\mr{Y}}$, which follows from the fact that the test interacts only once with $\P_{\mr{A}\mr{B}|\mr{X}\mr{Y}}$ (or, in other words, the test gets only a single copy of the box).}
			\begin{equation*}
				\mathrm{Pr}_{\textup{fail}}(\P_{\mr{A}\mr{B}|\mr{X}\mr{Y}}) = \frac{1}{n!} \sum_\pi \mathrm{Pr}_{\textup{fail}}(\P_{\mr{A}\mr{B}|\mr{X}\mr{Y}}\circ\pi) = \mathrm{Pr}_{\textup{fail}} \left(  \frac{1}{n!} \sum_\pi \P_{\mr{A}\mr{B}|\mr{X}\mr{Y}}\circ\pi \right) \;.
			\end{equation*}
			The box $\frac{1}{n!} \sum_\pi \P_{\mr{A}\mr{B}|\mr{X}\mr{Y}}\circ\pi$ is permutation invariant and therefore we can consider only permutation invariant boxes without loss of generality. 
			
			Next we define the following probabilities. Let $\mathrm{Pr}_{\textup{fail}\land c_z}(\tau_{\mr{A}\mr{B}C|\mr{X}\mr{Y}Z})$ be the probability that the second part of the box, $\tau_{C|Z}$, is used with the input $z$ and the output is $c_z$ and that the first part of the box, $\tau_{\mr{A}\mr{B}|\mr{X}\mr{Y}}$, fails the test $\mathcal{T}$ at the same time. That is, 
			\[
				\mathrm{Pr}_{\textup{fail}\land c_z}(\tau_{\mr{A}\mr{B}C|\mr{X}\mr{Y}Z}) = \mathrm{Pr}_{\textup{fail}}(\tau_{\mr{A}\mr{B}|\mr{X}\mr{Y}}) \cdot \tau_{C|Z}(c_z|z) \;.
			\]
			In a similar way we define $\mathrm{Pr}_{\textup{fail}|c_z}(\tau_{\mr{A}\mr{B}C|\mr{X}\mr{Y}Z})$ to be the probability that $\tau_{\mr{A}\mr{B}|\mr{X}\mr{Y}}$ fails the test $\mathcal{T}$ \emph{given} that $c_z$ is the output of $\tau_{C|Z}$ when used with the input $z$. We have 
			\[
				\mathrm{Pr}_{\textup{fail}|c_z}(\tau_{\mr{A}\mr{B}C|\mr{X}\mr{Y}Z}) = \frac{\mathrm{Pr}_{\textup{fail}\land c_z}(\tau_{\mr{A}\mr{B}C|\mr{X}\mr{Y}Z})}{\tau_{C|Z}(c_z|z)} \leq \frac{\mathrm{Pr}_{\textup{fail}}(\tau_{\mr{A}\mr{B}|\mr{X}\mr{Y}})}{\tau_{C|Z}(c_z|z)}
			\]
			since $\mathrm{Pr}_{\textup{fail}\land c_z}(\tau_{\mr{A}\mr{B}C|\mr{X}\mr{Y}Z}) \leq \mathrm{Pr}_{\textup{fail}}(\tau_{\mr{A}\mr{B}|\mr{X}\mr{Y}})$ always holds.
			
			Lemma \ref{lem:dF_post_selection} implies that $\tau_{C|Z}(c_z|z)\geq\frac{1}{(n+1)^{l(m-1)}}$ and that $\mathrm{Pr}_{\textup{fail}|c_z}(\tau_{\mr{A}\mr{B}C|\mr{X}\mr{Y}Z})=\mathrm{Pr}_{\textup{fail}}(\P_{\mr{A}\mr{B}|\mr{X}\mr{Y}})$ (given that the output was $c_z$, the resulting box is $\P_{\mr{A}\mr{B}|\mr{X}\mr{Y}}$). All together we get $\mathrm{Pr}_{\textup{fail}}(\P_{\mr{A}\mr{B}|\mr{X}\mr{Y}}) \leq (n+1)^{l(m-1)} \mathrm{Pr}_{\textup{fail}}(\tau_{\mr{A}\mr{B}|\mr{X}\mr{Y}})$ as required. 
			\end{proof}

		\subsection{Diamond norm}\label{sec:dF_diam_norm}

			Theorem~\ref{thm:dF_conditional} allows for a simple treatment of cases that can be analysed using the notation of a test. In some information processing tasks this is not possible and different ways of utilising the reductions are needed. 
			In this section we consider the task of distinguishing two channels acting on boxes. The channels can describe, for example, a cryptographic protocol.\footnote{Let us briefly explain why the notation of a test considered in Section~\ref{sec:dF_post_selection} is not appropriate in the cryptographic setting. When considering tests, we were interested in events defined over $\mathcal{X}^n\times\mathcal{Y}^n\times\mathcal{A}^n\times\mathcal{B}^n$. Whether an output of a protocol (a key, for example) is secure to use cannot be defined as an event. Security depends on the \emph{process} of producing the key rather on the specific \emph{data} that was produced during the run of the protocol.}
			
			When considering quantum protocols the distinguishing advantage is given by the diamond norm \cite{kitaev1997quantum}. The distance between two channels $\mathcal{E}$ and $\mathcal{F}$ which act on quantum states $\rho_{AB}$ is given by $\| \mathcal{E}-\mathcal{F}\|_{\diamond}=\underset{\rho_{ABC}}{\max}\| \left(\mathcal{E}-\mathcal{F}\right)\otimes \idn \; \rho_{ABC} \|_1$ where $\rho_{ABC}$ is a purification of $\rho_{AB}$ and $\| \cdot \|_1$ is the trace distance. 
			Informally, the idea is that in order to distinguish two channels we are not only allowed to choose the input state to the channels, $\rho_{AB}$, but also keep to ourselves a purifying state  $\rho_C$. 
			
			Although the definition of the diamond norm includes a maximisation over all states $\rho_{ABC}$ it was proven, using the quantum post-selection theorem~\cite{christandl2009postselection}, that when considering permutation invariant channels it is sufficient to calculate the distance for a specific quantum de Finetti state. 
			Motivated by this, we give a similar bound on a distance analogous to the diamond norm for channels which act on boxes (instead of quantum states). 
			
			In the following, we denote by $\mathcal{P}$ the set of all boxes $\P_{\mr{A}\mr{B}|\mr{X}\mr{Y}}$ and by $\mathcal{K}$ the set of all probability distributions $\mathrm{P}_K$ over $\{0,1\}^t$ for some $t\in\mathbb{N}$.
			We consider channels of the form $\mathcal{E}:\mathcal{P}\rightarrow \mathcal{K}$ which interact with boxes $\P_{\mr{A}\mr{B}|\mr{X}\mr{Y}}$ and output a classical bit string $k\in\{0,1\}^t$ of some length $t\geq 0$ with some probability $\mathrm{P}_K(k)$. 
			The connection between the channel and the box is illustrated in Figure \ref{fig:channel}.\footnote{Figure \ref{fig:channel} is almost identical to Figure~\ref{fig:dF_test_interaction}, describing a test. The difference between the two scenarios lies in the quantity that we wish to bound; see the previous footnote.}
			
			\begin{figure}
			\begin{center}
				\begin{tikzpicture}[scale=1.25]
				\begin{scope}[shift={(-0.5,-0.5)}]
					\node at (1,0.5) {$\P_{\mr{A}\mr{B}C|\mr{X}\mr{B}Z}$};
					\draw  (0,0) rectangle (2,1);
					
					\node at (-0.5,0.85) {$\mr{x},\mr{y}$};
					\draw[->] (-1,0.75) -- (0,0.75);
					\node at (-0.5,0.35) {$\mr{a},\mr{b}$};
					\draw[<-] (-1,0.25) -- (0,0.25);
				
					\node at (0.75,-0.7) {$z$};
					\draw[<-] (0.75,0) -- (0.75,-0.5);
					\node at (1.25,-0.7) {$c$};
					\draw[->] (1.25,0) -- (1.25,-0.5);
					
					\draw (-1.5,0) rectangle (-1,1);
					\node at (-1.25,0.5) {$\mathcal{E}$};
					\draw[->] (-1.5,0.25) -- (-2.2,0.25);
					\node at (-3.2,0.25) {$k=\mathcal{E}(\mr{a},\mr{b},\mr{x},\mr{y})$};
				\end{scope}
							  	
				\end{tikzpicture}
			\end{center}
			\caption{The channel $\mathcal{E}\otimes \idn$ acts on an extension $\P_{\mr{A}\mr{B}C|\mr{X}\mr{B}Z}$ of  $\P_{\mr{A}\mr{B}|\mr{X}\mr{Y}}$ and outputs a classical string $k\in\{0,1\}^t$ according to the probability $\mathrm{E}_K(k)$.}
			\label{fig:channel}
			\end{figure}

			The probability distribution of the output depends on the channel $\mathcal{E}$ itself and is given by the following definition.
			\begin{defn}
				The probability that a channel $\mathcal{E}$ outputs a string $k\in\{0,1\}^{t}$ when interacting with $\P_{\mr{A}\mr{B}|\mr{X}\mr{Y}}$ is 
				\[
					\mathrm{E}_K(k) = \sum_{\mr{x},\mr{y}} \mathrm{Pr}_{\mathcal{E}}(\mr{x},\mr{y}) \sum_{\substack{\mr{a},\mr{b} : \\ \mathcal{E}(\mr{a},\mr{b},\mr{x},\mr{y})=k}} \P_{\mr{A}\mr{B}|\mr{X}\mr{Y}}(\mr{a},\mr{b}|\mr{x},\mr{y})
				\]
				where $\mathrm{Pr}_{\mathcal{E}}(x)$ is the probability that $\mathcal{E}$ inputs $\mr{x},\mr{y}$ to $\P_{\mr{A}\mr{B}|\mr{X}\mr{Y}}$ and $\mathcal{E}(\mr{a},\mr{b},\mr{x},\mr{y})$ is the function according to which the output of the channel is determined. Analogously, 
				\[
					\mathrm{E}_{K|C}(k|c) = \sum_{\mr{x},\mr{y}} \mathrm{Pr}_{\mathcal{E}}(\mr{x},\mr{y}) \sum_{\substack{\mr{a},\mr{b} : \\ \mathcal{E}(\mr{a},\mr{b},\mr{x},\mr{y})=k}} \P_{\mr{A}\mr{B}|\mr{X}\mr{Y}C}(\mr{a},\mr{b}|\mr{x},\mr{y},c) \;.
				\] 
			\end{defn}

			\begin{defn}\label{def:diamond-norm}
				The distance between two channels $\mathcal{E},\mathcal{F}:\mathcal{P}\rightarrow \mathcal{K}$ according to the diamond norm is 
				\[
					\| \mathcal{E}-\mathcal{F}\|_{\diamond}=\underset{\P_{\mr{A}\mr{B}C|\mr{X}\mr{Y}Z}}{\max}\| \left(\mathcal{E}-\mathcal{F}\right)\otimes \idn(\P_{\mr{A}\mr{B}C|\mr{X}\mr{Y}Z} )\|_1 \;,
				\]
				where the maximisation is over all boxes $\P_{\mr{A}\mr{B}|\mr{X}\mr{Y}}$ and all possible extensions of them and
				\[
				\begin{split}
					\mathcal{E}\otimes \idn(\P_{\mr{A}\mr{B}C|\mr{X}\mr{Y}Z}) &= \mathcal{E}\otimes \idn(\P_{\mr{A}\mr{B}|\mr{X}\mr{Y}C}\cdot \mathrm{P}_{C|Z}) \\
					&= \mathrm{E}_{K|C}\cdot \mathrm{P}_{C|Z} \;.
				\end{split}	
				\]
				$\mathcal{F}\otimes \idn(\P_{\mr{A}\mr{B}C|\mr{X}\mr{Y}Z})$ is defined in a similar way.
			\end{defn}

			Similarly to the concept of a permutation invariant test presented in Definition~\ref{def:permutation-invariant-test}, we define a permutation invariant channel:
			\begin{defn}\label{def:permutation-invariant-channel}
				A channel $\mathcal{E}$ is permutation invariant if for all boxes $\P_{\mr{A}\mr{B}|\mr{X}\mr{Y}}$ and all permutations $\pi$ we have
				\[
				 	\mathcal{E}(\P_{\mr{A}\mr{B}|\mr{X}\mr{Y}})=\mathcal{E}(\P_{\mr{A}\mr{B}|\mr{X}\mr{Y}}\circ\pi) \;.
				\]
			\end{defn}
				
			Using the de Finetti reduction, Theorem~\ref{thm:dF_conditional}, we prove the following theorem.
			\begin{thm}\label{thm:dF_diamond_norm_bound} 
				For any two permutation invariant channels $\mathcal{E},\mathcal{F}:\mathcal{P}\rightarrow \mathcal{K}$
				\begin{equation}\label{eq:dF_diamond_norm_bound}
					\| \mathcal{E}-\mathcal{F}\|_{\diamond} \leq (n+1)^{l(m-1)} \underset{\tau_{\mr{A}\mr{B}C|\mr{X}\mr{Y}Z}}{\max}\| \left(\mathcal{E}-\mathcal{F}\right)\otimes \idn(\tau_{\mr{A}\mr{B}C|\mr{X}\mr{Y}Z} )\|_1
				\end{equation}
				where $\tau_{\mr{A}\mr{B}C|\mr{X}\mr{Y}Z}$ is a non-signalling extension of the de Finetti box $\tau_{\mr{A}\mr{B}|\mr{X}\mr{Y}}$ where given in Equation~\eqref{eq:dF_general_explicit}. 
			\end{thm}
			
			Theorem~\ref{thm:dF_diamond_norm_bound} tells us that when looking to bound the diamond norm for permutation invariant channels, one does not need to optimise over all possible boxes (as in Definition~\ref{def:diamond-norm}) but can consider only extensions of de Finetti boxes\footnote{Note, however, that the extension $\tau_{\mr{A}\mr{B}C|\mr{X}\mr{Y}Z}$ itself cannot be written as a convex combination of IID boxes, only its marginal $\tau_{\mr{A}\mr{B}|\mr{X}\mr{Y}}$ is a de Finetti box. Furthermore, $\tau_{\mr{A}\mr{B}|\mr{X}\mr{Y}}$ may be signalling in general, as before.} without loss of generality. This gives us another example as to why a de Finetti reduction is a \emph{reduction to IID} technique. 
			As in the case of Theorem~\ref{thm:dF_test_bound} if one is able to find an exponentially small  upper bound on $\| \left(\mathcal{E}-\mathcal{F}\right)\otimes \idn(\tau_{\mr{A}\mr{B}C|\mr{X}\mr{Y}Z} )\|_1$, an exponentially small upper bound on $\| \mathcal{E}-\mathcal{F}\|_{\diamond}$ follows. That is, the polynomial pre-factor $(n+1)^{l(m-1)}$ does not affect the asymptotic behaviour.
			
			The proof of Theorem \ref{thm:dF_diamond_norm_bound} builds on the following lemma. 
			\begin{lem}\label{lem:trace-distance-lem}
				For every two permutation invariant channels $\mathcal{E},\mathcal{F}:\mathcal{P}\rightarrow \mathcal{K}$ where $\mathrm{P}_K$ is a probability distribution over $k\in\{0,1\}^t$ for some $t>0$, and all $\P_{\mr{A}\mr{B}C|\mr{X}\mr{Y}Z}$,
				\[
					 \| \left(\mathcal{E}-\mathcal{F}\right)\otimes \idn(\P_{\mr{A}\mr{B}C|\mr{X}\mr{Y}Z} )\|_1 \leq (n+1)^{l(m-1)} \| \left(\mathcal{E}-\mathcal{F}\right)\otimes \idn(\tau^{\P_{\mr{A}\mr{B}C|\mr{X}\mr{Y}Z}}_{\mr{A}\mr{B}C|\mr{X}\mr{Y}Z} )\|_1
				\]
				where $\tau^{\P_{\mr{A}\mr{B}C|\mr{X}\mr{Y}Z}}_{\mr{A}\mr{B}C|\mr{X}\mr{Y}Z}$ is a non-signalling extension of $\tau_{\mr{A}\mr{B}|\mr{X}\mr{Y}}$ which depends on the specific box $\P_{\mr{A}\mr{B}C|\mr{X}\mr{Y}Z}$.
			\end{lem}
			The proof of the lemma follows by using Lemma~\ref{lem:dF_post_selection} in order to  construct a specific convex decomposition of $\tau_{\mr{A}\mr{B}|\mr{X}\mr{Y}}$ from a convex decomposition of $\P_{\mr{A}\mr{B}|\mr{X}\mr{Y}}$. A detailed proof is given in Appendix~\ref{appsec:diamond_norm_proof}.

			Theorem \ref{thm:dF_diamond_norm_bound} now easily follows from Lemma \ref{lem:trace-distance-lem}:
			\begin{proof}[Proof of Theorem \ref{thm:dF_diamond_norm_bound}]
				Using Lemma \ref{lem:trace-distance-lem}, 
				\[
				\begin{split}
					\| \mathcal{E} - \mathcal{F} \|_{\diamond} &= \underset{\P_{\mr{A}\mr{B}C|\mr{X}\mr{Y}Z}}{\max}\| \left(\mathcal{E}-\mathcal{F}\right)\otimes \idn(\P_{\mr{A}\mr{B}C|\mr{X}\mr{Y}Z} )\|_1\\
					&\leq (n+1)^{l(m-1)} \underset{\tau^{\P_{\mr{A}\mr{B}C|\mr{X}\mr{Y}Z}}_{\mr{A}\mr{B}C'|\mr{X}\mr{Y}Z}}{\max}\| \left(\mathcal{E}-\mathcal{F}\right)\otimes \idn(\tau^{\P_{\mr{A}\mr{B}C|\mr{X}\mr{Y}Z}}_{\mr{A}\mr{B}C'|\mr{X}\mr{Y}Z})\|_1\\
					&\leq (n+1)^{l(m-1)} \underset{\tau_{\mr{A}\mr{B}C|\mr{X}\mr{Y}Z}}{\max}\| \left(\mathcal{E}-\mathcal{F}\right)\otimes \idn(\tau_{\mr{A}\mr{B}C|\mr{X}\mr{Y}Z})\|_1\\
				\end{split}
				\]
				where $\tau_{\mr{A}\mr{B}C|\mr{X}\mr{Y}Z}$ is a non-signalling extension of $\tau_{\mr{A}\mr{B}|\mr{X}\mr{Y}}$. 
			\end{proof}

	\section{Impossibility results}\label{sec:dF_imp_res}
		
		Before concluding this chapter, let us discuss the directions in which one could hope to further develop the technique of device-independent de Finetti reductions. We do so by presenting several impossibility results with regards to different variants of Theorem~\ref{thm:dF_conditional}.
		
		\subsubsection*{Restricted de Finetti box}
		
			First, as explained in the above sections, our de Finetti box, given in Equation~\eqref{eq:dF_general_explicit}, is a \emph{signalling} box. 
			Clearly, this raises some difficulties when coming to use the different theorems presented in this chapter.\footnote{Though this does not make them useless; see Chapter~\ref{ch:par_rep_showcase}.} 
			Ideally, we would have wished to have a de Finetti reduction in which the de Finetti box $\tau_{\mr{A}\mr{B}|\mr{X}\mr{Y}}$ can be quantum or non-signalling when starting with a quantum or non-signalling box  $\P_{\mr{A}\mr{B}|\mr{X}\mr{Y}}$. 
			That is, we wish to find reductions of the form (with some $c$ polynomial\footnote{Weaker statements, e.g., with a pre-factor sub-exponential in $n$, may also be of interest in certain applications.} in $n$):
			\begin{equation}\label{eq:strong_dF_reduc}
				\P_{\mr{A}\mr{B}|\mr{X}\mr{Y}}^{\text{quant}} \leq c \cdot \tau_{\mr{A}\mr{B}|\mr{X}\mr{Y}}^{\text{quant}} 
				\qquad ; \qquad 
				\P_{\mr{A}\mr{B}|\mr{X}\mr{Y}}^{\text{ns}} \leq c \cdot \tau_{\mr{A}\mr{B}|\mr{X}\mr{Y}}^{\text{ns}} \;,
			\end{equation}
			where $\P_{\mr{A}\mr{B}|\mr{X}\mr{Y}}^{\text{quant}}$ and $\tau_{\mr{A}\mr{B}|\mr{X}\mr{Y}}^{\text{quant}}$ are quantum boxes and, similarly, $\P_{\mr{A}\mr{B}|\mr{X}\mr{Y}}^{\text{ns}}$ and~$\tau_{\mr{A}\mr{B}|\mr{X}\mr{Y}}^{\text{ns}}$ are non-signalling boxes. 
			
			Sadly, such reductions cannot be true when considering general permutation invariant boxes $\P_{\mr{A}\mr{B}|\mr{X}\mr{Y}}^{\text{quant}}$ and $\P_{\mr{A}\mr{B}|\mr{X}\mr{Y}}^{\text{ns}}$. 
			One way to see that this is the case is by considering the task of parallel repetition of games (which acts as one of our showcases; see Section~\ref{sec:pre_par_rep}). 	
			Reductions as those in Equation~\eqref{eq:strong_dF_reduc} will imply  very strong parallel repetition results. Indeed, if, e.g., $\P_{\mr{A}\mr{B}|\mr{X}\mr{Y}}^{\text{quant}} \leq c \cdot \tau_{\mr{A}\mr{B}|\mr{X}\mr{Y}}^{\text{quant}}$ holds for any permutation invariant quantum box $\P_{\mr{A}\mr{B}|\mr{X}\mr{Y}}^{\text{quant}}$, then it follows that, for \emph{any} game, 
			\begin{equation}\label{eq:strong_par_rep_dF}
				w\left(\P_{\mr{A}\mr{B}|\mr{X}\mr{Y}}^{\text{quant}}\right) \leq c \cdot w\left(\tau_{\mr{A}\mr{B}|\mr{X}\mr{Y}}^{\text{quant}}\right) = \mathrm{poly}(n) \cdot \omega^n \;,
			\end{equation}
			where $w\left(\circ\right)$ is the winning probability of the considered box in the repeated game,~$\omega$ is the winning probability of the optimal quantum strategy in a single game, and $\mathrm{poly}(n)$ is some polynomial of $n$, possibly depending on the alphabet of the RVs $A$, $B$, $X$, and $Y$.		
			However, there are examples of games (in the classical, quantum, and non-signalling case) for which a strong decrease in the winning probability with the number of games played $n$, as in Equation~\eqref{eq:strong_par_rep_dF}, does not hold; recall Section~\ref{sec:pre_par_rep}. Thus, reductions as in Equation~\eqref{eq:strong_dF_reduc} cannot be true. 
			
			Knowing that Equation~\eqref{eq:strong_dF_reduc} is not more than a wishful thinking, one could hope for the next best thing, i.e., an approximate version of the reduction. Concretely, we are interested in reductions of the form
			\begin{equation}\label{eq:approx_dF_reduc}
				\P_{\mr{A}\mr{B}|\mr{X}\mr{Y}}^{\text{quant}} \leq c \cdot \tau_{\mr{A}\mr{B}|\mr{X}\mr{Y}}^{\text{approx-quant}} 
				\qquad ; \qquad 
				\P_{\mr{A}\mr{B}|\mr{X}\mr{Y}}^{\text{ns}} \leq c \cdot \tau_{\mr{A}\mr{B}|\mr{X}\mr{Y}}^{\text{approx-ns}} \;,
			\end{equation}
			where $\tau_{\mr{A}\mr{B}|\mr{X}\mr{Y}}^{\text{approx-quant}}$ is an \emph{approximately}-quantum de Finetti box and  $\tau_{\mr{A}\mr{B}|\mr{X}\mr{Y}}^{\text{approx-ns}}$ is an \emph{approximately}-non-signalling one. 
			By approximately-quantum (and analogously for the non-signalling case) we mean that the de Finetti box can be written as
			\begin{equation*}
				\tau_{\mr{A}\mr{B}|\mr{X}\mr{Y}}^{\text{approx-quant}} = \int \left(\O_{AB|XY}^{\text{quant}}\right)^{\otimes n} \mathrm{d}\O_{AB|XY}^{\text{quant}} + \int \left(\O_{AB|XY}^{\text{non-quant}}\right)^{\otimes n} \mathrm{d}\O_{AB|XY}^{\text{non-quant}} \;,
			\end{equation*}
			where $\mathrm{d}\O_{AB|XY}^{\text{quant}}$ and $\mathrm{d}\O_{AB|XY}^{\text{non-quant}}$ are measures over quantum and non-quantum single-round boxes, respectively, and $\int \mathrm{d}\O_{AB|XY}^{\text{non-quant}}$ is, say, exponentially small in $n$ and/or assigns weight only to boxes $\O_{AB|XY}^{\text{non-quant}}$ which are close to quantum boxes, under some distance measure.\footnote{The hope here is that by adding the additional weight on non-quantum or signalling boxes one could account for the ``gap'' between Equation~\eqref{eq:strong_par_rep_dF} and the known parallel repetition results.}
			
			Parallel repetition results can, again, be used to show that such reductions cannot hold in general, at least in the non-signalling case. Here the reason lies in the observation that the reductions in Equation~\eqref{eq:approx_dF_reduc} are independent of the choice of \emph{distribution} over the inputs $\mathcal{X}^n$ and $\mathcal{Y}^n$ (while they may depend on the \emph{alphabet} of the inputs). Thus, they would imply general parallel repetition results which hold for any distribution over the inputs to the parallel boxes. As there are games for which such non-signalling parallel repetition results do not hold~\cite{holmgren2017counterexample}, at best $\P_{\mr{A}\mr{B}|\mr{X}\mr{Y}}^{\text{ns}} \leq c \cdot \tau_{\mr{A}\mr{B}|\mr{X}\mr{Y}}^{\text{approx-ns}}$ cannot be true in general.
			
			By this we learn that we ought to consider reductions that also include the input distribution $\P_{\mr{X}\mr{Y}}$:
			\begin{align}
				&\P_{\mr{X}\mr{Y}} \P_{\mr{A}\mr{B}|\mr{X}\mr{Y}}^{\text{quant}} \leq c \cdot \P_{\mr{X}\mr{Y}} \tau_{\mr{A}\mr{B}|\mr{X}\mr{Y}}^{\text{approx-quant}}\;, \label{eq:approx_quant_quest} \\
				&\P_{\mr{X}\mr{Y}} \P_{\mr{A}\mr{B}|\mr{X}\mr{Y}}^{\text{ns}} \leq c \cdot \P_{\mr{X}\mr{Y}} \tau_{\mr{A}\mr{B}|\mr{X}\mr{Y}}^{\text{approx-ns}} \;. \label{eq:approx_ns_quest}
			\end{align}
			The case of $\P_{\mr{X}\mr{Y}}=\Q_{XY}^{\otimes n}$ is of special interest. For such distributions, two results are known.
			In Section~\ref{sec:approx_ns_marginals} we prove a result in the \emph{flavour} of Equation~\eqref{eq:approx_ns_quest} using the de Finetti reduction given as Theorem~\ref{thm:dF_conditional}.
			The result, which originally appeared as part of~\cite{arnon2016non}, is stated informally as Theorem~\ref{thm:approx_ns_mar_informal}. Roughly speaking, it says that observed data that is sampled using a permutation invariant non-signalling parallel box looks \emph{as if} it was sampled using an approximately non-signalling IID box.
		
			In~\cite{lancien2016parallel} a reduction similar to that of Equation~\eqref{eq:approx_ns_quest} was proven by combining the de Finetti reduction in Theorem~\ref{thm:dF_conditional} together with another de Finetti-type theorem, presented in~\cite{lancien2017flexible}.
			Their theorem can be written as follows:\footnote{We present only the bipartite case; \cite[Theorem~4.3]{lancien2016parallel} is stated for an arbitrary number of parties.}
			\begin{thm}[Theorem~4.3 in~\cite{lancien2016parallel}]
					For any non-signalling permutation invariant parallel box $\P_{\mr{A}\mr{B}|\mr{X}\mr{Y}}$ and distribution $\Q_{XY}$ 
					\begin{equation}\label{eq:constrained_dF}
						\Q_{XY}^{\otimes n} \P_{\mr{A}\mr{B}|\mr{X}\mr{Y}} \leq \int \tilde{\mathrm{F}} \left( \O_{ABXY}\right)^{2n} \O_{ABXY}^{\otimes n} \mathrm{d}\O_{ABXY} \;,
					\end{equation}
					where 
					\[
						\tilde{\mathrm{F}} \left( \O_{ABXY}\right) = \min \left\{ \max_{\mathrm{R}_{A|X}} \mathrm{F}\left( \Q_{XY} \mathrm{R}_{A|X}, \O_{AXY} \right) ,\; \max_{\mathrm{R}_{B|Y}} \mathrm{F}\left( \Q_{XY} \mathrm{R}_{B|Y}, \O_{BXY} \right) \right\}
					\]
					for $\mathrm{F}$ the fidelity.
			\end{thm}
			To see that Equation~\eqref{eq:constrained_dF} is in the spirit of Equation~\eqref{eq:approx_ns_quest} note that $\tilde{\mathrm{F}} \left( \O_{ABXY}\right)$ is some measure of how far $\O_{ABXY}$ is from $\Q_{XY}\tilde{\O}_{AB|XY}$ for a non-signalling box~$\tilde{\O}_{AB|XY}$. 
			Recall that the fidelity is small when the distributions are far from one another; thus, $ \tilde{\mathrm{F}} \left( \O_{ABXY}\right)^{2n}$ assures that only negligible weight is assigned to distributions~$\O_{ABXY}$ originating from highly signalling boxes (or with marginals~$\O_{XY}$ far from~$\Q_{XY}$). 
						
			We conjecture that reductions similar to Equation~\eqref{eq:approx_quant_quest}, relevant for quantum boxes, should also hold. Yet, up to date there are no proofs in this direction (the difficulty in deriving such a statement is discussed in Chapter~\ref{ch:par_rep_showcase}).

		\subsubsection*{Extension to an adversary}
		
			Another direction in which one may wish to extend our de Finetti reductions is relevant for device-independent cryptographic protocols. 
			To explain what we aim for, let us first discuss the quantum variant of Theorem~\ref{thm:dF_diamond_norm_bound}, also called the post-selection technique, developed in~\cite{christandl2009postselection}.\footnote{\cite{christandl2009postselection} presented the first de Finetti reduction, i.e., an inequality relation between permutation invariant systems and de Finetti systems (all previous de Finetti-type theorems gave other types of relations between the two systems). The term ``de Finetti reduction'' was not used at that time and the authors chose the name ``post-selection technique'' as they first proved the quantum analogue of  Lemma~\ref{lem:dF_post_selection}.}
			The post-selection theorem implies that for any two permutation invariant quantum channels, $\mathcal{E}$ and $\mathcal{F}$, acting on quantum states $\rho_{\mr{Q_A}\mr{Q_B}}\in\mathcal{S}(\hilb_{Q_AQ_B}^{\otimes n})$ for some bipartite Hilbert space $\hilb_{Q_AQ_B}$ of dimension~$d$, 
				\begin{equation}\label{eq:quant_post_selection}
					\| \mathcal{E}-\mathcal{F}\|_{\diamond} \leq (n+1)^{d^2-1} 
					 \| \left(\mathcal{E}-\mathcal{F}\right)\otimes \idn(\tau_{\mr{Q_A}\mr{Q_B}E} )\|_1
				\end{equation}
				where $\tau_{\mr{Q_A}\mr{Q_B}E}$ is a purification of a given de Finetti state.
				Equation~\eqref{eq:quant_post_selection} should be compared to Equation~\eqref{eq:dF_diamond_norm_bound}; while  Equation~\eqref{eq:dF_diamond_norm_bound} includes a maximisation over all possible non-signalling extensions of the de Finetti box, in Equation~\eqref{eq:quant_post_selection} we consider only a single purification. The reason is simple -- in the quantum case all purifications of a state are equivalent up to local unitaries. 
				Furthermore (and crucially for applications), there exists a purification of a de Finetti state that has a very special form. To purify   
				\[
					\tau_{\mr{Q_A}\mr{Q_B}} = \int \left(\sigma_{Q_AQ_B}\right)^{\otimes n} \mathrm{d}\sigma_{Q_AQ_B}
				\]
				we can first purify the states $\sigma_{Q_AQ_B}$ to get
				\[
					\tau_{\mr{Q_A}\mr{Q_B}E'} = \int \left(\sigma_{Q_AQ_BE'}\right)^{\otimes n} \mathrm{d}\sigma_{Q_AQ_BE'}
				\]
				and then purify the state $\tau_{\mr{Q_A}\mr{Q_B}E'}$ using an additional system $E''$ to account for the convex combination of the pure states~$\left(\sigma_{Q_AQ_BE'}\right)^{\otimes n}$. 
				This defines us the pure state~$\tau_{\mr{Q_A}\mr{Q_B}E'E''}$. Denoting $E=E'E''$ we get  our pure $\tau_{\mr{Q_A}\mr{Q_B}E}$.
				
				In the cryptographic setting the quantum register $E$ is considered to belong to the adversary. Hence, any information about the structure of the system kept in it could be useful when analysing security. 
				Equation~\eqref{eq:quant_post_selection} in combination with the observation regarding the structure of the purification, $\tau_{\mr{Q_A}\mr{Q_B}E'E''}$,  we learn that the main task when proving security is to analyse the IID case, as in Chapter~\ref{ch:iid_assumption} (see~\cite{christandl2009postselection, renner2010simplifying} for the detailed explanation). That is, the quantum de Finetti reduction can be used as a reduction to IID in quantum cryptography. 
				
				In contrast, in general, it is impossible to prove a modified version of Theorem~\ref{thm:dF_diamond_norm_bound} in which the extension~$\tau_{\mr{A}\mr{B}C|\mr{X}\mr{Y}Z}$ of our de Finetti box~$\tau_{\mr{A}\mr{B}|\mr{X}\mr{Y}}$ will be as structured as the quantum state~$\tau_{\mr{Q_A}\mr{Q_B}E}$.
				 In particular, even if we can start with a de Finetti reduction where both $\P_{\mr{A}\mr{B}|\mr{X}\mr{Y}}$ and $\tau_{\mr{A}\mr{B}|\mr{X}\mr{Y}}$ are non-signalling,\footnote{In the presence of certain types of symmetries (in addition to permutation invariance) one can derive such de Finetti reductions; see~\cite{arnon2013finetti}.} it is impossible to derive a theorem which would imply that the analysis of device-independent cryptography in the presence of a non-signalling adversary can be reduced to the analysis under the IID assumption. 
				 This is due to the impossibility result of~\cite{arnon2012limits}, which asserts that, while exponential privacy amplification in the presence of a non-signalling adversary is possible under the IID assumption~\cite{hanggi2009ns}, it is impossible when the IID assumption is dropped. 
				 						
		\subsubsection*{Other de Finetti-type theorems}

				A final remark is with regards to the more common type of de Finetti theorem, in which one bounds the trace distance between an \emph{$n$-exchangeable} system and a de Finetti one.
				More specifically, let us first consider the classical case, i.e., a system is a probability distribution.
				$\P_{A_1,\dots,A_k}$ is permutation invariant if it is invariant under any permutation of $A_1,\dots,A_k$ (as before). We say that $\P_{A_1,\dots,A_k}$ is $n$-exchangeable, for $n\geq k$, if it is a marginal of some permutation invariant $\P_{A_1,\dots,A_n}$. 
				In~\cite{diaconis1980finite} a bound on the \emph{distance} between an $n$-exchangeable probability distribution and a de Finetti distribution was proven.\footnote{In this language, the original result of de Finetti~\cite{deFinetti69} stated that all infinitely-exchangeable distributions (i.e., distributions that are $n$-exchangeable for \emph{any} $n\geq k$) are equal to distributions of the form of a convex combination of IID distributions.} 
				Results of this type were also proven for quantum states~\cite{konig2005finetti,christandl2007one} and non-signalling boxes~\cite{christandl2009finite}.
				
				Let us focus on the non-signalling case~\cite{christandl2009finite}. There, a conditional probability distribution $\P_{A_1,\dots,A_n|X_1,\dots,X_n}$ is said to be non-signalling if the box cannot be used to signal from any subset of parties $I\subset [n]$ to the rest of the parties~$[n] \setminus I$.
				Permutation invariance is defined with respect to permutations $\pi:[n]\rightarrow [n]$. 
				Similarly to the classical case described above, $\P_{A_1,\dots,A_k|X_1,\dots,X_k}$ is~$n$-exchangeable, for $n\geq k$, when it is the marginal of a permutation invariant non-signalling box $\P_{A_1,\dots,A_n|X_1,\dots,X_n}$. 
				We then have the following bound~\cite[Theorem~3]{christandl2009finite} (using the above notation):
				\sloppy
				\begin{thm}[\cite{christandl2009finite}]\label{thm:dF_exchange}
					For any permutation invariant non-signalling box $\P_{A_1,\dots,A_n|X_1,\dots,X_n}$ and any $k<n$ there exists a de Finetti box~$\tau_{A_1,\dots,A_k|X_1,\dots,X_k}$ such that 
					\[
						\left|  \P_{A_1,\dots,A_k|X_1,\dots,X_k} - \tau_{A_1,\dots,A_k|X_1,\dots,X_k}  \right| \leq \min \left\{ \frac{2k |\mathcal{X}||\mathcal{A}|^{|\mathcal{X}|}}{n}, \; \frac{k(k-1)|\mathcal{X}|}{n} \right\} \;.
					\]
				\end{thm}
				
				\sloppy
				The crucial thing to note here is that the boxes $ \P_{A_1,\dots,A_k|X_1,\dots,X_k}$ and $\P_{A_1,\dots,A_n|X_1,\dots,X_n}$ are a very special type of parallel boxes: the non-signalling conditions must hold for any division of the indices in $[n]$. This implies that the for any $i,j\in[n]$, $A_i$ is independent of the inputs $X_j$ for $j\neq i$.
				Theorems such as Theorem~\ref{thm:dF_exchange} cannot be proven for general parallel boxes since they study exchangeable boxes, which inherently require the ability to consider the marginals of the boxes.

\chapter{Reductions to IID: sequential interaction}\label{ch:reductions_seq}


	
	Many device-independent protocols proceed in rounds and, hence, require devices with which the honest parties can interact sequentially, i.e., one round after the other. A particular example for such protocols is our showcase dealing with device-independent quantum cryptography. 
	When analysing the showcase under the IID assumption (Section~\ref{sec:crypt_under_iid}), we observed that the quantum AEP, given as Theorem~\ref{thm:quant_aep}, plays a crucial role in the proof of soundness. Specifically, the quantum AEP allowed us to bound the total amount of the relevant smooth entropy using a bound on the von Neumann entropy calculated for a single round of the protocol. 
	
	\sloppy
	The focus of this chapter is the so called ``entropy accumulation theorem'' (EAT)~\cite{dupuis2016entropy, dupuis2018entropy}. 
	The EAT is a generalisation of the AEP to a scenario in which, instead of the raw data being produced by an IID process, it is produced by certain sequential quantum processes of interest.\footnote{We remark that the EAT, as the quantum AEP, is only relevant when assuming that everything can be described within the quantum formalism. In particular, it cannot be used when talking about, e.g., cryptographic protocols in the presence of a non-signalling (super-quantum) adversary.}  
	In particular, similarly to the AEP, the EAT allows one to bound the total amount of the considered smooth entropy using the same bound on the von Neumann entropy calculated for the IID analysis. In this sense, the EAT can be seen as a \emph{reduction to IID}~--- with the aid of the EAT the analysis done under the IID assumption using the AEP is directly extended to the one relevant for \emph{multi-round sequential boxes}.

	Below, we motivate, present, and explain the EAT in the form relevant for device-independent quantum information processing~\cite{arnon2018practical}. The EAT is later used in the analysis of our showcase in Chapter~\ref{ch:crypto_showcase}.
	For the most general statement of the EAT and its proof the interested reader is referred to~\cite{dupuis2016entropy,dupuis2018entropy}.
	
	\section{Sequential quantum processes}\label{sec:seq_qunt_procc}
	
		We are interested in multi-round quantum sequential boxes (with communication between the rounds; see Definition~\ref{def:quant_seq_box_w_comm}) fulfilling certain conditions. The simplest way of describing the relevant conditions is by looking at the sequential quantum process defining the boxes, i.e., the process that results in the input-output distribution of the boxes.
		
	\begin{figure}
				\centering
				\includegraphics[width=\textwidth]{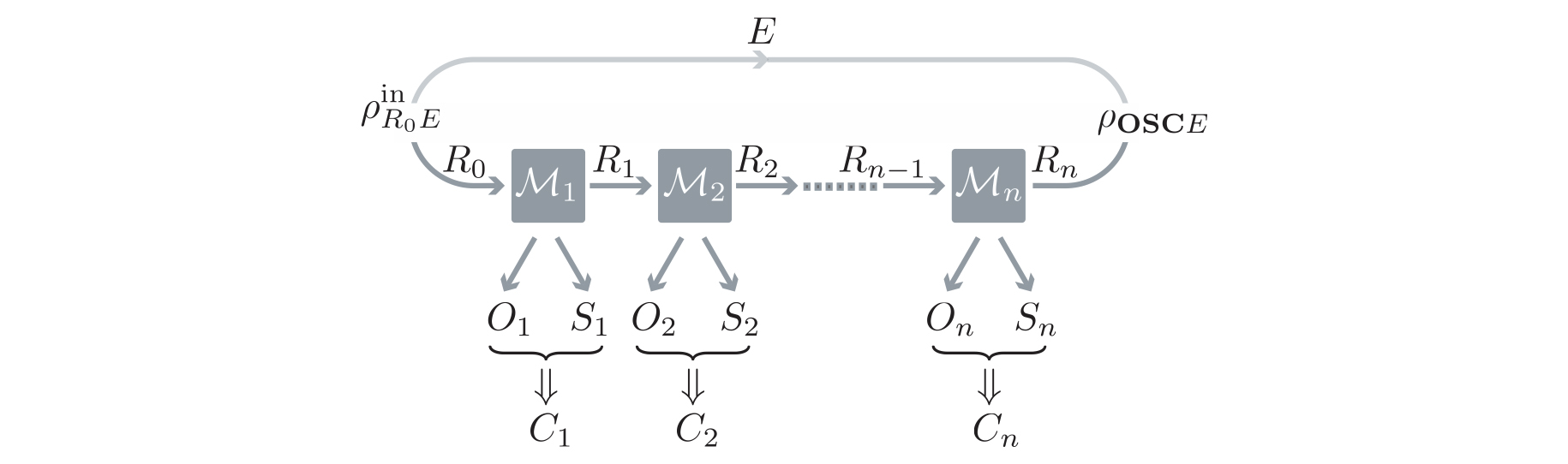}
				\caption{Sequential quantum process. The initial state $\rho^{\text{in}}_{R_0E}$ is transformed to the final one $\rho_{\mr{O}\mr{S}\mr{C}E}$ by applying a sequence of maps on the marginal $\rho^{\text{in}}_{R_0}$. Each map $\mathcal{M}_i$ in the sequence outputs the registers $O_i$ and $S_i$, from which $C_i$ is created. The ``memory system'' $R_i$ is being passed to the next map as input.}
				\label{fig:seq_proc}
		\end{figure}
		
		Consider a sequential process as illustrated in Figure~\ref{fig:seq_proc}. 
		We start with some initial state $\rho^{\text{in}}_{R_0E}$; the distinction between $R_0$ and $E$ is such that $R_0$ is the part of the state which may change during the considered process while $E$ denotes the ``environment'' register, i.e., the part of the state not being modified. 
		The marginal $\rho^{\text{in}}_{R_0}$ undergoes a sequence of operations in which a sequence of (non-IID) registers $\mr{O}=O_1,\dots,O_n$ and $\mr{S}=S_1,\dots,S_n$ are being created. We treat the registers $\mr{O}$ as the ``output registers'' while $\mr{S}$ are the ``side-information registers''. 
		Our ultimate objective is to bound the conditional smooth entropies $H^{\varepsilon}_{\min}(\mr{O}|\mr{S}E)$ and $H^{\varepsilon}_{\max}(\mr{O}|\mr{S}E)$.
		
		To be able to bound the above entropies, some statistical data must be collected during the protocol. Specifically, we consider additional classical registers $\mr{C}=C_1,\dots,C_2$ holding the information relevant for the estimation phase performed in the protocol. For every round $i\in[n]$,  $C_i$ is derived by performing some action on the registers $O_i$ and $S_i$. For example, the value of $C_i$ can be the result of applying a function on some classical information included in  $O_i$ and $S_i$.
		
		To gain a bit of intuition regarding all the different registers, let us quickly consider the cryptographic setting (a more precise discussion can be found in Chapter~\ref{ch:crypto_showcase}). 
		When analysing cryptographic protocols one may make the following choices: $E$ acts as the register belonging to the adversary, $\mr{O}$ as the raw data which is supposed to be secret, $\mr{S}$ as the side-information leaked during the protocol (e.g., all classical information which is communicated between Alice and Bob), and~$\mr{C}$ -- the indicators of whether the test rounds were successful or not (e.g., $C_i=1$ when the $i$'th game was won).
		The goal is then to lower bound  $H^{\varepsilon}_{\min}(\mr{O}|\mr{S}E)$ and this should  be done by using the statistics kept in~$\mr{C}$. 
		
		The sequential process itself is formally defined by a sequence of quantum channels, namely, CPTP maps (Definition~\ref{def:cptp_map}), 
		\begin{equation}\label{eq:eat_chan_def_general}
			\mathcal{M}_i:R_{i-1}\rightarrow R_i O_i S_i C_i \;,
		\end{equation}
		for all $i\in[n]$.  
		As seen from Equation~\eqref{eq:eat_chan_def_general}, when we say that a process is \emph{sequential} we not only mean that the maps act one after the other, but also that, in each round~$i\in[n]$, the output $O_i$, the side information $S_i$, and the estimation data $C_i$ are being created by the map $\mathcal{M}_i$ applied in \emph{that} round. That is, $\mr{O}$ denotes a sequence of registers created one after the other and similarly for $\mr{S}$ and $\mr{C}$.\footnote{The reader may be concerned by the distinction between, e.g., $O_1$ and $O_2$ -- clearly, we can also consider a situation in which $\mathcal{M}_1$ does not output $O_1$ but transfers it to $\mathcal{M}_2$ that later outputs~$O_1O_2$. We will soon assume that the different registers fulfil certain conditions and then the distinction will become clear.} 
		The state of interest in the end of the process is given by\footnote{We will be interested below in bounding the smooth entropies evaluated on a state closely related to the final state $\rho_{\mr{O}\mr{S}\mr{C}E}$, namely, the final state conditioned on the event of not aborting the considered protocol; see Section~\ref{sec:eat_statement}.}
		\begin{equation}
			\rho_{\mr{O}\mr{S}\mr{C}E} = \left( \Tr_{R_n} \circ \mathcal{M}_n \circ \dots \circ \mathcal{M}_1 \right) \otimes \idn_{E} \; \rho^{\text{in}}_{R_0E} \;.
		\end{equation}

		We remark that in the device-independent setting the initial state $ \rho^{\text{in}}_{R_0E}$, the maps $\left\{\mathcal{M}_i\right\}_{i\in[n]}$, and the final state $\rho_{\mr{O}\mr{S}\mr{C}E}$ are unknown; we merely require that some quantum states and maps  $\left\{\mathcal{M}_i\right\}_{i\in[n]}$, describing the overall process, exist. 
		In particular, this implies that we do not restrict the content of the registers $\left\{R_i\right\}_{i\in[n]}$ and, thus, they may include information which is being passed from previous rounds to the next ones. That is, we can think of these registers as holding some quantum memory. This is in stark contrast to what happens when working under the IID assumption.

\section{Entropy accumulation theorem}
	
		As mentioned above, the EAT~\cite{dupuis2016entropy} acts as a generalisation of the quantum AEP to scenarios in which certain sequential processes are considered, rather than IID ones. 
		The EAT, as the name suggests, quantifies the amount of entropy accumulated during the considered quantum processes. Moreover, as in the case of the AEP,   the total amount of smooth entropies can be bounded by calculating certain von Neumann entropies (the precise statements are given below). By this, the EAT justifies the use of the von Neumann entropy in quantum information processing even outside of the IID regime. 
		While we discuss the EAT in the context of device-independent quantum information processing, we remark that the EAT is a general information-theoretic tool, which can also be applied in other contexts.

		
		\subsection{Conceptual difficulties to overcome}\label{sec:eat_conc_diff}
		
			Before stating the theorem itself, let us explain the conceptual difficulties that arise when seeking for an ``AEP-style'' theorem in non-IID scenarios. Specifically, we would like to understand what is the form of the theorem we are aiming for and what is non-trivial about it. 
			To keep this section concise we focus on the smooth min-entropy; the same statements are relevant for the smooth max-entropy as well.

			Our goal is to have a theorem resembling the quantum AEP appearing as Theorem~\ref{thm:quant_aep}.
			That is, we look for a statement of the form
			\begin{equation}\label{eq:eat_general_form}
				 H^{\varepsilon}_{\min}(\mr{O}|\mr{S}E) \geq nt - \mu\sqrt{n} \;.
			\end{equation}
			for some $t$ and $\mu$ (independent of $n$ but otherwise unrestricted). 
			
			The EAT aims at providing a lower-bound on $H^{\varepsilon}_{\min}(\mr{O}|\mr{S}E)$ which scales linearly with the number of rounds $n$ (to first order in $n$, i.e., up to finite statistic effects, as in the AEP).  
			As $\mr{O}=O_1,\dots,O_n$ and $\mr{S}=S_1,\dots,S_n$ are being created in a sequential manner, we intuitively wish to say that in each round $i\in[n]$ we accumulate an additional constant amount of entropy due to the production of $O_i$ (while taking into account $S_i$ and $E$) until, in the end of the process, the total amount of entropy is linear in $n$.
			Consider, however, a sequential process in which $S_1,\dots,S_{n-1}$ are all empty (i.e., do not reveal any information about the outputs) while the side-information $S_n$ produced by the last map $\mathcal{M}_n$ includes all of the outputs $O_1,\dots,O_n$. Clearly, even though we may have accumulated entropy in the rounds $i\in[n-1]$, all of it is lost after $S_n$ is leaked in the last round $n$.
			This implies that we can only hope to prove a statement like the one given in Equation~\eqref{eq:eat_general_form} under some restrictions on the considered sequential processes. 
			
			A more fundamental difficulty to overcome is the following. 
			In the case of the AEP, i.e., when considering IID processes, or boxes, it is clear what $t$, appearing in Equation~\eqref{eq:eat_general_form}, is -- it is a quantity describing the \emph{single-round box} defining the IID box. (And, as it turns out, this quantity is the relevant conditional von Neumann entropy evaluated on a single-round box; recall Section~\ref{sec:quant_aep}).
			
			Moving to the sequential processes, or multi-round boxes, it is not obvious at all which quantity $t$ should describe. We would like to find a quantity related to some ``single-round property'', but how can we even define such a thing in a meaningful way? 
			Since the behaviour of the box in each round may depend on everything that happened in previous rounds (see Definition~\ref{def:quant_seq_box_w_comm}), we cannot directly define a multi-round box in terms of single-round ones. 
			To put it differently, when holding a multi-round device, there is no physical system that we can point to and treat as an isolated subsystem. Thus, $t$ cannot refer to such a system as in the IID case. 
			
			Keeping in mind the conceptual difficulties that one needs to overcome when phrasing the theorem, we are now ready to discuss the EAT on a more concrete level. In particular, the following section unveils the resolutions of the issues presented above. 
		
		\subsection{Prerequisites of the theorem}\label{sec:eat_prereq}
		
			Before presenting the EAT, we need to define two objects to which the theorem refers~-- EAT channels and tradeoff functions. The ``correct'' definition of these objects is what allows us to overcome the conceptual difficulties discussed above. 
			Furthermore, when coming to use the EAT, the choice, or construction, of these objects is what allows one to derive a strong bound on the considered entropy. Thus, understanding the prerequisites of the theorem is of great importance.
		
			\subsubsection{EAT channels}
			
				As mentioned in the previous section, entropy does not accumulate in any general sequential process. Therefore, we must restrict our attention to processes which fulfil certain conditions. Specifically, we work with processes defined via the following type of maps, called ``EAT channels'':
			
				\begin{defn}[EAT channels]\label{def:eat_channels}
					Quantum channels $\left\{\mathcal{M}_i:R_{i-1}\rightarrow R_i O_i S_i C_i\right\}_{i\in[n]}$ are said to be EAT channels if the following requirements hold:					
					\begin{enumerate}
						\item $\left\{O_i\right\}_{i\in[n]}$ are finite dimensional quantum systems of dimension $d_{O}$ and $\left\{C_i\right\}_{i\in[n]}$ are finite-dimensional classical systems (RV).  
							$\left\{S_i\right\}_{i\in[n]}$ and $\left\{R_i\right\}_{i\in[n]}$ are arbitrary quantum systems. 
						
						\item For any $i\in[n]$ and any input state $\sigma_{R_{i-1}}$, the output state $\sigma_{R_i O_i S_i} =  \mathcal{M}_i  \left(\sigma_{R_{i-1}} \right)$ has the property that the classical value $C_i$ can be measured from the marginal $\sigma_{O_iS_i}$ without changing the state. 
						That is, for the map $\mathcal{T}_i :O_iS_i\rightarrow O_iS_iC_i$ describing the process of deriving $C_i$ from $O_i$ and $S_i$, it holds that $\Tr_{C_i}\circ\mathcal{T}_i \left(\sigma_{O_iS_i}\right) = \sigma_{O_iS_i} $.
						
						\item For any initial state 
						$\rho_{R_0E}^{\text{in}}$, the final state $\rho_{\mr{O}\mr{S}\mr{C}E} = \left( \Tr_{R_n} \circ \mathcal{M}_n \circ \dots \circ \mathcal{M}_1 \right) \otimes \idn_{E} \; \rho^{\text{in}}_{R_0E}$ 
						fulfils the Markov-chain conditions (Defintion~\ref{eq:markov_chain_def})
						\begin{equation}\label{eq:req_markov_cond}
							O_1,\dotsc ,O_{i-1} \leftrightarrow S_1,\dotsc, S_{i-1}, E \leftrightarrow S_i 
						\end{equation}
							for all $i\in[n]$. 
					\end{enumerate}
				\end{defn}
				
				In words, Equation~\eqref{eq:req_markov_cond} states that in each round, the previous outcomes $O_1,\dotsc ,O_{i-1}$ are independent of the future side-information $S_i$ given all the past side information $S_1,\dotsc, S_{i-1}, E$. That is, the side-information of any given round does not reveal new information about previous outcomes.  
				When coming to use the EAT, one is free to choose the different systems as one wishes. By choosing $O_i$ and $S_i$ properly the required Markov chain condition can be satisfied by sequential protocols such as device-independent quantum key distribution, as will be shown in Chapter~\ref{ch:crypto_showcase}.\footnote{In some cases the obvious choices for $O_i$ and $S_i$ are such that  Equation~\eqref{eq:req_markov_cond} does not hold. Still, sometimes, one can overcome the problem by considering related protocols in which the Markov-chain conditions are ``enforced''. This is done, for example, in~\cite{arnon2017device}.}

				Equation~\eqref{eq:req_markov_cond} acts as the additional constraint on the sequential process which allows us to avoid processes in which entropy does not accumulate.\footnote{One can easily verify that the problematic process described in Section~\ref{sec:eat_conc_diff} does not fulfil the Markov-chain conditions.}
				We remark that the above requirements, and Equation~\eqref{eq:req_markov_cond} in particular,  give sufficient, but perhaps not necessary, conditions for the entropy to accumulate. That is, there might be sets of weaker or incomparable conditions that can also be used to show that entropy accumulates. 

				\subsubsection{Tradeoff functions}\label{sec:trade_off_func_def}
			
				As explained in Section~\ref{sec:eat_conc_diff}, since we cannot directly define a multi-round box in terms of single-round ones, it is not clear what quantity should replace $t$ in Equation~\eqref{eq:eat_general_form}. 
				The tradeoff functions, defined below, give an adequate way to quantify the amount of entropy which is accumulated in a single step of the process, i.e., in an application of just one channel, and by this allow us to define $t$ in a meaningful way. 
				We first present the formal definition of the functions and then explain. 
				
				\begin{defn}[Tradeoff functions]\label{def:tradeoff_funcs}
					Let $\left\{\mathcal{M}_i\right\}_{i\in[n]}$ be a family of EAT channels and $\mathcal{C}$ denote the common alphabet of $C_1,\dots,C_n$. 	
					A differentiable and convex function $f_{\min}$ from the set of probability distributions $p$ over $\mathcal{C}$ to the real numbers is called a \emph{min-tradeoff function} for $\left\{\mathcal{M}_i\right\}_{i\in[n]}$ if it satisfies\footnote{The infimum and supremum over the empty set are defined as plus and minus infinity, respectively.} 
					\begin{equation}\label{eq:min_tradeoff_def}
						f_{\min}(p) \leq \inf_{\sigma_{R_{i-1}R'}:\mathcal{M}_i(\sigma)_{C_i}=p} H\left( O_i | S_i R' \right)_{\mathcal{M}_i(\sigma)} \;
					\end{equation}
					for all $i\in [n]$, where the infimum is taken over all purifications of input states of $\mathcal{M}_i$ for which the marginal on $C_i$ of the output state is the probability distribution $p$.   
					
					Similarly, a differentiable and concave function $f_{\max}$ from the set of probability distributions $p$ over $\mathcal{C}$ to the real numbers is called a \emph{max-tradeoff function} for $\{\mathcal{M}_i\}$ if it satisfies 
					\begin{equation}\label{eq:max_tradeoff_def}
						f_{\max}(p) \geq \sup_{\sigma_{R_{i-1}R'}:\mathcal{M}_i(\sigma)_{C_i}=p} H\left( O_i | S_i R' \right)_{\mathcal{M}_i(\sigma)} \;
					\end{equation}
					for all $i\in [n]$, where the supremum is taken over all purifications of input states of $\mathcal{M}_i$ for which the marginal on $C_i$ of the output state is the probability distribution~$p$.  
				\end{defn}
				
				\begin{figure}
					\centering
					\includegraphics[width=\textwidth]{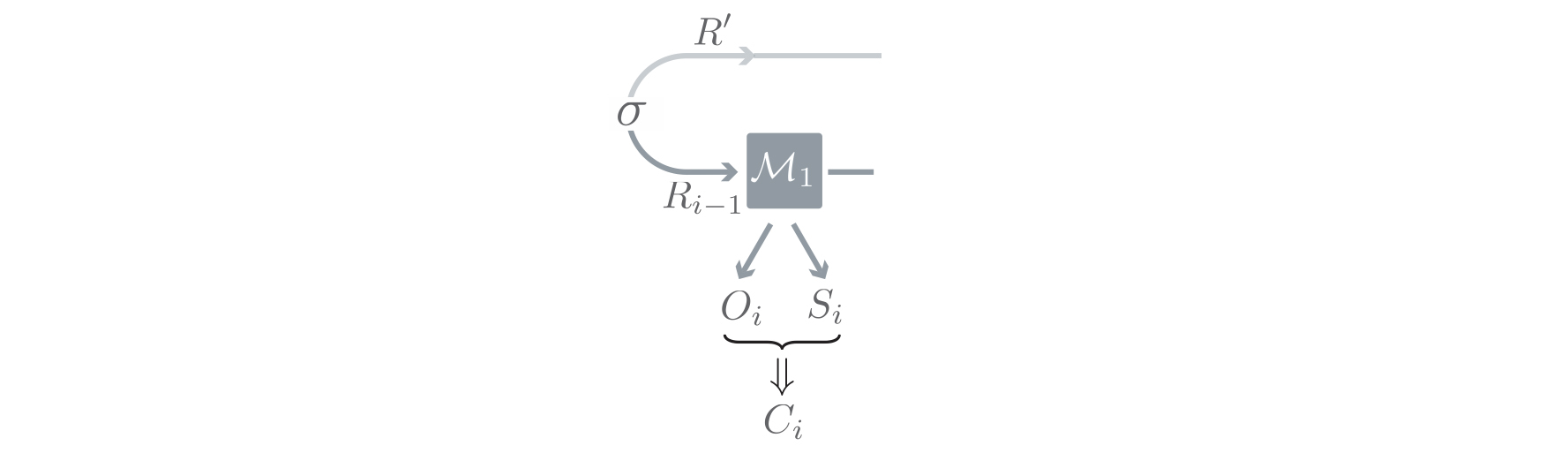}
					\caption{A single step in the sequential process. The initial state is $\sigma_{R_{i-1}R'}$; $\sigma_{R_{i-1}}$ is the input of the map $\mathcal{M}_i$ while $R'$ acts as the environment register and is not affected by the map (similarly to  $E$ in Figure~\ref{fig:seq_proc}). The map produces the registers $O_i$ and $S_i$, from which $C_i$ can be inferred.}
					\label{fig:sing_chan}
				\end{figure}
				
				Figure~\ref{fig:sing_chan} illustrates the considered scenario -- a single step in the sequential process (compare to Figure~\ref{fig:seq_proc}). 
				For any possible input state $\sigma_{R_{i-1}}\in\mathcal{S}(\hilb_{R_{i-1}})$ of the map $\mathcal{M}_i$, we denote by $\sigma_{R_{i-1}R'}\in \mathcal{S}(\hilb_{R_{i-1}}\otimes\hilb_{R'})$ its purification. Note that the register $R'$ is not being affected by the map (similarly to  $E$ in Figure~\ref{fig:seq_proc}). 
				$O_i$~and $S_i$ denote the output and side-information registers of the output state $\mathcal{M}_i(\sigma)$. $C_i$ can be inferred from  $O_i$ and $S_i$ as before.
				
				To comprehend these so called tradeoff functions, let us first discuss the set over which we perform the optimisations in Equations~\eqref{eq:min_tradeoff_def} and~\eqref{eq:max_tradeoff_def}:
				\begin{equation}\label{eq:compatible_states_set}
					\Sigma(p) = \left\{\sigma_{R_{i-1}R'}:\mathcal{M}_i(\sigma)_{C_i}=p\right\} \;.
				\end{equation}
				$\mathcal{M}_i(\sigma_{R_{i-1}})$ is the output state of the map and $\mathcal{M}_i(\sigma)_{C_i}$ is its marginal over $C_i$. Recall that the classical registers $C_i$ are used to collect statistics during the run of the considered protocol. Hence, $\mathcal{M}_i(\sigma)_{C_i}$ can be seen as a probability distribution over $\mathcal{C}$. The condition $\mathcal{M}_i(\sigma)_{C_i}=p$ therefore restricts the set of considered states~-- $\Sigma(p)$ only includes states $\sigma$ that exhibit the statistics defined by the probability distribution $p$. If there are no such states then  $\Sigma(p)$ is empty.
				
				As an example, consider a protocol in which the CHSH game is being played in each round and $C_i$ records whether the game was won ($C_i=1$) or lost ($C_i=0$). Denoting by $\omega$ the probability that $\sigma$ wins the game, we can write
				\begin{equation}\label{eq:p_to_w_illustration}
					\mathcal{M}_i(\sigma)_{C_i} = \begin{pmatrix} \omega & 0  \\  0  & 1-\omega   \end{pmatrix} \;.
				\end{equation}
				$\Sigma(p)$ then includes all of the states for which $\omega = p(1)$. 
				For a probability distribution with $p(1)=1$, for example, the set $\Sigma(p)$ is empty, since there are no quantum states which can be used to play the CHSH game with probability $\omega=1$.
				
				Given the above, the infimum/supremum of $H\left( O_i | S_i R' \right)_{\mathcal{M}_i(\sigma)} $ over the set $\Sigma(p)$
				 describes the worst-case\footnote{By ``worst-case'' we mean lowest or largest, depending on whether we are working with min- or max- tradeoff functions. This will become clearer when discussing the EAT itself.} conditional von Neumann entropy in a \emph{single round, restricted to states with the correct marginal over $C_i$.}

				To get some intuition as to why the tradeoff functions in Definition~\ref{def:tradeoff_funcs} give an adequate way of quantifying the amount of entropy accumulated in a single step of the process, let us present two ``alternative'' definitions that one could try to use and refute them with the help of simple classical examples. 
				
				In both examples we consider classical processes in which each channel $\mathcal{M}_i$ outputs a single bit $O_i$ without any side information $S_i$ about it; the system~$E$ is empty as well.  Every bit $O_i$ may depend on the ones produced previously.
				We would like to extract randomness out of $\mr{O}$ and thus aim to calculate $H_{\min}^{\varepsilon} \left( \mr{O} \right)$, which tightly describes the amount of extractable randomness~\cite{renner2008security,tomamichel2013hierarchy}. 
				How much randomness does a single round contribute to the extractable randomness \emph{given} that we already accounted for the randomness of the previous rounds?

				One possible guess is the conditional von-Neumann entropy: 
				\begin{equation}\label{eq:min_trad_intu_guess_one}
					H(O_i|O_1,\dots,\O_{i-1}) = - \mathbb{E}_{o_1,\dots,o_i}  \log\Pr(o_i|o_1,\dots,o_{i-1}) \;.
				\end{equation}
				The von Neumann entropy fulfils the chain rule and so we have $\sum_i H(O_i|O_1,\dots,\O_{i-1}) = H(\mr{O})$. Unfortunately, the smooth min-entropy $H_{\min}^{\varepsilon} \left( \mr{O} \right)$ can be arbitrarily lower than $H(\mr{O})$.
				An example for a sequential process in which this is the case is as follows: $O_1$ is uniform while, for all $i\in[n]\setminus\{1\}$, 
				\[
					O_i = \begin{cases}
									0 & O_1 = 0 \\
									\text{uniform} & \text{otherwise.}
								\end{cases}
				\]
				Direct calculation of $H(\mr{O})$ gives $H(\mr{O})=1+(n-1)/2$. 
				The min-entropy, however, depends on the most probable value of $\mr{O}$ rather than its expected value. One can easily check that $H_{\min}(\mr{O})=1$, which implies that the extractable randomness is independent of $n$. 
				Thus, $H(\mr{O})$  is too optimistic~--- it suggests that one can get arbitrarily more randomness than we can possibly extract from this process.

				Let us try a worst-case version of the min-entropy instead: 
				\begin{equation}\label{eq:min_trad_intu_guess_two}
					H_{\min}^{w.c.}= -  \log\max_{o_1,\dots,o_i} \Pr(o_i|o_1,\dots,o_{i-1})\;.
				\end{equation}
				While this option at least does not result in a contradiction (in contrast to the one above), it is too pessimistic.  
				To see this, consider IID RVs $\mr{O}$, where each $O_i$ is a Bernoulli random variable with expectation $p < 1/2$.
				Then,  Equation~\eqref{eq:min_trad_intu_guess_two} tells us that we can extract $-\log(1-p)$ randomness per round.
				However, it follows from the EAP that $h(p)>-\log(1-p)$ randomness can be extracted per round, for sufficiently large $n$.

				The following quantity lies between those given in Equations~\eqref{eq:min_trad_intu_guess_one} and~\eqref{eq:min_trad_intu_guess_two}:  
				\[
					\min_{o_1,\dots,o_{i-1}}H(O_i|O_1=o_1,\dots,O_{i-1}=o_{i-1}) \;.
				\]
				This quantity describes the von Neumann entropy of~$O_i$, evaluated for the worst case values of $O_1,\dots,O_{i-1}$. 
				Going back to the two processes considered above, one can easily verify that this choice gives the ``correct'' amount of extractable randomness in both cases.
				The min-tradeoff function defined above is the quantum analogue of this.

				The tradeoff functions are not uniquely defined by Equations~\eqref{eq:min_tradeoff_def} and~\eqref{eq:max_tradeoff_def}. The equations merely pose a constraint on the functions. 
				That is, a min-tradeoff function can be chosen to be any differentiable convex function satisfying the condition\footnote{The value of the functions at points $p$ for which $\Sigma(p)$ is the empty set is unconstrained and can be chosen freely (while keeping the function differentiable and convex).} given in Equation~\eqref{eq:min_tradeoff_def}, i.e., it is upper-bounded by $\inf_{\Sigma(p)} H\left( O_i | S_i R' \right)_{\mathcal{M}_i(\sigma)}$. 
				Similarly, a max-tradeoff function is any differentiable concave function lower-bounded by $\sup_{\Sigma(p)} H\left( O_i | S_i R' \right)_{\mathcal{M}_i(\sigma)}$. 
				To get the tightest bounds on the smooth entropies using the EAT one should construct tradeoff functions in the tightest way possible, ideally matching the exact value of the worst-case von Neumann entropy given in Equations~\eqref{eq:min_tradeoff_def} and~\eqref{eq:max_tradeoff_def}.
				In device-independent cryptographic protocols based on the CHSH game, for example, Lemma~\ref{lem:single_round_secrecy} can be used to construct a tight min-tradeoff function; this will be done in Chapter~\ref{ch:crypto_showcase}.

\subsection{Statement of the theorem}\label{sec:eat_statement}
	
			After presenting the prerequisites of the EAT, we are now ready to discuss the statement of the theorem. 
			
			\subsubsection{Conditioning on not aborting}\label{sec:eat_cond_on_omega}
			
				Consider a sequential protocol, i.e., one which proceeds in rounds; the development of the quantum state throughout the protocol can be described by a sequential process. In the end of the protocol, the honest parties can choose whether to abort the protocol or not. For example, if Alice and Bob run a device-independent cryptographic protocols and observe that the device does not win the game in sufficiently many games, they conclude that the device might be malicious and abort the protocol. 
				Our goal is to bound the smooth entropies of the outputs when the protocol \emph{does not abort}. 
				
				Whether the protocol aborts or not depends on the observed data produced during the execution of the protocol and, specifically, on the value assigned to $\mr{C}$. Thus, the event of not aborting the protocol, denoted by $\Omega$, is defined to be a subset of $\mathcal{C}^n$. 
				The most common way of choosing the set $\Omega$ is such that whether $\mr{c}=c_1, \dotsc, c_n \in \mathcal{C}^n$ belongs to $\Omega$ or not depends on its ``frequencies''. 
				Formally, for any $\mr{c}\in\mathcal{C}^n$, denote by $\mathrm{freq}_{\mr{c}}$ the probability distribution over $\mathcal{C}$ defined by 
					\begin{equation}\label{eq:freq_definition}
						\mathrm{freq}_{\mr{c}}(\tilde{c}) = \frac{| \left\{ i | c_i = \tilde{c} \right\} |}{n}
					\end{equation}
					for $\tilde{c}\in\mathcal{C}$.
				We define a set $\hat{\Omega}$ that includes all the probability distributions approved by the protocol, i.e., the desired frequencies $\mathrm{freq}_{\mr{c}}$ for which the protocol does not abort.				
				Then, we can write the event of not aborting in terns of the desired frequencies:
				\begin{equation*}
					\Omega = \left\{ \mr{c} : \mathrm{freq}_{\mr{c}} \in \hat{\Omega}  \right\} \subseteq \mathcal{C}^n \;.
				\end{equation*}
				Note that one can also start by choosing the set $\Omega$ describing the event of not aborting the protocol. Then, $\hat{\Omega}$ can be chosen to be any set fulfilling\footnote{It will become clear from the statement of the EAT that one should choose a minimal convex set $\hat{\Omega}$ that includes the frequencies considered in $\Omega$. It is perhaps instructive to observe that, for a finite $n$, $\Omega$ is a finite set; $\hat{\Omega}$, on the other hand, includes infinitely many probability distributions.}
				\[
					\left\{ \mathrm{freq}_{\mr{c}} : \mr{c} \in \Omega \right\} \subseteq \hat{\Omega}  \;.
				\]
				Focusing on permutation invariant sets $\Omega$, in the sense that $\mr{c}\in\Omega$ if and only if $\pi(\mr{c})\in\Omega$ for all permutations $\pi$ of the $n$ indices,  defining $\hat{\Omega}$ via $\Omega$ is practically the same as defining $\Omega$ via $\hat{\Omega}$.

				Let us present a simple example of the above definitions and sets. Let $\mathcal{C}=\{0,1\}$ and consider, e.g., 
				\begin{equation}\label{eq:freq_string_exmp}
					\mr{c} = 01101000110100111011 \;.
				\end{equation}
				To write $\mathrm{freq}_{\mr{c}}$ we count the number of zeros and ones in the above string and get, according to Equation~\eqref{eq:freq_definition}, the probability distribution over $\{0,1\}$ defined by
				\[
					\mathrm{freq}_{\mr{c}}(0) = \frac{9}{20} \quad \;; \quad \mathrm{freq}_{\mr{c}}(1) = \frac{11}{20} \;.
				\]
				We can now consider a protocol which \emph{does not abort} whenever the observed statistics are such that the fraction of ones is greater than half. This leads to
				\begin{align*}
					\hat{\Omega} &= \left\{ p : p(1) > \frac{1}{2} \right\} \\
					\Omega &= \left\{ \mr{c} : \mathrm{freq}_{\mr{c}} \in \hat{\Omega} \right\} = \left\{ \mr{c} : \mathrm{freq}_{\mr{c}}(1) > \frac{1}{2} \right\} 
				\end{align*}
				and, in particular, for $\mr{c}$ appearing in Equation~\eqref{eq:freq_string_exmp}, $\mr{c}\in\Omega$.

				\subsubsection{The theorem}

				We first give the formal statement of the EAT and then explain. 
				\begin{thm}[EAT]\label{thm:eat}
					Let $\mathcal{M}_i:R_{i-1}\rightarrow R_i O_i S_i  C_i$ for $i\in [n]$ be EAT channels, 
					$\rho$ be the final state, 
					$\Omega$ an event  defined over $\mathcal{C}^n$, 
					$p_\Omega$ the probability of $\Omega$ in $\rho$, 
					and $\rho_{|\Omega}$ the final state conditioned on $\Omega$.
					Let~$\varepsilon \in (0,1)$.
					 
					For $\hat{\Omega} = \{  \mathrm{freq}_{\mr{c}} : \mr{c} \in \Omega \}$ convex,\footnote{We consider only \emph{convex} sets $\hat{\Omega}$ (as was done in~\cite{arnon2018practical}). One can convince oneself that choosing a convex  $\hat{\Omega}$ is the sensible thing to do. For example, a set $\hat{\Omega}$ including all frequencies, or probability distributions, for which $p(1)\in[a,b]$ for some constants $0\leq a,b \leq 1$ is convex. If, nevertheless, one wishes to consider arbitrary sets $\Omega$, which are not defined via a convex $\hat{\Omega}$,  then this comes at the cost of considering only affine tradeoff functions (instead of convex/concave functions as in Definition~\ref{def:tradeoff_funcs}); see~\cite{dupuis2016entropy} for the original claim. It is not clear that there are scenarios in which $\Omega$ cannot be defined with an underlying convex set $\hat{\Omega}$ and, at the same time, applying the EAT with adequate affine tradeoff functions does not result in a trivial statement. Hence, the convexity of $\hat{\Omega}$ should not be seen as a restriction.}
					$f_{\min}$ a min-tradeoff function for $\{\mathcal{M}_i\}_{i\in[n]}$, 
					and any $t\in \mathbb{R}$ such that $f_{\min}\left( \mathrm{freq}_{\mr{c}} \right) \geq t$ for any $\mathrm{freq}_{\mr{c}} \in \hat{\Omega}$, 
					\begin{equation}\label{eq:eat_min_bound}
						H_{\min}^{\varepsilon} \left( \mr{O}|\mr{S}E \right)_{\rho_{|\Omega}} > n t - \mu\sqrt{n} \;,
					\end{equation}
					where 
					\begin{equation}\label{eq:eat_min_bound_second_ord}
						\mu = 2\left(\log(1+2 d_{O}) + \lceil \|  \triangledown f_{\min} \|_{\infty} \rceil \right)\sqrt{1-2\log (\varepsilon \cdot p_\Omega)}\;,
					\end{equation}
					$d_{O}$ the dimension of the systems $O_i$, and $\|  \triangledown f_{\min} \|_{\infty}$ is the infinity norm of the gradient of $f_{\min}$. 
					
					Similarly, for $\hat{\Omega} = \{  \mathrm{freq}_{\mr{c}} : \mr{c} \in \Omega \}$ convex,  
					$f_{\max}$ a max-tradeoff function 
					and any $t\in \mathbb{R}$ such that $f_{\max}\left(\mathrm{freq}_{\mr{c}} \right) \leq t$ for any $\mathrm{freq}_{\mr{c}} \in \hat{\Omega}$, 
					\begin{equation}\label{eq:eat_max_bound}
						H_{\max}^{\varepsilon} \left( \mr{O} |\mr{S}E \right)_{\rho_{|\Omega}} < n t + \mu\sqrt{n} \;,
					\end{equation}
					with 
					\begin{equation}
						\mu = 2\left(\log(1+2 d_{O}) + \lceil \|  \triangledown f_{\max} \|_{\infty} \rceil \right)\sqrt{1-2\log (\varepsilon \cdot p_\Omega)}\;.
					\end{equation}
				\end{thm}
				
				Let us parse the statement of the theorem while focusing on the smooth min-entropy for the moment. 
				Equation~\eqref{eq:eat_min_bound} has exactly the form that we were aiming for: it gives a lower-bound on the conditional smooth min-entropy, evaluated on the state $\rho_{|\Omega}$ in the end of the protocol and conditioned on not aborting, where the first order term is linear in $n$ and the second, describing finite statistic effects, scales like~$\sqrt{n}$. 
				
				The constant $t$, governing the entropy rate $H_{\min}^{\varepsilon} \left( \mr{O}|\mr{S}E \right)_{\rho_{|\Omega}}/n$ when $n\rightarrow\infty$, is defined via the min-tradeoff function in the following way. The min-tradeoff function~$f_{\min}$ assigns to each probability distribution $p$, or frequency, a number describing the minimal amount of conditional von Neumann entropy which is compatible with the probability distribution $p$ (recall Definition~\ref{eq:min_tradeoff_def}). 
				We now consider all frequencies $\mathrm{freq}_{\mr{c}}$ (i.e., probability distributions) which are accepted by the protocol. The theorem asserts that $t$ should be chosen as the minimal value of~$f_{\min}$ over this set of accepted frequencies.  
				That is,\footnote{The reader may be concerned that for finite $n$ all frequencies belonging to $\mathrm{freq}_{\mr{c}}\in \hat{\Omega}$ actually lead to empty sets $\Sigma(\mathrm{freq}_{\mr{c}})$, defined in Equation~\eqref{eq:compatible_states_set} and hence $t$ can be arbitrary. Note however that the tradeoff functions are defined over the set of \emph{all} probability distributions, not only over the possible frequencies. Since the tradeoff functions must be differential convex/concave functions with a finite gradient, the points in which the functions are constrained by Equations~\eqref{eq:min_tradeoff_def} and~\eqref{eq:max_tradeoff_def} also constrain the values of the functions at the points $\mathrm{freq}_{\mr{c}} \in \hat{\Omega}$.}
				\[	
					t = \inf \left\{ f_{\min}\left( \mathrm{freq}_{\mr{c}} \right)  : \mathrm{freq}_{\mr{c}} \in \hat{\Omega} \right\}\;.
				\]
				Since the min-tradeoff function is practically the worst-case conditional von Neumann entropy, we get that, asymptotically, $H_{\min}^{\varepsilon} \left( \mr{O}|\mr{S}E \right)_{\rho_{|\Omega}}/n$ is equal to the lowest von Neumann entropy of a single-round (in the sense defined by the min-tradeoff function) that is compatible with the statistics observed in the protocol.

\begin{figure}
					\centering
					\begin{tikzpicture}
						\begin{axis}[
							height=7cm,
							width=11cm,
							xlabel=$\omega$,
							ylabel=$f_{\min}(\omega)$,
							xmin=0.75,
							xmax=0.853553,
							ymax=1,
							ymin=0,
						     xtick={0.82},
						     xticklabels={$\omega_T$},
					         ytick={0.53},
					          yticklabels={$t$},
							legend style={at={(0.21,0.95)},anchor=north,legend cell align=left,font=\footnotesize} 
						]
						
						
						\coordinate (tpoint) at (axis cs:0.82,0.53);
						
						\path[name path=axisone] (axis cs:0.75,0) -- (axis cs:0.82,0);
						\path[name path=axistwo] (axis cs:0.82,0) -- (axis cs:0.853553,0);
					
						\addplot[blue, thick, smooth, name path global = partone] coordinates {
						(0.75, 0.) (0.752071, 0.0120347) (0.754142, 0.0242374) (0.756213, 0.0366113) (0.758284, 0.0491598) (0.760355, 0.0618861) (0.762426, 0.074794) (0.764497, 0.0878872) (0.766569, 0.10117) (0.76864, 0.114646) (0.770711, 0.128319) (0.772782, 0.142196) (0.774853, 0.156279) (0.776924, 0.170575) (0.778995, 0.185089) (0.781066, 0.199826) (0.783137, 0.214793) (0.785208, 0.229996) (0.787279, 0.245441) (0.78935, 0.261137) (0.791421, 0.277091) (0.793492, 0.293311) (0.795563, 0.309806) (0.797635, 0.326586) (0.799706, 0.343661) (0.801777, 0.361042) (0.803848, 0.378741) (0.805919, 0.396771) (0.80799, 0.415147) (0.810061, 0.433884) (0.812132, 0.452998) (0.814203, 0.47251) (0.816274, 0.49244) (0.818345, 0.51281) (0.82, 0.53) 
						};
						
						\addplot[blue, thick, smooth, name path global = parttwo] coordinates {
						 	(0.82, 0.53) (0.820416, 0.533648) (0.822487, 0.554982) (0.824558, 0.576846) (0.82663, 0.599279) (0.828701, 0.622324) (0.830772, 0.646033) (0.832843, 0.670469) (0.834914, 0.695705) (0.836985, 0.721832) (0.839056, 0.748965) (0.841127, 0.777251) (0.843198, 0.806888) (0.845269, 0.838156) (0.84734, 0.871481) (0.849411, 0.907587) (0.851482, 0.948007) (0.853553, 1.)	
						};
						
						\addplot[black,smooth,dotted] coordinates {
							(0.75,0.53) (0.853553, 0.53)
						};
						
						\addplot[black,smooth,dotted] coordinates {
							(0.82,0) (0.82, 1)
						};

						\addplot [
					        thick,
					        color=red,
					        fill=red, 
					        fill opacity=0.1
					    ]
					    fill between[
					        of=partone and axisone,
					        soft clip={domain=0:1},
					    ];
					    
					    \addplot [
					        thick,
					        color=green,
					        fill=green, 
					        fill opacity=0.1
					    ]
					    fill between[
					        of=parttwo and axistwo,
					        soft clip={domain=0:1},
					    ];

						\end{axis}  
						
						\filldraw[draw=black]  (tpoint) circle(0.07);
						
					\end{tikzpicture}
					
					\caption{First order term from the min-tradeoff function. We consider a protocol which does not abort if the fraction of games won is above $\omega_T$. The value of $t$, appearing in Equation~\eqref{eq:eat_min_bound}, should be chosen to be the lowest value that the min-tradeoff function $f_{\min}$ assigns to accepted winning probabilities, i.e., the black point in the plot.}
					\label{fig:f_min_value_for_t_illust}
				\end{figure}
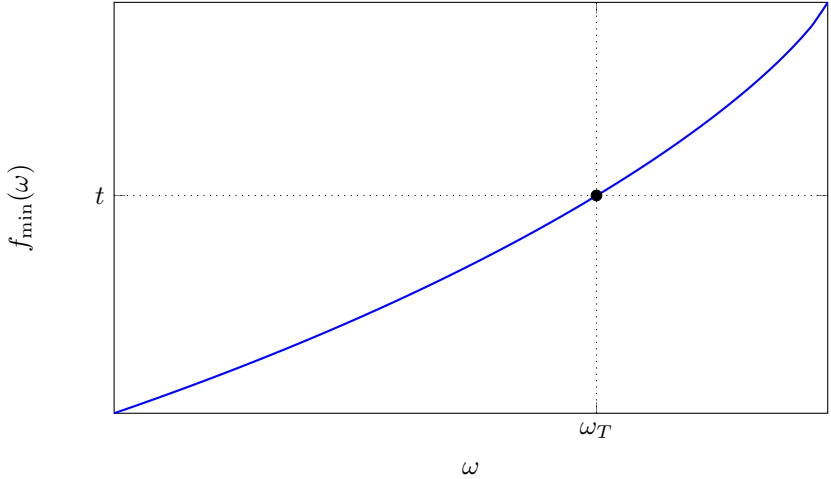

				As an example, consider a device-independent protocol that assigns $C_i=1$ when the $i$'th game is won and $C_i=0$ otherwise and which does not abort as long as the fraction of games won $(\sum_i C_i )/ n$ is above some threshold $\omega_T$. 
				The min-tradeoff function is defined over probability distributions $p$ over $\mathcal{C}=\{0,1\}$. Thus, we can also think of it as a function over winning probabilities $\omega$ via the relation $p(1)=\omega$ and $p(0)=1-\omega$ (see Equation~\eqref{eq:p_to_w_illustration}). 
				Assuming that the min-tradeoff function is increasing with $\omega$ (as expected to be; see Lemma~\ref{lem:single_round_secrecy}), the lowest value that $f_{\min}$ assigns to accepted frequencies is $f_{\min}(\omega_T)$ and hence this should be the value of $t$; this is illustrated in Figure~\ref{fig:f_min_value_for_t_illust}.

				The above discussion refers to the first order term in Equation~\eqref{eq:eat_min_bound}. Let us now briefly discuss the second order term. 
				Similarly to the AEP, the second order term scales like $\sqrt{n}$, which is the optimal scaling. 
				The constant $\mu$, defined in Equation~\eqref{eq:eat_min_bound_second_ord}, depends on the different parameters and constants. In particular, it depends on the dimension of the output systems $O_i$ and the gradient of the tradeoff functions. To get a good second order term one should therefore choose the possible values that can be assigned to the registers $O_i$ and the tradeoff functions such that the quantities of interest can be bounded in a good manner. 			
				In particular, to control the gradient of the tradeoff function one can ``cut'' when the gradient becomes too large and linearise the function at that point.  An example is given in Section~\ref{sec:crypto_min_tradeoff}.
				Improved second order terms for the EAT were derived in~\cite{dupuis2018entropy}.  

				A last remark is with regards to the statement of the EAT for the smooth max-entropy. When interested in the smooth max-entropy, the registers describing the environment, i.e., $E$ in Equations~\eqref{eq:req_markov_cond} and~\eqref{eq:eat_max_bound} as well as $R'$ in Equation~\eqref{eq:max_tradeoff_def}, can be dropped. 
				The fact that $R'$ can be dropped from Equation~\eqref{eq:max_tradeoff_def} was already noted in~\cite[Remark~4.2]{dupuis2016entropy}; the reason is that for the calculation of the supremum one can always assume that the system on $R'$ is in product with the rest of the systems.
				To see that $E$ can be dropped from Equations~\eqref{eq:markov_chain_def} and~\eqref{eq:eat_max_bound} note that the EAT must hold for any initial state $\rho_{R_0E}^{\text{in}}$ and, hence, in particular to a tensor product state $\rho_{R_0E}^{\text{in}}=\rho_{R_0}\otimes\rho_E$, for which the conditional smooth max-entropy is maximal.

				To conclude this chapter, we summarise the reasons for why the EAT can be seen as an extension of the AEP for non-IID processes:
				\begin{enumerate}
					\item Similarly to the AEP, the EAT tells us that for large enough $n$ the smooth entropies are equal to the \emph{von Neumann entropy of a single-round times the number of rounds}. 
				
					\item The observed frequencies from the entire process are used when calculating the entropy accumulated in a single step of the process, \emph{as if} all steps contribute equally and independently of each other. This is analogous to the analysis done in the IID case, as we saw in Section~\ref{sec:using_IID_assump}.
					
					\item The asymptotic bounds on the smooth entropies derived using the EAT are equal to those derived using the AEP under the IID assumption. As a result, the bounds resulting form the application of the EAT are tight (assuming that the constructed tradeoff functions are tight). The second order terms of both theorems are also similar, as the scaling of both is $\sqrt{n}$.
				\end{enumerate}
				All the above justifies the use of the EAT as a ``reduction to IID'' technique.


\chapter{Showcase: non-signalling parallel repetition}\label{ch:par_rep_showcase}

	In this chapter we consider the showcase of non-signalling parallel repetition, introduced in Section~\ref{sec:pre_par_rep}, and show how threshold theorems derived under the IID assumption can be extended to threshold theorems for general strategies, using a \emph{reduction to~IID}.
	
	We focus on the case of non-signalling players.\footnote{Most steps of our proof can be used as is when considering classical and quantum players as well. There is one lemma, however, which we do not know how to modify so it can capture the classical and quantum case. We explain the difficulty later on.} That is, the \emph{only} restriction on the players is that they are not allowed to communicate.  
	Considering the non-signalling case is interesting for several reasons. A first reason is to minimise the set of assumptions to the mere necessary. 
		Minimising the set of assumptions can be useful in cryptography when one wishes to get the strongest result possible, i.e., one where the attack strategies of malicious parties are only restricted minimally (as in \cite{hanggi2009quantum,masanes2009universally,masanes2014full} for example). 
	In theoretical physics, non-signalling correlations enable the study of generalised theories possibly beyond quantum theory.  
	It is also important to mention that, due to their linearity, the non-signalling constraints are often easier to analyse than the quantum or the classical constraints. Therefore, even if additional constraints hold, focusing on the non-signalling ones serves as a way to get first insights into a given problem.
		
	Our theorem deals with complete-support games; these are game in which the distribution $\Q_{XY}$ over the questions has complete support, i.e., for all $x,y$, $\Q_{XY}(x,y)>0$.\footnote{When considering games with only two players, the requirement for complete support can be dropped but, even though we focus on two-player games, we do not discuss this here; see~\cite{arnon2016non} for the details.}
	The main result presented in this chapter can be informally stated as follows.\footnote{See Theorem~\ref{thm:final_threshold_theorem} for the formal statement.}
	\begin{thm}[Informal]\label{thm:parallel_rep_informal}
		For any complete-support game, a threshold theorem for general non-signalling strategies follows from a threshold theorem for non-signalling IID strategies. 
		Furthermore, given a game with optimal non-signalling winning probability $1-\alpha$, the resulting threshold theorem states that, for any $\beta>0$, the probability to win more than a fraction $1-\alpha+\beta$ of $n$ games is exponentially small in $n\beta^2$, as in the IID case. 
	\end{thm}
	
	\sloppy
	The result previously appeared in~\cite{arnon2016non}. 
	We remark that while the non-signalling  threshold theorem of~\cite{buhrman2013parallel} was known prior to~\cite{arnon2016non}, the dependence on $\beta$ did not match that of the IID case, which is optimal (as follows from the optimal formulation of the Chernoff bound).
	Following~\cite{arnon2016non}, the work of~\cite{holmgren2017counterexample} showed that parallel repetition does \emph{not} hold for general games without complete-support in the non-signalling case. Thus, Theorem~\ref{thm:parallel_rep_informal} is as general as it gets. 
	
	Proving parallel repetition via a reduction to IID has the advantage that, as in the IID case, the proof is oblivious to the number of players and structure of the considered game. This leads to a general theorem applicable to all games (with complete support) that is, arguably, simpler than other proof techniques.

	When considering a strategy for the repeated game there is one type of symmetry which one can take advantage of -- since the repeated game is permutation invariant the same symmetry can be assumed to hold for the optimal strategies, without loss of generality.
	Permutation-invariant strategies are strategies which are indifferent to the ordering of the questions given by the referee. That is, the probability of answering a specific set of questions correctly does not depend on the ordering of the questions (see Section~\ref{sec:threshold_theorem} below for the formal definitions). 
	Once we restrict our attention to permutation-invariant strategies, de Finetti theorems presents themselves as a natural tool to leverage for the analysis. Indeed, our proof builds on the de Finetti reduction discusses in Chapter~\ref{ch:reductions_par}, which acts as a reduction to IID in our analysis of parallel repetition. 
	
	The chapter is arranged as follows. In Section~\ref{sec:par_rep_chall_goal} we explain the main challenge when proving parallel repetition and threshold theorems using techniques employed by works predating~\cite{arnon2016non} and why de Finetti theorems were not used in the context of parallel repetition before~\cite{arnon2016non}. 
	In Section~\ref{sec:approx_ns_marginals} we give the main technical statements needed to prove our non-signalling threshold theorem. The different observations and lemmas of Section~\ref{sec:approx_ns_marginals} may be of independent interest when analysing non-signalling parallel boxes and therefore we perform the analysis without referring to multi-player games. 
	The explicit threshold theorem and its proof are given in Section~\ref{sec:threshold_theorem}. 
	 As in the rest of the thesis, we focus on the case of two parties for simplicity; we refer to~\cite{arnon2016non} for the proofs in the case of more than two players as well as a couple of extensions of our theorem to games without complete support. 
	
	\section{Main challenge and goal}\label{sec:par_rep_chall_goal}

		The main difficulty in proving a parallel repetition result comes from the, almost arbitrary, correlations between the different questions-answers pairs in the players' strategy for the repeated $\G_{1-\alpha+\beta}^n$: as the players get all the $n$ questions together they can answer them in a correlated way. In most of the known parallel repetition results (e.g., \cite{raz1998parallel,holenstein2007parallel,rao2011parallel,buhrman2013parallel}) the main idea of the proof is to bound the winning probability for some of the questions, \emph{conditioned} on winning the game in several other coordinates. 
		However, as the strategy itself introduces correlations between the different tuples of questions, a large amount of technical work is devoted to dealing with the effect of conditioning on the event of winning the previous questions.

		As mentioned above (and formally stated in Section~\ref{sec:threshold_theorem}), due to the permutation invariance of $\G_{1-\alpha+\beta}^n$ one can study only permutation invariant strategies without loss of generality. 
		Once we restrict our attention to permutation-invariant strategies, de Finetti theorems seem like a natural tool to leverage for the analysis.
		In the context of games and strategies, de Finetti theorems suggest one may be able to reduce the analysis of general permutation-invariant strategies to the analysis of a de Finetti strategy, i.e., a convex combination of IID strategies; recall Chapter~\ref{ch:reductions_par}. 
		As presented in Section~\ref{sec:par_rep_under_iid}, the behaviour of IID strategies is trivial under parallel repetition. Hence, a reduction to IID using a de Finetti-type theorem could significantly simplify the analysis of parallel repetition theorems and threshold theorems.  
		
		\sloppy
		Yet, de Finetti theorems were not used in this context prior to~\cite{arnon2016non}, and for a good reason. The many versions of quantum de Finetti theorems (e.g., \cite{renner2007symmetry,christandl2009postselection}) could not have been used as they depend on the dimension of the underlying quantum strategies, while in the quantum multi-player game setting one does not wish to restrict the dimension. 
		Non-signalling de Finetti theorems, as in~\cite{barrett2009finetti,christandl2009finite}, were also not applicable for non-signalling parallel repetition theorems, as they restrict almost completely the type of allowed correlations in the strategies for the repeated game by assuming very strict non-signalling constraints between the different repetitions, i.e., between the different questions-answers pairs. 
		
		In the proof presented in the next sections, we use the de Finetti reduction presented in Chapter~\ref{ch:reductions_par}, which imposes no assumptions at all regarding the structure of the strategies (apart from permutation invariance), and is therefore applicable in the context of parallel repetition. 
		This allows us to devise a proof technique which is completely different from the  proofs of parallel repetition results predating~\cite{arnon2016non}.\footnote{Following~\cite{arnon2016non},~\cite{lancien2016parallel} presented another, conceptually similar but technically different, proof of non-signalling parallel repetition based on de Finetti reductions.} In particular, at least in the non-signalling case presented here, the conditioning problem described above disappears completely and the number of players does not play a role in the proof structure. 
		
		As explained in Section~\ref{sec:dF_reductions}, the de Finetti strategy that one ought to consider when using our de Finetti reduction assigns some weight to \emph{signalling} IID strategies.\footnote{As discussed in Section~\ref{sec:dF_imp_res}, this is inevitable.}
		Most of the effort is therefore directed to, informally, showing that when starting with a permutation-invariant non-signalling strategy the de Finetti strategy must assign only a small weight to signalling IID strategies. Formally, a similar in spirit but somewhat different statement is proven; see Theorem~\ref{thm:approx_ns_mar_informal} below.
		Without further ado, let us get into the proof of the exact statements in the following section.
		
	\section{Approximately non-signalling marginals}\label{sec:approx_ns_marginals}
	

		Consider a parallel box $\P_{\mr{A}\mr{B}|\mr{X}\mr{Y}}$ (as defined in Section~\ref{sec:parallel_mr_boxes}) and some complete-support distribution $\Q_{XY}$, i.e., for all $\tilde{x}\in\mathcal{X}$ and $\tilde{y}\in\mathcal{Y}$, $\Q(\tilde{x},\tilde{y})\neq 0$.		
		Sample $\mr{x}$, $\mr{y}$, $\mr{a}$, and $\mr{b}$ according to $\Q_{\mr{X}\mr{Y}}^{\otimes n} \P_{\mr{A}\mr{B}|\mr{X}\mr{Y}}$.
		We assume in this section that all possible inputs $\tilde{x},\tilde{y}$ appear in the observed data $\mr{x}$ and $\mr{y}$ (that is, there exists $i\in[n]$ for which $(x_i,y_i)=(\tilde{x},\tilde{y})$. 
		For a complete-support distribution $\Q_{XY}$, the probability that this is \emph{not} the case is exponentially small in $n$ and we will account for it later.

		Next, let $\O_{ABXY}^{\textup{freq}(\mr{a},\mr{b},\mr{x},\mr{y})}$ be the distribution derived from the frequencies in the observed data via
		\begin{equation*}
			\O_{ABXY}^{\textup{freq}(\mr{a},\mr{b},\mr{x},\mr{y})} (\tilde{a}\tilde{b}\tilde{x}\tilde{y}) = \frac{\Big|\left\{i : (a_i,b_i,x_i,y_i) = (\tilde{a},\tilde{b},\tilde{x},\tilde{y}) \right\}\Big|}{n} 
		\end{equation*}
		and define
		\begin{equation}\label{eq:freq_single_round_dF}
			\O_{AB|XY}^{\textup{freq}(\mr{a},\mr{b},\mr{x},\mr{y})} = \frac{\O_{ABXY}^{\textup{freq}(\mr{a},\mr{b},\mr{x},\mr{y})}}{\Q_{XY}} \;.
		\end{equation}
		
		Without the complete-support requirement on $\Q_{XY}$ it does not even make sense to talk about a fully defined $\O_{AB|XY}^{\textup{freq}(\mr{a},\mr{b},\mr{x},\mr{y})}$, i.e., a conditional probability distribution which is defined for all $x\in\mathcal{X}$ and $y\in\mathcal{Y}$. Indeed, $\O_{AB|XY}^{\textup{freq}(\mr{a},\mr{b},\mr{x},\mr{y})}$ can only be defined for $x\in\mathrm{Supp}(\mathcal{X})$ and $y\in\mathrm{Supp}(\mathcal{Y})$ due to the estimation process (at least when assuming that all inputs appear in the observed data, which happens with high probability for large enough $n$). 
		If one is willing to consider conditional probability distributions which are allowed to not assign values to certain inputs then $\O_{AB|XY}^{\textup{freq}(\mr{a},\mr{b},\mr{x},\mr{y})}$ regains its meaning. In the context of non-signalling boxes, these  conditional probability distributions were termed ``sub-non-signalling'' boxes in~\cite{lancien2016parallel}; sub-non-signalling boxes fulfil the subset of the non-signalling conditions which apply for the defined inputs. 
			In the case of two parties, it is known that there is always a way to ``complete'' a sub-non-signalling box to a non-signalling box, defined over all inputs~\cite{ito2010polynomial,lancien2016parallel}.

		The current section deals with the following question: given that we start with a non-signalling box $\P_{\mr{A}\mr{B}|\mr{X}\mr{Y}}$, what is the probability that the \emph{single-round box}~$\O_{AB|XY}^{\textup{freq}(\mr{a},\mr{b},\mr{x},\mr{y})}$ is signalling?
		Under the IID assumption, i.e., when $\P_{\mr{A}\mr{B}|\mr{X}\mr{Y}}=\O_{AB|XY}^{\otimes n}$, this question is natural and can be easily answered. In that case,~$\O_{AB|XY}^{\textup{freq}(\mr{a},\mr{b},\mr{x},\mr{y})}$ can simply be seen as an estimation of the ``real box'', or marginal,~$\O_{AB|XY}$. In particular, according to Sanov's theorem (Lemma~\ref{lem:sanovs_theorem}), as~$n\rightarrow\infty$, we have~$\O_{AB|XY}^{\textup{freq}(\mr{a},\mr{b},\mr{x},\mr{y})}=\O_{AB|XY}$ almost surely.
		Thus, for a non-signalling IID box~$\P_{\mr{A}\mr{B}|\mr{X}\mr{Y}}$, $\O_{AB|XY}^{\textup{freq}(\mr{a},\mr{b},\mr{x},\mr{y})}$ must be non-signalling as ~$n\rightarrow\infty$.

		Our goal is to show that roughly the same is true for permutation invariant non-signalling parallel boxes  $\P_{\mr{A}\mr{B}|\mr{X}\mr{Y}}$ when boxes such as $\O_{AB|XY}^{\textup{freq}(\mr{a},\mr{b},\mr{x},\mr{y})} $ take the role of the marginals, which are not properly defined for parallel boxes.
		The theorem can be stated informally as follows:\footnote{For the formal statement see Theorem~\ref{thm:strong_condition}.}
		\begin{thm}[Informal]\label{thm:approx_ns_mar_informal}
			Let $\P_{\mr{A}\mr{B}|\mr{X}\mr{Y}}$ be a permutation invariant non-signalling parallel box and $\O_{AB|XY}^{\textup{freq}(\mr{a},\mr{b},\mr{x},\mr{y})}$ be the single-round box defined via the observed data sampled using $\Q_{XY}^{\otimes n}\P_{\mr{A}\mr{B}|\mr{X}\mr{Y}}$, as in Equation~\eqref{eq:freq_single_round_dF}. Then, for sufficiently large $n$, $\O_{AB|XY}^{\textup{freq}(\mr{a},\mr{b},\mr{x},\mr{y})}$ is close to a non-signalling single-round box with high probability.
			This also implies that the observed data can be seen as if, with high probability, it was sampled using an IID box~$\O_{AB|XY}^{\otimes n}$ with $\O_{AB|XY}$  close to a non-signalling single-round box.
		\end{thm}

		To prove the theorem we utilise the concept of a test, discussed in Section~\ref{sec:dF_fail_prob_test}. Roughly speaking, we define a \emph{signalling test} $\mathcal{T}$, interacting with a parallel box, which \emph{accepts} whenever the box $\O_{AB|XY}^{\textup{freq}(\mr{a},\mr{b},\mr{x},\mr{y})}$ is highly signalling and \emph{rejects} whenever the box is close to a non-signalling box. 
		With the aid of the variant of the de Finetti reduction phrased as Theorem~\ref{thm:dF_test_bound} and a rather simple signalling game (defined in Section~\ref{sec:sig_game}) we prove that the probability that the test accepts when interacting with a permutation invariant non-signalling parallel box is small.
		We follow this proof idea in the succeeding sections.

		\subsection{Single-round boxes from frequencies}\label{sec:box_from_freq}
			
			To ease notation we denote $\mathsf{data} = \mr{a},\mr{b},\mr{x},\mr{y}$ when it is clear from the context which $\mr{a},\mr{b},\mr{x},\mr{y}$ are considered.
			Every observed $\mathsf{data}$ is split into two non-overlapping parts, $\mathsf{data}_1$ and $\mathsf{data}_2$. Specifically, let\footnote{For simplicity we assume $n$ is even; otherwise replace $n/2$ by $\lceil n/2 \rceil$ and modify everything else accordingly.}
			\begin{equation}\label{eq:data_partition}
			\begin{split}
				&\mathsf{data}_1 = a_1,\dots,a_{\frac{n}{2}},b_1,\dots,b_{\frac{n}{2}},x_1,\dots,x_{\frac{n}{2}},y_1,\dots,y_{\frac{n}{2}} \;, \\
				&\mathsf{data}_2 = a_{\frac{n}{2}+1},\dots,a_n,b_{\frac{n}{2}+1},\dots,b_n,x_{\frac{n}{2}+1},\dots,x_n,y_{\frac{n}{2}+1},\dots,y_n \;.
			\end{split}
			\end{equation}
			$\mathsf{data}$ (and hence also $\mathsf{data}_1$ and $\mathsf{data}_2$) is sampled according to $\Q_{\mr{X}\mr{Y}}^{\otimes n} \P_{\mr{A}\mr{B}|\mr{X}\mr{Y}}$, where~$\P_{\mr{A}\mr{B}|\mr{X}\mr{Y}}$ is a non-signalling permutation invariant parallel box. Note the following:
			\begin{enumerate}
				\item $\P_{\mr{A}\mr{B}|\mr{X}\mr{Y}}$ may be signalling between the different rounds $i\in[n]$ (i.e., for a given party, the output of one round may depend on the input of other rounds). Therefore, even though $\mathsf{data}_1$ and $\mathsf{data}_2$ are each defined only by part of the observed data, they may depend on the entire data. 
				\item Due to permutation invariance, it does not matter which indices $i\in[n]$ belong to each part of the data. We could as well define $\mathsf{data}_1$ and $\mathsf{data}_2$ by splitting the data according to whether the index $i$ is even or odd (for example). For any partition of the data, $\mathsf{data}_1$ and $\mathsf{data}_2$ are distributed in the same way. Hence, the choice of partition made in Equation~\eqref{eq:data_partition} is arbitrary and all other choices give rise to the same results.
			\end{enumerate}
			
			We define two single-round boxes from the observed data, similarly to what was done in Equation~\eqref{eq:freq_single_round_dF}:
			\begin{equation}
			\begin{split}
					&\O_{ABXY}^{\textup{freq}(\mathsf{data}_1)} (\tilde{a}\tilde{b}\tilde{x}\tilde{y}) = \frac{\Big|\left\{i\in[n/2] : (a_i,b_i,x_i,y_i) = (\tilde{a},\tilde{b},\tilde{x},\tilde{y}) \right\}\Big|}{n/2} \;, \\
					&\O_{ABXY}^{\textup{freq}(\mathsf{data}_2)} (\tilde{a}\tilde{b}\tilde{x}\tilde{y}) = \frac{\Big|\left\{|i\in\{n/2+1,\dots,n\} : (a_i,b_i,x_i,y_i) = (\tilde{a},\tilde{b},\tilde{x},\tilde{y})  \right\}\Big|}{n/2} 
				\end{split}
			\end{equation}
			and, for $t\in\{0,1\}$,
			\begin{equation}\label{eq:partial_freq_dF}
				\O_{AB|XY}^{\textup{freq}(\mathsf{data}_t)} = \frac{	\O_{ABXY}^{\textup{freq}(\mathsf{data}_t)}}{\Q_{XY}} \;.
			\end{equation}
			As mentioned above, for $\O_{AB|XY}^{\textup{freq}(\mathsf{data}_1)}$ and $\O_{AB|XY}^{\textup{freq}(\mathsf{data}_2)}$ to be defined for all inputs we assume that all inputs $(x,y)$ appear in both $\mathsf{data}_1$ and  $\mathsf{data}_2$.

		\subsection{Signalling test}
		
			Below we consider distributions $\O_{ABXY}=\Q_{XY}\O_{AB|XY}$. One can then consider different marginals of  $\O_{ABXY}$. For example, $\O_{BY}$ is simply defined by $\O_{BY}(b,y)=\sum_{a,x} \O_{ABXY}(a,b,x,y)$. Note that the marginals of $\O_{ABXY}$ are all well-defined even if~$\O_{AB|XY}$ itself is signalling. 
			
			We wish to define a signalling test. To this end, let us first define a signalling \emph{measure} over single-round boxes:
			\begin{defn}\label{def:sign_measure_dF}
				Let $\O_{AB|XY}$ be a single-round box defined over $\mathcal{A}$, $\mathcal{B}$, $\mathcal{X}$, and~$\mathcal{Y}$, $\Q_{XY}$ a distribution over the inputs of the single-round box, and $\O_{ABXY}=\Q_{XY}\O_{AB|XY}$.
				The amount of signalling from Alice to Bob using the inputs $(x,y)$ and Bob's output~$b$ is given by
				\begin{align*}
					&\mathrm{Sig}^{(A\rightarrow B, x, y, b)} \left( \O_{AB|XY} \right) 
					=  \O_{BY}(b,y) \left[ \O_{X|BY}(x|b,y) - \Q_{X|Y}(x|y) \right] \;.
				\end{align*} 	
				Similarly, the amount of signalling from Bob to Alice using the inputs $(x,y)$, distributed according to $\Q_{XY}$, and Alice's output $a$, distributed according to $\O_{A|X=x,Y=y}$, is given by
				\begin{align*}
					\mathrm{Sig}^{(B\rightarrow A, x, y, a)} \left( \O_{AB|XY} \right) 
					=  \O_{AX}(a,x) \left[ \O_{Y|AX}(y|a,x) - \Q_{Y|X}(y|x) \right] \;.
				\end{align*} 	
				The box $\O_{AB|XY}$ is non-signalling if and only if 
				\begin{equation}\label{eq:sign_zero}
					\mathrm{Sig}^{(A\rightarrow B, x, y, b)} \left( \O_{AB|XY} \right) = \mathrm{Sig}^{(B\rightarrow A, x, y, a)} \left( \O_{AB|XY} \right)  = 0 \;.
				\end{equation}
			\end{defn}
			
			To see that the above definition makes sense as a signalling measure first notice that, when positive,  $\O_{X|BY}(x|b,y) - \Q_{X|Y}(x|y)$ can be understood as quantifying Bob's advantage in guessing Alice's input $x$ when observing $b$, compared to his prior information~$\Q_{X|Y}(x|y)$ about her input. 
			For a uniform distribution over $\mathcal{X}\times\mathcal{Y}$, $\Q_{X|Y}(x|y)=0$ for all $x$ and $y$. Then, the non-signalling requirement means that Bob cannot infer Alice's input from his output (as otherwise Alice could signal Bob), that is,~$\O_{X|BY}(x|b,y)=0$ as well. The above is a generalisation of this requirement to non-uniform distributions $\Q_{XY}$.
			On the more technical level~--- the non-signalling conditions (Definition~\ref{def:ns_box}) can be equivalently written as
			\begin{equation}\label{eq:ns_cond_par_rep}
				\begin{split}
					&\forall b,x,y \quad \sum_{\tilde{a}} \O_{AB|XY}(\tilde{a},b|x,y) = \sum_{\tilde{x}} \Q_{X|Y}(\tilde{x}|y) \sum_{\tilde{a}} \O_{AB|XY}(\tilde{a},b|\tilde{x},y) \;; \\
				&\forall a,x,y \quad  \sum_{\tilde{b}} \O_{AB|XY}(a,\tilde{b}|x,y) = \sum_{\tilde{y}} \Q_{Y|X}(\tilde{y}|x) \sum_{\tilde{b}} \O_{AB|XY}(a,\tilde{b}|a,\tilde{y}) \;.
				\end{split}
			\end{equation}
			One can verify that, for complete support $\Q_{XY}$, these conditions are equivalent to Equation~\eqref{eq:sign_zero}.

			All statements proven below regarding our signalling measure hold for signalling in both directions, i.e., from Alice to Bob and from Bob to Alice. We present all the statements and proofs in terms of signalling from Alice to Bob; to derive the same statements for signalling from Bob to Alice one can simply replace the parties (their inputs and outputs) with one another.
			
			We use the following definition to measure the distance between two single-round boxes.\footnote{ More commonly in the literature, one considers a definition in which $\mathbb{E}_{(x,y)}$ is replaced by $\max_{x,y}$. We use Definition~\ref{def:trace_distance_box_dF} since it allows us to apply Sanov's theorem later on.} 
			\begin{defn}\label{def:trace_distance_box_dF}
				The distance between $\mathrm{K}_{AB|XY}$ and $\mathrm{R}_{AB|XY}$ is defined as
				\[
					\big|\mathrm{K}_{AB|XY}-\mathrm{R}_{AB|XY}\big|_1 =  \mathbb{E}_{(x,y)\in \mathcal{X}\times\mathcal{Y}} \sum_{(a,b) \in \mathcal{A}\times\mathcal{B}} \big| \mathrm{K}_{AB|XY}(a,b|x,y) - \mathrm{R}_{AB|XY}(a,b|x,y) \big| \;.
				\]
			\end{defn}

			The following lemma shows that our measure of signalling is continuous. That is, if two strategies are close to one another according to Definition \ref{def:trace_distance_box_dF} then their signalling values are also close. The proof is given in Appendix \ref{appsec:ns_averages_proofs}.
			\begin{lem}\label{lem:continuity_sig}
				Let $\O_{AB|XY}^1$ and $\O_{AB|XY}^2$ be two single-round boxes such that 
				\[
					\big|\O_{AB|XY}^1-\O_{AB|XY}^2\big|_1 \leq \epsilon \;.
				\]
				Then, for all $a$, $b$, $x$, and $y$,
				\begin{align*}
					&\big|\mathrm{Sig}^{(A\rightarrow B, x, y, b)}(\O_{AB|XY}^1) - \mathrm{Sig}^{(A\rightarrow B, x, y, b)}(\O_{AB|XY}^2)\big| \leq 2\varepsilon 
				\end{align*}
			\end{lem}
			
			We can now define our \emph{signalling test}. 
			The test interacts with the parallel box~$\P_{\mr{A}\mr{B}|\mr{X}\mr{Y}}$ by sampling $\mathsf{data}$ according to $\Q_{XY}^{\otimes n}\P_{\mr{A}\mr{B}|\mr{X}\mr{Y}}$ and then checking whether~$\O_{AB|XY}^{\textup{freq}(\mathsf{data}_2)}$ is sufficiently signalling. Formally:
			\begin{defn}\label{def:sig_test} 
				Let $\zeta,\epsilon>0$ be parameters satisfying $\zeta \geq 7\epsilon$. 
				For any $x$, $y$, and $b$, a signalling test is defined by\footnote{If $\mathsf{data}_1$ does not include an index in which the inputs are $(x,y)$ then the test $\mathcal{T}^{(A\rightarrow B,x,y,b)}$ rejects by definition (recall Definition~\ref{def:sign_measure_dF}).}
				\begin{equation}\label{eq:test_definition}
					\mathcal{T}^{(A\rightarrow B,x,y,b)} (\P_{\mr{A}\mr{B}|\mr{X}\mr{Y}})= 
					\begin{cases}
				    		1& \text{if } \mathrm{Sig}^{(A\rightarrow B,x,y,b)}\left( \O_{AB|XY}^{\textup{freq}(\mathsf{data}_2)}\right) \geq \zeta - 2\epsilon\\
				    		0& \text{otherwise} \;,
					\end{cases} \\
				\end{equation}
				where $\O_{AB|XY}^{\textup{freq}(\mathsf{data}_2)}$ is defined as in Equation~\eqref{eq:partial_freq_dF}.
				
				Let $\passtest$ denote the event that the signalling test $\mathcal{T}^{(A\rightarrow B,x,y,b)}$ passes.
				The probability of the test passing when interacting with $\P_{\mr{A}\mr{B}|\mr{X}\mr{Y}}$ is given by
				\begin{equation*}
					\mathrm{Pr}_{\mathsf{data}\sim\P_{\mr{A}\mr{B}\mr{X}\mr{Y}}} \left[ \passtest \right]  = \sum_{\mr{x},\mr{y}} \Q_{XY}^{\otimes n}(\mr{x},\mr{y}) \sum_{\substack{\mr{a},\mr{b} : \\ \mathrm{Sig}^{(A\rightarrow B,x,y,b)}\left( \O_{AB|XY}^{\textup{freq}(\mathsf{data}_2)}\right) \\ \quad \geq \zeta - 2\epsilon}} \P_{\mr{A}\mr{B}|\mr{X}\mr{Y}}(\mr{a},\mr{b}|\mr{x},\mr{y}) \;.
				\end{equation*}
			\end{defn}
			
			The signalling test above is defined with $\mathrm{Sig}^{(A\rightarrow B,x,y,b)}\left( \O_{AB|XY}^{\textup{freq}(\mathsf{data}_2)}\right)$, rather than its absolute value, since this will be the only case relevant for our analysis; see Appendix~\ref{appsec:pr_sen_analy}.

			When considering IID boxes $\O_{AB|XY}^{\otimes n}$, the signalling test $\mathcal{T}^{(A\rightarrow B,x,y,b)}$ is reliable~--- if $\mathrm{Sig}^{(A\rightarrow B,x,y,b)}\left( \O_{AB|XY} \right)\geq \zeta$ the test will detect it with high probability, i.e. the test will accept with high probability, and if $\O_{AB|XY}$ is non-signalling then the test will reject with high probability. 
			It follows, in particular, that if signalling is detected by the test in $\O_{AB|XY}^{\textup{freq}(\mathsf{data}_2)}$, $\O_{AB|XY}^{\textup{freq}(\mathsf{data}_1)}$ is also signalling with high probability. 
			This holds also when considering the de Finetti box as in Definition~\ref{defn:dF_box}.
			
			To make the statement precise, let us define two sets of single-round boxes for every signalling test $\mathcal{T}^{(A\rightarrow B,x,y,b)}$.
			The first set is given by
			\begin{equation*}
			\begin{split}
				\sigma^{(A\rightarrow B,x,y,b)} = \big\{ \O_{AB|XY} : \forall \bar{\O}_{AB|XY} & \text{ s.t. } |\O_{AB|XY} - \bar{\O}_{AB|XY}|_1 \leq \epsilon \\
				&\Rightarrow \mathrm{Sig}^{(A\rightarrow B, x, y, b)}(\bar{\O}_{AB|XY}) \geq \zeta \big\} \;.
			\end{split}
			\end{equation*}
			Using the continuity of the signalling measure, Lemma~\ref{lem:continuity_sig}, we observe that
			\begin{equation}\label{eq:not_in_sigma}
				 \O_{AB|XY} \notin \sigma^{(A\rightarrow B,x,y,b)} \Rightarrow \mathrm{Sig}^{(A\rightarrow B, x, y, b)}(\O_{AB|XY}) \leq \zeta + 2\epsilon \;.
			\end{equation}
			
\begin{figure}
			\begin{center}
				\begin{tikzpicture}[scale=0.9]
				
					\draw (0,0) node {$\mathrm{Sig}^{A\rightarrow B,x,y,b )}\left( \O_{AB|XY}\right)$};
					\draw  [<->] (2,0) -- (12,0);
					
					\draw[fill] (3,0) circle [radius=0.025];
			
				\node [below] at (3,0) {0};
					
					\draw[fill] (4.5,0) circle [radius=0.025];
			
				\node [below] at (4.5,0) {$\nu$};
					
					\draw[fill] (7.5,0) circle [radius=0.025];
			
				\node [below] at (7.5,0) {$\zeta$};
					
					\draw[fill] (9,0) circle [radius=0.025];
			
				\node [below] at (9,0) {$\zeta+2\epsilon$};

					\draw[decorate,decoration={brace,mirror,amplitude=10pt}] (4.6,-0.5) -- (12,-0.5) node [midway,yshift=-0.75cm] {$\mathrm{Sig}$ of $\O\in\Sigma^{(A\rightarrow B,x,y,b)}$};
					\draw[decorate,decoration={brace,amplitude=10pt}] (2.10,0.25) -- (8.90,0.25) node [midway,yshift=0.75cm] {$\mathrm{Sig}$ of $\O\notin\sigma^{(A\rightarrow B,x,y,b)}$};
					\draw[decorate,decoration={brace,mirror,amplitude=10pt}] (3.10,-0.5) -- (4.5,-0.5) node [midway,yshift=-0.75cm] {constant gap};
				
				\end{tikzpicture}
			\par\end{center}
			\caption{Visualisation of the sets $\sigma^{(A\rightarrow B,x,y,b)}$ and $\Sigma^{(A\rightarrow B,x,y,b)}$.} \label{fig:sig_sets}
			\end{figure}
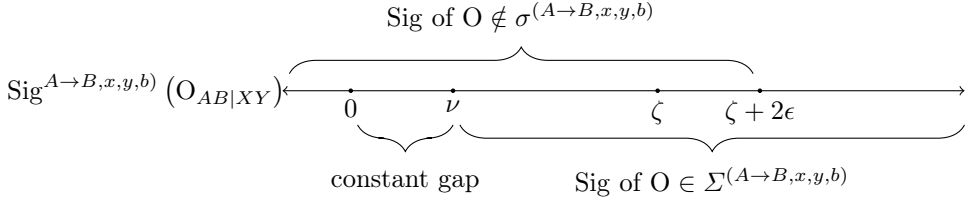

			The second set is defined to be
			\begin{equation*}
				\begin{split}
					\Sigma^{(A\rightarrow B,x,y,b)} = \big\{ \O_{AB|XY} :  \exists \bar{\O}_{AB|XY}  \text{ s.t. } & |\O_{AB|XY} - \bar{\O}_{AB|XY}|_1 \leq \epsilon \\
					& \land  \mathrm{Pr}_{\mathsf{data}\sim\bar{\O}_{AB|XY}^{\otimes n}}[\passtest] >\delta \big\} \;,
				\end{split}
			\end{equation*}
			where  $\delta=\delta\left(\frac{n}{2},\epsilon\right)=\left(\frac{n}{2}+1\right)^{|\mathcal{A}||\mathcal{B}||\mathcal{X}||\mathcal{Y}|-1}e^{-n\epsilon^2/4}$. 
			Since the signalling test is reliable when acting on IID boxes, one can easily show that
			\begin{equation}\label{eq:Sigma_alternative_form}
				 \O_{AB|XY} \in 	\Sigma^{(A\rightarrow B,x,y,b)} \Rightarrow \mathrm{Sig}^{(A\rightarrow B, x, y, b)}(\O_{AB|XY}) > \nu
			\end{equation}
			for any $0<\nu<\zeta-6\epsilon$. 
			This is stated and proven as Lemma~\ref{lem:ns_Sigma_alternative} in Appendix~\ref{appsec:ns_averages_proofs}.
			The sets and the relevant constants are illustrated in Figure~\ref{fig:sig_sets}.
			
			We use below the following notation:
			\begin{itemize}
				\item  $\insigma$ denotes the event that $\O_{AB|XY}^{\textup{freq}(\mathsf{data}_1)}\in \sigma^{(A\rightarrow B,x,y,b)}$.
				\item $\inSigma$ denotes the event that $\O_{AB|XY}^{\textup{freq}(\mathsf{data}_1)}\in \Sigma^{(A\rightarrow B,x,y,b)}$.
				\item ``For all signalling test $\mathcal{T}^{(A\rightarrow B,x,y,b)}$...'' should be understood as ``for all $x$, $y$, and $b$, defining a signalling test $\mathcal{T}^{(A\rightarrow B,x,y,b)}$,..." and similarly for other quantifiers.
			\end{itemize}
			 Furthermore, to avoid confusion, we explicitly denote the probability distributions on which we evaluate the probability of the above events. 
			
			As shown in Appendix~\ref{appsec:ns_averages_proofs},  the following lemma holds for a de Finetti box:						
			\begin{lem} \label{lem:de_finetti_prop}
				Let $\tau_{\mr{A}\mr{B}\mr{X}\mr{Y}}=\Q_{XY}^{\otimes n} \tau_{\mr{A}\mr{B}|\mr{X}\mr{Y}}$, where $\tau_{\mr{A}\mr{B}|\mr{X}\mr{Y}}$ is a de Finetti box. 
				For every signalling test $\mathcal{T}^{(A\rightarrow B,x,y,b)}$,
				\begin{enumerate}
					\item $\mathrm{Pr}_{\mathsf{data}\sim\tau_{\mr{A}\mr{B}\mr{X}\mr{Y}}} \left[ \lnot\inSigma \land \passtest  \right] \leq \delta $
					\item $\mathrm{Pr}_{\mathsf{data}\sim\tau_{\mr{A}\mr{B}\mr{X}\mr{Y}}} \left[  \insigma \land \lnot\passtest \right] \leq \delta $, 
				\end{enumerate}
				where $\delta=\delta\left(\frac{n}{2},\epsilon\right)=\left(\frac{n}{2}+1\right)^{|\mathcal{A}||\mathcal{B}||\mathcal{X}||\mathcal{Y}|-1}e^{-n\epsilon^2/4}$.
			\end{lem}

			A similar lemma can be proven for \emph{permutation invariant parallel boxes}, using the de Finetti reduction of Theorem~\ref{thm:dF_conditional}:			
			\begin{lem}\label{lem:events_prob_reduction}
				Given a parallel box $\P_{\mr{A}\mr{B}|\mr{X}\mr{Y}}$ let $\P_{\mr{A}\mr{B}\mr{X}\mr{Y}}=\Q_{XY}^{\otimes n} \P_{\mr{A}\mr{B}|\mr{X}\mr{Y}}$.
				For every permutation-invariant box $\P_{\mr{A}\mr{B}|\mr{X}\mr{Y}}$ and every $\mathcal{T}^{(A\rightarrow B,x,y,b)}$:
				\begin{enumerate}
					\item \label{it:reduction1}$\mathrm{Pr}_{\mathsf{data}\sim\P_{\mr{A}\mr{B}\mr{X}\mr{Y}}} \left[  \lnot\inSigma \land \passtest  \right] \leq c\delta$ 
					\item \label{it:reduction2} $\mathrm{Pr}_{\mathsf{data}\sim\P_{\mr{A}\mr{B}\mr{X}\mr{Y}}} \left[  \insigma \land \lnot\passtest \right] \leq c\delta$ ,
				\end{enumerate}
				where $c=(n+1)^{|\mathcal{X}||\mathcal{Y}|\left(|\mathcal{A}||\mathcal{B}|-1\right)}$ and $\delta$ is as in Lemma~\ref{lem:de_finetti_prop}.
			\end{lem}
			
			\begin{proof}
				We prove both of the claims together. Denote the relevant event by $E(\mathsf{data})$ and note that for both events we can write 
				\[
					\mathrm{Pr}_{\mathsf{data}\sim\P_{\mr{A}\mr{B}\mr{X}\mr{Y}}} \left[ E(\mathsf{data}) =1 \right] = \sum_{\substack{\mathsf{data} | \\ E(\mathsf{data})=1 }} \P_{\mr{A}\mr{B}\mr{X}\mr{Y}}(\mathsf{data}) \;.
				\]
				From Theorem~\ref{thm:dF_conditional} we get $\P_{\mr{A}\mr{B}\mr{X}\mr{Y}}(\mathsf{data}) \leq c \cdot \tau_{\mr{A}\mr{B}\mr{X}\mr{Y}}(\mathsf{data})$ and therefore 
				\begin{equation*}
					\begin{split}
						\mathrm{Pr}_{\mathsf{data}\sim\P_{\mr{A}\mr{B}\mr{X}\mr{Y}}} \left[ E(\mathsf{data}) =1 \right] &= \sum_{\substack{\mathsf{data} | \\ E(\mathsf{data})=1 }} \P_{\mr{A}\mr{B}\mr{X}\mr{Y}}(\mathsf{data}) \\
						& \leq c \cdot \sum_{\substack{\mathsf{data} | \\ E(\mathsf{data})=1 }} \tau_{\mr{A}\mr{B}\mr{X}\mr{Y}}(\mathsf{data}) \\
						& = c \cdot \mathrm{Pr}_{\mathsf{data}\sim\tau_{\mr{A}\mr{B}\mr{X}\mr{Y}}} \left[ E(\mathsf{data}) =1 \right] \;.	
					\end{split}
				\end{equation*}
				Combining this with Lemma \ref{lem:de_finetti_prop} proves the lemma. 
			\end{proof}
												
		\subsection{Guessing game}\label{sec:sig_game}
		
			The previous section discussed the relations between $\O_{AB|XY}^{\textup{freq}(\mathsf{data}_1)}$ and $\O_{AB|XY}^{\textup{freq}(\mathsf{data}_2)}$ in terms of the probabilities of certain events which depend on these averaged single-round boxes. All statements made so far were general, in the sense that they hold for any permutation-invariant parallel box $\P_{\mr{A}\mr{B}|\mr{X}\mr{Y}}$.
			In the current section we focus on permutation-invariant \emph{non-signalling} parallel boxes $\P_{\mr{A}\mr{B}|\mr{X}\mr{Y}}$. 
			Our goal is to show that for non-signalling boxes $\P_{\mr{A}\mr{B}|\mr{X}\mr{Y}}$, the averaged boxes $\O_{AB|XY}^{\textup{freq}(\mathsf{data}_1)}$ and~$\O_{AB|XY}^{\textup{freq}(\mathsf{data}_2)}$ cannot be too signalling as we were set to prove.
			
			To this end, we construct a guessing game for every signalling test $\mathcal{T}^{(A\rightarrow B, x, y, b)}$. In the game, a referee gives Alice and Bob $n/2$ questions, distributed according to~$\Q_{XY}^{\otimes n/2}$. Bob's goal is to output an index $j\in\left[n/2\right]$ for which $(x_j,y_j)=(x,y)$ (while Alice does not need to output anything).
			Alice and Bob are allowed to use any non-signalling box to win the game.
			Clearly, a non-signalling box should not allow Bob to learn anything about Alice's input from his outputs. Bob's best strategy is thus to guess an index $j$ for which $y_j=y$. The probability that his guess is correct is~$\Q_{X|Y}(x|y)$. 
			If Alice and Bob are able to win the game with higher probability then the used box must be signalling.\footnote{This motivates our signalling measure given in Definition~\ref{def:sign_measure_dF}.}
			
			The following lemma asserts that, for non-signalling $\P_{\mr{A}\mr{B}|\mr{X}\mr{Y}}$, \emph{conditioned} on the signalling test detecting a lot of signalling in $\O_{AB|XY}^{\textup{freq}(\mathsf{data}_2)}$, the probability that~$\O_{AB|XY}^{\textup{freq}(\mathsf{data}_1)}$ is highly signalling is bounded away from 1. 
			Intuitively, we would have expected that if signalling  is detected in $\mathsf{data}_2$ then $\mathsf{data}_1$ should exhibit signalling  practically with certainty. The lemma (roughly) shows that, when starting with non-signalling boxes, this is not the case. 
			
			
			Before starting, we remind the reader that we assume in this section that all pairs of questions appear in $\mathsf{data}_1$ and $\mathsf{data}_2$.
			For the lemmas and proofs below it is important to remember that all the probabilities are conditioned on $\mathsf{data}_1$.
			To ease notation we do not explicitly write it.
			\begin{lem}\label{lem:weak_condition}
				Let $\epsilon\in[0,1]$ and $n$ be such that 
				\begin{equation}\label{eq:number_of_repet_bound}
					\frac{n}{\ln(n)} > 20|\mathcal{X}||\mathcal{Y}||\mathcal{A}||\mathcal{B}| \frac{\ln(2/\epsilon)}{\epsilon^{2}}\;,
				\end{equation} 
				and $\P_{\mr{A}\mr{B}|\mr{X}\mr{Y}}$ a non-signalling parallel box. For any  signalling test $\mathcal{T}^{(A\rightarrow B, x, y, b)}$ denote by $\P_{\mr{A}\mr{B}\mr{X}\mr{Y}|\mathcal{T}=1}$ the probability distribution $\mathrm{P}_{\mr{A}\mr{B}\mr{X}\mr{Y}}$ conditioned on the event $\mathcal{T}^{(A\rightarrow B, x, y, b)}\left( \P_{\mr{A}\mr{B}|\mr{X}\mr{Y}} \right)=1$, whenever such a conditional probability distribution is defined. Then,
				\begin{equation}\label{eq:weak_lemma_eq}
					\mathrm{Pr}_{\mathsf{data}\sim\P_{\mr{A}\mr{B}\mr{X}\mr{Y}|\mathcal{T}=1}} \left[\inSigma  \right]  < 1 - \sqrt{c\delta}\;,
				\end{equation}
				for $c$ and $\delta$ as in Lemma~\ref{lem:events_prob_reduction}.
			\end{lem}
			\begin{proof}
				We denote~$\mr{x}_{\mathsf{data}_1}=x_1,\dots,x_{n/2}$ and  $\mr{y}_{\mathsf{data}_1}= y_1,\dots,y_{n/2}$.
			
				\sloppy
				For every signalling test $\mathcal{T}^{(A\rightarrow B, x, y, b)}$ and inputs for Bob $\mr{y}_{\mathsf{data}_1}$ such that~$\Pr_{\mathsf{data}\sim\mathrm{P}_{\mr{A}\mr{B}\mr{X}\mr{Y}}} \left[ \passtest | \mr{y}_{\mathsf{data}_1}\right] \neq 0$ we construct a guessing game. Our goal is to derive a contradiction by showing that if Equation \eqref{eq:weak_lemma_eq} is not true, then the guessing game can be won with probability higher than the optimal non-signalling winning probability. 
					
				The guessing game is as explained above. A referee gives Bob the inputs $\mr{y}_{\mathsf{data}_1}$ and Alice gets $\mr{x}_{\mathsf{data}_1}$ distributed according to $\Q_{XY}(x|y)$. 
				Bob's goal is to guess an index $j\in[n/2]$ such that $(x_j,y_j) = (x , y)$ (we assume that such exists). 
			
				If the parties share a non-signalling box $\P_{\mr{A}\mr{B}|\mr{X}\mr{Y}}$ then Bob's marginals are the same for all $\mr{x}_{\mathsf{data}_1}$. Therefore, his outputs do not give him any information about the inputs that Alice got from the referee. The best non-signalling strategy for the guessing game is therefore to choose, uniformly at random, an index $j$ for which~$y_j=y$. The winning probability is then given by $W_{\mathrm{ns}} = \Q_{X|Y}(x|y)<1$.\footnote{Note that while Bob's inputs, $\mr{y}_{\mathsf{data}_1}= y_1,\dots,y_{n/2}$, are fixed in a specific instance of the guessing game, Alice's inputs are still distributed according to the prior $\Q_{XY}(x|y)$.}
			
				We now show that if the parties share $\P_{\mr{A}\mr{B}|\mr{X}\mr{Y}}$ for which 
				\begin{equation}\label{eq:contr_assump}
					\mathrm{Pr}_{\mathsf{data}\sim\P_{\mr{A}\mr{B}\mr{X}\mr{Y}|\mathcal{T}=1}} \left[\inSigma | \mr{y}_{\mathsf{data}_1}\right]  \geq 1 - \sqrt{c\delta}
				\end{equation}
				then they can win the above guessing game with probability higher than the optimal non-signalling winning probability~$W_{\mathrm{ns}}$.
				
				The idea is as follows. The parties share many identical copies of $\P_{\mr{A}\mr{B}|\mr{X}\mr{Y}}$. They use the inputs given by the referee as  $\mr{x}_{\mathsf{data}_1}$ and $\mr{y}_{\mathsf{data}_1}$ in all of the copies and choose, using shared randomness, the rest of the inputs, associated with $\mathsf{data}_2$, in each copy (i.e., there are different inputs for $\mathsf{data}_2$ for each copy). 
				They use the copies of~$\P_{\mr{A}\mr{B}|\mr{X}\mr{Y}}$ with the described inputs. 
				Bob then looks for the first copy of $\P_{\mr{A}\mr{B}|\mr{X}\mr{Y}}$ in which the event $\passtest$ holds~-- such a copy exists as long as\footnote{To see this note that since the box is non-signalling between Alice and Bob, Bob can check in which copy the test passes even before Alice uses her input. Therefore, the probability to pass the test is independent of Alice's inputs and hence must be non-zero for any of them.} $\mathrm{Pr}_{\mathsf{data}\sim\mathrm{P}_{\mr{A}\mr{B}\mr{X}\mr{Y}}} \left[\passtest | \mr{y}_{\mathsf{data}_1} \right] \neq 0$; he can find it since he knows all the inputs in $\mathsf{data}_2$ (as they were chosen using shared randomness).\footnote{Recalling Definitions~\ref{def:sign_measure_dF} and~\ref{def:sig_test}, we see that only $\O_{BXY}^{\textup{freq}(\mathsf{data}_2)}$ is needed in order to check whether the signalling test passes or not. Thus, Bob indeed has all the relevant information and he can locally check whether the test passes or not.} 
				Alice does not need to know in which copy the test holds.
				Using the chosen copy, Bob chooses a random index~$j\in[n/2]$ such that~$y_j=y$ and~$b_j=b$.

				Let us show that, as long as $\mathrm{Pr}_{\mathsf{data}\sim\mathrm{P}_{\mr{A}\mr{B}\mr{X}\mr{Y}}} \left[\passtest|\mr{y}_{\mathsf{data}_1}\right] \neq 0$, this box achieves a winning probability which is higher than $W_{\mathrm{ns}}$.  
				For the chosen copy, the event $\passtest$ holds and hence $\mathsf{data}_1$ can be seen as data distributed according to $n/2$ identical copies of~$\O_{AB|XY}^{\textup{freq}(\mathsf{data}_1)}$, which is with high probability in $\Sigma_{(i,b, x , y)}$ according to Equation~\eqref{eq:contr_assump}. 
				Using Equation~\eqref{eq:Sigma_alternative_form} this implies 
				\begin{equation}\label{eq:weak_cond1_plus}
					\mathrm{Pr}_{\mathsf{data}\sim\P_{\mr{A}\mr{B}\mr{X}\mr{Y}|\mathcal{T}=1}} \left[\mathrm{Sig}^{(A \rightarrow B,b, x , y)}(\O_{AB|XY}^{\textup{freq}(\mathsf{data}_1)}) > \nu | \mr{y}_{\mathsf{data}_1} \right]   \geq 1 - \sqrt{c\delta} \;,
				\end{equation}
				where $\nu>0$ is any parameter satisfying $\nu < \zeta - 6\epsilon$ (recall Equation~\eqref{eq:Sigma_alternative_form}).
				
				Using Definition~\ref{def:sign_measure_dF} we know that if indeed 
				\[
					\mathrm{Sig}^{(A \rightarrow B,b, x , y)}(\O_{AB|XY}^{\textup{freq}(\mathsf{data}_1)}) > \nu 
				\]
				 then $\O_{BY}^{\textup{freq}(\mathsf{data}_1)}(b, y)>0$ and 
				\begin{align}
					 \O_{X|BY}^{\textup{freq}(\mathsf{data}_1)}( x | b, y)  &> \frac{\nu}{\O_{BY}^{\textup{freq}(\mathsf{data}_1)}(b, y)} + \Q_{X|Y}(x |y) \label{eq:weak_cond2_plus}  \\
					&= \frac{\nu}{\O_{BY}^{\textup{freq}(\mathsf{data}_1)}(b, y)} + W_{\mathrm{ns}} \;. \nonumber
				\end{align}
				That is, by choosing an index for which $b_j=b$ Bob increase the winning probability. 
				
				On the other hand, if $\mathrm{Sig}^{(A \rightarrow B,b, x , y)}(\O_{AB|XY}^{\textup{freq}(\mathsf{data}_1)}) \leq \nu $, which can happen with probability $\sqrt{c\delta}$, then Bob may decrease his winning probability. In the worst case the winning probability is~0.
				Therefore, for the chosen copy (for which the test passed) we get the following winning probability
				\begin{equation} \label{eq:win_cond_on_test}
					W \geq  (1 - \sqrt{c\delta}) \left( \frac{\nu}{\O_{BY}^{\textup{freq}(\mathsf{data}_1)}(b, y)} + W_{\mathrm{ns}} \right) +  \sqrt{c\delta} \cdot 0 \;.
				\end{equation}

				Thus, $W>W_{\mathrm{ns}}$ for 
				\begin{equation}\label{eq:rep_num_condition}
					\nu > \frac{\sqrt{c\delta}}{1-\sqrt{c\delta}} W_{\mathrm{ns}} \geq \frac{\sqrt{c\delta}}{1-\sqrt{c\delta}} W_{\mathrm{ns}} \cdot \O_{BY}^{\textup{freq}(\mathsf{data}_1)}(b, y)  \;.
				\end{equation}
				Using $W_{\mathrm{ns}} \cdot \O_{BY}^{\textup{freq}(\mathsf{data}_1)}(b, y) \leq 1$ and $\sqrt{c\delta} \leq (n+1)^{|\mathcal{X}||\mathcal{Y}||\mathcal{A}||\mathcal{B}|}e^{-n\epsilon^2/8}$, we see that as long as $n/\ln(n) > 20|\mathcal{X}||\mathcal{Y}||\mathcal{A}||\mathcal{B}| \epsilon^{-2}\ln(2/\epsilon)$ we have
				\[
					\frac{\sqrt{c\delta} \; W_{\mathrm{ns}}  \O_{BY}^{\textup{freq}(\mathsf{data}_1)}(b, y)} {1-\sqrt{c\delta}} < \epsilon \;.
				\]
				 Assuming $\zeta \geq 7\epsilon$, there is a choice of $\nu$ that satisfies both Equation~\eqref{eq:rep_num_condition} and the earlier condition that $\nu < \zeta - 6\epsilon$. 
				
				We get that Equation \eqref{eq:contr_assump} must not hold for any $\mr{y}_{\mathsf{data}_1}$ and hence cannot hold also when we omit the conditioning on $\mr{y}_{\mathsf{data}_1}$. The lemma therefore follows.
			\end{proof}

			The bound given in Equation \eqref{eq:weak_lemma_eq} is weak for two reasons. First, $\mathsf{data}$ is distributed according to the conditional distribution $\P_{\mr{A}\mr{B}\mr{X}\mr{Y}|\mathcal{T}=1}$ and not according to $\P_{\mr{A}\mr{B}\mr{X}\mr{Y}}$ itself. Second, it only tells us that 
			$\mathrm{Pr}_{\mathsf{data}\sim\P_{\mr{A}\mr{B}\mr{X}\mr{Y}|\mathcal{T}=1}} \big[\O_{AB|XY}^{\textup{freq}(\mathsf{data}_1)} \notin \Sigma^{(A\rightarrow B,x,y,b)}  \big]  \geq \sqrt{c\delta}$, i.e., the probability that $\O_{AB|XY}^{\textup{freq}(\mathsf{data}_1)}$ has a small value of signalling is higher than $\sqrt{c\delta}$. 
			We show how the statement can be amplified using Lemma~\ref{lem:events_prob_reduction}, which utilised our de Finetti reduction.

			\begin{thm}\label{thm:strong_condition}
				Let $\P_{\mr{A}\mr{B}|\mr{X}\mr{Y}}$ be a permutation-invariant non-signalling parallel box and $n$ such that Equation \eqref{eq:rep_num_condition} is satisfied. Then for any  signalling test $\mathcal{T}^{(A\rightarrow B, x, y, b)}$ such that $\Q_{XY}(x,y)\neq 0$ and $\Q_{X|Y}(x|y)\neq 1$ and conditioned on the event of all questions $(x,y)$ appearing in $\mathsf{data}_1$ and $\mathsf{data}_2$,
				\begin{equation}\label{eq:max_prob_sig_val}
					\mathrm{Pr}_{\mathsf{data}\sim\P_{\mr{A}\mr{B}\mr{X}\mr{Y}}} \left[ \mathrm{Sig}^{(A\rightarrow B, x, y, b)}\left(\O_{AB|XY}^{\textup{freq}(\mathsf{data})}\right) > \zeta + 2\epsilon \right] \leq 4\sqrt{c\delta}\;.
				\end{equation}
				Similarly, for any  signalling test $\mathcal{T}^{(B\rightarrow A, x, y, a)}$,
				\begin{equation}\label{eq:max_prob_sig_val_bob}
					\mathrm{Pr}_{\mathsf{data}\sim\P_{\mr{A}\mr{B}\mr{X}\mr{Y}}} \left[ \mathrm{Sig}^{(B\rightarrow A, x, y, b)}\left(\O_{AB|XY}^{\textup{freq}(\mathsf{data})}\right) > \zeta + 2\epsilon \right] \leq 4\sqrt{c\delta}\;.
				\end{equation}
			\end{thm}
			\begin{proof}
				From Lemma \ref{lem:events_prob_reduction} part \ref{it:reduction1} we get
				\[
					\mathrm{Pr}_{\mathsf{data}\sim\P_{\mr{A}\mr{B}\mr{X}\mr{Y}}} \left[\passtest \right] > \sqrt{c\delta} \Rightarrow \mathrm{Pr}_{\mathsf{data} \sim \P_{\mr{A}\mr{B}\mr{X}\mr{Y}|\mathcal{T}=1}} \left[ \lnot\inSigma \right] \leq \sqrt{c\delta}
				\]
				and this can be rewritten as 
				\[
					\mathrm{Pr}_{\mathsf{data}\sim\P_{\mr{A}\mr{B}\mr{X}\mr{Y}}} \left[\passtest\right] > \sqrt{c\delta} \Rightarrow \mathrm{Pr}_{\mathsf{data} \sim \P_{\mr{A}\mr{B}\mr{X}\mr{Y}|\mathcal{T}=1}} \left[ \inSigma \right]  \geq 1 - \sqrt{c\delta} \;.
				\]
				According to Lemma \ref{lem:weak_condition}, this implies 
				\[
					\mathrm{Pr}_{\mathsf{data}\sim\P_{\mr{A}\mr{B}\mr{X}\mr{Y}}} \left[\passtest \right] > \sqrt{c\delta} \Rightarrow \P_{\mr{A}\mr{B}|\mr{X}\mr{Y}} \text{ is signalling} \;.
				\]
				Therefore it must be that
				\begin{equation}\label{eq:test_prob1}
					\mathrm{Pr}_{\mathsf{data}\sim\P_{\mr{A}\mr{B}\mr{X}\mr{Y}}} \left[ \passtest \right] \leq \sqrt{c\delta}\;
				\end{equation} 
				or alternatively, 
				\begin{equation}\label{eq:test_prob2}
					\mathrm{Pr}_{\mathsf{data}\sim\P_{\mr{A}\mr{B}\mr{X}\mr{Y}}} \left[ \lnot\passtest \right] \geq 1 - \sqrt{c\delta}\;
				\end{equation} 
				
				Next, combining Lemma \ref{lem:events_prob_reduction} part \ref{it:reduction2} with Equation \eqref{eq:test_prob2} we get 
				\[
					\mathrm{Pr}_{\mathsf{data} \sim \mathrm{P}_{\mr{A}\mr{B}\mr{X}\mr{Y}|\mathcal{T}=0}} \left[\insigma \right] \leq \sqrt{c\delta} \;.
				\]
				Using Equation \eqref{eq:test_prob1} we get
				\[
					\mathrm{Pr}_{\mathsf{data} \sim \P_{\mr{A}\mr{B}\mr{X}\mr{Y}}} \left[ \insigma \right]  \leq 2\sqrt{c\delta}  \;.
				\]
			
				Using the definition of the set $\sigma^{(A \rightarrow B, x, y, b)}$ and Equation~\eqref{eq:not_in_sigma} we get that
				\begin{equation*}
					\mathrm{Pr}_{\mathsf{data}\sim\P_{\mr{A}\mr{B}\mr{X}\mr{Y}}} \left[ \mathrm{Sig}^{(A\rightarrow B, x, y, b)}\left(\O_{AB|XY}^{\textup{freq}(\mathsf{data}_1)}\right) > \zeta + 2\epsilon \right] \leq 2\sqrt{c\delta}\;.
				\end{equation*}
				Permutation invariance implies that $\mathsf{data}_1$ and $\mathsf{data}_2$ are distributed in the same way. Therefore, we also have
				\begin{equation*}
					\mathrm{Pr}_{\mathsf{data}\sim\P_{\mr{A}\mr{B}\mr{X}\mr{Y}}} \left[ \mathrm{Sig}^{(A\rightarrow B, x, y, b)}\left(\O_{AB|XY}^{\textup{freq}(\mathsf{data}_2)}\right) > \zeta + 2\epsilon \right] \leq 2\sqrt{c\delta}\;.
				\end{equation*}
				By definition, 
				\[	
					\O_{AB|XY}^{\textup{freq}(\mathsf{data})} = \frac{1}{2} \O_{AB|XY}^{\textup{freq}(\mathsf{data}_1)} + \frac{1}{2} \O_{AB|XY}^{\textup{freq}(\mathsf{data}_2)} 
				\]
				and, thus, using the linearity of the signalling measure, for any fixed observed data
				\begin{align*}
					 \mathrm{Sig}^{(A\rightarrow B, x, y, b)}\left(\O_{AB|XY}^{\textup{freq}(\mathsf{data})}\right) =& \frac{1}{2}  \mathrm{Sig}^{(A\rightarrow B, x, y, b)}\left(\O_{AB|XY}^{\textup{freq}(\mathsf{data}_1)}\right) \\
					 &+ \frac{1}{2}  \mathrm{Sig}^{(A\rightarrow B, x, y, b)}\left(\O_{AB|XY}^{\textup{freq}(\mathsf{data}_2)}\right) \;.
				\end{align*}
				Equation~\eqref{eq:max_prob_sig_val} follows by combining the above equations.
				Switching Alice and Bob in all lemmas above, Equation~\eqref{eq:max_prob_sig_val_bob} follows in the exact same way. 
			\end{proof}

			Theorem~\ref{thm:strong_condition} tells us that if $\P_{\mr{A}\mr{B}|\mr{X}\mr{Y}}$ is a permutation-invariant non-signalling parallel box then the probability that $\O_{AB|XY}^{\textup{freq}(\mathsf{data})}$ is highly signalling (in any direction and using any inputs and outputs) is exponentially small in the number of games. 
			De facto, this means that we can think of the observed data sampled from a non-signalling parallel box \emph{as if} it came from an IID box defined via a single-round box that is approximately non-signalling, with high probability. 
			This can be used to infer properties of non-signalling parallel boxes. In Section~\ref{sec:threshold_theorem} we show that Theorem~\ref{thm:strong_condition} implies a threshold theorem almost directly. 
			
			Apart from the applications of Theorem~\ref{thm:strong_condition}, we see it as an abstract mathematical statement about the observed data produced by non-signalling parallel boxes. 
			Deriving a similar statement for quantum boxes is also of interest (and, in particular, will imply a threshold theorem for all quantum games). 
			The main difficulty in deriving a quantum analogue of Theorem~\ref{thm:strong_condition} lies in finding a ``non-quantumness'' measure which, ideally, can be performed locally  by one of the parties (as in our guessing game in the proof of Lemma~\ref{lem:weak_condition}).
			
	\section{Threshold theorem}\label{sec:threshold_theorem}
	
		This section is devoted to deriving the following threshold theorem:
		\begin{thm}\label{thm:final_threshold_theorem}
			For any complete-support two-player game $\G$ whose optimal non-signalling winning probability is $w_{\mathrm{ns}} = 1-\alpha$, there exist  $\mathcal{C}(\G)$ such that for every $0<\beta\leq\alpha$ and large enough $n$, the probability that non-signalling players win more than a fraction $1-\alpha+\beta$ of the $n$ questions in the threshold game $\G^n_{1-\alpha+\beta}$ is at most $\exp \left[ - \mathcal{C}(\G) n \beta^2 \right]$.
		\end{thm} 
		
		That is, for sufficiently many repetitions the probability to win more than a fraction $1-\alpha+\beta$ of the $n$ games is exponentially small. 
		A sufficient condition on the number of repetitions for the bound in the theorem to hold is stated in~Equation \eqref{eq:number_of_repet_bound}, and a choice of constants made around Equation \eqref{eq:choice_of_constants}, for a more precise bound.
		
	
		The proof builds on Theorem~\ref{thm:strong_condition} and is rather simple. 
		If $\O_{AB|XY}^{\textup{freq}(\mathsf{data})}$ is not too signalling (for any signalling test), then its winning probability in a single game cannot be much higher than the winning probability of the optimal \emph{non-signalling} strategy for $\G$. 
		Furthermore, the number of games won in any given observed data can be read directly from the winning probability of $\O_{AB|XY}^{\textup{freq}(\mathsf{data})}$. Thus, by analysing $\O_{AB|XY}^{\textup{freq}(\mathsf{data})}$ we are actually analysing the number of games won.
		The combination of these two observations gives the final theorem. 
		We follow these steps below.
		
		\subsection{Winning probability of approximately non-signalling strategies}\label{sec:lin_prog_ns_wp}
			
			The linearity of the non-signalling conditions (Definition~\ref{def:ns_box}) and the winning probability in a game (Equation~\eqref{eq:game_win_prob}) allow us to phrase the  optimisation problem of finding the optimal non-signalling winning probability in any game $\G$ as a linear program~\cite{schrijver1998theory}. 
			For complete-support game we can use the following linear program over the variables $\O_{AB|XY}(a,b|x,y)$:
			\begin{subequations} \label{eq:linear_non_relaxed}
			\begin{align}
				\max \quad &\sum_{a,b,x,y} \Q_{XY}(xy) R(a,b,x,y) \O_{AB|XY}(a,b|x,y) \label{eq:linear1_obj} \\
				\text{s.t.} \quad &   
					\mathrm{Sig}^{(A\rightarrow B, x, y, b)} \left(\O_{AB|XY}\right) = 0 &\forall x,y,b \label{eq:linear1_ns_a} \\
					& \mathrm{Sig}^{(B\rightarrow A, x, y, b)} \left(\O_{AB|XY}\right) = 0 &\forall x,y,a \label{eq:linear1_ns_b} \\
				&\sum_{a,b} \O_{AB|XY}(a,b|x,y) = 1  &\forall x,y \label{eq:linear1_nornmal} \\
				& \O_{AB|XY}(a,b|x,y) \geq 0 &\forall a,b,x,y \label{eq:linear1_positive}
			\end{align} 
			\end{subequations}
			
			The objective function, Equation \eqref{eq:linear1_obj}, is exactly the winning probability in the game when using strategy $\O_{AB|XY}$. Equations \eqref{eq:linear1_nornmal} and \eqref{eq:linear1_positive} are the normalisation and positivity constraints on the strategy~$\O_{AB|XY}$.
			In Equations \eqref{eq:linear1_ns_a} and~\eqref{eq:linear1_ns_a} all the non-signalling constraints are listed, as follows  for complete-support games from Equation~\eqref{eq:sign_zero}.\footnote{\cite{arnon2016non} includes an explanation of the implications of the linear program \eqref{eq:linear_non_relaxed} to games with incomplete support.}

			Our goal is to upper-bound the winning probability of $\O_{AB|XY}^{\textup{freq}(\mathsf{data})}$, which may be slightly signalling. The optimal winning probability of strategies which are slightly signalling can be written as a linear program similar to the one above, by relaxing the constraint in Equations~\eqref{eq:linear1_ns_a} and~\eqref{eq:linear1_ns_b} so that it allows for some signalling.
			Specifically, keeping in mind Equation~\eqref{eq:max_prob_sig_val}, we are interested in the following  program:
			\begin{equation} \label{eq:perturbation}
			\begin{aligned}
				\max \quad &\sum_{a,b,x,y} \Q_{XY}(xy) R(a,b,x,y) \O_{AB|XY}(a,b|x,y) \\
				\text{s.t.} \quad &  
				\mathrm{Sig}^{(A\rightarrow B, x, y, b)} \left(\O_{AB|XY}\right)  \leq \zeta+2\epsilon &\forall x,y,b  \\
					& \mathrm{Sig}^{(B\rightarrow A, x, y, b)} \left(\O_{AB|XY}\right)  \leq \zeta+2\epsilon &\forall x,y,a \\
				&\sum_{a,b} \O_{AB|XY}(a,b|x,y) = 1  &\forall x,y  \\
				& \O_{AB|XY}(a,b|x,y) \geq 0 &\forall a,b,x,y
			\end{aligned}
			\end{equation}

			Program~\eqref{eq:perturbation} can be seen as a perturbation of Program~\eqref{eq:linear_non_relaxed}. Their optimal values are therefore related to one another; the exact relation can be derived by studying how sensitive the objective function is to small modifications of the constraints. This process is called ``sensitivity analysis of linear programs'' and we follow it in Appendix~\ref{appsec:pr_sen_analy} for the programs of interest. 
			As a result, we get that strategies~$\O_{AB|XY}$ with
			\begin{equation*}
				\begin{split}
					&\mathrm{Sig}^{(A\rightarrow B, x, y, b)}\left(\O_{AB|XY}\right) \leq  \zeta + 2\epsilon \\
					&\mathrm{Sig}^{(B\rightarrow A, x, y, a)}\left(\O_{AB|XY}\right) \leq  \zeta + 2\epsilon \;,
				\end{split}
			\end{equation*}
			for all $x$, $y$, $a$, and $b$  achieve winning probability $w \left( \O_{AB|XY} \right)$ such that
			\begin{equation}\label{eq:aprox_ns_up_bound}
				w \left( \O_{AB|XY} \right) \leq 1 - \alpha + (\zeta + 2\epsilon )  d \;,
			\end{equation}
			where $1-\alpha$ is the optimal winning probability of a \emph{non-signalling} strategy, i.e., it is the solution of Program~\eqref{eq:linear_non_relaxed}, and $ d =|\mathcal{X}||\mathcal{Y}|\left(|\mathcal{A}|+|\mathcal{B}|\right)$ is the number of the non-signalling constraints in the linear programs above.

		\subsection{Final result}
		
			The results of Section~\ref{sec:approx_ns_marginals} are applicable when considering permutation invariant strategies $\P_{\mr{A}\mr{B}|\mr{X}\mr{Y}}$ (recall Definition~\ref{def:permutation}).
			As the repeated game $\G^n_{1-\alpha+\beta}$ is by itself permutation invariant we can restrict the strategies of the players to be permutation invariant without loss of generality.\footnote{This is not to say that all strategies are permutation invariant but only that the optimal strategy can be assumed to be permutation invariant. It is perhaps interesting to note that, more commonly, the optimal strategies are taken to be, without loss of generality, deterministic in proofs of classical parallel repetition and pure in proofs of quantum parallel repetition. Here we are choosing to focus on permutation invariant strategies instead.}
			This is shown in the next lemma. 
			
			\begin{lem}\label{lem:prem_wlog}
				For every strategy $\P_{\mr{A}\mr{B}|\mr{X}\mr{Y}}$ for the repeated game $\G^n_{1-\alpha+\beta}$ there exists a permutation-invariant strategy $\tilde{\P}_{\mr{A}\mr{B}|\mr{X}\mr{Y}}$ such that $w\left( \P_{\mr{A}\mr{B}|\mr{X}\mr{Y}}\right) =w\left( \tilde{\P}_{\mr{A}\mr{B}|\mr{X}\mr{Y}}\right)$.
			\end{lem}
			\begin{proof}
				Given $\P_{\mr{A}\mr{B}|\mr{X}\mr{Y}}$ define its permutation-invariant version to be 
				\[
					\tilde{\P}_{\mr{A}\mr{B}|\mr{X}\mr{Y}} = \frac{1}{n!} \sum_{\pi}  \P_{\mr{A}\mr{B}|\mr{X}\mr{Y}}\circ\pi \;.
				\]
				The winning probability of the game is linear in the strategy, therefore we have 
				\begin{equation}\label{eq:winning_prob_linear}
					w\left( \tilde{\P}_{\mr{A}\mr{B}|\mr{X}\mr{Y}} \right) = w\left( \frac{1}{n!} \sum_{\pi}  \P_{\mr{A}\mr{B}|\mr{X}\mr{Y}}\circ\pi \right) =  \frac{1}{n!} \sum_{\pi} w\left( \P_{\mr{A}\mr{B}|\mr{X}\mr{Y}}\circ\pi \right) \;.
				\end{equation}
				
				Since the questions in the repeated game are chosen in an IID manner and the winning condition is checked for each game separately, the winning probability is indifferent to the ordering of the questions-answers pairs. As $\pi$ permutes the questions and answers together we have $w\left( \P_{\mr{A}\mr{B}|\mr{X}\mr{Y}}\circ\pi \right) = w\left( \P_{\mr{A}\mr{B}|\mr{X}\mr{Y}} \right)$.  
				Thus, we get $w\left( \tilde{\P}_{\mr{A}\mr{B}|\mr{X}\mr{Y}} \right)=w\left( \P_{\mr{A}\mr{B}|\mr{X}\mr{Y}} \right)$. 
			\end{proof}

			We can now combine everything we have learned in the previous sections in order to derive the final results. 
			As before, we denote by $d$ the number of non-signalling conditions appearing in the linear programs above, i.e., $ d =|\mathcal{X}||\mathcal{Y}|\left(|\mathcal{A}|+|\mathcal{B}|\right)$.

			\begin{lem}\label{lem:threshold}
				Let $w(G) = 1-\alpha$ be the optimal winning probability of a non-signalling strategy in $\G$. Let $0<\beta\leq\alpha$ be some constant and $n$ a sufficiently large integer such that Equation \eqref{eq:rep_num_condition} is satisfied. Then for any non-signalling strategy~$\P_{\mr{A}\mr{B}|\mr{X}\mr{Y}}$ of the threshold game $\G^n_{1-\alpha-\beta}$, 
				\[
					\mathrm{Pr}_{\mathsf{data}\sim\P_{\mr{A}\mr{B}\mr{X}\mr{Y}}} \left[w\left(\O_{AB|XY}^{\textup{freq}(\mathsf{data})}\right) > 1 - \alpha + \beta \right] \leq 6d\sqrt{c\delta} \;.
				\]
			\end{lem}
			
			\begin{proof}
				We denote the event of all inputs appearing in $\mathsf{data}_1$ and $\mathsf{data}_2$ by $\agq$.
				Furthermore, let $\zeta,\epsilon>0$ be such that $d(\zeta+2\epsilon) \leq \beta$, $\epsilon \leq \min_{x,y} \Q_{XY}(xy)$ and $7\epsilon \leq \zeta \leq 1$.
				
				If all questions $(x,y)$ appear at least once in $\mathsf{data}_1$ and $\mathsf{data}_2$, i.e., the event $\agq$ holds, then we can use Equation~\eqref{eq:aprox_ns_up_bound} in combination with Theorem \ref{thm:strong_condition} and get
				\begin{align*}
					&\mathrm{Pr}_{\mathsf{data}\sim \P_{\mr{A}\mr{B}\mr{X}\mr{Y}}}  \left[ w\left(\O_{AB|XY}^{\textup{freq}(\mathsf{data})}\right) > 1-\alpha + \beta \big| \agq \right] \\
					&\qquad \leq \mathrm{Pr}_{\mathsf{data}\sim \P_{\mr{A}\mr{B}\mr{X}\mr{Y}}} \Big[ \exists a,b,x,y \text{ s.t. }  \mathrm{Sig}^{(A\rightarrow B, x, y, b)}\left(\O_{AB|XY}^{\textup{freq}(\mathsf{data})}\right) \leq  \zeta + 2\epsilon \\
					& \hspace{145pt} \text{ or } \mathrm{Sig}^{(B\rightarrow A, x, y, a)}\left(\O_{AB|XY}^{\textup{freq}(\mathsf{data})}\right) \leq  \zeta + 2\epsilon \big| \agq \Big]  \\
					&\qquad \leq  d\cdot 4\sqrt{c\delta} \;. 
				\end{align*}

				The probability that the event $\agq$ does not hold is upper bounded by 
				\begin{align*}
					2 |\mathcal{X}||\mathcal{Y}|\left(1-\min_{x,y} \Q_{XY}(x,y)\right)^{n/2} 
					&\leq 2 |\mathcal{X}||\mathcal{Y}|e^{-\min_{x,y} \Q_{XY}(x,y)n/2} \\
					&\leq 2  |\mathcal{X}||\mathcal{Y}|e^{-\epsilon n/2} \\
					&\leq 2 d\delta
				\end{align*}
				and therefore all together we have
				\[
					\mathrm{Pr}_{\mathsf{data}\sim \P_{\mr{A}\mr{B}\mr{X}\mr{Y}}} \left[ w\left(\O_{AB|XY}^{\textup{freq}(\mathsf{data})}\right) > 1-\alpha + \beta \right] \leq 6d\sqrt{c\delta}\;. \qedhere 
				\]
			\end{proof}
			
			Our threshold theorem, Theorem \ref{thm:final_threshold_theorem}, follows from Lemma \ref{lem:threshold}:
			\begin{proof}[Proof of Theorem \ref{thm:final_threshold_theorem}]
				
				Let  $f$ denote the fraction of coordinates in which the players win the game in the observed data. 
				Note that $f$ is equals exactly $w\left(\O_{AB|XY}^{\textup{freq}(\mathsf{data})}\right)$ by the definition of $\O_{AB|XY}^{\textup{freq}(\mathsf{data})}$. 
				 Lemma  \ref{lem:threshold} therefore implies 
				 \begin{equation}\label{eq:complete_statement}
				 		\mathrm{Pr}_{\mathsf{data}\sim\P_{\mr{A}\mr{B}\mr{X}\mr{Y}}} \left[ f > 1 - \alpha + \beta \right] \leq 6d\sqrt{c\delta} \;.
				 \end{equation}

%
				
				Plugging the values of the parameters $d$, $c$, and $\delta$, we see that Equation~\eqref{eq:complete_statement} can be written, for an appropriately defined $\hat{C}(\G,n)$, as 
				\begin{align}
					\mathrm{Pr}_{\mathsf{data}\sim \P_{\mr{A}\mr{B}\mr{X}\mr{Y}}}  \left[ f > 1 - \alpha + \beta \right] & \leq \hat{C}(\G,n) \exp [-n \epsilon^2 / 8] \nonumber \\
					& = \mathrm{poly}(n)  \exp [-n \epsilon^2 / 8]  \;. \label{eq:complete_statement_poly}
				\end{align}
				Our goal now is to show that, actually, it must be possible to replace $\hat{C}(\G,n)$ with a constant smaller than 1 and by this drop the polynomial pre-factor. We do this using a step that appeared in~\cite[Proof of Theorem~3.1]{lancien2016parallel}.\footnote{The part of the proof starting at this point onward did not appear in the proof of the threshold theorem of~\cite{arnon2016non}.  We follow here the last part of the proof of the threshold theorem presented in~\cite{lancien2016parallel}, which appeared after~\cite{arnon2016non}, and can be used to improve the result of~\cite{arnon2016non}.}
				
				To this end, denote by $\omega_{\text{opt}}(\G^n_{1-\alpha+\beta})$ the optimal winning probability in the threshold game $\G^n_{1-\alpha+\beta}$.
				For any $n$, let $\tilde{C}_n$ be the constant for which the tight bound 
				\begin{equation}\label{eq:complete_bound_tight}
					\omega_{\text{opt}}(\G^n_{1-\alpha+\beta}) = \tilde{C}_n  \exp [-n \epsilon^2 / 8]
				\end{equation}
				holds. In particular, this means that there exists a strategy $\P_n$ that achieves the above winning probability.
				
				Assume by contradiction that there exists $N_0$ such that $\tilde{C}_{N_0} > 1$. Thus, there exists a strategy $\P_{N_0}$ achieving $\tilde{C}_{N_0}  \exp [-N_0 \epsilon^2 / 8]$, with $\tilde{C}_{N_0} > 1$, in the game~$\G^{N_0}_{1-\alpha+\beta}$.
				
				Let $N_1$ be sufficiently large, so that Equation~\eqref{eq:number_of_repet_bound} holds for $n=N_0N_1$ and consider the threshold game $\G^{N_0N_1}_{1-\alpha+\beta}$. 
				On the one hand, using $N_1$ independent copies of $\P_{N_0}$ achieves winning probability of $\left(\tilde{C}_{N_0}\right)^{N_1}  \exp [-N_0N_1 \epsilon^2 / 8]$ and thus 
				\begin{equation*}
					\omega_{\text{opt}}(\G^{N_0N_1}_{1-\alpha+\beta}) \geq \left(\tilde{C}_{N_0}\right)^{N_1}  \exp [-N_0N_1 \epsilon^2 / 8] \;.
				\end{equation*}
				On the other hand, Equation~\eqref{eq:complete_statement_poly} must hold for $n=N_0N_1$:
				\begin{equation*}
					\omega_{\text{opt}}(\G^{N_0N_1}_{1-\alpha+\beta}) \leq  \mathrm{poly}\left(N_0N_1\right)  \exp [-N_0N_1 \epsilon^2 / 8] \;.
				\end{equation*}
				To reconcile both bounds, we must have $\left(\tilde{C}_{N_0}\right)^{N_1} \leq \mathrm{poly}\left(N_0N_1\right)$ for all sufficiently large $N_1$. Thus, $\tilde{C}_{N_0}\leq 1$, which leads to a contradiction. 
				
				We get that for \emph{all} sufficiently large $n$, $\tilde{C}_n\leq 1$. In combination with Equation~\eqref{eq:complete_bound_tight} we therefore have 
				\[
					\omega_{\text{opt}}(\G^n_{1-\alpha+\beta}) \leq  \exp [-n \epsilon^2 / 8] \;. \qedhere
				\]
			\end{proof}

			To get a better feeling of the result, without trying to optimise it, one can make the following choices. Let $\epsilon = \frac{\beta}{10d}$, $\zeta=8\epsilon$ and $\nu=\epsilon$ (assuming $\min_{x,y} \Q_{XY}(x,y) > \frac{\beta}{10d}$). 
			Using these choices, our proof holds for $n$ and $\beta$ such that 
			\begin{equation*}
				\frac{n}{\ln(n)}>20|\mathcal{A}||\mathcal{B}||\mathcal{X}||\mathcal{Y}|\frac{\ln(20d/\beta)}{(\beta/10d)^2}
			\end{equation*}
			with the following constants in Theorem \ref{thm:final_threshold_theorem}:
			\begin{equation}\label{eq:choice_of_constants}
				\mathcal{C}(\G) = (30d)^{-2} = \left(30 |\mathcal{X}||\mathcal{Y}|\left(|\mathcal{A}|+|\mathcal{B}| \right) \right)^{-2} \;.
			\end{equation}
			The theorem then reads
			\begin{equation*}
				\mathrm{Pr}_{\mathsf{data}\sim \P_{\mr{A}\mr{B}\mr{X}\mr{Y}}}  \left[ f > 1 - \alpha + \beta \right] \leq \exp\left[-n\beta^2\left(30 |\mathcal{X}||\mathcal{Y}|\left(|\mathcal{A}|+|\mathcal{B}| \right) \right)^{-2}\right] \;.
			\end{equation*}
			A different choice of parameters can improve the dependence of the constants on the game $\G$.

		
	\section{Open questions}\label{sec:pr_open_ques}
	
		In this chapter we considered the question of parallel repetition of games when the players are allowed to use any non-signalling strategy.
		The most interesting direction for future work is the development of a similar proof technique, based on de Finetti reductions or other forms of reductions to IID, for classical and quantum parallel repetition. 
		In the case of classical games, parallel repetition results for general games with more than two parties are  unknown. For quantum games, even the case of two-player games is not completely solved. (Recall Section~\ref{sec:par_rep_intro_thresh_thm} for further information). 
		Since our proof captures all types of games and any number of players (see~\cite{arnon2016non}), a similar proof technique for classical and quantum games will solve some open questions.	
		
		To understand what is the main challenge when trying to extend the proof to classical and quantum case, note the following.
		 In the standard proofs of parallel repetition theorems, i.e., proofs following the approach of~\cite{raz1998parallel} most of the difficulties arise due to the effect of conditioning on the event of winning some of the game repetitions. As this event is one that depends on the structure of the game and strategy and we have no control over them, conditioning can introduce arbitrary correlations between the questions used in different repetitions of the game, a major source of difficulty for the remainder of the argument. 
		 In our proof we also need to analyse the effect of conditioning on a certain event, the event of the non-signalling test accepting, and this is done in Lemma \ref{lem:weak_condition}. 
		 However, the key advantage of our approach is that the test has a very specific structure, and in particular conditioning on the test passing can be done locally by the players in a way that respects the non-signalling constraints. As a result it is almost trivial to deal with the conditioning in the remainder of the proof. 
		 This shift from conditioning on an uncontrolled event, success in the game, to a highly controlled one, a non-signalling test that we design ourselves, is a key simplification that we expect to play an important role in any extension of our method to  classical or quantum strategies.
		  
		By finding appropriate ``non-classicality'' and ``non-quantumness'' measures which can replace our signalling measure in Definition \ref{def:sign_measure_dF} one may be able to adapt the proof to the multi-player classical and quantum cases as well.  
		Unfortunately, it is not clear which measure can be used by the players, preferably locally, to determine if their systems are classical or quantum. In other words, the main difficulty is finding a measure for which Lemma \ref{lem:weak_condition} can be proven. 
		The rest of the proofs should follow easily for most ``non-classicality'' and ``non-quantumness'' measures of one-game strategies.

\chapter{Showcase: device-independent quantum cryptography}\label{ch:crypto_showcase}

	In this chapter we consider the showcase of device-independent quantum cryptography and show how the security proof of device-independent cryptographic protocols can be performed via a \emph{reduction to IID}. 
	We introduce a general framework for obtaining proofs of device-independent security for a broad range of cryptographic tasks. For the sake of explicitness, we focus in this chapter on the task of device-independent quantum key distribution (DIQKD).\footnote{Since the initial announcement of our work~\cite{arnon2016simple}, our framework has already been applied to a variety of additional tasks, including conference key agreement~\cite{ribeiro2017fully}, randomness expansion~\cite{arnon2016simple} and privatization~\cite{kessler2017device}, as well as randomness generation with sub-linear quantum resources~\cite{bamps2017device}.}
	
	The main result that we present can be phrased in the following informal way (the formal theorem is stated as Theorem~\ref{thm:QKD_security}):
	\begin{thm}[Informal]
		Security of DIQKD in the most general case follows from security under  the IID assumption. 
		Moreover, the dependence of the key rate on the number of rounds, $n$, is the same as the one in the IID case, up to terms that scale like $1/\sqrt{n}$.
	\end{thm}
	The theorem establishes the a priori surprising fact that general quantum adversaries are no stronger than an adversary restricted to IID attacks, even in the device-independent setting. 
	This allows us to give simple and modular security proofs of DIQKD and to extend tight results known for DIQKD under the IID assumption to the most general setting thus deriving essentially optimal key rates and noise tolerance.\footnote{This is crucial for experimental implementations of device-independent protocols. Our quantitive results have been applied to the analysis of the first experimental implementation of a protocol for randomness generation in the fully device-independent framework~\cite{liu2017high}.}
	
	Our technique takes advantage of the sequential nature of the protocol, as well as the specific way in which classical statistics are collected by users of the protocol, and makes use of the entropy accumulation theorem (EAT), discussed as part of Chapter~\ref{ch:reductions_seq}. 
	The analysis and results of this chapter previously appeared in~\cite{arnon2016simple,arnon2018practical}.

	The chapter is arranged as follows. 
	We first explain in Section~\ref{sec:crypt_showcase_main_cha} what is the main challenge when proving the security of device-independent quantum cryptographic protocols, such as DIQKD.
	Section~\ref{sec:crypt_showcase_ent_acc} deals with the analysis of the main subroutine of most device-independent protocols. Then, the security proof of DIQKD is given in Section~\ref{sec:diqkd_proof}.
	As we are about to encounter many parameters and variables throughout the proofs, we list them in Appendix~\ref{appsec:qkd_param_summ} for convenience. 
		
	\section{Main challenge and goal}\label{sec:crypt_showcase_main_cha}

		The central task when proving security of cryptographic protocols consists in bounding the information that an adversary, called Eve, may obtain about certain values generated by the protocol, which are supposed to be secret. 
		In the case of QKD, for example, the relevant output of the protocol is the raw data $K$, and proving security is essentially equivalent\footnote{From that point onward standard classical post-processing steps, e.g., error correction and privacy amplification, suffice to prove the security of the protocol; recall Section~\ref{sec:qkd_main_task}.} to establishing a lower bound on the smooth conditional min-entropy $H_{\min}^{\varepsilon}(K|E)$, where $E$ is Eve's quantum system, which can be initially correlated to the device producing $K$.
		The quantity $H_{\min}^{\varepsilon}(K|E)$ determines the maximal length of the secret key that can be created by the protocol. Hence, proving security amounts to establishing a lower bound on $H_{\min}^{\varepsilon}(K|E)$.
		Evaluating the smooth min-entropy $H_{\min}^{\varepsilon}(K|E)$ of a large system is often difficult, especially in the device-independent setting where not much is known about the way $K$ is produced and the system $E$ is out of our control. 
		
		\sloppy
		The IID assumption, discussed in Chapter~\ref{ch:iid_assumption}, is 	commonly used to simplify the task of calculating  $H_{\min}^{\varepsilon}(K|E)$. The analysis of the smooth min-entropy under the IID assumption was sketched in Section~\ref{sec:crypt_under_iid};  in that case the total smooth min-entropy can be easily related to the sum of the von Neumann entropies in each round separately, using the quantum asymptotic equipartition property (Section~\ref{sec:quant_aep}). A bound on the entropy accumulated in one round can usually be derived using the expected winning probability in the game played in that round (as appeared in Section~\ref{sec:randomness_single_round}), which in turn can be easily estimated during the protocol in the IID case using standard Chernoff-type bounds.  
		A long line of  works~\cite{acin2006bell,acin2006efficient,scarani2006secrecy,acin2007device,masanes2009universally,pironio2009device,hanggi2010efficient,hanggi2010device,masanes2011secure,masanes2014full}  considered the security of device-independent quantum and non-signalling cryptography under the IID assumption. 
		Most relevant for our work are the results of~\cite{pironio2009device}, where security of a DIQKD protocol was proven in the asymptotic limit, i.e., when the device is used $n\rightarrow\infty$ times, and under the IID assumption. Their protocol is based on the CHSH inequality~\cite{clauser1969proposed}, and their analysis shows that it achieves the best possible rates under these assumptions.

		Unfortunately, even though quite convenient for the analysis, the IID assumption is a very strong one in the DI scenario.
		In particular, under such an assumption the device cannot use any internal memory (i.e., its actions in one round cannot depend on the previous rounds) or even display time-dependent behaviour (due to inevitable imperfections for example). 	
		Without this assumption, however, very little is known about the structure of the untrusted device and hence also about its output (as the device might correlate the different rounds in an almost arbitrary way). As a consequence, DIQKD security proofs~\cite{reichardt2013classical,vazirani2014fully,miller2014robust} that estimated $H_{\min}^{\varepsilon}(K|E)$ directly for the most general case had to use complicated techniques and statistical analysis compared to the IID case. 
		This led to security statements which are of limited relevance for practical experimental implementations; they are applicable only in an unrealistic regime of parameters, e.g., small amount of tolerable noise and large number of signals.

		To overcome the above difficulty we take the approach of \emph{reductions to IID} in the analysis presented in the following sections. In particular, we leverage the sequential nature of our DIQKD protocol to prove its security by reducing the analysis of multi-round sequential boxes to that of IID boxes as discussed in Chapter~\ref{ch:reductions_seq}. Specifically, we use the EAT presented in Section~\ref{sec:eat_statement} to establish that entropy accumulates additively throughout the multiple rounds of the protocol and use it to bound the total amount of smooth min-entropy $H_{\min}^{\varepsilon}(K|E)$.\footnote{The security proof presented in~\cite{miller2014robust} is similar in spirit (but technically very different) to the one presented here. It bounds the total amount of smooth min-entropy generated in the protocol in a round-by-round fashion but the entropy accumulated in a single round is not the von Neumann entropy.}
		
		This results in a proof technique with several benefits. 
		Firstly, since the analysis of the IID case is rather simple and modular (as it builds mainly on the analysis of a single-round box) a security proof via a reduction to IID ends up being simple and modular by itself. For example, if one wishes to consider a DIQKD protocol based on a game other than the CHSH game, the sole significant modification of the security proof is the analysis of a single-round box (see Sections~\ref{sec:randomness_single_round} and~\ref{sec:crypto_min_tradeoff}). 
		Secondly, due to the optimality of the EAT (at least to first order in $n$), we are able to extend tight results known for, e.g., DIQKD, under the IID assumption, to the most general setting. This yields the best rates known for any protocol for a device-independent cryptographic task. 
		Thirdly, performing a finite-size analysis is no harder than performing the asymptotic one as all dependency on $n$ is either trivial or already incorporated in the EAT.

		We are now ready to embark on the mission of proving the security of our DIQKD protocol, described in Section~\ref{sec:diqkd-protocol} (see also Protocol~\ref{pro:diqkd_chsh} below).

	\section{Device-independent entropy accumulation}\label{sec:crypt_showcase_ent_acc}
	
		The current section is devoted to the analysis of the entropy accumulation protocol presented as Protocol~\ref{pro:randomness_generation}. 
		The entropy accumulation protocol acts as the main building block of many device-independent cryptographic protocols. 
		It is used to generate the raw data for Alice and Bob by playing a non-local game $n$ times in sequence using an untrusted device $D$.
		We remark that even though we call the entropy accumulation protocol a ``protocol'',  one should see it more as a mathematical tool which allows us to use the machinery of the EAT rather than an actual protocol to be implemented.\footnote{In particular, in a setting with two distinct parties, Alice and Bob, communication is required to actually implement Protocol~\ref{pro:randomness_generation}. We ignore this as it is not relevant for the analysis.}  
		The relevance of the protocol stems from the fact that the final state at the end of the protocol, on which a smooth min-entropy is evaluated, is closely related to the final state in the actual protocol to be executed (e.g., our DIQKD protocol).  
		
		\begin{algorithm}[t]
				\caption{CHSH-based entropy accumulation protocol}
				\label{pro:randomness_generation}
				\begin{algorithmic}[1]
					\STATEx \textbf{Arguments:} 
						\STATEx\hspace{\algorithmicindent} $D$ -- untrusted device of two components that can play CHSH repeatedly
						\STATEx\hspace{\algorithmicindent} $n \in \mathbb{N}_+$ -- number of rounds
						\STATEx\hspace{\algorithmicindent} $\gamma \in (0,1]$ -- expected fraction of test rounds 
				
						\STATEx\hspace{\algorithmicindent} $\omega_{\mathrm{exp}}$ -- expected winning probability in an honest implementation    
						\STATEx\hspace{\algorithmicindent} $\delta_{\mathrm{est}} \in (0,1)$ -- width of the confidence interval for parameter estimation
						
					\STATEx
					
					\STATE For every round $i\in[n]$ do Steps~\ref{prostep:choose_Ti}-\ref{prostep:calculate_Ci_EA}:
						\STATE\hspace{\algorithmicindent} Alice and Bob choose a random $T_i\in\{0,1\}$ such that $\Pr(T_i=1)=\gamma$.  \label{prostep:choose_Ti}
						\STATE\hspace{\algorithmicindent} If $T_i=0$, Alice and Bob choose $(X_i,Y_i)=(0,2)$ and otherwise $X_i,Y_i\in \{0,1\}$ uniformly at random.  
						\STATE\hspace{\algorithmicindent} Alice and Bob use $D$ with $X_i,Y_i$ and record their outputs as $A_i$ and $B_i$ respectively. \label{prostep:measurement} 
						\STATE\hspace{\algorithmicindent}If $T_i=0$ then Bob updates $B_i$ to $B_i = \perp$, and they set $W_i=\perp$. If $T_i=1$ they set $W_i =w\left(A_i,B_i,X_i,Y_i\right)$.\label{prostep:calculate_Ci_EA}
				\STATE Alice and Bob abort if $\sum_{j:T_j=1} W_j < \left(\omega_{\mathrm{exp}}\gamma - \delta_{\mathrm{est}}\right) \cdot n\;$. \label{prostep:abort_general_EA}
				\end{algorithmic}
				\end{algorithm}

		Our primary task is to lower-bound the amount of smooth min-entropy generated by playing the $n$ games. This lower-bound can then be used as the starting point of security proofs of device-independent cryptographic protocols, such as DIQKD. 
		The informal statement is given below (for the explicit formulation see Theorem~\ref{thm:main_generation_chsh}):
		\begin{thm}[Informal]\label{thm:main_generation_informal}
			Fix a choice of parameters for Protocol~\ref{pro:randomness_generation}. Then there exist constants $c_1,c_2>0$ such that the following holds. 
			Let $D$ be any device and $\rho_{|\Omega}$ the state generated using Protocol~\ref{pro:randomness_generation}, conditioned on the protocol not aborting. 
			Then for any $\varepsilon_1,\varepsilon_2\in(0,1)$, either the protocol aborts with probability greater than $1-\varepsilon_1$ or
				\begin{equation}\label{eq:main_thm-inf}
					 H^{\varepsilon_2}_{\min} \left( \mr{A}\mr{B} | \mr{X}\mr{Y}\mr{T} E \right)_{\rho_{|\Omega}} > c_1 n - c_2\sqrt{n \log(1/\varepsilon_1\varepsilon_2)} \;.
				\end{equation}
		\end{thm}

		The registers $\mr{A}\mr{B}$ in Equation~\eqref{eq:main_thm-inf} contain the classical outputs generated by the device during the protocol. The registers $\mr{X}\mr{Y}\mr{T}$ hold the classical information exchanged during the protocol, that may be leaked to the adversary. $E$ is a quantum register that describes the adversary's quantum system. Thus, Equation~\eqref{eq:main_thm-inf} gives a precise bound on the amount of the smooth min-entropy present in the users' outputs at the end of the protocol, conditioned on all information available to the adversary. 
	
		We give below explicit formulas for computing the constants $c_1$ and $c_2$ that appear in Equation~\eqref{eq:main_thm-inf} as a function of the parameters of the protocol. 
		Importantly, the constant $c_1$ that governs the leading-order term equals the optimal constant, i.e., the same leading constant that would be obtained under the IID assumption, which by the asymptotic equipartition property (Theorem~\ref{thm:quant_aep}) is the von Neumann entropy accumulated in one round of the protocol. 
		Furthermore, our analysis provides control over the constant~$c_2$ in front of the second-order term. Such control is necessary for any application where finite values of $n$ need to be considered, such as in  quantum cryptography, where the values of $n$ achieved in practice remain relatively small.\footnote{See e.g. Figure~\ref{fig:qkd_rates_n_mod}, where one can see that finite-size effects can play an important role up to even moderately large values of $n\approx 10^{10}$.}

		As we show below, Theorem~\ref{thm:main_generation_informal} can be proven by reducing the general sequential scenario to the IID one using the EAT. 
		To use the EAT, we first need to construct the relevant objects, i.e., the EAT-channels and the min-tradeoff functions defined in Section~\ref{sec:eat_prereq}. 	This is done in the following two sections. A lower-bound on the smooth min-entropy is then proven in Section~\ref{sec:ent_acc_rate}.
		
		\subsection{EAT channels}\label{sec:crypto_eat_chan}
		
			Protocol~\ref{pro:randomness_generation} proceeds in rounds and can therefore be presented by an application of a sequence of quantum channels (recall Section~\ref{sec:seq_qunt_procc}). 
			In this section we define the considered channels and prove that they are EAT-channels, according to Definition~\ref{def:eat_channels}.
			Note that one has some freedom in \emph{choosing} the channels to work with (i.e., the channels are not completely defined by the protocol itself). We choose our particular channels so that all the prerequisites of the EAT are fulfilled and, at the same time, the final bound on the smooth min-entropy can be converted to a bound on the smooth min-entropy in our DIQKD protocol (see Section~\ref{sec:diqkd_soundness}, and Lemma~\ref{lem:smooth_bound_qkd} in particular, for details).
			
			Every EAT channel $\mathcal{M}_i$ describes one round of the protocol, where one round includes Steps~\ref{prostep:choose_Ti}-\ref{prostep:calculate_Ci_EA} of Protocol~\ref{pro:randomness_generation}. 
			For every $i\in\{0\}\cup[n]$, the (unknown) quantum state of the device $D$ shared by Alice and Bob after round $i$ of the protocol is denoted by $\rho_{Q_AQ_B}^i$. We denote the register holding this state by $R_i$. In particular, $R_0 = Q_AQ_B$ at the start of the protocol. 
			At Step~\ref{prostep:measurement} in Protocol~\ref{pro:randomness_generation}, the quantum state of the devices is changed from $\rho_{Q_AQ_B}^{i-1}$ in $R_{i-1}$ to $\rho_{Q_AQ_B}^{i}$ in $R_i$ by the use of the device. To be a bit more precise, the quantum state is changed in two stages. First, the relevant measurement of Step~\ref{prostep:measurement} is done (where it is assumed that the measurements of the different components are in tensor product). Then, after $A_i$ and~$B_i$ are recorded, the different components of the device are allowed to communicate. Thus, some further changes can be made to the post-measurement state even based on the memory of all components together (recall Section~\ref{sec:mr_seq_with_comm}).
		
			In the notation of Chapter~\ref{ch:reductions_seq}, we make the following choices:
			\begin{equation}\label{eq:register_ident_eat_crypt}
				\begin{split}
					O_i &= A_iB_i \\
					S_i &= X_i Y_i T_i \\
					C_i &= W_i \\
					R_i &= R_i \\
					E &= E \;.
				\end{split}
			\end{equation}
			Our EAT channels are then  
			\[
				\mathcal{M}_i:R_{i-1}\rightarrow R_i A_i B_i X_i Y_i T_i  W_i
			\]
			 defined by the CPTP map describing the $i$-th round of Protocol~\ref{pro:randomness_generation}, as implemented by the untrusted device $D$. That is,  the channel describes the random choices of~$T_i$, $X_i$, and $Y_i$, the quantum operations made by the device, and the production of~$A_i$, $B_i$, and $W_i$. 
			 
			 Since the operations of $D$ are unknown, our EAT channels are not completely explicit. The important thing is merely that we know that some quantum channels describing the operation of the device  \emph{exist}. The lack of knowledge regarding the channels does not raise any problems when applying the EAT but it does make the task of deriving good min-tradeoff functions more challenging (compared to the scenario of a characterised device). This difficulty, however, is inherent to device-independent information processing tasks and has nothing to do with the proof technique; see Section~\ref{sec:crypto_min_tradeoff} below for further details. 
			
			We prove that the described channels can act as our EAT channels. 
			\begin{lem}\label{lem:eat_chan_crypt}
				The channels $\left\{\mathcal{M}_i:R_{i-1}\rightarrow R_i A_i B_i X_i Y_i T_i W_i\right\}_{i\in[n]}$ defined by the \textup{CPTP} map describing the $i$-th round of Protocol~\ref{pro:randomness_generation}, as implemented by the untrusted device $D$ are EAT channels according to Definition~\ref{def:eat_channels} and the identification made in Equation~\eqref{eq:register_ident_eat_crypt}.
			\end{lem}
			\begin{proof}
				To prove that the constructed channels $\left\{\mathcal{M}_i\right\}_{i\in[n]}$ are EAT channels we need to show that the three conditions stated in Definition~\ref{def:eat_channels} are fulfiled.
				
				\begin{enumerate}
						\item $\left\{O_i\right\}_{i\in[n]}=\left\{A_iB_i\right\}_{i\in[n]}$, $\left\{S_i\right\}_{i\in[n]}=\left\{X_i Y_i T_i\right\}_{i\in[n]}$, and $\left\{C_i\right\}_{i\in[n]}=\left\{W_i\right\}_{i\in[n]}$ are all finite-dimensional classical systems. 
							$\left\{R_i\right\}_{i\in[n]}$ are arbitrary quantum systems. 
							Finally, we have $d_{O}=d_{A_i}\cdot d_{B_i}=2\cdot 3 = 6$.

						\item For any $i\in[n]$ and any input state $\sigma_{R_{i-1}}$, $W_i$ is a function of the classical values $A_i$, $B_i$, $X_i$, and $Y_i$. Hence, the marginal $\sigma_{O_iS_i}=\sigma_{A_i B_i X_i Y_i T_i }$ of the output state is unchanged when deriving $W_i$ from it. (In other words, we can ``measure'' $\sigma_{A_i B_i X_i Y_i T_i }$ to get the value of $W_i$ repeatedly without disturbing $\sigma_{A_i B_i X_i Y_i T_i}$). 
					
						\item For any initial state 
						$\rho_{R_0E}^{\text{in}}$ and the resulting final state $\rho_{\mr{O}\mr{S}\mr{C}E} = \rho_{\mr{A}\mr{B}\mr{X}\mr{Y}\mr{T}\mr{W}E}$, the Markov-chain conditions
						\begin{equation*}
							(AB)_1,\dotsc ,(AB)_{i-1} \leftrightarrow (XYT)_1,\dotsc, (XYT)_{i-1}, E \leftrightarrow (XYT)_i 
						\end{equation*}
							trivially hold for all $i\in[n]$ since, according to Protocol~\ref{pro:randomness_generation}, $X_i$, $Y_i$, and $T_i$ are chosen independently from everything else.  \qedhere
					\end{enumerate}
			\end{proof}
			
			When defining the EAT channels we identified $O_i$ with $A_iB_i$. Looking ahead, this means that we are about to derive a bound on $H_{\min}^{\varepsilon}(\mr{A}\mr{B}|\mr{X}\mr{Y}\mr{T}E)$. For the analysis of DIQKD, however, a bound on $H_{\min}^{\varepsilon}(\mr{A}|\mr{X}\mr{Y}\mr{T}E)$ is needed. 
			Why not set $O_i=A_i$ instead of $O_i=A_iB_i$? 
			The reason is that the definition of an EAT channel requires that $C_i$ can be derived from $O_iS_i$ alone. Thus, choosing $O_i=A_i$, $S_i=X_i Y_i T_i$, and $C_i=W_i$ would render the above lemma wrong. 
			The reader may then ask -- why not choose $O_i=A_i$ and $S_i=B_iX_i Y_i T_i$?
			With this choice, however, the required Markov-chain conditions read 
			\[
				A_1,\dotsc ,A_{i-1} \leftrightarrow (BXYT)_1,\dotsc, (BXYT)_{i-1}, E \leftrightarrow (BXYT)_i 
			\]
			and these do not hold for an arbitrary initial states since nothing restricts $B_i$ from being correlated to, e.g., $A_{i-1}$, when considering untrusted devices.\footnote{Consider for example a device in which the initial state $\rho_{Q_{A_1}Q_{B_1}Q_{A_2}Q_{B_2}} = \ket{\Phi}\bra{\Phi}_{Q_{A_1}Q_{B_2}}\otimes\ket{\Phi}\bra{\Phi}_{Q_{A_2}Q_{B_1}}$, i.e., the systems over $Q_{A_1}$ and $Q_{B_2}$ are entangled. Thus, $A_1$ and $B_2$ may be correlated even given $B_1X_1Y_1T_1$. In this case the Markov-chain conditions do not hold since the side-information $B_2$ reveals information regarding the past output $A_1$.} Hence, the lemma would not hold for this choice as well. 
			We therefore stick with the choices made in Equation~\eqref{eq:register_ident_eat_crypt} and relate $H_{\min}^{\varepsilon}(\mr{A}\mr{B}|\mr{X}\mr{Y}\mr{T}E)$ to $H_{\min}^{\varepsilon}(\mr{A}|\mr{X}\mr{Y}\mr{T}E)$ in Section~\ref{sec:diqkd_soundness}.
					
			Now that our EAT channels are defined, the next step is to construct a min-tradeoff function for them. This is done in the next section.
	
		\subsection{Min-tradeoff function}\label{sec:crypto_min_tradeoff}
		
			When working with the EAT the most important task is to devise a good tradeoff function, as defined in Definition~\ref{def:tradeoff_funcs}. As mentioned in Section~\ref{sec:eat_statement}, this is where the ``physics kicks in''. 
			We are aiming for a lower-bound on the smooth min-entropy and hence in  need of a min-tradeoff function $f_{\min}$. 
			That is, we need to construct a convex differential function for which, for all $i\in[n]$,
			\begin{equation}\label{eq:min_trad_cond_crypt}
				f_{\min}(p) \leq \inf_{\sigma_{R_{i-1}R'}:\mathcal{M}_i(\sigma)_{W_i}=p} H\left( A_iB_i | X_i Y_i T_i R' \right)_{\mathcal{M}_i(\sigma)} \;,
			\end{equation}
			where $p$ is a probability distribution over $\mathcal{W} = \{\perp,0,1\}$ and $\left\{\mathcal{M}_i\right\}_{i\in[n]}$ are the EAT channels defined in the previous section. 
			
			To understand the task at hand, let us first focus on the set of states
			\[
				\Sigma(p) = \left\{\sigma_{R_{i-1}R'}:\mathcal{M}_i(\sigma)_{W_i}=p\right\} \;.
			\]
			on which the infimum is evaluated. 
			First observe that, due to the structure of our channels, the distributions over $X_i$, $Y_i$, and $T_i$	are \emph{fixed} for any $\sigma\in\Sigma(p)$. That is, even though we take an infimum over many possible input states for the channels, and even though the actions of the untrusted device are not characterised, the values of $X_i$, $Y_i$, and $T_i$ are always chosen according to Protocol~\ref{pro:randomness_generation}.\footnote{A different model for the sequential process could have been one in which the initial quantum state \emph{itself} includes the registers $\mr{X}$ and $\mr{Y}$ and the channel is defined such that a measurement is performed on those registers to get the inputs (and then use the device in the protocol). When starting with maximally mixed states over $\mr{X}\mr{Y}$ the entire sequential process is exactly the same as the one described by our EAT channels. However, when coming to construct a min-tradeoff function with this (somewhat strange) alternative choice of channels, we see that the set $\Sigma(p)$ can include states in which, e.g., $X_i=0$ with probability 1 (since we need to consider \emph{all possible} input states). In the context of Bell inequalities, this is similar to dropping the ``free choice assumption''. Clearly, if this had been the case, the only min-tradeoff function one could construct is the constant function $f_{\min}(p)=0$ for all $p$, which is trivial and useless.}
			This implies, in particular, that for probability distributions $p$ with $p(\perp)\neq 1-\gamma$ the set $\Sigma(p)$ is empty.

			Given the above, for any $p$ over $\{\perp,0,1\}$, the set $\Sigma(p)$ includes only states  $\sigma$ for which 
			\begin{equation}\label{eq:p_to_w_illustration}
					\mathcal{M}_i(\sigma)_{W_i} 
					= \begin{pmatrix} 
					p(\perp) & 0 & 0 \\ 
					0 & p(0) & 0  \\
					0 &  0  & p(1)   \end{pmatrix}
					=\begin{pmatrix} 
					1-\gamma & 0 & 0 \\ 
					0 & \gamma (1-\omega)  & 0  \\
					0 &  0  & \gamma \omega  \end{pmatrix} \;,
				\end{equation}
				where we identify $\omega$ with the winning probability of the state $\sigma$ in the CHSH game (when using the measurements defining the channel $\mathcal{M}_i$). 
				For this reason, we can, slightly informally,\footnote{Formally, we will need to extend the function to all probability distributions $p$ (even those with $p(\perp)\neq 1-\gamma$). We can extended the function in any way we wish, while keeping it convex and differentiable.} see the function $f_{\min}$ as defined over a single variable $\omega\in[0,1]$. 
				
				In total, we can understand the set $\Sigma(p)=\Sigma(p(\omega))$ as the set including all states~$\sigma$ that can be used to win the CHSH game with probability $\omega$. It is this information about the relevant input states $\sigma$ that allows us to construct a min-tradeoff function fulfilling Equation~\eqref{eq:min_trad_cond_crypt}. In fact, given the above observation, the main ingredient needed to construct a valid (and tight) min-tradeoff function is Lemma~\ref{lem:single_round_secrecy}, which was discussed in the context of \emph{single-round boxes} (Chapter~\ref{ch:single_round_box}). This clarifies why the presented proof technique can be seen as a \emph{reduction to~IID}.
				
				We are ready to embark on the construction of the min-tradeoff function. 
				
				\begin{lem}\label{lem:min_tradeoff_crypt}
					Let\footnote{We define the functions $g$ and  $f_{min}$ only in the regime in which the protocol does not abort, i.e., $ p(1)/\gamma\geq 3/4$.}
					\begin{equation*}
						g(p) =  \begin{cases}
							 1 - h\left( \frac{1}{2} + \frac{1}{2}\sqrt{16 \; \frac{p(1)}{\gamma} \left(\frac{p(1)}{\gamma}-1\right) +3}  \right)&  \frac{p(1)}{\gamma}\in\left[\frac{3}{4},\frac{2+\sqrt{2}}{4}\right] \\
							1 & \frac{p(1)}{\gamma}\in\left[\frac{2+\sqrt{2}}{4},1\right]\;,
							\end{cases} 
					\end{equation*}
					and
					\begin{equation}\label{eq:f_min_choice}
						f_{\min}\left(p,p_{\mathrm{cut}}\right) = \begin{cases}
						g\left(p\right)&  p(1) \leq p_{\mathrm{cut}}(1) \;  \\
						\begin{split}
							&\frac{\mathrm{d}}{\mathrm{d}p(1)} g(p)\big|_{p_{\mathrm{cut}}}  \cdot p(1) \\
							&\quad +  g(p_{\mathrm{cut}}) -	\frac{\mathrm{d}}{\mathrm{d}p(1)} g(p)\big|_{p_{\mathrm{cut}}} \cdot p_{\mathrm{cut}}(1)
						\end{split} 
						& p(1)> p_{\mathrm{cut}}(1)\; \;. 
						\end{cases}
					\end{equation}
					Then, for any probability distribution $p_{\mathrm{cut}}$ over $\{\perp,0,1\}$, $f_{\min}\left(p,p_{\mathrm{cut}}\right)$ is a min-tradeoff functions for the EAT channels from Lemma~\ref{lem:eat_chan_crypt}.
				\end{lem}

 				Before proving the lemma, let us parse the above lengthy equations. 
 				The function~$g$ is basically the  single-round bound presented in Lemma~\ref{lem:single_round_secrecy}, where we replace~$\omega$ with$~\frac{p(1)}{\gamma}$ and trivially extend the function to the regime of winning probabilities above the optimal quantum winning probability $\frac{2+\sqrt{2}}{4}$. 
				Notice that for an arbitrary~$p$, the correct relation between $p$ and $\omega$ is given by $\omega=\frac{p(1)}{p(0)+p(1)}$. However, as explained above, due to the definition of the channels $\mathcal{M}_i$, the set $\Sigma(p)$ is empty for $p$ with $p(0)+p(1)\neq \gamma$. This implies that there are no constraints on the value of the min-tradeoff function for such $p$'s and we are free to define it as we wish in this regime. Thus, we are not going to run into problems even though the value of the function~$g$ does not seem to have any ``physical meaning'' for~$p$ with~$p(0)+p(1)\neq \gamma$.\footnote{Alternatively, one could replace $p(1)/\gamma$ with $p(1)/(p(0)+p(1))$, which is more meaningful, in the definition of the function $g$. However, since the $\mathrm{d}f_{\min}/\mathrm{d}p(1)$ will affect the final smooth min-entropy bound, using $p(1)/\gamma$ leads to better quantitive results.}
				
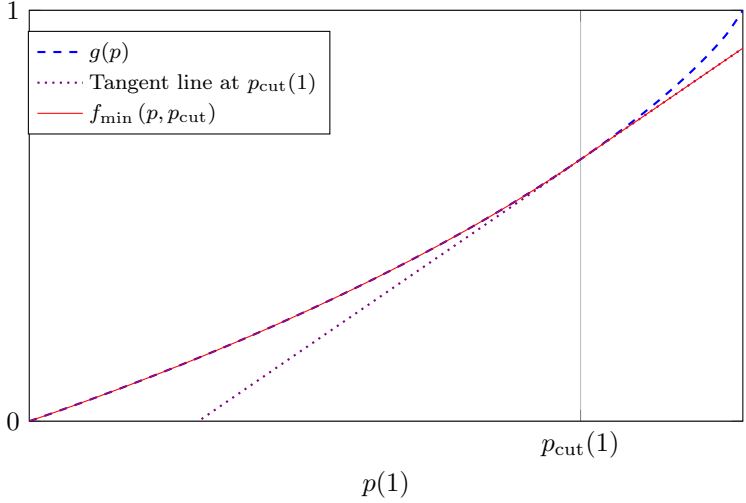
\begin{figure}
					\centering
					\begin{tikzpicture}
						\begin{axis}[
							height=7cm,
							width=11cm,
							grid=major,
							xlabel=$p(1)$,
							xmin=0.75,
							xmax=0.853553,
							ymax=1,
							ymin=0,
						     xtick={0.83},
						     xticklabels={$p_{\mathrm{cut}}(1)$},
					          ytick={0,1},
							legend style={at={(0.21,0.95)},anchor=north,legend cell align=left,font=\footnotesize} 
						]
						
					
						\addplot[blue,thick,smooth,dashed] coordinates {
						(0.75, 0.) (0.752071, 0.0120347) (0.754142, 0.0242374) (0.756213, 0.0366113) (0.758284, 0.0491598) (0.760355, 0.0618861) (0.762426, 0.074794) (0.764497, 0.0878872) (0.766569, 0.10117) (0.76864, 0.114646) (0.770711, 0.128319) (0.772782, 0.142196) (0.774853, 0.156279) (0.776924, 0.170575) (0.778995, 0.185089) (0.781066, 0.199826) (0.783137, 0.214793) (0.785208, 0.229996) (0.787279, 0.245441) (0.78935, 0.261137) (0.791421, 0.277091) (0.793492, 0.293311) (0.795563, 0.309806) (0.797635, 0.326586) (0.799706, 0.343661) (0.801777, 0.361042) (0.803848, 0.378741) (0.805919, 0.396771) (0.80799, 0.415147) (0.810061, 0.433884) (0.812132, 0.452998) (0.814203, 0.47251) (0.816274, 0.49244) (0.818345, 0.51281) (0.820416, 0.533648) (0.822487, 0.554982) (0.824558, 0.576846) (0.82663, 0.599279) (0.828701, 0.622324) (0.830772, 0.646033) (0.832843, 0.670469) (0.834914, 0.695705) (0.836985, 0.721832) (0.839056, 0.748965) (0.841127, 0.777251) (0.843198, 0.806888) (0.845269, 0.838156) (0.84734, 0.871481) (0.849411, 0.907587) (0.851482, 0.948007) (0.853553, 1.)	
						};
						\addlegendentry{$g(p)$}
						
						\addplot[violet,thick,smooth,dotted] coordinates {
						(0.75, -0.282055) (0.752071, -0.258259) (0.754142, -0.234463) (0.756213, -0.210667) (0.758284, -0.186871) (0.760355, -0.163075) (0.762426, -0.139279) (0.764497, -0.115484) (0.766569, -0.0916877) (0.76864, -0.0678918) (0.770711, -0.0440959) (0.772782, -0.0203) (0.774853, 0.00349582) (0.776924, 0.0272917) (0.778995, 0.0510876) (0.781066, 0.0748834) (0.783137, 0.0986793) (0.785208, 0.122475) (0.787279, 0.146271) (0.78935, 0.170067) (0.791421, 0.193863) (0.793492, 0.217659) (0.795563, 0.241455) (0.797635, 0.26525) (0.799706, 0.289046) (0.801777, 0.312842) (0.803848, 0.336638) (0.805919, 0.360434) (0.80799, 0.38423) (0.810061, 0.408026) (0.812132, 0.431821) (0.814203, 0.455617) (0.816274, 0.479413) (0.818345, 0.503209) (0.820416, 0.527005) (0.822487, 0.550801) (0.824558, 0.574597) (0.82663, 0.598393) (0.828701, 0.622188) (0.830772, 0.645984) (0.832843, 0.66978) (0.834914, 0.693576) (0.836985, 0.717372) (0.839056, 0.741168) (0.841127, 0.764964) (0.843198, 0.788759) (0.845269, 0.812555) (0.84734, 0.836351) (0.849411, 0.860147) (0.851482, 0.883943) (0.853553, 0.907739)
						};
						\addlegendentry{Tangent line at $p_{\mathrm{cut}}(1)$}
						
						\addplot[red,smooth] coordinates {
						(0.75, 0.) (0.752071, 0.0120347) (0.754142, 0.0242374) (0.756213, 0.0366113) (0.758284, 0.0491598) (0.760355, 0.0618861) (0.762426, 0.074794) (0.764497, 0.0878872) (0.766569, 0.10117) (0.76864, 0.114646) (0.770711, 0.128319) (0.772782, 0.142196) (0.774853, 0.156279) (0.776924, 0.170575) (0.778995, 0.185089) (0.781066, 0.199826) (0.783137, 0.214793) (0.785208, 0.229996) (0.787279, 0.245441) (0.78935, 0.261137) (0.791421, 0.277091) (0.793492, 0.293311) (0.795563, 0.309806) (0.797635, 0.326586) (0.799706, 0.343661) (0.801777, 0.361042) (0.803848, 0.378741) (0.805919, 0.396771) (0.80799, 0.415147) (0.810061, 0.433884) (0.812132, 0.452998) (0.814203, 0.47251) (0.816274, 0.49244) (0.818345, 0.51281) (0.820416, 0.533648) (0.822487, 0.554982) (0.824558, 0.576846) (0.82663, 0.599279) (0.828701, 0.622324) (0.830772, 0.645984) (0.832843, 0.66978) (0.834914, 0.693576) (0.836985, 0.717372) (0.839056, 0.741168) (0.841127, 0.764964) (0.843198, 0.788759) (0.845269, 0.812555) (0.84734, 0.836351) (0.849411, 0.860147) (0.851482, 0.883943) (0.853553, 0.907739)
						};
						\addlegendentry{$f_{\min}\left(p,p_{\mathrm{cut}}\right) $}
								
						\end{axis}  
					\end{tikzpicture}
					
					\caption{The construction of the min-tradeoff function $f_{\min}$ appearing in Equation~\eqref{eq:f_min_choice}. The plot shows the values of the min-tradeoff function restricted to a slice $p(0)+p(1)= \text{constant}$. }
							\label{fig:f_min_crypt}
					\end{figure}

				The function $f_{\min}$ in Equation~\eqref{eq:f_min_choice} is governed by $g$ and can be understood as follows. Fix a probability distribution $p_{\mathrm{cut}}\in[0,1]$. For $p$ with $p(1) \leq p_{\mathrm{cut}}(1)$, $f_{\min}$ is identical to $g$. Otherwise,~$f_{\min}$ is a linear function (when restricting ourselves to a slice $p(0)+p(1)= \text{constant}$) defined via the value and the tangent of $g$ at the point~$p_{\mathrm{cut}}(1)$. That is, we ``cut and glue'' the function at point $p_{\mathrm{cut}}$. 
				By doing so, we make sure that $f_{\min}$ is a convex and differentiable function, as required by Definition~\ref{def:tradeoff_funcs} while restraining its gradient, which will later affect the bound on the smooth min-entropy (via Equation~\eqref{eq:eat_min_bound_second_ord}).
				This construction of $f_{\min}$ is illustrated in Figure~\ref{fig:f_min_crypt}.
				
					\begin{proof}[Proof of Lemma~\ref{lem:min_tradeoff_crypt}.]
						We start by using the chain rule of the von Neumann entropy,
						\[
							H\left( A_i B_i | X_i Y_i T_i R' \right)_{\mathcal{M}_i(\sigma)} \geq H\left( A_i | X_i Y_i T_i R' \right)_{\mathcal{M}_i(\sigma)}.
						\]
						
						Due to the bipartite requirement on the untrusted device $D$ used to implement the protocol, the actions of Alice's device are independent of Bob's choice of $Y_i$ as well as of and $T_i$.\footnote{We assume that the value of $T_i$ is exchanged over a classical authenticated channel to which the device $D$ does not have access. In particular, Alice's part of the device is independent from the value of $T_i$ given $X_i$.} We thus have 
						\[
							H\left( A_i | X_i Y_i T_i R'   \right)_{\mathcal{M}_i(\sigma)} = H\left(A_i | X_i R'   \right)_{\mathcal{M}_i(\sigma)} \;.
						\]	
						
						Combined with Lemma~\ref{lem:single_round_secrecy} we get that for any state $\sigma$ with winning probability~$\omega$ in the CHSH game,
						\begin{equation}\label{eq:entropy_bound_for_min_tradeoff}
							H\left( A_i B_i | X_i Y_i T_i R' \right) \geq 1 - h\left( \frac{1}{2} + \frac{1}{2}\sqrt{16\omega \left(\omega-1\right) +3}  \right) \;.
						\end{equation}
						
						For probability distributions $p$ with $p(0)+p(1)\neq \gamma$, the set of states fulfilling $\mathcal{M}_i(\sigma)_{W_i}=p$ is empty and the condition on the min-tradeoff function given in Equation~\ref{eq:min_trad_cond_crypt} becomes trivial. Hence, for the construction of the min-tradeoff function we can restrict our attention to $p$ with $p(0)+p(1)= \gamma$. For such $p$'s one can write $\omega=\frac{p(1)}{p(0)+p(1)}=\frac{p(1)}{\gamma}$.
						All together we learn that \emph{for all} $p$ with $\frac{p(1)}{\gamma}\geq \frac{3}{4}$,
						\begin{equation}\label{eq:one_box_entropy_final}
						\begin{split}
							\inf_{\sigma_{R_{i-1}R'}:\mathcal{M}_i(\sigma)_{W_i}=p} H&\left( A_i B_i |  X_i Y_i T_i R' \right)_{\mathcal{M}_i(\sigma)} \geq \\
							&1 - h\left( \frac{1}{2} + \frac{1}{2}\sqrt{16\;\frac{p(1)}{\gamma} \left(\frac{p(1)}{\gamma}-1\right) +3}  \right) \;.
							\end{split}
						\end{equation}
	
						Define a function $g$ by  
						\begin{equation}\label{eq:def-g}
						g(p) \,=\,  \begin{cases}
								1 - h\left( \frac{1}{2} + \frac{1}{2}\sqrt{16\; \frac{p(1)}{\gamma} \left(\frac{p(1)}{\gamma}-1\right) +3}  \right) &  \frac{p(1)}{\gamma}\in\left[\frac{3}{4},\frac{2+\sqrt{2}}{4}\right] \\
								1& \frac{p(1)}{\gamma}\in\left[\frac{2+\sqrt{2}}{4},1\right] \;.
								\end{cases}
						\end{equation}
					
						From Equation~\eqref{eq:one_box_entropy_final} it follows that any choice of $f_{\min}$ that is differentiable and satisfies $f_{\min}(p) \leq g(p)$ for all $p$ will satisfy Equation~\eqref{eq:min_trad_cond_crypt}. 

						For $\frac{p(1)}{\gamma}=\frac{2+\sqrt{2}}{4}$ the derivative of $g$ is infinite. 
						Looking ahead, for the final bound on the smooth min-entropy derived using the EAT to be meaningful, $f_{\min}$ should be chosen such that $\| \triangledown f_{\min} \|_{\infty}$ is finite. 
						To assure that this is the case we choose $f_{\min}$ by ``cutting'' the function $g$ and ``gluing'' it to a linear function at some point $p_{\mathrm{cut}}$, while keeping the function differentiable. By doing this we ensure that the gradient of $f_{\min}$ is bounded, at the cost of losing a bit of entropy for $p$ with $p(1)> p_{\mathrm{cut}}(1)$.\footnote{The point $p_{\mathrm{cut}}$ can later be chosen such that the derived smooth entropy bounds are optimised.} 
						Towards this, denote 
						\begin{equation}\label{eq:derivative} 
							a(p_{\mathrm{cut}})= \frac{\mathrm{d}}{\mathrm{d}p(1)} g(p)\big|_{p_{\mathrm{cut}}}  \qquad\text{and} \qquad b(p_{\mathrm{cut}}) =  g(p_{\mathrm{cut}})-a(p_{\mathrm{cut}})\cdot p_{\mathrm{cut}}(1). 
						\end{equation}
					
						We then make the following choice for the min-tradeoff function $f_{\min}$ (see Figure~\ref{fig:f_min_crypt}):
						\begin{equation}\label{eq:f_min_choice}
							f_{\min}\left(p,p_{\mathrm{cut}}\right) = \begin{cases}
								g\left(p\right) &  p(1) \leq p_{\mathrm{cut}}(1) \\
								a(p_{\mathrm{cut}}) \cdot p(1) + b(p_{\mathrm{cut}}) & p(1)> p_{\mathrm{cut}}(1)\;
							\end{cases}
						\end{equation}
						From the definition of $a$ and $b$ in Equation~\eqref{eq:derivative}, this function is convex, differentiable, and fulfils the condition given in Equation~\eqref{eq:min_trad_cond_crypt}. $f_{\min}$ can therefore be rightfully called a min-tradeoff function. 
						Furthermore, by definition, for any choice of $p_{\mathrm{cut}}$ it holds that $\| \triangledown f_{\min}(\cdot,p_{\mathrm{cut}}) \|_{\infty} \leq a(p_{\mathrm{cut}})$. \qedhere
						
					\end{proof}
		
	\subsection{Smooth min-entropy rate}\label{sec:ent_acc_rate}
		
			After constructing the EAT channels and min-tradeoff function in the previous sections, we are ready to apply Theorem~\ref{thm:eat} to derive our lower-bound on the conditional smooth min-entropy generated by the entropy accumulation protocol, Protocol~\ref{pro:randomness_generation}. 
			
			We use the following notation. 
			The event of \emph{not aborting} the protocol is given~by
			\begin{equation}\label{eq:good_event_def}
				\Omega =  \Big\{\mr{w} : \sum_{j: T_j=1} w_j \geq \left(\omega_{\mathrm{exp}}\gamma - \delta_{\mathrm{est}}\right) \cdot n\;\Big\}\;.
			\end{equation}
			For any initial state $\rho_{Q_AQ_BE}^{\text{in}}$, the final state in the end of the protocol is denoted by $\rho=\rho_{\mathbf{ABXYTW}E}$ 	and the final state conditioned on not aborting the entropy accumulation protocol is  $\rho_{|\Omega}$.

			As shown in Theorem~\ref{thm:main_generation_chsh} below, The smooth min-entropy rate is governed by the following functions, where $h$ is the binary entropy and~$\gamma,p(1)\in(0,1]$:
			\begin{align}
				&g(p) =  \begin{cases}
						 1 - h\left( \frac{1}{2} + \frac{1}{2}\sqrt{16 \; \frac{p(1)}{\gamma} \left(\frac{p(1)}{\gamma}-1\right) +3}  \right)&  \frac{p(1)}{\gamma}\in\left[\frac{3}{4},\frac{2+\sqrt{2}}{4}\right] \\
						1 & \frac{p(1)}{\gamma}\in\left[\frac{2+\sqrt{2}}{4},1\right]\;,
						\end{cases}\notag\\
				&f_{\min}\left(p,p_{\mathrm{cut}}\right) = \begin{cases}
				g\left(p\right)&  p(1) \leq p_{\mathrm{cut}}(1) \;  \\
				\frac{\mathrm{d}}{\mathrm{d}p(1)} g(p)\big|_{p_{\mathrm{cut}}}  \cdot p(1)+  g(p_{\mathrm{cut}}) -	\frac{\mathrm{d}}{\mathrm{d}p(1)} g(p)\big|_{p_{\mathrm{cut}}} \cdot p_{\mathrm{cut}}(1) & p(1)> p_{\mathrm{cut}}(1)\;,
				\end{cases} \nonumber \\
				&\mu(p,p_{\mathrm{cut}},\varepsilon_{\mathrm{s}},\varepsilon_{\mathrm{e}}) =  
				f_{\min}\left(p, p_{\mathrm{cut}}\right)  \nonumber\\
					&\hspace{90pt} - \frac{1}{\sqrt{n}}2\left( \log 13 + \frac{\mathrm{d}}{\mathrm{d}p(1)} g(p)\big|_{p_{\mathrm{cut}}}  \right)\sqrt{1-2 \log (\varepsilon_{\mathrm{s}} \cdot \varepsilon_{\mathrm{e}})}\;,
				 \nonumber\\
				&\mu_{\mathrm{opt}}(\varepsilon_{\mathrm{s}}, \varepsilon_{\mathrm{e}}) = \max_{\frac{3}{4} < \frac{p_{\mathrm{cut}}(1)}{\gamma} < \frac{2+\sqrt{2}}{4}} \; \mu(\omega_{\mathrm{exp}}\gamma - \delta_{\mathrm{est}},p_{\mathrm{cut}},\varepsilon_{\mathrm{s}},\varepsilon_{\mathrm{e}})\;. \label{eq:eta_opt}
			\end{align}

			\begin{thm}\label{thm:main_generation_chsh}
			Let $D$ be any device, $\rho$ the state generated by running Protocol~\ref{pro:randomness_generation},~$\Omega$ the event that the protocol does not abort (as defined in Equation~\eqref{eq:good_event_def}), and~$\rho_{|\Omega}$ the state conditioned on $\Omega$.
			Then, for any $\varepsilon_{\mathrm{EA}},\varepsilon_{\mathrm{s}}\in (0,1)$, either the protocol aborts with probability greater than $1-\varepsilon_{\mathrm{EA}}$ or
				\begin{equation}\label{eq:main_thm_di_ent_acc}
					 H^{\varepsilon_{\mathrm{s}}}_{\min} \left( \mr{A}\mr{B} | \mr{X}\mr{Y}\mr{T} E \right)_{\rho_{|\Omega}} > n\cdot \mu_{\mathrm{opt}}(\varepsilon_{\mathrm{s}},\varepsilon_{\mathrm{EA}}) \;,
				\end{equation}
				where  $\mu_{\mathrm{opt}}$ is defined in Equation~\eqref{eq:eta_opt}. 
			\end{thm}

			\begin{proof}
				
				We wish to apply the EAT, stated as Theorem~\ref{thm:eat}. 
				To this end, denote by $\mathrm{freq}_{\mr{w}}(\tilde{w}) = \frac{| \left\{ i | w_i = \tilde{w} \right\} |}{n}$ the frequency defined by the raw data $\mr{w}$ (recall Equation~\eqref{eq:freq_definition}) and observe the following:
				\begin{enumerate}
				
					\item The EAT channels $\{\mathcal{M}_i\}_{i\in[n]}$ constructed in Section~\ref{sec:crypto_eat_chan} faithfully describe the protocol and the device $D$, in the sense that the final state of the protocol,~$\rho$, can be written as
						\begin{equation}\label{eq:final_state_before_abort}
							\rho_{\mathbf{ABXYTW}E} = \left( \Tr_{R_n}\circ \mathcal{M}_n \circ \dots \circ \mathcal{M}_1\right) \otimes \mathbb{I}_E \; \rho_{Q_AQ_BE}^{\text{in}} \;.
						\end{equation} 
					
					\item The set $ \hat{\Omega} = \{ p: p(1) \geq \omega_{\mathrm{exp}}\gamma - \delta_{\mathrm{est}} \} $ is convex and $\{  \mathrm{freq}_{\mr{w}} : \mr{w} \in \Omega \} \subseteq  \hat{\Omega}$ (recall Section~\ref{sec:eat_cond_on_omega}).
					
					\item According to Lemma~\ref{lem:min_tradeoff_crypt}, $f_{\min}\left(p,p_{\mathrm{cut}}\right)$ is a min-tradeoff function for the considered EAT channels, for any $p_{\mathrm{cut}}$ with $\frac{3}{4}<\frac{p_{\mathrm{cut}}(1)}{\gamma}<\frac{2+\sqrt{2}}{4} $.
					
					\item For any $p_{\mathrm{cut}}$ with $\frac{3}{4}<\frac{p_{\mathrm{cut}}(1)}{\gamma}<\frac{2+\sqrt{2}}{4} $, the value $t=f_{\min}\left(\omega_{\mathrm{exp}}\gamma - \delta_{\mathrm{est}},p_{\mathrm{cut}}\right)$ satisfies $f_{\min}\left( \mathrm{freq}_{\mr{w}},p_{\mathrm{cut}} \right) \geq t$ for any $\mathrm{freq}_{\mr{w}} \in \hat{\Omega}$.
					
					\item $d_O=d_{A_iB_i}=6$ and $ \|  \triangledown f_{\min}(\cdot ,p_{\mathrm{cut}}) \|_{\infty}  = a(p_{\mathrm{cut}})$ for any $p_{\mathrm{cut}}$ with $\frac{3}{4}<\frac{p_{\mathrm{cut}}(1)}{\gamma}<\frac{2+\sqrt{2}}{4} $. 
					
				\end{enumerate}
				
				Using the EAT (Theorem~\ref{thm:eat}) in combination with the above observations 
				we conclude that  for any  $p_{\mathrm{cut}}$ with $\frac{3}{4}<\frac{p_{\mathrm{cut}}(1)}{\gamma}<\frac{2+\sqrt{2}}{4} $, either the protocol aborts with probability greater than $1-\varepsilon_{\mathrm{EA}}$, or
				\begin{equation}\label{eq:EAT_statement}
					H^{\varepsilon_{\mathrm{s}}}_{\min} \left(  \mr{A}\mr{B} | \mr{X}\mr{Y}\mr{T} E \right)_{\rho_{|\Omega}} > n f_{\min}\left( \omega_{\mathrm{exp}}\gamma - \delta_{\mathrm{est}},p_{\mathrm{cut}} \right) - \sqrt{n} \zeta(p_{\mathrm{cut}})  \;,
				\end{equation}
				for $\zeta(p_{\mathrm{cut}},\varepsilon_{\mathrm{s}},\varepsilon_{\mathrm{EA}}) = 2\left( \log 13 + a(p_{\mathrm{cut}}) \right)\sqrt{1-2 \log (\varepsilon_{\mathrm{s}} \cdot\varepsilon_{\mathrm{EA}})}$.
				 To obtain the optimal rate we maximise $H^{\varepsilon_{\mathrm{s}}}_{\min} \left(  \mr{A}\mr{B} | \mr{X}\mr{Y}\mr{T} E \right)_{\rho_{|\Omega}} $ over $p_{\mathrm{cut}}$. 
				 Denote $\mu(p,p_{\mathrm{cut}},\varepsilon_{\mathrm{s}},\varepsilon_{\mathrm{EA}}) =  f_{\min}\left(p, p_{\mathrm{cut}}\right) - \frac{1}{\sqrt{n}}\zeta(p_{\mathrm{cut}},\varepsilon_{\mathrm{s}},\varepsilon_{\mathrm{EA}})$ and let
				\[
					\mu_{\mathrm{opt}} (\varepsilon_{\mathrm{s}},\varepsilon_{\mathrm{EA}})= \max_{\frac{3}{4} < \frac{p_{\mathrm{cut}}(1)}{\gamma} < \frac{2+\sqrt{2}}{4}} \mu(\omega_{\mathrm{exp}}\gamma - \delta_{\mathrm{est}},p_{\mathrm{cut}},\varepsilon_{\mathrm{s}},\varepsilon_{\mathrm{EA}}) \;.
				\]	
				Plugging this into Equation~\eqref{eq:EAT_statement} the theorem follows. \qedhere
				
			\end{proof}

			Importantly, the theorem tells us that the first order term of the smooth min-entropy is linear in $n$. Moreover, asymptotically, the entropy rate is simply given by the min-tradeoff function $f_{\min}\left(p,p_{\mathrm{cut}}\right)$. This is why it was crucial to construct an optimal min-tradeoff function in Section~\ref{sec:crypto_min_tradeoff}.

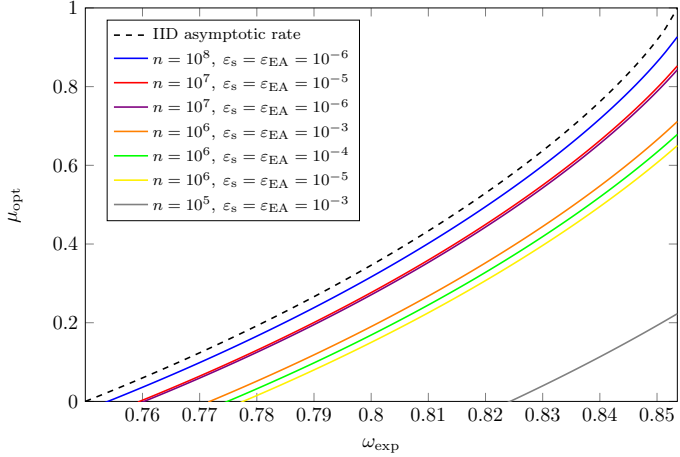
\begin{figure}
			\centering
			\begin{tikzpicture}[scale=0.75]
				\begin{axis}[
					height=8.5cm,
					width=12cm,
					xlabel=$\omega_{\mathrm{exp}}$,
					ylabel=$\mu_{\mathrm{opt}}$,
					xmin=0.75,
					xmax=0.853553,
					ymax=1,
					ymin=0,
				     xtick={0.76,0.77,0.78,0.79,0.80,0.81,0.82,0.83,0.84,0.85},
			          ytick={0,0.2,0.4,0.6,0.8,1},
					legend style={at={(0.25,0.97)},anchor=north,legend cell align=left,font=\footnotesize} 
				]
				
			
				\addplot[black,thick,smooth,dashed] coordinates {
				(0.75, 0.) (0.752071, 0.0120347) (0.754142, 0.0242374) (0.756213, 0.0366113) (0.758284, 0.0491598) (0.760355, 0.0618861) (0.762426, 0.074794) (0.764497, 0.0878872) (0.766569, 0.10117) (0.76864, 0.114646) (0.770711, 0.128319) (0.772782, 0.142196) (0.774853, 0.156279) (0.776924, 0.170575) (0.778995, 0.185089) (0.781066, 0.199826) (0.783137, 0.214793) (0.785208, 0.229996) (0.787279, 0.245441) (0.78935, 0.261137) (0.791421, 0.277091) (0.793492, 0.293311) (0.795563, 0.309806) (0.797635, 0.326586) (0.799706, 0.343661) (0.801777, 0.361042) (0.803848, 0.378741) (0.805919, 0.396771) (0.80799, 0.415147) (0.810061, 0.433884) (0.812132, 0.452998) (0.814203, 0.47251) (0.816274, 0.49244) (0.818345, 0.51281) (0.820416, 0.533648) (0.822487, 0.554982) (0.824558, 0.576846) (0.82663, 0.599279) (0.828701, 0.622324) (0.830772, 0.646033) (0.832843, 0.670469) (0.834914, 0.695705) (0.836985, 0.721832) (0.839056, 0.748965) (0.841127, 0.777251) (0.843198, 0.806888) (0.845269, 0.838156) (0.84734, 0.871481) (0.849411, 0.907587) (0.851482, 0.948007) (0.853553, 1)	
				};
				\addlegendentry{IID asymptotic rate}
				
				\addplot[blue,thick,smooth] coordinates {
				(0.753051, -0.00519263) (0.755102, 0.00675606) (0.757153, 0.0188591) (0.759204, 0.0311305) (0.761255, 0.0435732) (0.763306, 0.0561908) (0.765357, 0.0689867) (0.767409, 0.0819646) (0.76946, 0.0951283) (0.771511, 0.108482) (0.773562, 0.12203) (0.775613, 0.135776) (0.777664, 0.149726) (0.779715, 0.163884) (0.781766, 0.178256) (0.783817, 0.192847) (0.785868, 0.207663) (0.787919, 0.222711) (0.78997, 0.237996) (0.792021, 0.253526) (0.794072, 0.269309) (0.796123, 0.285352) (0.798175, 0.301665) (0.800226, 0.318255) (0.802277, 0.335134) (0.804328, 0.352312) (0.806379, 0.369801) (0.80843, 0.387613) (0.810481, 0.405761) (0.812532, 0.424261) (0.814583, 0.44313) (0.816634, 0.462384) (0.818685, 0.482044) (0.820736, 0.502133) (0.822787, 0.522674) (0.824838, 0.543696) (0.82689, 0.56523) (0.828941, 0.587313) (0.830992, 0.609986) (0.833043, 0.633297) (0.835094, 0.657305) (0.837145, 0.682076) (0.839196, 0.707695) (0.841247, 0.734266) (0.843298, 0.761921) (0.845349, 0.790836) (0.8474, 0.821253) (0.849451, 0.853531) (0.851502, 0.888251) (0.853553, 0.926527)
				};
				\addlegendentry{$n=10^8,\; \varepsilon_{\mathrm{s}}=\varepsilon_{\mathrm{EA}}=10^{-6}$}
			
				\addplot[red,thick,smooth] coordinates {
				(0.759204, -0.00167058) (0.761255, 0.01049) (0.763306, 0.0228199) (0.765357, 0.0353224) (0.767409, 0.0480009) (0.76946, 0.0608589) (0.771511, 0.0739003) (0.773562, 0.0871289) (0.775613, 0.100549) (0.777664, 0.114164) (0.779715, 0.12798) (0.781766, 0.142001) (0.783817, 0.156232) (0.785868, 0.170679) (0.787919, 0.185347) (0.78997, 0.200242) (0.792021, 0.21537) (0.794072, 0.230739) (0.796123, 0.246355) (0.798175, 0.262226) (0.800226, 0.278361) (0.802277, 0.294768) (0.804328, 0.311457) (0.806379, 0.328437) (0.80843, 0.345721) (0.810481, 0.363319) (0.812532, 0.381245) (0.814583, 0.399513) (0.816634, 0.418137) (0.818685, 0.437136) (0.820736, 0.456528) (0.822787, 0.476333) (0.824838, 0.496574) (0.82689, 0.517277) (0.828941, 0.538472) (0.830992, 0.56019) (0.833043, 0.582471) (0.835094, 0.605358) (0.837145, 0.628901) (0.839196, 0.653163) (0.841247, 0.678215) (0.843298, 0.704149) (0.845349, 0.731076) (0.8474, 0.759141) (0.849451, 0.788541) (0.851502, 0.81955) (0.853553, 0.852586)
				};
				\addlegendentry{$n=10^7,\; \varepsilon_{\mathrm{s}}=\varepsilon_{\mathrm{EA}}=10^{-5}$}
				
				\addplot[violet,thick,smooth] coordinates {
				(0.759204, -0.00635168) (0.761255, 0.00576902) (0.763306, 0.0180583) (0.765357, 0.0305193) (0.767409, 0.0431556) (0.76946, 0.0559705) (0.771511, 0.0689678) (0.773562, 0.0821515) (0.775613, 0.0955254) (0.777664, 0.109094) (0.779715, 0.122862) (0.781766, 0.136833) (0.783817, 0.151014) (0.785868, 0.165409) (0.787919, 0.180023) (0.78997, 0.194863) (0.792021, 0.209936) (0.794072, 0.225246) (0.796123, 0.240803) (0.798175, 0.256613) (0.800226, 0.272684) (0.802277, 0.289025) (0.804328, 0.305646) (0.806379, 0.322556) (0.80843, 0.339766) (0.810481, 0.357288) (0.812532, 0.375135) (0.814583, 0.39332) (0.816634, 0.411858) (0.818685, 0.430766) (0.820736, 0.450063) (0.822787, 0.469767) (0.824838, 0.489902) (0.82689, 0.510493) (0.828941, 0.531567) (0.830992, 0.553158) (0.833043, 0.575301) (0.835094, 0.598039) (0.837145, 0.621421) (0.839196, 0.645505) (0.841247, 0.670362) (0.843298, 0.696076) (0.845349, 0.722754) (0.8474, 0.750533) (0.849451, 0.779594) (0.851502, 0.810189) (0.853553, 0.842692)
				};
				\addlegendentry{$n=10^7,\; \varepsilon_{\mathrm{s}}=\varepsilon_{\mathrm{EA}}=10^{-6}$}
			
				\addplot[orange,thick,smooth] coordinates {
				(0.771511, -0.0015508) (0.773562, 0.0110039) (0.775613, 0.0237357) (0.777664, 0.0366482) (0.779715, 0.0497451) (0.781766, 0.0630305) (0.783817, 0.0765085) (0.785868, 0.0901835) (0.787919, 0.10406) (0.78997, 0.118143) (0.792021, 0.132438) (0.794072, 0.14695) (0.796123, 0.161685) (0.798175, 0.176649) (0.800226, 0.191849) (0.802277, 0.207291) (0.804328, 0.222982) (0.806379, 0.238932) (0.80843, 0.255147) (0.810481, 0.271637) (0.812532, 0.288412) (0.814583, 0.305482) (0.816634, 0.322858) (0.818685, 0.340553) (0.820736, 0.35858) (0.822787, 0.376952) (0.824838, 0.395687) (0.82689, 0.414801) (0.828941, 0.434314) (0.830992, 0.454247) (0.833043, 0.474624) (0.835094, 0.495471) (0.837145, 0.516818) (0.839196, 0.538701) (0.841247, 0.561158) (0.843298, 0.584235) (0.845349, 0.607986) (0.8474, 0.632476) (0.849451, 0.657782) (0.851502, 0.683999) (0.853553, 0.711249)
				};
				\addlegendentry{$n=10^6,\; \varepsilon_{\mathrm{s}}=\varepsilon_{\mathrm{EA}}=10^{-3}$}
				
				\addplot[green,thick,smooth] coordinates {
				(0.773562, -0.00798184) (0.775613, 0.00458303) (0.777664, 0.0173252) (0.779715, 0.0302482) (0.781766, 0.0433558) (0.783817, 0.0566522) (0.785868, 0.0701415) (0.787919, 0.083828) (0.78997, 0.0977165) (0.792021, 0.111812) (0.794072, 0.126119) (0.796123, 0.140644) (0.798175, 0.155392) (0.800226, 0.170369) (0.802277, 0.185583) (0.804328, 0.201039) (0.806379, 0.216745) (0.80843, 0.23271) (0.810481, 0.24894) (0.812532, 0.265447) (0.814583, 0.282238) (0.816634, 0.299326) (0.818685, 0.31672) (0.820736, 0.334434) (0.822787, 0.35248) (0.824838, 0.370873) (0.82689, 0.389629) (0.828941, 0.408766) (0.830992, 0.428302) (0.833043, 0.44826) (0.835094, 0.468663) (0.837145, 0.489538) (0.839196, 0.510916) (0.841247, 0.53283) (0.843298, 0.555322) (0.845349, 0.578436) (0.8474, 0.602229) (0.849451, 0.626763) (0.851502, 0.652119) (0.853553, 0.678392)
				};
				\addlegendentry{$n=10^6,\; \varepsilon_{\mathrm{s}}=\varepsilon_{\mathrm{EA}}=10^{-4}$}
				
				\addplot[yellow,thick,smooth] coordinates {
				(0.775613, -0.0122004) (0.777664, 0.00039393) (0.779715, 0.0131662) (0.781766, 0.0261199) (0.783817, 0.039259) (0.785868, 0.0525874) (0.787919, 0.0661095) (0.78997, 0.0798296) (0.792021, 0.0937525) (0.794072, 0.107883) (0.796123, 0.122226) (0.798175, 0.136788) (0.800226, 0.151574) (0.802277, 0.166591) (0.804328, 0.181845) (0.806379, 0.197342) (0.80843, 0.213091) (0.810481, 0.2291) (0.812532, 0.245376) (0.814583, 0.26193) (0.816634, 0.27877) (0.818685, 0.295908) (0.820736, 0.313355) (0.822787, 0.331124) (0.824838, 0.349227) (0.82689, 0.36768) (0.828941, 0.386499) (0.830992, 0.405701) (0.833043, 0.425307) (0.835094, 0.445338) (0.837145, 0.465818) (0.839196, 0.486775) (0.841247, 0.50824) (0.843298, 0.530249) (0.845349, 0.552841) (0.8474, 0.576066) (0.849451, 0.599978) (0.851502, 0.624644) (0.853553, 0.650146)
				};
				\addlegendentry{$n=10^6,\; \varepsilon_{\mathrm{s}}=\varepsilon_{\mathrm{EA}}=10^{-5}$}
				
				\addplot[gray,thick,smooth] coordinates {
				(0.822787, -0.0102655) (0.824838, 0.00363856) (0.82689, 0.0177499) (0.828941, 0.0320736) (0.830992, 0.0466152) (0.833043, 0.0613805) (0.835094, 0.0763757) (0.837145, 0.0916072) (0.839196, 0.107082) (0.841247, 0.122808) (0.843298, 0.138792) (0.845349, 0.155044) (0.8474, 0.171572) (0.849451, 0.188385) (0.851502, 0.205496) (0.853553, 0.222914)
				};
				\addlegendentry{$n=10^5,\; \varepsilon_{\mathrm{s}}=\varepsilon_{\mathrm{EA}}=10^{-3}$}
				
				\end{axis}  
			\end{tikzpicture}
			
			\caption{$\mu_{\mathrm{opt}}(\omega_{\mathrm{exp}})$ for $\gamma=1$ and several choices of $n$, $\varepsilon_{\mathrm{EA}},$ and the smoothing parameter $\varepsilon_{\mathrm{s}}$. $\delta_{\mathrm{est}}=10^{-2}$ in the curve with $n=10^5$ and $\delta_{\mathrm{est}}=10^{-3}$  in all other curves. Note that for the errors of the protocols to be meaningful the number of rounds $n$ should be at least of order $\delta_{\mathrm{est}}^{-2}$.  $\varepsilon_{\mathrm{EA}}$ and $\varepsilon_{\mathrm{s}}$ affect the soundness error in the DIQKD protocol considered in Section~\ref{sec:diqkd_proof} and therefore should be chosen to be relatively small. The dashed line shows the optimal asymptotic ($n\rightarrow\infty$) rate under the IID assumption.}
			\label{fig:eta_rates}
			\end{figure}

			The rate $\mu_{\mathrm{opt}}$ is plotted in Figure~\ref{fig:eta_rates} as a function of the expected winning probability $\omega_{\mathrm{exp}}$ in the CHSH game for $\gamma=1$ and several choices of values for the parameters $\varepsilon_{\mathrm{EA}}$, $\delta_{\mathrm{est}}$, and $n$ (while optimising over all other parameters). 
			For comparison, we also plot  in Figure~\ref{fig:eta_rates} the asymptotic rate ($n\rightarrow\infty$) under the IID assumption. In this case, the quantum asymptotic equipartition property implies that the optimal rate is the Shannon entropy accumulated in one round of the protocol (recall Section~\ref{sec:quant_aep}). This rate, appearing as the dashed line in Figure~\ref{fig:eta_rates}, is an upper bound on the smooth min-entropy that can be accumulated. One can see that as the number of rounds in the protocol increases the rate $\mu_{\mathrm{opt}}$ approaches this optimal rate.

			\section{Device-independent quantum key distribution}\label{sec:diqkd_proof}
	
		Our DIQKD protocol is stated as Protocol~\ref{pro:diqkd_chsh}. 
		In the first part of the protocol Alice and Bob use their devices to produce the raw data, similarly to what is done in the entropy accumulation protocol, Protocol~\ref{pro:randomness_generation}, analysed in the previous section.
		In the second part of the protocol Alice and Bob apply classical post-processing steps to produce their final keys from the raw data. The classical post-processing consists of error correction, parameter estimation, and privacy amplification; all discussed in detail in Section~\ref{sec:diqkd-protocol}.
		
		Apart from the classical post-processing, the main difference between the entropy accumulation protocol and the DIQKD protocol is the way we set Bob's outputs. 
		In Protocol~\ref{pro:randomness_generation}, Bob's outputs are being set to $\perp$ in all rounds for which $T_i=0$, i.e., in the generation rounds. 
		In contrast, when dealing with QKD Bob needs to keep the outputs produced in the generation rounds so that he could create a key identical to Alice's key. To make the distinction explicit we denote Bob's outputs in Protocol~\ref{pro:diqkd_chsh} with a tilde,~$\tilde{\mr{B}}$. 
		We will get back to this point later and explain why the distinction is relevant for our analysis.

		\begin{algorithm}[t]
			\caption{CHSH-based DIQKD protocol}
			\label{pro:diqkd_chsh}
			\begin{algorithmic}[1]
				\STATEx \textbf{Arguments:} 
					\STATEx\hspace{\algorithmicindent} $D$ -- untrusted device of two components that can play CHSH repeatedly
					\STATEx\hspace{\algorithmicindent} $n \in \mathbb{N}_+$ -- number of rounds
					\STATEx\hspace{\algorithmicindent} $\gamma \in (0,1]$ -- expected fraction of test rounds 
			
					\STATEx\hspace{\algorithmicindent} $\omega_{\mathrm{exp}}$ -- expected winning probability in an honest implementation    
					\STATEx\hspace{\algorithmicindent} $\delta_{\mathrm{est}} \in (0,1)$ -- width of the confidence interval for parameter estimation
					
					\STATEx\hspace{\algorithmicindent} $\mathrm{EC}$ -- error correction protocol that leaks $\mathrm{leak_{EC}}$ bits and has completeness and soundness error probabilities $\varepsilon^c_{\mathrm{EC}}$ and $\varepsilon_{\mathrm{EC}}$ respectively
					 \STATEx\hspace{\algorithmicindent} $\mathrm{PA}$ -- privacy amplification protocol with error probability $\varepsilon_{\mathrm{PA}}$
					
				\STATEx
				
				\STATE For every round $i\in[n]$ do Steps~\ref{prostep:choosing_est_test}-\ref{prostep:use_device_qkd}:
					\STATE\hspace{\algorithmicindent} Alice and Bob choose a random $T_i\in\{0,1\}$ such that $\Pr(T_i=1)=\gamma$.  \label{prostep:choosing_est_test}
					\STATE\hspace{\algorithmicindent} If $T_i=0$, Alice and Bob choose $(X_i,Y_i)=(0,2)$ and otherwise $X_i,Y_i\in \{0,1\}$ uniformly at random.  
					\STATE\hspace{\algorithmicindent} Alice and Bob use $D$ with $X_i,Y_i$ and record their outputs as $A_i$ and $\tilde{B}_i$ respectively. \label{prostep:use_device_qkd}
				
				\STATEx
				
				\STATE \textbf{Error correction:} Alice and Bob apply the error correction protocol $\mathrm{EC}$. If $\mathrm{EC}$ aborts they abort the protocol. Otherwise, they obtain raw keys denoted by $K_A$ and $K_B$. \label{prostep:ec}
				\STATE \textbf{Parameter estimation:} Using $\tilde{\mr{B}}$ and $K_B$, Bob sets $W_i = w_{\text{CHSH}}\left({K_B}_i,\tilde{B}_i,X_i,Y_i\right)$ for the test rounds and $W_i = \perp$ otherwise. He aborts if $\sum_{j:T_j=1} W_j < \left(\omega_{\mathrm{exp}}\gamma - \delta_{\mathrm{est}}\right) \cdot n;$. \label{prostep:abort_chsh_qkd}
				\STATE \textbf{Privacy amplification:} Alice and Bob apply the privacy amplification protocol $\mathrm{PA}$ on $K_A$ and $K_B$ to create their final keys $\tilde{K}_A$ and $\tilde{K}_B$ of length $\ell$. \label{prostep:pa}	
			\end{algorithmic}
		\end{algorithm}

		Our main goal in the following sections is to prove the security (according to Definition~\ref{def:security_QKD}) of Protocol~\ref{pro:diqkd_chsh}:
		\begin{thm}\label{thm:QKD_security}
			For any choice of parameters, the DIQKD protocol given in Protocol~\ref{pro:diqkd_chsh} is $(\varepsilon_{\mathrm{QKD}}^s,\varepsilon_{\mathrm{QKD}}^c,)$-secure according to Definition~\ref{def:security_QKD}, with $\varepsilon^s_{\mathrm{QKD}} \leq  2\varepsilon_{\mathrm{EC}} + \varepsilon_{\mathrm{PA}} + \varepsilon_{\mathrm{s}} + \varepsilon_{\mathrm{EA}}$, $\varepsilon^c_{\mathrm{QKD}} \leq \varepsilon^c_{EC} + \varepsilon_{\mathrm{EA}}^c + \varepsilon_{\mathrm{EC}}$, and for key length
			\begin{equation}\label{eq:key_length_def}
			\begin{split}
				\ell = &\; n \cdot \mu_{\mathrm{opt}}\left(\varepsilon_{\mathrm{s}}/4,\varepsilon_{\mathrm{EA}} + \varepsilon_{\mathrm{EC}}\right) - \mathrm{leak_{EC}} \\
				&- 3 \log\left(1-\sqrt{1-(\varepsilon_{\mathrm{s}}/4)^2}\right) - \gamma n  \\
				& - \sqrt{n} 2\log(7)\sqrt{1-2\log \left( \varepsilon_{\mathrm{s}}/4 \cdot \left(\varepsilon_{\mathrm{EA}} + \varepsilon_{\mathrm{EC}}\right) \right)}   -2\log\left(\varepsilon^{-1}_{\mathrm{PA}}\right) \;,
			\end{split}
			\end{equation}
			where $\mu_{\mathrm{opt}}$ is specified in Equation~\eqref{eq:eta_opt}. 
		\end{thm}
		
		The resulting key rates, $\ell/n$, are discussed and plotted in Section~\ref{sec:key_rates} for different choices of parameters.
	
		In the following sections we are set to prove Theorem~\ref{thm:QKD_security}. 
		The theorem follows from the completeness of the protocol, stated as Lemmas~\ref{lem:QKD_complete},  and its soundness, stated as Lemma~\ref{lem:QKD_sound}. 
	
		\subsection{Completeness}
		
			We seek to prove that Protocol~\ref{pro:diqkd_chsh} is complete, i.e., that there exists an honest implementation of the device $D$ that leads to a negligible probability of the protocol aborting. 
			We remark that in order for the protocol to be relevant in practice, completeness has to be proven with respect to a \emph{realistic} honest implementation that can be realised in experiments (or, at the least, believed to be feasible in the future).
			The honest implementation that we consider is the standard one and is described in Section~\ref{sec:honest_qkd_imp}. In short, the honest device makes IID measurements on an IID quantum state $\rho=\sigma^{\otimes n}$. The state and measurements are such that the winning probability achieved in the CHSH game in a single round is $\omega_{\mathrm{exp}}$ that can be chosen freely.\footnote{ For any $\omega_{\mathrm{exp}}$ there are many devices that fit this description; an explicit example can be found in Section~\ref{sec:honest_qkd_imp}.}
			
		
			The following lemma gives the relation between  the probability  $\varepsilon^c_{\mathrm{QKD}}$ that the protocol aborts for an honest implementation of the device $D$ and the other parameters of the protocol.
			\begin{lem}\label{lem:QKD_complete}
				Protocol~\ref{pro:diqkd_chsh} is complete with completeness error 
				\[
					\varepsilon^c_{\mathrm{QKD}} \leq \varepsilon^c_{EC} + \varepsilon_{\mathrm{EC}} + \varepsilon^c_{EA}\;,
				\]
				where 	$\varepsilon^c_{EA} \leq \exp(- 2 n \delta_{\mathrm{est}}^2 )$ and $\varepsilon^c_{EC}$ and $\varepsilon_{\mathrm{EC}}$ are two independent parameters of the error correction protocol. 
			\end{lem}
			
			\begin{proof}
				We wish to upper-bound the probability that Protocol~\ref{pro:diqkd_chsh} aborts when running using the honest implementation. 
				There are two steps in which Alice and Bob can abort Protocol~\ref{pro:diqkd_chsh}: 
				\begin{enumerate}
					\item The protocol may abort after the error correction step (Step~\ref{prostep:ec}). This happens with probability $\varepsilon^c_{EC}$.
					\item Assuming the protocol did not abort in Step~\ref{prostep:ec}, it may abort after the parameter estimation step (Step~\ref{prostep:abort_chsh_qkd}).
						Recall that Bob performs parameter estimation using  $K_B$ and $\tilde{\mr{B}}$, i.e., he checks whether sufficiently many games were won when looking at his data $K_B$ and $\tilde{\mr{B}}$.
						There are two scenarios which lead to the protocol aborting after parameter estimation: 
						\begin{enumerate}
							\item Error correction was successful, i.e., $K_B=K_A$, but not sufficiently many games were won when comparing $K_A$ and $\tilde{\mr{B}}$. 
								When utilising the honest implementation, $W_i$ are IID RVs with $\mathbb{E}\left[ W_i\right] = \omega_{\mathrm{exp}}\gamma$.  Therefore, we can use Hoeffding's inequality to bound the probability of such an event:
							 	\begin{equation}\label{eq:completeness_error_EA}
									\varepsilon^c_{EA} =  \Pr \left[ \sum_{j:T_j\neq \perp} W_j \leq \left(\omega_{\mathrm{exp}}\gamma - \delta_{\mathrm{est}}\right) \cdot n\ \right] \leq \exp(- 2 n \delta_{\mathrm{est}}^2 )  \;.
								\end{equation}
							\item Error correction was not successful, i.e., $K_B\neq K_A$ (but $\mathrm{EC}$ did not abort) and not sufficiently many games were won when comparing at $K_B$ and~$\tilde{\mr{B}}$. This happens with probability at most $\varepsilon_{EC}$.
						\end{enumerate} 
				\end{enumerate}
				The lemma follows by using the above in combination with the union bound. \qedhere
			\end{proof}

		\subsection{Soundness}\label{sec:diqkd_soundness}

			To establish soundness first note that, by definition, as long as Protocol~\ref{pro:diqkd_chsh} does not abort it produces a key of length $\ell$. Therefore it remains to verify correctness (Definition~\ref{def:qkd_correctness}), which depends on the error correction step, and security (Definition~\ref{defn:qkd_secrecy}), which is based on the privacy amplification step. 
			
			To prove security we start by assuming that the error correction step is successful and lower-bound the smooth min-entropy of the quantum state shared between Alice and Bob right before the privacy amplification step. 
			The main ingredient in the proof is the lower-bound on the smooth min-entropy established in Theorem~\ref{thm:main_generation_chsh}. 
			Most effort in proving security is devoted to relating the state considered in the entropy accumulation protocol (to which Theorem~\ref{thm:main_generation_chsh} refers) and the state in the end of the DIQKD protocol.

			To be more precise, let $\overset{\approx}{\Omega}$ denote the event of Protocol~\ref{pro:diqkd_chsh} not aborting \emph{and} the $\mathrm{EC}$ protocol being successful, and let $\tilde{\rho}_{\mr{A}\tilde{\mr{B}}\mr{X}\mr{Y}\mr{T}OE|\overset{\approx}{\Omega}}$ be the state at the end of the protocol,\footnote{$O$ denotes the classical information sent from Alice to Bob during error correction; see Section~\ref{sec:diqkd-protocol}.} conditioned on this event.
			Success of the privacy amplification step relies on the smooth min-entropy $H^{\varepsilon_{\mathrm{s}}}_{\min} ( \mr{A} |\mr{X}\mr{Y}\mr{T} O E )_{\tilde{\rho}_{|\overset{\approx}{\Omega}}}$ being sufficiently large.  Lemma~\ref{lem:smooth_bound_qkd} connects this quantity to 
			 $H^{\frac{\varepsilon_{\mathrm{s}}}{4}}_{\min} ( \mr{A}\mr{B} | \mr{X}\mr{Y}\mr{T}  E )_{\rho_{|\Omega}}$, on which a lower bound is provided by Theorem~\ref{thm:main_generation_chsh}. 
			
			\begin{lem}\label{lem:smooth_bound_qkd}
				For any device $D$, let $\tilde{\rho}$ be the state generated in Protocol~\ref{pro:diqkd_chsh} right before the privacy amplification step, Step~\ref{prostep:pa}. Let $\tilde{\rho}_{|\overset{\approx}{\Omega}}$ be the state conditioned on not aborting the protocol and success of the $\mathrm{EC}$ protocol. Then, for any $\varepsilon_{\mathrm{EA}},\varepsilon_{\mathrm{EC}},\varepsilon_{\mathrm{s}}\in (0,1)$, either the protocol aborts with probability greater than $1-\varepsilon_{\mathrm{EA}} - \varepsilon_{\mathrm{EC}}$ or
				\begin{equation}\label{eq:final_min_entropy_bound_qkd}
				\begin{split}
					H^{\varepsilon_{\mathrm{s}}}_{\min} \left( \mr{A} | \mr{X} \mr{Y} \mr{T} O E \right)_{\tilde{\rho}_{|\overset{\approx}{\Omega}}} \geq n \cdot \mu_{\mathrm{opt}}\left(\varepsilon_{\mathrm{s}}/4,\varepsilon_{\mathrm{EA}} + \varepsilon_{\mathrm{EC}}\right) - \mathrm{leak_{EC}}  \\
					-  3 \log\left(1-\sqrt{1-(\varepsilon_{\mathrm{s}}/4)^2}\right) - \gamma n \\
					 - \sqrt{n} 2\log7\sqrt{1-2\log \left( \varepsilon_{\mathrm{s}}/4 \cdot \left(\varepsilon_{\mathrm{EA}} + \varepsilon_{\mathrm{EC}}\right) \right)} \;.
				\end{split}
				\end{equation}
			\end{lem}
			
			\begin{proof}
			
				Consider the following events:
				\begin{enumerate}
					\item $\Omega$: the event of not aborting in the entropy accumulation protocol, Protocol~\ref{pro:randomness_generation}. This happens when the Bell violation, calculated using Alice and Bob's outputs and inputs, is sufficiently high. 
					\item $\widetilde{\Omega}$: Suppose Alice and Bob run  Protocol~\ref{pro:randomness_generation}, and then execute the $\mathrm{EC}$ protocol. The event $\widetilde{\Omega}$ is defined by $\Omega$ \emph{and} $K_B = \mathbf{A}$.	
					\item $\overset{\approx}{\Omega}$: the event of not aborting the DIQKD protocol, Protocol~\ref{pro:diqkd_chsh}, \emph{and} $K_B = \mathbf{A}$. 
				\end{enumerate}
				The state $\rho_{|\widetilde{\Omega}}$ then denotes the state at the end of Protocol~\ref{pro:randomness_generation} conditioned on $\widetilde{\Omega}$. 
			
				As we are only interested in the case where the $\mathrm{EC}$ protocol outputs the correct guess of Alice's bits, that is $K_B = \mathbf{A}$ (which happens with probability $1-\varepsilon_{\mathrm{EC}}$), we have 
				$\tilde{\rho}_{\mr{A} \mr{X}\mr{Y}\mr{T}  E|\overset{\approx}{\Omega}} = \rho_{\mr{A}  \mr{X}\mr{Y}\mr{T} E|\widetilde{\Omega}}$ (note that $\tilde{\mr{B}}$ and $\mr{B}$ were traced out from $\tilde{\rho}$ and $\rho$ respectively). Hence, 
				\begin{equation}\label{eq:marginals_equiv}
					H^{\varepsilon_{\mathrm{s}}}_{\min} \left( \mr{A} | \mr{X} \mr{Y} \mr{T}  E \right)_{\tilde{\rho}_{|\overset{\approx}{\Omega}}} = H^{\varepsilon_{\mathrm{s}}}_{\min} \left( \mr{A} | \mr{X} \mr{Y} \mr{T}  E \right)_{\rho_{|\widetilde{\Omega}}} \;.
				\end{equation}
			
				Using the chain rule given in~\cite[Lemma 6.8]{tomamichel2015quantum} together with Equation~\eqref{eq:marginals_equiv} we get  that 
				\begin{align}
					H^{\varepsilon_{\mathrm{s}}}_{\min} \left( \mr{A} | \mr{X} \mr{Y} \mr{T} O E \right)_{\tilde{\rho}_{|\overset{\approx}{\Omega}}} &\geq H^{\varepsilon_{\mathrm{s}}}_{\min} \left( \mr{A} | \mr{X} \mr{Y} \mr{T}  E \right)_{\tilde{\rho}_{|\overset{\approx}{\Omega}}} - \mathrm{leak_{EC}} \nonumber \\
					&= H^{\varepsilon_{\mathrm{s}}}_{\min} \left( \mr{A} | \mr{X} \mr{Y} \mr{T}  E \right)_{\rho_{|\widetilde{\Omega}}} - \mathrm{leak_{EC}} \;, \label{eq:entropy_chain_1}
				\end{align}
				where $\mathrm{leak_{EC}}$ denotes the amount of information leaked during error correction.
				
				\sloppy
				To apply Theorem~\ref{thm:main_generation_chsh} it remains to relate $H^{\varepsilon_{\mathrm{s}}}_{\min} \left( \mr{A} | \mr{X} \mr{Y} \mr{T}  E \right)_{\rho_{|\widetilde{\Omega}}}$ to $H^{\varepsilon'_{\mathrm{s}}}_{\min} \left( \mr{AB} | \mr{X} \mr{Y} \mr{T}  E \right)_{\rho_{|\widetilde{\Omega}}}$ for some~$\varepsilon'_{\mathrm{s}}$. For this we first write
				\begin{align*}
					H^{\varepsilon_{\mathrm{s}}}_{\min} \left( \mr{A} | \mr{X} \mr{Y} \mr{T}  E \right)_{\rho_{|\widetilde{\Omega}}} &\geq H^{\frac{\varepsilon_{\mathrm{s}}}{4}}_{\min} \left( \mr{AB} | \mr{X} \mr{Y} \mr{T}  E \right)_{\rho_{|\widetilde{\Omega}}} - H^{\frac{\varepsilon_{\mathrm{s}}}{4}}_{\max} \left( \mr{B} | \mr{A} \mr{X} \mr{Y} \mr{T}  E \right)_{\rho_{|\widetilde{\Omega}}}  \\
					& \quad \quad - 3 \log\left(1-\sqrt{1-(\varepsilon_{\mathrm{s}}/4)^2}\right) \\
					&\geq H^{\frac{\varepsilon_{\mathrm{s}}}{4}}_{\min} \left( \mr{AB} | \mr{X} \mr{Y} \mr{T}  E \right)_{\rho_{|\widetilde{\Omega}}} - H^{\frac{\varepsilon_{\mathrm{s}}}{4}}_{\max} \left( \mathbf{B} | \mathbf{T} E  \right)_{\rho_{|\widetilde{\Omega}}} \\
					 & \quad \quad - 3 \log\left(1-\sqrt{1-(\varepsilon_{\mathrm{s}}/4)^2}\right) \;,
				\end{align*}
				where the first inequality is due to the chain rule~\cite[Equation~(6.57)]{tomamichel2015quantum} and the second is due to strong sub-additivity of the smooth max-entropy.

				One can now apply the EAT to upper bound $H^{\frac{\varepsilon_{\mathrm{s}}}{4}}_{\max} \left( \mr{B} | \mr{ T}  E \right)_{\rho_{|\widetilde{\Omega}}}$ in the following way. We use Theorem~\ref{thm:eat} with the replacements $\mr{O} \rightarrow \mr{B}, \; \mr{S}\rightarrow \mr{T}, \; E \rightarrow E$. The Markov conditions $B_{1,\dotsc, i-1} \leftrightarrow T_{1,\dotsc, i-1} E \leftrightarrow T_i$ then trivially hold and the condition on the max-tradeoff function reads
				\[
					f_{\max}(p) \geq \sup_{\sigma_{R_{i-1}R'}:\mathcal{M}_i(\sigma)_{W_i}=p} H\left(  B_i | T_i R' \right)_{\mathcal{M}_i(\sigma)} \;.
				\]
				By the definition of the EAT channels $\{\mathcal{M}_i\}_{i\in [n]}$, $B_i \neq \perp$ only for $T_i = 1$, which happens with probability~$\gamma$.\footnote{This is why we made the distinction between $B_i$ in the entropy accumulation protocol and $\tilde{B}_i$ in the DIQKD protocol.}
				Hence, for any state $\sigma_{R_{i-1}R'}$ we have, 
				\begin{align*}
					 H\left(  B_i | T_i R' \right)_{\mathcal{M}_i(\sigma)} \leq  H\left(  B_i | T_i \right)_{\mathcal{M}_i(\sigma)} \leq \gamma 
				\end{align*}
				and the max-tradeoff function is simply $f_{\max}(p) = \gamma$ for any $p$ (and thus $\|  \triangledown f_{\max} \|_\infty =0$). Applying\footnote{Here a slightly more general version of the EAT than the one given in Section~\ref{sec:eat_statement} is needed, in which the event $\Omega$ can be defined via $A,B,X,Y$ and not only $C$; see~\cite{dupuis2016entropy} for the details.} Theorem~\ref{thm:eat} with this choice of $f_{\max}$ we get
				\begin{equation}\label{eq:bound_max_entropy}
					H^{\frac{\varepsilon_{\mathrm{s}}}{4}}_{\max} \left( \mr{B} | \mr{ T}  E \right)_{\rho_{|\widetilde{\Omega}}} < \gamma n + \sqrt{n} 2\log7\sqrt{1-2\log \left( \varepsilon_{\mathrm{s}}/4 \cdot \left(\varepsilon_{\mathrm{EA}} + \varepsilon_{\mathrm{EC}}\right) \right)} \;.
				\end{equation}

				Combing the equations above we get that
				\begin{equation*}
				\begin{split}
					H^{\varepsilon_{\mathrm{s}}}_{\min} \left( \mr{A} | \mr{X} \mr{Y} \mr{T} O E \right)_{\tilde{\rho}_{|\overset{\approx}{\Omega}}} \geq H^{\frac{\varepsilon_{\mathrm{s}}}{4}}_{\min} \left( \mr{AB} | \mr{X} \mr{Y} \mr{T}  E \right)_{\rho_{|\widetilde{\Omega}}} - \mathrm{leak_{EC}} \\
					 - 3 \log\left(1-\sqrt{1-(\varepsilon_{\mathrm{s}}/4)^2}\right)   - \gamma n\\
					  - \sqrt{n} 2\log7\sqrt{1-2\log \left( \varepsilon_{\mathrm{s}}/4 \cdot \left(\varepsilon_{\mathrm{EA}} + \varepsilon_{\mathrm{EC}}\right) \right)}   \;.
				\end{split}
				\end{equation*}
				
				Finally, note that by applying the EAT on $\rho_{|\widetilde{\Omega}}$, as in Theorem~\ref{thm:main_generation_chsh}, we have that either $1-\Pr(\widetilde{\Omega})\geq 1-\varepsilon_{\mathrm{EA}} - \varepsilon_{\mathrm{EC}}$, or 
				\begin{equation*}
					H^{\frac{\varepsilon_{\mathrm{s}}}{4}}_{\min} ( \mr{AB} | \mr{X} \mr{Y} \mr{T}  E )_{\rho_{|\widetilde{\Omega}}} > n \cdot \mu_{\mathrm{opt}}\left(\varepsilon_{\mathrm{s}}/4,\varepsilon_{\mathrm{EA}} + \varepsilon_{\mathrm{EC}}\right) \;.
				\end{equation*}
				
				The last two equations together give us the desired bound on $H^{\varepsilon_{\mathrm{s}}}_{\min} \left( \mr{A} | \mr{X} \mr{Y} \mr{T} O E \right)_{\tilde{\rho}_{|\overset{\approx}{\Omega}}}$: either the protocol aborts with probability greater than $1-\varepsilon_{\mathrm{EA}} - \varepsilon_{\mathrm{EC}}$ or 
				\begin{equation*}
				\begin{split}
					H^{\varepsilon_{\mathrm{s}}}_{\min} \left( \mr{A} | \mr{X} \mr{Y} \mr{T} O E \right)_{\tilde{\rho}_{|\overset{\approx}{\Omega}}} \geq n \cdot \mu_{\mathrm{opt}}\left(\varepsilon_{\mathrm{s}}/4,\varepsilon_{\mathrm{EA}} + \varepsilon_{\mathrm{EC}}\right) - \mathrm{leak_{EC}}  \\
					-  3 \log\left(1-\sqrt{1-(\varepsilon_{\mathrm{s}}/4)^2}\right) - \gamma n  \\
					- \sqrt{n} 2\log7\sqrt{1-2\log \left( \varepsilon_{\mathrm{s}}/4 \cdot \left(\varepsilon_{\mathrm{EA}} + \varepsilon_{\mathrm{EC}}\right) \right)} \;. \qedhere
				\end{split}
				\end{equation*}
			\end{proof}
			
			Using Lemma~\ref{lem:smooth_bound_qkd}, we prove that  Protocol~\ref{pro:diqkd_chsh} is sound.
			
			\begin{lem}\label{lem:QKD_sound}
				For any device $D$ let $\tilde{\rho}$ be the state generated using Protocol~\ref{pro:diqkd_chsh}. Then either the protocol aborts with probability greater than $ 1 - \varepsilon_{EA} - \varepsilon_{EC}$ or it is $(\varepsilon_{\mathrm{EC}} + \varepsilon_{\mathrm{PA}} + \varepsilon_{\mathrm{s}})$-correct-and-secret while producing keys of length $\ell$, as defined in Equation~\eqref{eq:key_length_def}.
			\end{lem}
			
			\begin{proof}
				Denote all the classical public communication during the protocol by $J=\mr{X} \mr{Y} \mr{T} O S$ where $S$ is the seed used in the privacy amplification protocol $\mathrm{PA}$. Denote the final state of Alice, Bob, and Eve at the end of Protocol~\ref{pro:diqkd_chsh}, \emph{conditioned on not aborting}, by $\tilde{\rho}_{\tilde{K}_A \tilde{K}_B JE|\dot{\Omega}}$.
				
				We consider two cases. First assume that the $\mathrm{EC}$ protocol was not successful (but did not abort). Then Alice and Bob's final keys might not be identical. This happens with probability at most~$\varepsilon_{\mathrm{EC}}$. 
				
				Otherwise, assume the $\mathrm{EC}$ protocol was successful, i.e., $K_B = \mr{A}$. In that case, Alice and Bob's keys must be identical also after the final privacy amplification step. That is, conditioned on $K_B = \mr{A}$, $\tilde{K}_A=\tilde{K}_B$.
					
				We continue to show that in this case the key is also secret. The secrecy depends only on the privacy amplification step, and for universal hashing a secure key is produced as long as
				\begin{equation*}
							\ell = H^{\varepsilon_{\mathrm{s}}}_{\min} ( \mr{A} | \mr{X} \mr{Y} \mr{T} O E ) -2\log\frac{1}{\varepsilon_{\mathrm{PA}}} 
				\end{equation*} 
				holds (recall Section~\ref{sec:diqkd-protocol}). 
				Hence, a uniform and independent key of length $\ell$ as in Equation~\eqref{eq:key_length_def} is produced by the privacy amplification step unless the smooth min-entropy is not high enough (i.e., the bound in Equation~\eqref{eq:final_min_entropy_bound_qkd} does not hold)
				or the privacy amplification protocol was not successful, which happens with probability at most $\varepsilon_{\mathrm{PA}} + \varepsilon_{\mathrm{s}}$.
					
				According to Lemma~\ref{lem:smooth_bound_qkd}, either the protocol aborts with probability greater than $1-\varepsilon_{\mathrm{EA}}-\varepsilon_{\mathrm{EC}}$, or the entropy is sufficiently high for us to have (recall Definition~\ref{defn:quant_proof_extr})
				\begin{equation*}
					\| \tilde{\rho}_{\tilde{K}_A JE|\dot{\Omega}} - \rho_{U_l} \otimes \tilde{\rho}_{JE} \|_1 \leq \varepsilon_{\mathrm{PA}} + \varepsilon_{\mathrm{s}}  \;.
				\end{equation*} 
				
				Combining both cases above the lemma follows. \qedhere

			\end{proof}

		\subsection{Key rate analysis}\label{sec:key_rates}

			Theorem~\ref{thm:QKD_security} establishes a relation between the length $\ell$ of the secure key produced by our protocol and the different error terms. As this relation, given in Equation~\eqref{eq:key_length_def}, is somewhat hard to visualise, we analyse the key rate $r=\ell/n$ for some specific choices of parameters and compare it to the key rates achieved in device-dependent QKD with finite resources~\cite{scarani2008quantum,scarani2008security} and DIQKD with infinite resources and a restricted set of attacks~\cite{pironio2009device}. 
			
			The key rate depends on the amount of leakage of information due to the error correction step, which in turn depends on the honest implementation of the protocol (recall Section~\ref{sec:diqkd-protocol}). We use the honest IID implementation described in Section~\ref{sec:honest_qkd_imp} and choose the honest state of each round to be the two-qubit Werner state $\rho_{Q_AQ_B} = (1-\nu) \ket{\phi^+}\bra{\phi^+} + \nu\mathbb{I}/4$ (and the measurements are as described in Section~\ref{sec:honest_qkd_imp}). 
			The quantum bit error rate is then $Q=\frac{\nu}{2}$ and the expected winning probability is $\omega_{\mathrm{exp}}=\frac{2+\sqrt{2}(1-2Q)}{4}$. 
			
			We emphasise that this is only a choice of the \emph{honest} implementation and it does not in any way restrict the actions of the adversary (and, in particular, the types of imperfections in the device). Furthermore, the analysis done below can be adapted to any other honest implementation of interest.

			\subsubsection{Leakage due to error correction}\label{sec:leakage_ec_calc}
			
				To calculate the rates we first need to explicitly upper bound the leakage of information due to the error correction protocol, $\mathrm{leak_{EC}}\;$. As shown in Equation~\eqref{eq:ec_leakage}, this can be done by evaluating $H_{0}^{\varepsilon'_{\mathrm{EC}}}(\mr{A}|\tilde{\mr{B}}\mr{X}\mr{Y}\mr{T})$ on Alice and Bob's state in an honest IID implementation of the protocol, described in Section~\ref{sec:honest_qkd_imp}. 
				
				\sloppy
				For this we first use the following relation between $H_{0}^{\varepsilon}$ and $H_{\max}^{\varepsilon'}$~\cite[Lemma 18]{tomamichel2011leftover}:
				\[
					H_{0}^{\varepsilon'_{\mathrm{EC}}}(\mr{A}|\tilde{\mr{B}}\mr{X}\mr{Y}\mr{T}) \leq H_{\max}^{\frac{\varepsilon'_{\mathrm{EC}}}{2}}\left(\mr{A}|\tilde{\mr{B}}\mr{X}\mr{Y}\mr{T}\right) + \log \left( 8/\varepsilon'^2_{\mathrm{EC}} + 2/\left(2-\varepsilon'_{\mathrm{EC}}\right)\right) \;.
				\]
				
				The quantum asymptotic equipartition property, given as Theorem~\ref{thm:quant_aep}, tells us that 
				\[
					H_{\max}^{\frac{\varepsilon'_{\mathrm{EC}}}{2}}\left(\mr{A}|\tilde{\mr{B}}\mr{X}\mr{Y}\mr{T}\right) \leq n H(A_i|\tilde{B}_iX_i Y_i T_i) + \sqrt{n} \delta (\varepsilon'_{\mathrm{EC}}, \tau) \;,
				\]
				for $\tau =2 \sqrt{2^{H_{\max}(A_i|\tilde{B}_iX_i Y_i T_i)}}+1$ and $\delta (\varepsilon'_{\mathrm{EC}}, \tau) = 4\log \tau \sqrt{2 \log \left(8/ {\varepsilon'}^2_{\mathrm{EC}} \right)}$. 
				
				For the honest implementation of Protocol~\ref{pro:diqkd_chsh} $H_{\max}(A_i|\tilde{B}_iX_i Y_i T_i)=1$  and 
				\begin{align*}
					H(A_i|\tilde{B}_iX_i Y_i T_i) = &\Pr(T_i=0) \cdot H(A_i|\tilde{B}_i X_i Y_i, T_i=0) + \\
					&\Pr(T_i=1) \cdot H(A_i|\tilde{B}_iX_i Y_i ,T_i=1) \\
					=&\left( 1-\gamma \right) \cdot H(A_i|\tilde{B}_i X_i Y_i, T_i=0) + \\
					&\gamma \cdot H(A_i|\tilde{B}_iX_i Y_i ,T_i=1) \\
					=& \left( 1-\gamma \right)  h(Q) + \gamma h(\omega_{\mathrm{exp}}) \;,
				\end{align*}
				where the first equality follows from the definition of conditional entropy and the second from the way $T_i$ is chosen in Protocol~\ref{pro:diqkd_chsh}. 
				The last equality holds since for the generation rounds the error rate (i.e., the probability that $A_i$ and $\tilde{B}_i$ differ) in the honest case is $Q$ and for the test rounds  Bob can guess $A_i$  with probability $\omega_{\mathrm{exp}}$ given $\tilde{B}_i$, $X_i$, and $Y_i$. 
				
				We thus have 
				\[
				\begin{split}
					H_{0}^{\varepsilon'_{\mathrm{EC}}}\left(\mr{A}|\tilde{\mr{B}}\mr{X}\mr{Y}\mr{T}\right) \leq & \; n \left[\left( 1-\gamma \right)  h(Q) + \gamma h(\omega_{\mathrm{exp}}) \right] \\
					& + \sqrt{n} 4\log \left(2\sqrt{2} +1\right) \sqrt{2 \log \left(8/ {\varepsilon'}^2_{\mathrm{EC}} \right)}  \\
					& + \log \left( 8/\varepsilon'^2_{\mathrm{EC}} + 2/\left(2-\varepsilon'_{\mathrm{EC}}\right)\right) \;.
					\end{split}
				\]
				Plugging this into Equation~\eqref{eq:ec_leakage} we get 
				\begin{equation}\label{eq:explicit_leak_ec}
					\begin{split}
						\mathrm{leak_{EC}} \leq & \; n \left[\left( 1-\gamma \right)  h(Q) + \gamma h(\omega_{\mathrm{exp}}) \right] \\
						& + \sqrt{n} 4\log \left(2\sqrt{2} +1\right) \sqrt{2 \log \left(8/ {\varepsilon'}^2_{\mathrm{EC}} \right)}\\
						& + \log \left( 8/\varepsilon'^2_{\mathrm{EC}} + 2/\left(2-\varepsilon'_{\mathrm{EC}}\right)\right) + \log\left(\frac{1}{\varepsilon_{\mathrm{EC}}}\right) \;.
					\end{split}
				\end{equation}

			\subsubsection{Key rate curves}\label{sec:qkd_curves}

				In Appendix~\ref{sec:better_rate} a slightly modified protocol is considered in which, instead of fixing the number of rounds in the protocol, only the expected number of rounds is fixed. The completeness and soundness proofs follow the same lines as the proofs above, as detailed in Appendix~\ref{sec:better_rate} and do not include additional crucial insights. 
				The modification of the protocol improves the dependency of the key rate on the probability of a test round $\gamma$\footnote{The second order term of the smooth min-entropy rate given in Lemma~\ref{lem:smooth_bound_qkd} scales with $\gamma$, roughly, as $1/\gamma$, while in Appendix~\ref{sec:better_rate} the dependency is roughly $1/\sqrt{\gamma}$. The modified analysis can be seen as a ``patch'' used to overcome the non-optimal dependency of the EAT given in Theorem~\ref{thm:eat} on the testing probability in the considered protocols. This issue was overcome in a more recent version of the EAT~\cite{dupuis2018entropy}.} 
				The analysis presented in the appendix leads to the key rates for the modified protocol and these are the rates presented here. 
				Putting the technical details aside, the reader may simply think of $\bar{n}$ below as taking the place of the number of rounds $n$ used so far.
				
				In an asymptotic analysis ($\bar{n}\rightarrow\infty$) it is well understood that the soundness and completeness errors $\varepsilon_{\mathrm{QKD}}^s,\varepsilon_{\mathrm{QKD}}^c$ should tend to zero as $\bar{n}$ increases. However, in the non-asymptotic scenario considered here these errors are always finite. We therefore fix some values for them which are considered to be realistic and relevant for actual applications. We choose the parameters such that the security parameters are at least as good (and in general even better) as in~\cite{scarani2008quantum}, such that a fair comparison can be made. All other parameters are chosen in a consistent way while (roughly) optimising the key rate.
				
				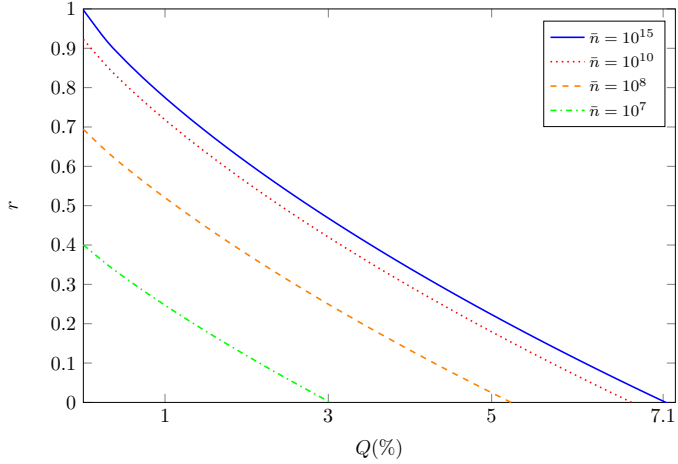
\begin{figure}
				\centering
				\begin{tikzpicture}[scale=0.75]
					\begin{axis}[
						height=8.5cm,
						width=12cm,
						xlabel=$Q(\%)$,
						ylabel=$r$,
						xmin=0,
						xmax=0.145,
						ymax=1,
						ymin=0,
					     xtick={0.02,0.06,0.1,0.142},
					     xticklabels={$1$, $3$, $5$, $7.1$},
				          ytick={0,0.1,0.2,0.3,0.4,0.5,0.6,0.7,0.8,0.9,1},
						legend style={at={(0.88,0.97)},anchor=north,legend cell align=left,font=\footnotesize} 
					]
					
				
					\addplot[blue,thick,smooth] coordinates {
						(1.*10^-10, 0.997348) (0.00584, 0.91723) (0.01168, 0.854049) (0.01752, 0.797357) (0.02336, 0.744869) (0.0292, 0.695502) (0.03504, 0.648617) (0.04088, 0.603795) (0.04672, 0.560736) (0.05256, 0.519218) (0.0584, 0.479068) (0.06424, 0.44015) (0.07008, 0.402351) (0.07592, 0.365578) (0.08176, 0.329752) (0.0876, 0.294806) (0.09344, 0.260681) (0.09928, 0.227327) (0.10512, 0.194699) (0.11096, 0.162758) (0.1168, 0.131467) (0.12264, 0.100796) (0.12848, 0.0707151) (0.13432, 0.0411983) (0.14016, 0.0122219) (0.146, -0.0162357)	
					};
					\addlegendentry{$\bar{n}=10^{15}$}
					
					\addplot[red,thick,smooth,dotted] coordinates {
					(1.*10^-10, 0.922753) (0.00584, 0.85326) (0.01168, 0.794151) (0.01752, 0.740126) (0.02336, 0.689641) (0.0292, 0.641886) (0.03504, 0.596356) (0.04088, 0.552704) (0.04672, 0.510679) (0.05256, 0.470087) (0.0584, 0.430779) (0.06424, 0.392632) (0.07008, 0.355544) (0.07592, 0.319433) (0.08176, 0.284226) (0.0876, 0.249862) (0.09344, 0.216287) (0.09928, 0.183453) (0.10512, 0.15132) (0.11096, 0.11985) (0.1168, 0.0890098) (0.12264, 0.0587699) (0.12848, 0.0291029) (0.13432, -0.0000156698) 
					};
					\addlegendentry{$\bar{n}=10^{10}$}
				
					\addplot[orange,thick,smooth,dashed] coordinates {
					(1.*10^-10, 0.694276) (0.00584, 0.636597) (0.01168, 0.586384) (0.01752, 0.539418) (0.02336, 0.494819) (0.0292, 0.452125) (0.03504, 0.411042) (0.04088, 0.371362) (0.04672, 0.332928) (0.05256, 0.295616) (0.0584, 0.259328) (0.06424, 0.22398) (0.07008, 0.189504) (0.07592, 0.15584) (0.08176, 0.122936) (0.0876, 0.090747) (0.09344, 0.0592343) (0.09928, 0.028362) (0.10512, -0.00190154) 
					};
					\addlegendentry{$\bar{n}=10^{8}$}
					
					\addplot[green,thick,smooth,dashdotted] coordinates {
					(1.*10^-10, 0.400565) (0.00584, 0.350245) (0.01168, 0.305868) (0.01752, 0.264116) (0.02336, 0.224246) (0.0292, 0.186061) (0.03504, 0.149213) (0.04088, 0.113476) (0.04672, 0.078735) (0.05256, 0.0448971) (0.0584, 0.011889) (0.06424, -0.02035) 
					};
					\addlegendentry{$\bar{n}=10^{7}$}

					\end{axis}  
				\end{tikzpicture}
				
				\caption{The expected key rate $r=\ell/\bar{n}$ as a function of the quantum bit error rate $Q$ for several values of the expected number of rounds $\bar{n}$ (see the main text and Appendix~\ref{sec:better_rate}). For $\bar{n}=10^{15}$ the curve essentially coincides with the curve for the IID asymptotic case~\cite[Equation~(12)]{pironio2009device}. 
					The following values for the error terms were chosen: $\varepsilon_{\mathrm{EC}}=10^{-10}, \; \varepsilon_{\mathrm{QKD}}^s=10^{-5}$ and $\varepsilon_{\mathrm{QKD}}^c=10^{-2}$.}
				\label{fig:qkd_rates_Q_mod}
				\end{figure}

				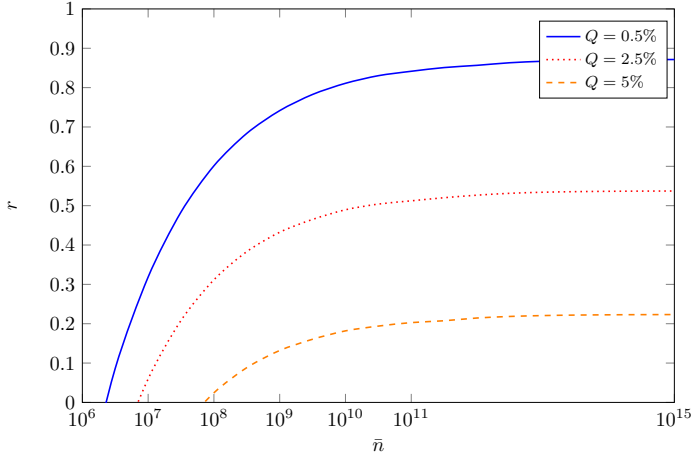
\begin{figure}
				\centering
				\begin{tikzpicture}[scale=0.75]
					\begin{axis}[
						height=8.5cm,
						width=12cm,
						xlabel=$\bar{n}$,
						ylabel=$r$,
						xmin=6,
						xmax=15,
						ymax=1,
						ymin=0,
					     xtick={6,7,8,9,10,11,15},
					     xticklabels={ $10^6$, $10^7$, $10^8$, $10^9$,$10^{10}$,$10^{11}$,$10^{15}$},
				          ytick={0,0.1,0.2,0.3,0.4,0.5,0.6,0.7,0.8,0.9,1},
						legend style={at={(0.88,0.97)},anchor=north,legend cell align=left,font=\footnotesize} 
					]
					
				
					\addplot[blue,thick,smooth] coordinates {
					(6, -0.2442) (13/2, 0.0866932) (7, 0.318636) (15/2, 0.483919) (8, 0.601281) (17/2, 0.683763) (9, 0.741528) (19/2, 0.782662) (10, 0.810984) (21/2, 0.830336) (11, 0.841544) (23/2, 0.850964) (12, 0.856629) (25/2, 0.863243) (13, 0.867006) (27/2, 0.869136) (14, 0.870339) (29/2, 0.871017) (15, 0.871399)
					};
					\addlegendentry{$Q = 0.5\%$}
					
					\addplot[red,thick,smooth,dotted] coordinates {
					(6, -0.459711) (13/2, -0.152595) (7, 0.059624) (15/2, 0.208344) (8, 0.311673) (17/2, 0.383259) (9, 0.432467) (19/2, 0.465061) (10, 0.489571) (21/2, 0.504051) (11, 0.512275) (23/2, 0.520235) (12, 0.526662) (25/2, 0.531429) (13, 0.53412) (27/2, 0.535636) (14, 0.536489) (29/2, 0.536969) (15, 0.53724)
					};
					\addlegendentry{$Q = 2.5\%$}
				
					\addplot[orange,thick,smooth,dashed] coordinates {
					(6, -0.683142) (13/2, -0.398978) (7, -0.203553) (15/2, -0.0685348) (8, 0.02432) (17/2, 0.0884077) (9, 0.131977) (19/2, 0.160513) (10, 0.181837) (21/2, 0.193957) (11, 0.202814) (23/2, 0.207513) (12, 0.214511) (25/2, 0.218461) (13, 0.220688) (27/2, 0.221941) (14, 0.222646) (29/2, 0.223043) (15, 0.223266)
					};
					\addlegendentry{$Q = 5\%$}

					\end{axis}  
				\end{tikzpicture}
				\caption{The expected key rate $r=\ell/\bar{n}$ as a function of the expected number of rounds $\bar{n}$ (see the main text and Appendix~\ref{sec:better_rate}) for several values of the quantum bit error rate $Q$. For $Q=0.5\%$, $2.5\%$, and $5\%$ the achieved key rates are approximatly $r=87\%$, $53\%$, and $22\%$ respectively.
					The following values for the error terms were chosen: $\varepsilon_{\mathrm{EC}}=10^{-10}, \; \varepsilon_{\mathrm{QKD}}^s=10^{-5}$ and $\varepsilon_{\mathrm{QKD}}^c=10^{-2}$.}
				\label{fig:qkd_rates_n_mod}
				\end{figure}

				In Figure~\ref{fig:qkd_rates_Q_mod} the expected key rate $r=\ell/\bar{n}$ is plotted as a function of the quantum bit error rate $Q$ for several values of  the expected number of rounds $\bar{n}$. 
				 For $\bar{n}=10^{15}$ the curve essentially coincides with the rate achieved in the asymptotic IID case~\cite{pironio2009device}. Since the latter was shown to be optimal~\cite{pironio2009device} it provides an upper bound on the key rate and the amount of tolerable noise. 
				Hence, for large enough $\bar{n}$ our rates become optimal and the protocol can tolerate up to the maximal error rate $Q=7.1\%$. 
				For comparison, the previously established explicit rates~\cite{vazirani2014fully} are well below the lowest curve presented in Figure~\ref{fig:qkd_rates_Q_mod}, even when the number of signals goes to infinity, with a maximal noise tolerance of 1.6\%.

				In Figure~\ref{fig:qkd_rates_n_mod}, $r$ is plotted as a function of $\bar{n}$ for several values of $Q$. As can be seen from the figure, the achieved rates are significantly higher than those achieved in previous works. Moreover, they are practically comparable to the key rates achieved in device-\emph{dependent} QKD (see Figure~1 in~\cite{scarani2008quantum}). The main difference between the curves for the device-dependent case  and the independent one is the minimal value of $\bar{n}$ which is required for a positive key rate. (That is, for the protocols considered in~\cite{scarani2008quantum} one can get a positive key rate with less rounds.)

	\section{Open questions}\label{sec:diqkd_open_ques}
		
		To end the chapter, we list some future work directions and open questions specific for the showcase of quantum cryptography.
		
		\subsubsection{Experimental realistions}
		
			The results presented in this chapter provide the theoretical groundwork for experimental implementations of device-independent cryptographic protocols.
			The quantitive results imply that the first proof of principle experiments, with small distances and small rates, are within reach with today's state-of-the-art technology, which  recently enabled the violation of Bell inequalities in a loophole-free way~\cite{hensen2015loophole,shalm2015strong,giustina2015significant} (a necessity for device-independent cryptography).
			Indeed,  Theorem~\ref{thm:main_generation_chsh} has already been applied to the analysis of the first experimental implementation of a protocol for randomness generation in the fully device-independent framework~\cite{liu2017high}.
			The next major challenge in experimental implementations is a field demonstration of a DIQKD protocol. This would provide the strongest cryptographic experiment ever realised. 
			
			As can be seen from Figures~\ref{fig:eta_rates},~\ref{fig:qkd_rates_Q_mod}, and~\ref{fig:qkd_rates_n_mod}, implementing a DIQKD protocol is more challenging than implementing a randomness generation protocol~--- positive key rates require higher number of signals and lower noise levels. 
			It therefore becomes increasingly relevant to achieve the best possible dependence of the rate curves on the number of rounds $\bar{n}$, even for very small values of $\bar{n}$. 
			As can be seen from the figures our rate curves approach (and essentially coincide) with the optimal curves as the number of rounds increases. This is the case since our first-order term of the key rate is tight. 
						
			However, one thing that can perhaps still be further optimised is the dependency on the number of rounds, or in other words, how fast the curves approach the optimal curve. Although this seems like a minor issue, it can make actual implementations more feasible. 
			The explicit dependency on $\bar{n}$ given in Equation~\eqref{eq:key_length_def} is already close to optimal. Still:
			\begin{enumerate}
				\item The numerical analysis used to plot the curves can be made somewhat better for the range of $\bar{n}=10^4-10^6$. 
				\item Throughout the proof we made certain non-optimal choices that almost surely effect the second order term. For example, the use of the chain rules for the smooth entropies can be done in a tighter way (compare the general form of the chain rules~\cite[Equation (6.57)]{tomamichel2015quantum} to the way it was applied here). A tighter analysis of similar steps was used in~\cite{tomamichel2017largely} for the case of device-\emph{dependent} QKD protocols and may be of use here as well. 
			\end{enumerate}

			
		\subsubsection{Possible extensions}
			
			The optimality of our key rates is only with respect to the structure of the considered protocol, which is the standard (and only, as far as we are aware) DIQKD protocol studied in the literature. 
			It is interesting to come up with new DIQKD protocols and see if they lead to key rates with higher first-order terms.  
			Apart from the theoretical curiosity, protocols with better asymptotic key rates can, of course, help us reach an experimental implementation.

			Due to the modularity of our analysis, it can at large be directly applied to the analysis of other protocols.  The main challenge is to come up with interesting protocols. 
			We discuss two possible directions to consider.
			
			On the ``quantum side'' of the protocol, one may modify the protocol by considering different Bell inequalities.
			Even more, one can construct protocols in which more information than the violation of a single inequality is used: Alice and Bob may use the collected statistics to evaluate several quantities and decide accordingly whether to abort or not; see for example the related work~\cite{nieto2018device}. 
			To apply our proof to other Bell inequalities and additional statistical information one should find a good bound on the min-tradeoff function, as done in Equation~\eqref{eq:one_box_entropy_final} for the CHSH inequality. 
			For many Bell inequalities such bounds are known, but for the min-entropy instead of the von Neumann entropy. In most cases using a bound on the min-entropy will result in far from optimal rate curves. Therefore, to adapt our protocol in this direction one should probably first bound the min-tradeoff function using the von Neumann entropy directly.\footnote{This should not be dismissed as can be seen from the following state of affairs. \cite{nieto2018device} reports an advantage in terms of the \emph{min-entropy} when considering the full statistics instead of merely the violation if the CHSH inequality. 
			Comparing the bound on the min-entropy from the full statistics to the bound on the von Neumann entropy from the violation alone, both evaluated on the quantum states produced by the honest implementations, we find that it is still better to use the bound on the von Neumann entropy as we do here. 
			Thus, to truly see if an advantage can be gained by considering the full statistics, one should aim to a direct bound on the von Neumann entropy.}

			On the ``classical side'' of the protocol, different classical post-processing steps can be considered. 
			It is known that, asymptotically, considering protocols with other one-way classical post-processing cannot lead to an improvement over our protocol~\cite{devetak2005distillation}. 
			Hence, the interesting to check is whether there are protocols with \emph{two-way} classical post-processing protocols that lead to an improvement of the first-order term of the key rates.

		\subsubsection{Bounding the von Neumann entropy}
		
			As mentioned above, the main task to perform when modifying the protocol and the analysis is to lower-bound the min-tradeoff function. Getting a tight bound on the von Neumann entropy, and hence on the min-tradeoff function, does not seem to be an easy task. 
			Is there a numerical technique that allows one to get good bounds on the von Neumann entropy for general Bell inequalities? 
			
			Interestingly, as it turns out, good numerical tools are known for a couple of similar quantities:
			\begin{enumerate}
				\item When considering the min-entropy instead of the von Neumann entropy one can use SDP hierarchies to get (not necessarily tight) lower-bounds~\cite{navascues2008convergent}. 
				\item In the device-\emph{dependent} case, a recent development~\cite{winick2017reliable} presents a numerical technique to lower-bound the von Neumann entropy and by this derive better key rates for QKD. 
			\end{enumerate}
			
			We hope that it is possible to devise a general technique (numerical or analytical) to calculate good lower bounds on the von Neumann entropy relevant for our case. Such a technique, in combination with our work, will allow us to ``enumerate'' over all possible protocols and  calculate their key rates when looking for better protocols.
			If it is not possible to devise a general tool, it will at least be interesting to understand why this is the case. Is there a fundamental mathematical reason behind the complexity of the problem?

\chapter{Outlook}\label{ch:outlook}


	The development and application of the concept of reductions to IID, taking the form of de Finetti theorems, flourished in ``standard'' quantum information processing in the last decade and more. 
	The tools used, unfortunately, were not applicable when considering device-independent information processing tasks, where the devices being analysed are uncharacterised. 
	The reductions presented in the thesis, namely the de Finetti reduction (Chapter~\ref{ch:reductions_par}) and the entropy accumulation theorem (Chapter~\ref{ch:reductions_seq}), are the first to be applicable in the device-independent setting. As such, they have opened the possibility of a significantly simpler analysis of device-independent information processing tasks.
		
	Among the advantages of applying the approach of reductions to IID in the device-independent setting, compared to directly analysing the most general case, are tighter quantitive results and modular proofs. 
	The thesis' showcases, used to exemplify the usage of the reductions, indeed report such benefits. 
	Our proof of non-signalling parallel repetition (Chapter~\ref{ch:par_rep_showcase}) is automatically valid for any complete-support game with any number of players and achieves an exponential decrease that matches that of IID strategies. 
	Our security proof for device-independent quantum key distribution (Chapter~\ref{ch:crypto_showcase}) achieves tight key rates, as under the IID assumption, that are significantly better than all prior results and can be easily adapted to other related protocols.

	With this in mind, it is interesting to investigate how the presented reductions or variants thereof can be used in the analysis of other tasks. Let us discuss a partial list of questions and possible future work that we find intriguing.\footnote{We list here questions that are not directly related to the showcases considered in the thesis. For concrete open questions regarding parallel repetition (e.g., extensions of the results) and device-independent quantum key distribution (such as possible improvements and experimental implementations) see Sections~\ref{sec:pr_open_ques} and~\ref{sec:diqkd_open_ques}, respectively.}

	\subsubsection{Two-party device-independent quantum cryptography}
	
		In the cryptographic protocols discussed in the thesis we considered two honest and cooperating parties, Alice and Bob. Two-party cryptography, on the other hand, refers to cryptographic protocols in which Alice and Bob do not trust each other. 
		When considering device-independent two-party cryptography the dishonest party (which can be either Alice or Bob) takes the role of the adversary and hence is allowed to prepare the device used to implement the protocol. 
		~\cite{fu2018local} and~\cite{ribeiro2018device} present examples for such protocols.
		
		The above mentioned works study the security of the protocols under the IID assumption (or a closely related assumption). Clearly, it is interesting to see if the analysis can be extended to capture the most general adversarial scenario, which, in the case of these protocols, includes the use of sequential boxes. Applying a reduction to IID can be beneficial here. 
		Unfortunately, it is not clear whether the entropy accumulation theorem, in its current form, can be of use in such protocols.
		The reason is that the Markov-chain conditions stated in Equation~\eqref{eq:req_markov_cond} do not hold, at least when considering the most obvious choices of random variables.\footnote{In the case of two-party cryptography, the natural choice to make when trying to use the entropy accumulation theorem is one in which the~$\mr{O}$ systems belong to the honest party and the~$\mr{S}$ systems to the dishonest party. One can easily come up with boxes that do not fulfil  Equation~\eqref{eq:req_markov_cond} with these choices.}
		
		In some cases, one can overcome the problem by considering ``imaginary'' protocols, closely related to the ``real'' protocol,  in which the Markov-chain conditions do hold. The idea is then to reduce the problem of proving the security of the real protocol to that of the imaginary one and perform the analysis of the imaginary protocol using the entropy accumulation theorem. 
		
		Such a proof technique is used in~\cite{arnon2017device}. There, the protocol of interest is a device-independent entanglement certification protocol and its analysis requires an upper bound on the smooth max-entropy, rather than a lower bound on the smooth min-entropy as in cryptographic scenarios.\footnote{In the considered scenarios the two quantities are not dual to one another; see~\cite{arnon2017device} for the details.} Thus, the steps used in~\cite{arnon2017device} are not directly applicable to two-party device-independent cryptography. 
		It is interesting to see if similar ideas can be useful in cryptographic scenarios as well. 
		
		Alternatively, one could also try to prove a different variant of the entropy accumulation theorem in which the Markov-chain conditions are replaced by some other restrictions on the sequential process, which are fulfilled by two-party cryptographic protocols. (Finding such conditions is interesting by itself).
		As discussed in Section~\ref{sec:eat_conc_diff}, some conditions on the process must appear in the theorem, since entropy does not accumulate in any sequential process. 
		While the Markov-chain conditions are sufficient, we currently have no reason to believe that they are necessary; it might as well be that some weaker or incomparable conditions also suffice.

	\subsubsection{Parallel device-independent quantum cryptography}
	
		Another type of cryptographic protocols to which the presented reductions to IID are not applicable in a trivial manner are ones in which the most general analysis should be done with quantum parallel boxes. 
		An example is the parallel device-independent quantum key distribution protocol of~\cite{jain2017parallel}, in which all the non-local games are played in parallel with the device (as in the parallel repetition question).
		While~\cite{jain2017parallel} includes a security proof that goes beyond the IID scenario, it achieves quantitively weak results. 
		This raises the fundamental question of whether parallel adversaries, i.e., adversaries that can create parallel boxes, are stronger than sequential and IID adversaries (which are proven to have the same strength by our work). 
		To learn the answer to this question there is a need to supply tight key rates for parallel device-independent quantum key distribution protocols.

		Utilising a reduction to IID instead of analysing the general case directly, as in~\cite{jain2017parallel},  will almost surely lead to stronger, perhaps even tight, results.
		Alas, the known reductions are not directly applicable here.
		The entropy accumulation theorem is not useful in this case since it is restricted to sequential boxes and here one ought to analyse parallel boxes. 
		The de Finetti reduction, while suitable for parallel boxes, is a priori not applicable  here since the de Finetti box does not include the adversary and is not a quantum box; see Section~\ref{sec:dF_imp_res}.
		It is therefore interesting to investigate whether the analysis can somehow be manipulated so that the known techniques can be utilised to prove security of parallel device-independent quantum cryptography or, otherwise, whether other types of reductions, more adequate for such scenarios, can be developed.

	\subsubsection{Device-independent tomography}
	
		One of the applications of the ``original'' quantum de Finetti reduction (also called the post-selection technique)~\cite{christandl2009postselection} is a technique for a reliable quantum  state tomography~\cite{christandl2012reliable}. The technique is said to be reliable since it reports not just an estimation of the quantum state but also a confidence region around the estimated state, which acts as a meaningful ``error bar''. This is of crucial importance as the other more standard approaches, such as the maximum-likelihood optimisation and least-square-error estimation, suffer from systematic errors~\cite{schwemmer2015systematic}. 
		
		Recently, the device-independent equivalents of the maximum-likelihood optimisation and least-square-error estimation were considered in~\cite{lin2018device}. The goal of such device-independent tomographic techniques is to report an estimated quantum box from the observed finite statistics. 
		Apart from systematic errors, device-independent tomographic procedures as above are also at risk of providing a non-quantum box, since up to date it is unknown how to perform optimisation problems over the set of quantum boxes. 
		In analogy to~\cite{christandl2012reliable}, applying our de Finetti reductions to achieve reliable device-independent tomography can therefore be of interest.

%

	

\appendix

\chapter{Additional proofs: de Finetti reductions}

	\section{Bounding the de Finetti box}\label{appsec:proof_dF_red}
		
		We use the notation used in Section~\ref{sec:dF_reductions}:
		\begin{enumerate}
				\item $|\mathcal{X}||\mathcal{Y}| = l$ and we identify each pair $(x,y)\in \mathcal{X}\times \mathcal{Y}$ with a label $j\in[l]$ by writing $(x,y)=j$.
				\item $|\mathcal{A}||\mathcal{B}| = m$ and we identify each pair $(a,b)\in \mathcal{A}\times \mathcal{B}$ with a label $k\in[m]$ by writing $(a,b)=k$.
				\item For all $j\in[l]$  and $k\in[m]$, $p^j_k\in[0,1]$ such that $\sum_k p^j_k =1$ for all $j$. 
				\item For all $j\in[l]$  and $k\in[m]$, $c^j_k=1-\sum_{t<k} p^j_t$.
				\item  For all $\mr{x}$, $\mr{y}$, and $j\in[l]$, $n^j=|\left\{i : (x_i,y_i)=j\right\}|$, i.e., $n^j$ denotes the number of indices  of $(\mr{x},\mr{y})$ in which the type of inputs is $(x,y)=j$.
				\item For all $\mr{x}$, $\mr{y}$, $\mr{a}$, $\mr{b}$, $j\in[l]$, and $k\in[m]$, $n^j_k=|\left\{i : (x_i,y_i)=j \land (a_i,b_i)=k \right\}|$, i.e., $n^j_k$ denotes the number of indices of $(\mr{x},\mr{y},\mr{a},\mr{b})$ in which the type of inputs  is $(x,y)=j$ and the type of outputs is $(a,b)=k$.
			\end{enumerate}	
			and notice that:
			\begin{enumerate}
				\item For all  $j\in[l]$  and $k\in[m-1]$, $p^j_k\in[0,c^j_k]$ and $p^j_m = c^j_m$.
				\item For all  $j\in[l]$  and $k\in[m]$, $c^j_k = c^j_{k-1} - p^j_{k-1}$. 
				\item For all  $j\in[l]$, $n^j_m = n^j - \sum_{k=1}^{m-1} n^j_k$.
			\end{enumerate}

		\begin{customlemma}{\ref{lem:dF_up_bound}}
			For all $\mr{a}$, $\mr{b}$, $\mr{x}$, and $\mr{y}$, 
				\[
					\tau_{\mr{A}\mr{B}|\mr{X}\mr{Y}} (\mr{a}\mr{b}|\mr{x}\mr{y}) \geq \prod_{j=1}^l {n^j \choose {n^j_1, \dots, n^j_m}}^{-1} \frac{1}{(n^j + 1)^{m-1}} \;,
				\]
				where $\tau_{\mr{A}\mr{B}|\mr{X}\mr{Y}}$ is the de Finetti box defined by 
				\begin{equation*}
		\begin{split}
			&\tau_{\mr{A}\mr{B}|\mr{X}\mr{Y}} (\mr{a},\mr{b}|\mr{x},\mr{y}) = \int \O_{AB|XY}^{\otimes n} \mathrm{d}\O_{AB|XY} \\
			&\qquad  = \prod_{j=1}^l 
			\left[ \int_0^{c^j_1} \frac{\mathrm{d}p^j_1}{c^j_1} \left( p^j_1 \right)^{n^j_1} \right]
			\dots
			\left[ \int_0^{c^j_{m-1}} \frac{\mathrm{d}p^j_{m-1}}{c^j_{m-1}} \left( p^j_{m-1} \right)^{n^j_{m-1}} \right] 
			\cdot \left( p^j_m \right)^{n^j - \sum_{k=1}^{m-1} n^j_k} \;.
		\end{split}
			\end{equation*}
		\end{customlemma}
	
		In the proof of the lemma we use the following formula:
		\begin{equation} \label{eq:basic_formula}
			\begin{split}
				\forall c>0 \; \forall n,n'\in \mathbb{N}, n'\leq n\\
				\int_0^{c}  \frac{\mathrm{d}p}{c} \; p^{n'} (c-p)^{n-n'}  \ &= c^n \int_0^1 q^{n'} (1-q)^{(n-n')}  \mathrm{d}q  \\
				&= c^n \mathrm{B}(n-n'+1, n'+1)  \\
				&= c^n {n \choose n'}^{-1} \frac{1}{n+1} 
			\end{split}
		\end{equation}
		where $\mathrm{B}$ is the Beta function. We also need the following identity:
		\begin{equation}\label{eq:multinomial}
			 {{n- \sum_{t=1}^s n_t} \choose {n_{s+1}}} \cdot {n \choose {n_1,\dotsc ,n_s, n-\sum_{t=1}^s n_t}} = {n \choose {n_1, \dotsc , n_{s+1},n-\sum_{t=1}^{s+1} n_t}}
		\end{equation}

		\begin{proof}
		
			Abusing notation we denote below, for $t\in[2,m-1]$,\footnote{This is just a notation and $\prod_{k=1}^{t}$ should not be understood as the product operation. In particular, the order of terms is relevant since the different parameters are not independent of one another.}
			\[
				\left[ \int_0^{c^j_1} \frac{\mathrm{d}p^j_1}{c^j_1} \left( p^j_1 \right)^{n^j_1} \right]
			\dots
			\left[ \int_0^{c^j_{t}} \frac{\mathrm{d}p^j_{m-1}}{c^j_{t}} \left( p^j_{t} \right)^{n^j_{t}} \right] = \prod_{k=1}^{t} \left[ \int_0^{c^j_k} \frac{\mathrm{d}p^j_k}{c^j_k}\left( p^j_k \right)^{n^j_k} \right] \;.
			\]

			We start by proving the following, for all $j\in[l]$, by induction:
			\begin{equation} \label{eq:induction}
				\begin{split}
					\prod_{k=1}^{m-1} \left[ \int_0^{c^j_k} \frac{\mathrm{d}p^j_k}{c^j_k}\left( p^j_k \right)^{n^j_k} \right]   \left(c^j_{m-1}-p^j_{m-1} \right)^{n^j-\sum_{k=1}^{m-1} n^j_k} \geq \\
					 {n^j \choose {n^j_1, \dots n^j_{m-1} ,n^j - \sum_{k=1}^{m-1} n^j_k}}^{-1} \frac{1}{(n^j+1)^{m-1}}
				\end{split}
			\end{equation}
			
			\emph{Base case, $m=2$:} 
			\begin{align*}
				\int_0^{c^j_1}  \frac{\mathrm{d}p^j_1}{c^j_1} \; \left(p^j_1\right)^{n^j_1} \left[(c^j_1-p^j_1)\right]^{n^j-n^j_1} ={n^j \choose n^j_1}^{-1} \frac{1}{n^j+1}
			\end{align*}
			This follows from Equation \eqref{eq:basic_formula} while noting that for the first index we have $c^j_1=1$ by definition. 
			
			\emph{Induction hypothesis for $m-2$:}
			\begin{equation} \label{eq:induction_hypothesis}
				\begin{split}
					\prod_{k=1}^{m-2} \left[ \int_0^{c^j_k} \frac{\mathrm{d}p^j_k}{c^j_k}\left( p^j_k \right)^{n^j_k} \right]  \left(c^j_{m-2}-p^j_{m-2} \right)^{n^j-\sum_{k=1}^{m-2} n^j_k} \geq \\
					  {n^j \choose {n^j_1, \dots n^j_{m-2} ,n^j - \sum_{k=1}^{m-2} n^j_k}}^{-1} \frac{1}{(n^j+1)^{m-2}}
				\end{split}
			\end{equation}

			\emph{Inductive step:}
			\begin{align}
				& \prod_{k=1}^{m-1} \left[ \int_0^{c^j_k} \frac{\mathrm{d}p^j_k}{c^j_k}\left( p^j_k \right)^{n^j_k} \right]   \left(c^j_{m-1}-p^j_{m-1} \right)^{n^j-\sum_{k=1}^{m-1} n^j_k} = \nonumber \\
				& \prod_{k=1}^{m-2} \left[ \int_0^{c^j_k} \frac{\mathrm{d}p^j_k}{c^j_k}\left( p^j_k \right)^{n^j_k} \right] \int_0^{c^j_{m-1}} \frac{\mathrm{d}p^j_{m-1}}{c^j_{m-1}} \; \left(p^j_{m-1}\right)^{n^j_{m-1}}  \left(c^j_{m-1}-p^j_{m-1} \right)^{n^j-\sum_{k=1}^{m-2} n^j_k -n^j_{m-1}} = \label{step_1}\\
				\begin{split}
					& \prod_{k=1}^{m-2} \left[ \int_0^{c^j_k} \frac{\mathrm{d}p^j_k}{c^j_k}\left( p^j_k \right)^{n^j_k} \right] \times \\
					& \quad \qquad \qquad \times \left(c^j_{m-1}\right)^{n^j-\sum_{k=1}^{m-2} n^j_k } { n^j-\sum_{k=1}^{m-2} n^j_k  \choose n^j_{m-1}}^{-1} \frac{1}{n^j-\sum_{k=1}^{m-2} n^j_k +1} =  
				\end{split} \label{step_2} \\
				\begin{split}
					& { n^j-\sum_{k=1}^{m-2} n^j_k  \choose n^j_{m-1}}^{-1} \frac{1}{n^j-\sum_{k=1}^{m-2} n^j_k + 1}  \times \\
					&  \qquad \quad \times \prod_{k=1}^{m-2} \left[ \int_0^{c^j_k} \frac{\mathrm{d}p^j_k}{c^j_k}\left( p^j_k \right)^{n^j_k} \right]  \left(c^j_{m-2}-p^j_{m-2} \right)^{n^j-\sum_{k=1}^{m-2} n^j_k} \geq 
				\end{split} \label{step_4} \\
				\begin{split}
					& { n^j-\sum_{k=1}^{m-2} n^j_k  \choose n^j_{m-1}}^{-1} \frac{1}{n^j-\sum_{k=1}^{m-2} n^j_k + 1}  \times \\
					& \qquad \qquad \qquad \qquad \times {n^j \choose {n^j_1, \dots n^j_{m-2} ,n^j - \sum_{k=1}^{m-2} n^j_k}}^{-1} \frac{1}{(n^j+1)^{m-2}} \geq 
				\end{split} \label{step_5}\\
				& {n^j \choose {n^j_1, \dots n^j_{m}}}^{-1} \frac{1}{(n^j+1)^{m-1}} \nonumber \;.
			\end{align}
			where we used Equation \eqref{eq:basic_formula} to get from \eqref{step_1} to \eqref{step_2}, 
			$c^j_{m-1}=c^j_{m-2}-p^j_{m-2}$ to get from \eqref{step_2} to \eqref{step_4}, 
			the induction hypothesis \eqref{eq:induction_hypothesis} to get from \eqref{step_4} to \eqref{step_5} 
			and Equation \eqref{eq:multinomial} as well as $n^j-\sum_k=1^{m-2}+1 \geq n^j+1$ in the last line.
			
			Finally, for any $\mr{a}$, $\mr{b}$, $\mr{x}$, and $\mr{y}$,
			\begin{align*}
				& \tau_{\mr{A}\mr{B}|\mr{X}\mr{Y}} (\mr{a}\mr{b}|\mr{x}\mr{y}) = \nonumber \\
				&   \prod_{j=1}^l  \prod_{k=1}^{m-1} \left[ \int_0^{c^j_k} \frac{\mathrm{d}p^j_k}{c^j_k}\left( p^j_k \right)^{n^j_k} \right] \left( p^j_{m} \right)^{n^j-\sum_{k=1}^{m-1} n^j_k} = \\
				&  \prod_{j=1}^l   \prod_{k=1}^{m-1} \left[ \int_0^{c^j_k} \frac{\mathrm{d}p^j_k}{c^j_k}\left( p^j_k \right)^{n^j_k} \right] \left( c^j_m \right)^{n^j-\sum_{k=1}^{m-1} n^j_k} =\\
				&  \prod_{j=1}^l  \prod_{k=1}^{m-1} \left[ \int_0^{c^j_k} \frac{\mathrm{d}p^j_k}{c^j_k}\left( p^j_k \right)^{n^j_k} \right]   \left(c^j_{m-1}-p^j_{m-1} \right)^{n^j-\sum_{k=1}^{m-1} n^j_k} \geq  \\
				&   \prod_{j=1}^l  {n^j \choose {n^j_1, \dots n^j_{m}}}^{-1} \frac{1}{(n^j+1)^{m-1}} 
			\end{align*}
			where we used Equation \eqref{eq:induction} it the last step. \qedhere

		\end{proof}

	\section{Diamond norm reduction}\label{appsec:diamond_norm_proof}
	
		We prove Lemma~\ref{lem:trace-distance-lem}
			\begin{customlemma}{\ref{lem:trace-distance-lem}}
			For every two permutation invariant channels $\mathcal{E},\mathcal{F}:\mathcal{P}\rightarrow \mathcal{K}$ where $\mathrm{P}_K$ is a probability distribution over $k\in\{0,1\}^t$ for some $t>0$, and all $\P_{\mr{A}\mr{B}C|\mr{X}\mr{Y}Z}$,
				\[
					 \| \left(\mathcal{E}-\mathcal{F}\right)\otimes \idn(\P_{\mr{A}\mr{B}C|\mr{X}\mr{Y}Z} )\|_1 \leq (n+1)^{l(m-1)} \| \left(\mathcal{E}-\mathcal{F}\right)\otimes \idn(\tau^{\P_{\mr{A}\mr{B}C|\mr{X}\mr{Y}Z}}_{\mr{A}\mr{B}C|\mr{X}\mr{Y}Z} )\|_1
				\]
				where $\tau^{\P_{\mr{A}\mr{B}C|\mr{X}\mr{Y}Z}}_{\mr{A}\mr{B}C|\mr{X}\mr{Y}Z}$ is a non-signalling extension of $\tau_{\mr{A}\mr{B}|\mr{X}\mr{Y}}$ which depends on the specific box $\P_{\mr{A}\mr{B}C|\mr{X}\mr{Y}Z}$.
			\end{customlemma}
			
			\begin{proof}
					First, as in the proof of Theorem~\ref{thm:dF_test_bound}, since the channels are permutation invariant it is sufficient to consider boxes $\P_{\mr{A}\mr{B}|\mr{X}\mr{Y}}$ which are permutation invariant. 
				
				Given a specific box $\P_{\mr{A}\mr{B}C|\mr{X}\mr{Y}Z}$ we can see this extension as a set of convex decompositions of $\P_{\mr{A}\mr{B}|\mr{X}\mr{Y}}$, according to Lemma \ref{lem:extension}. That is, every possible input $z$ induces a specific decomposition $\{ ( p_{c_z}, \P^{c_z}_{\mr{A}\mr{B}|\mr{X}\mr{Y}}) \}_{c_z}$ such that $p_{c_z}=\mathrm{P}_{C|Z}(c_z|z)$ and $\P^{c_z}_{\mr{A}\mr{B}|\mr{X}\mr{Y}}(\mr{a},\mr{b}|\mr{x},\mr{y})=\P_{\mr{A}\mr{B}C|\mr{X}\mr{Y}Z}(\mr{a},\mr{b},c_z|\mr{x},\mr{y},z)$. Since this is a convex decomposition of $\P_{\mr{A}\mr{B}|\mr{X}\mr{Y}}$ we also have 
				\begin{equation} \label{eq:decomposition}
					\forall z \quad \sum_c p_c \cdot \P^{c}_{\mr{A}\mr{B}|\mr{X}\mr{Y}} = \P_{\mr{A}\mr{B}|\mr{X}\mr{Y}} \;.
				\end{equation}
				We now use the set of decompositions of $\P_{\mr{A}\mr{B}|\mr{X}\mr{Y}}$ to construct a set of decompositions of the de Finetti box $\tau_{\mr{A}\mr{B}|\mr{X}\mr{Y}}$. Combining Lemmas~\ref{lem:dF_post_selection},~\ref{lem:extension_condition} and~\ref{lem:extension} together, we know that there exists a non-signalling box $\mathrm{R}_{\mr{A}\mr{B}|\mr{X}\mr{Y}}$ such that 
				\[
				\begin{split}
					\tau_{\mr{A}\mr{B}|\mr{X}\mr{Y}} &= \frac{1}{(n+1)^{l(m-1)}}\P_{\mr{A}\mr{B}|\mr{X}\mr{Y}} + \left(1- \frac{1}{(n+1)^{l(m-1)}} \right)\mathrm{R}_{\mr{A}\mr{B}|\mr{X}\mr{Y}} \\
					&=\frac{1}{(n+1)^{l(m-1)}}\sum_c p_c \cdot \P^{c}_{\mr{A}\mr{B}|\mr{X}\mr{Y}} + \left(1- \frac{1}{(n+1)^{l(m-1)}} \right)\mathrm{R}_{\mr{A}\mr{B}|\mr{X}\mr{Y}} \;,
				\end{split}
				\]
				where the second equality is due to Equation \eqref{eq:decomposition}. For every $z$ this defines a decomposition $\{ ( \frac{1}{(n+1)^{l(m-1)}} \cdot p_{c_z}, \P^{c_z}_{\mr{A}\mr{B}|\mr{X}\mr{Y}} ) \}_{c_z} \cup \{ (1- \frac{1}{(n+1)^{l(m-1)}}, \mathrm{R}_{\mr{A}\mr{B}|\mr{X}\mr{Y}}) \} $ of~$\tau_{\mr{A}\mr{B}|\mr{X}\mr{Y}}$. That is, this defines an extension $\tau^{\P_{\mr{A}\mr{B}C|\mr{X}\mr{Y}Z}}_{\mr{A}\mr{B}C'|\mr{X}\mr{Y}Z}$ of  $\tau_{\mr{A}\mr{B}|\mr{X}\mr{Y}}$ where $C'=C\cup \{c'\}$. 
				
				This connection between the extensions $\P_{\mr{A}\mr{B}C|\mr{X}\mr{Y}Z}$ and $\tau^{\P_{\mr{A}\mr{B}C|\mr{X}\mr{Y}Z}}_{\mr{A}\mr{B}C'|\mr{X}\mr{Y}Z}$ allow us to get the following bound on the trace distance, from which the lemma follows:
				\begin{equation}\label{eq:dF_trace_dist_relation}
					\| \left(\mathcal{E}-\mathcal{F}\right)\otimes \idn(\tau^{\P_{\mr{A}\mr{B}C|\mr{X}\mr{Y}Z}}_{\mr{A}\mr{B}C'|\mr{X}\mr{Y}Z} )\|_1 \geq \frac{1}{(n+1)^{l(m-1)}} \| \left(\mathcal{E}-\mathcal{F}\right)\otimes \idn(\P_{\mr{A}\mr{B}C|\mr{X}\mr{Y}Z} )\|_1 \;.
				\end{equation}
					
				Equation~\eqref{eq:dF_trace_dist_relation} can be proven using the following sequence of steps. 		
				First, the diamond norm can be written in the following way.
				\begin{equation*}
					\begin{split}
						\| \mathcal{E}-\mathcal{F}\|_{\diamond} & =\underset{\P_{\mr{A}\mr{B}C|\mr{X}\mr{Y}Z}}{\max}\| \left(\mathcal{E}-\mathcal{F}\right)\otimes \idn(\P_{\mr{A}\mr{B}C|\mr{X}\mr{Y}Z} )\|_1\\
						&= \underset{\P_{\mr{A}\mr{B}C|\mr{X}\mr{Y}Z}}{\max}\| \mathrm{E}_{K|C}\cdot \mathrm{P}_{C|Z} - \mathrm{F}_{K|C}\cdot \mathrm{P}_{C|Z}\|_1\\
						&=\underset{\P_{\mr{A}\mr{B}C|\mr{X}\mr{Y}Z}}{\max} \frac{1}{2} \sum_k \max_z \sum_c \mathrm{P}_{C|Z}(c|z) \Big| \mathrm{E}_{K|C}(k|c)-\mathrm{F}_{K|C}(k|c) \Big| \\
						&= \underset{\P_{\mr{A}\mr{B}C|\mr{X}\mr{Y}Z}}{\max} \frac{1}{2} \sum_k \max_z \sum_c \mathrm{P}_{C|Z}(c|z) \times \\
						& \qquad \qquad \times \Big| \sum_{\mr{x},\mr{y}} \mathrm{Pr}_{\mathcal{E}}(\mr{x},\mr{y}) \sum_{\substack{\mr{a},\mr{b} : \\ \mathcal{E}(\mr{a},\mr{b},\mr{x},\mr{y})=k}} \P_{\mr{A}\mr{B}|\mr{X}\mr{Y}C}(\mr{a},\mr{b}|\mr{x},\mr{y},c)  - \\
						& \qquad \qquad  \qquad\sum_{\mr{x},\mr{y}} \mathrm{Pr}_{\mathcal{F}}(\mr{x},\mr{y}) \sum_{\substack{\mr{a},\mr{b} : \\ \mathcal{F}(\mr{a},\mr{b},\mr{x},\mr{y})=k}} \P_{\mr{A}\mr{B}|\mr{X}\mr{Y}C}(\mr{a},\mr{b}|\mr{x},\mr{y},c)  \Big|
					\end{split}
				\end{equation*}
				where the third equality is due to the explicit form of the trace distance previously given in \cite{masanes2009universally,hanggi2009quantum}. 
				
				This can then be used to write
				\begin{align*}
					&\| \left(\mathcal{E}-\mathcal{F}\right)\otimes \idn(\tau^{\P_{\mr{A}\mr{B}C|\mr{X}\mr{Y}Z}}_{\mr{A}\mr{B}C'|\mr{X}\mr{Y}Z} )\|_1 \\
					& \qquad = \frac{1}{2} \sum_k \max_z \sum_{c\in C'} \tau^{\P_{\mr{A}\mr{B}C|\mr{X}\mr{Y}Z}}_{C'|Z}(c|z)  \Big| \sum_{\mr{x},\mr{y}} \mathrm{Pr}_{\mathcal{E}}(\mr{x},\mr{y}) \sum_{\substack{\mr{a},\mr{b} : \\\mathcal{E}(\mr{a},\mr{b},\mr{x},\mr{y})=k}} \tau^{\P_{\mr{A}\mr{B}C|\mr{X}\mr{Y}Z}}_{\mr{A}\mr{B}|\mr{X}\mr{Y}C'}(\mr{a}\mr{b}|\mr{x}\mr{y}c)  \\ 
					& \qquad \qquad - \sum_{\mr{x},\mr{y}} \mathrm{Pr}_{\mathcal{F}}(\mr{x},\mr{y}) \sum_{\substack{\mr{a},\mr{b} : \\ \mathcal{F}(\mr{a},\mr{b},\mr{x},\mr{y})=k}} \tau^{\P_{\mr{A}\mr{B}C|\mr{X}\mr{Y}Z}}_{\mr{A}\mr{B}|\mr{X}\mr{Y}C'}(\mr{a}\mr{b}|\mr{x}\mr{y}c)  \Big|  \\
					&\qquad =\frac{1}{2} \sum_k \max_z \Big[ \sum_{c\in C} \tau^{\P_{\mr{A}\mr{B}C|\mr{X}\mr{Y}Z}}_{C'|Z}(c|z)  \Big| \sum_{\mr{x},\mr{y}} \mathrm{Pr}_{\mathcal{E}}(\mr{x},\mr{y}) \sum_{\substack{\mr{a},\mr{b} : \\ \mathcal{E}(\mr{a},\mr{b},\mr{x},\mr{y})=k}} \tau^{\P_{\mr{A}\mr{B}C|\mr{X}\mr{Y}Z}}_{\mr{A}\mr{B}|\mr{X}\mr{Y}C'}(\mr{a}\mr{b}|\mr{x}\mr{y}c)   \\
					& \qquad \qquad - \sum_{\mr{x},\mr{y}} \mathrm{Pr}_{\mathcal{F}}(\mr{x},\mr{y}) \sum_{\substack{\mr{a},\mr{b} : \\ \mathcal{F}(\mr{a},\mr{b},\mr{x},\mr{y})=k}} \tau^{\P_{\mr{A}\mr{B}C|\mr{X}\mr{Y}Z}}_{\mr{A}\mr{B}|\mr{X}\mr{Y}C'}(\mr{a}\mr{b}|\mr{x}\mr{y}c)  \Big|  \\
					& \qquad \qquad + \left(1- \frac{1}{(n+1)^{l(m-1)}} \right)  \Big| \sum_{\mr{x},\mr{y}} \mathrm{Pr}_{\mathcal{E}}(\mr{x},\mr{y}) \sum_{\substack{\mr{a},\mr{b} : \\ \mathcal{E}(\mr{a},\mr{b},\mr{x},\mr{y})=k}} \tau^{\P_{\mr{A}\mr{B}C|\mr{X}\mr{Y}Z}}_{\mr{A}\mr{B}|\mr{X}\mr{Y}C'}(\mr{a}\mr{b}|\mr{x}\mr{y}c')  \\
					& \qquad \qquad - \sum_{\mr{x},\mr{y}} \mathrm{Pr}_{\mathcal{F}}(\mr{x},\mr{y}) \sum_{\substack{\mr{a},\mr{b} : \\ \mathcal{F}(\mr{a},\mr{b},\mr{x},\mr{y})=k}} \tau^{\P_{\mr{A}\mr{B}C|\mr{X}\mr{Y}Z}}_{\mr{A}\mr{B}|\mr{X}\mr{Y}C'}(\mr{a}\mr{b}|\mr{x}\mr{y}c')  \Big| \Big] \displaybreak \\
					&\qquad \geq \frac{1}{2} \sum_k \max_z  \sum_{c\in C} \tau^{\P_{\mr{A}\mr{B}C|\mr{X}\mr{Y}Z}}_{C'|Z}(c|z)  \Big| \sum_{\mr{x},\mr{y}} \mathrm{Pr}_{\mathcal{E}}(\mr{x},\mr{y}) \sum_{\substack{\mr{a},\mr{b} : \\ \mathcal{E}(\mr{a},\mr{b},\mr{x},\mr{y})=k}} \tau^{\P_{\mr{A}\mr{B}C|\mr{X}\mr{Y}Z}}_{\mr{A}\mr{B}|\mr{X}\mr{Y}C'}(\mr{a}\mr{b}|\mr{x}\mr{y}c)  \\
					& \qquad \qquad  - \sum_{\mr{x},\mr{y}} \mathrm{Pr}_{\mathcal{F}}(\mr{x},\mr{y}) \sum_{\substack{\mr{a},\mr{b} : \\ \mathcal{F}(\mr{a},\mr{b},\mr{x},\mr{y})=k}} \tau^{\P_{\mr{A}\mr{B}C|\mr{X}\mr{Y}Z}}_{\mr{A}\mr{B}|\mr{X}\mr{Y}C'}(\mr{a}\mr{b}|\mr{x}\mr{y}c)  \Big|  \\
					&\qquad =\frac{1}{2} \sum_k \max_z  \sum_{c\in C} \frac{1}{(n+1)^{l(m-1)}}\cdot \mathrm{P}_{C|Z}(c|z) \\
					 & \qquad \qquad \cdot   \Big| \sum_{\mr{x},\mr{y}} \mathrm{Pr}_{\mathcal{E}}(\mr{x},\mr{y}) \sum_{\substack{\mr{a},\mr{b} : \\ \mathcal{E}(\mr{a},\mr{b},\mr{x},\mr{y})=k}} \P_{\mr{A}\mr{B}|\mr{X}\mr{Y}C}(\mr{a}\mr{b}|\mr{x}\mr{y}c)  \\ 
					& \qquad \qquad - \sum_{\mr{x},\mr{y}} \mathrm{Pr}_{\mathcal{F}}(\mr{x},\mr{y}) \sum_{\substack{\mr{a},\mr{b} : \\ \mathcal{F}(\mr{a},\mr{b},\mr{x},\mr{y})=k}} \P_{\mr{A}\mr{B}|\mr{X}\mr{Y}C}(\mr{a}\mr{b}|\mr{x}\mr{y}c)  \Big|  \\
					&\qquad = \frac{1}{(n+1)^{l(m-1)}} \| \left(\mathcal{E}-\mathcal{F}\right)\otimes \idn(\P_{\mr{A}\mr{B}C|\mr{X}\mr{Y}Z} )\|_1 \;.
				\end{align*}
				where in order to get the second equality we divide the sum over $C'=C\cup \{c'\}$ to the sum over $C$ and then additional part of the partition $c'$. The next inequality is then correct since
				\[
				\begin{split}
					 \left(1- \frac{1}{(n+1)^{l(m-1)}} \right) & \Big| \sum_{\mr{x},\mr{y}} \mathrm{Pr}_{\mathcal{E}}(\mr{x},\mr{y}) \sum_{\substack{\mr{a},\mr{b} : \\ \mathcal{E}(\mr{a},\mr{b},\mr{x},\mr{y})=k}} \tau^{\P_{\mr{A}\mr{B}C|\mr{X}\mr{Y}Z}}_{A|XC}(\mr{a}\mr{b}|\mr{x}\mr{y}c')  - \\ 
					 & \qquad \qquad \sum_{\mr{x},\mr{y}} \mathrm{Pr}_{\mathcal{F}}(\mr{x},\mr{y}) \sum_{\substack{\mr{a},\mr{b} : \\ \mathcal{F}(\mr{a},\mr{b},\mr{x},\mr{y})=k}} \tau^{\P_{\mr{A}\mr{B}C|\mr{X}\mr{Y}Z}}_{A|XC}(\mr{a}\mr{b}|\mr{x}\mr{y}c')  \Big| \geq 0 \;,
				\end{split}
				\]
				 and the two last equalities are due to the specific decomposition of $\tau_{\mr{A}\mr{B}|\mr{X}\mr{Y}}$ that we defined and the definition of the trace distance. 
				 This proves the lemma.\qedhere

			\end{proof}

\chapter{Additional proofs: non-signalling parallel repetition}

	\section{Signalling measure and test}\label{appsec:ns_averages_proofs}
	
		We present here proofs of lemmas relevant to Section~\ref{sec:approx_ns_marginals}. The proofs previously appeared in~\cite{arnon2016non}. 
		
		\begin{customlemma}{\ref{lem:continuity_sig}}
			Let $\O_{AB|XY}^1$ and $\O_{AB|XY}^2$ be two single-round boxes such that 
				\[
					\big|\O_{AB|XY}^1-\O_{AB|XY}^2\big|_1 \leq \epsilon \;.
				\]
				Then, for all $a$, $b$, $x$, and $y$,
				\begin{align*}
					&\big|\mathrm{Sig}^{(A\rightarrow B, x, y, b)}(\O_{AB|XY}^1) - \mathrm{Sig}^{(A\rightarrow B, x, y, b)}(\O_{AB|XY}^2)\big| \leq 2\varepsilon 
				\end{align*}
		\end{customlemma}

		\begin{proof}
			We prove a stronger result from which the lemma follows. We prove 
			\[
				  \sum_{b,x,y} \big|\mathrm{Sig}^{(A\rightarrow B, x, y, b)}\left( \O_{AB|XY}^1 \right) - \mathrm{Sig}^{(A\rightarrow B, x, y, b)}\left( \O_{AB|XY}^2 \right)\big|\leq 2\epsilon \;.
			\]

			To do so first note the following,
			\begin{align*}
				\big|\O_{AB|XY}^1-\O_{AB|XY}^2\big|_1 &= \mathbb{E}_{x,y} \sum_{a,b} \big| \O_{AB|XY}^1(a,b|x,y) - \O_{AB|XY}^2(a,b|x,y) \big| \\
				& \geq \mathbb{E}_{x,y} \sum_{b} \Big| \sum_{a} \left( \O_{AB|XY}^1(a,b|x,y) - \O_{AB|XY}^2(a,b|x,y) \right) \Big| \\
				& = \mathbb{E}_{x,y} \sum_{b} \big| \O_{B|XY}^1(b|x,y) - \O_{B|XY}^2(b|x,y) \big| \\
				& =  \sum_{b,x,y} \Q_{XY}(x,y) \big| \O_{B|XY}^1(b|x,y) - \O_{B|XY}^2(b|x,y) \big| \;,
			\end{align*}
			therefore if $\big|\O_{AB|XY}^1-\O_{AB|XY}^2\big|_1 \leq \epsilon$ then 
			\begin{equation}\label{eq:continuity_proof}
				\sum_{b,x,y} \Q_{XY}(x,y) \big| \O_{B|XY}^1(b|x,y) - \O_{B|XY}^2(b|x,y) \big|  \leq \epsilon \;.
			\end{equation}
		
			Next, using Definition~\ref{def:sign_measure_dF} and the discussion following it,
			\begin{align*}
				 \sum_{b,x,y}  \big|& \mathrm{Sig}^{(A\rightarrow B, x, y, b)}\left( \O_{AB|XY}^1\right) - \mathrm{Sig}^{(A\rightarrow B, x, y, b)}\left( \O_{AB|XY}^2 \right)\big|  \\ 
				& = \sum_{b,x,y} \Q_{XY}(x,y) \Big|  \O_{B|XY}^1(b|x,y) - \sum_{\tilde{x}} \Q_{X|Y}(\tilde{x}|y)\O_{B|XY}^1(b | \tilde{x},y) \\
				& \qquad \qquad - \O_{B|XY}^2(b|x,y) + \sum_{\tilde{x}} \Q_{X|Y}(\tilde{x}|y)\O_{B|XY}^2(b | \tilde{x},y) \Big| \\
				& =\sum_{b,x,y}  \Q_{XY}(x,y) \Big|  \O_{B|XY}^1(b|x,y) - \O_{B|XY}^2(b|x,y) \\
				& \qquad \qquad + \sum_{\tilde{x}} \Q_{X|Y}(\tilde{x}|y) \left( \O_{B|XY}^2(b | \tilde{x},y) - \O_{B|XY}^1(b | \tilde{x},y)\right) \Big| \\
				& \leq \sum_{b,x,y} \Q_{XY}(x,y) \Big|  \O_{B|XY}^1(b|x,y) - \O_{B|XY}^2(b|x,y)\Big| \\
				& \qquad \qquad + \sum_{b,x,y} \Q_{XY}(x,y) \Big|\sum_{\tilde{x}} \Q_{X|Y}(\tilde{x}|y) \left( \O_{B|XY}^2(b | \tilde{x},y) - \O_{B|XY}^1(b | \tilde{x},y)\right) \Big| \\
				& \leq \sum_{b,x,y} \Q_{XY}(x,y) \Big|  \O_{B|XY}^1(b|x,y) - \O_{B|XY}^2(b|x,y)\Big| \\
				& \qquad \qquad  +  \sum_{b,x,y} \sum_{\tilde{x}} \Q_{X|Y}(\tilde{x}|y) \Q_{XY}(x,y)  \Big| \O_{B|XY}^2(b | \tilde{x},y) - \O_{B|XY}^1(b | \tilde{x},y) \Big| \\
				&= \sum_{b,x,y} \Q_{XY}(x,y) \Big|  \O_{B|XY}^1(b|x,y) - \O_{B|XY}^2(b|x,y)\Big| \\
				& \qquad \qquad +  \sum_{b,y} \sum_{\tilde{x}} \Q_{X|Y}(\tilde{x}|y) \Q_{Y}(y)  \Big| \O_{B|XY}^2(b | \tilde{x},y) - \O_{B|XY}^1(b | \tilde{x},y) \Big| \\
				&= \sum_{b,x,y} \Q_{XY}(x,y) \Big|  \O_{B|XY}^1(b|x,y) - \O_{B|XY}^2(b|x,y)\Big| \\ 
				& \qquad \qquad  +  \sum_{b,x,y} \Q_{XY}(x,y)  \Big| \O_{B|XY}^2(b | x,y) - \O_{B|XY}^1(b | x,y) \Big| \\
				& \leq 2 \epsilon 
			\end{align*} 
		 	where the last inequality follows from Equation \eqref{eq:continuity_proof}. \qedhere
		\end{proof}


		\begin{lem}\label{lem:ns_Sigma_alternative}
			Let $\nu > 0$ be any parameter such that $\nu < \zeta - 6\epsilon$. Then for every $x$, $y$, and $b$,
			\[
			 	\forall \O_{AB|XY}\in\Sigma^{(A\rightarrow B,x,y,b)}, \quad \mathrm{Sig}^{(A\rightarrow B, x, y, b)}\left( \O_{AB|XY}\right) > \nu \;.
			\]		
		\end{lem}

		\begin{proof}
			Assume by contradiction that there exists $\O_{AB|XY}\in\Sigma^{(A\rightarrow B,x,y,b)}$ such that $\mathrm{Sig}^{(A\rightarrow B, x, y, b)}\left( \O_{AB|XY}\right) \leq \nu$. Since $\O_{AB|XY}\in\Sigma^{(A\rightarrow B,x,y,b)}$ there exists $\bar{\O}_{AB|XY}$ such that $ |\O_{AB|XY} - \bar{\O}_{AB|XY}|_1 \leq \epsilon$ and 
			\begin{equation}\label{eq:nu_assumption}
				\mathrm{Pr}_{\textup{data}\sim\bar{\O}_{AB|XY}^{\otimes n}} \left[ \passtest\right] > \delta\;.
			\end{equation}
			
			Using Lemma \ref{lem:continuity_sig} we get $\mathrm{Sig}^{(A\rightarrow B, x, y, b)}\left( \bar{\O}_{AB|XY}\right) \leq \nu + 2\epsilon$. 
			From Lemma \ref{lem:sanovs_theorem} we know that $ \mathrm{Pr}_{\textup{data}\sim\bar{\O}_{AB|XY}^{\otimes n}} \left[ |\bar{\O}_{AB|XY}^{\textup{freq}(\textup{data}_2)} -  \bar{\O}_{AB|XY}|_1 > \epsilon \right] \leq \delta$ and therefore, using Lemma \ref{lem:continuity_sig} again, 
			\[
				\mathrm{Pr}_{\textup{data}\sim\bar{\O}_{AB|XY}^{\otimes n}} \left[  \mathrm{Sig}^{(A\rightarrow B, x, y, b)}\left( \bar{\O}_{AB|XY}^{\textup{freq}(\textup{data}_2)} \right) > \nu + 4\epsilon  \right] \leq \delta \;.
			\]
			Since $\nu  < \zeta - 6\epsilon$ this implies 
			\[
				\mathrm{Pr}_{\textup{data}\sim\bar{\O}_{AB|XY}^{\otimes n}} \left[  \mathrm{Sig}^{(A\rightarrow B, x, y, b)}\left( \bar{\O}_{AB|XY}^{\textup{freq}(\textup{data}_2)} \right) > \zeta - 2\epsilon  \right] \leq \delta 
			\]
			and therefore, according to the definition of the test,
			\[
				\mathrm{Pr}_{\textup{data}\sim\bar{\O}_{AB|XY}^{\otimes n}} \left[ \passtest \right] \leq \delta \;,
			\]
			which contradicts Equation \eqref{eq:nu_assumption}.
		\end{proof}


	Next we would like to prove Lemma~\ref{lem:de_finetti_prop}. To do so, we first prove the same statement but for IID boxes:
		\begin{lem}\label{lem:reliable_test}
			Assume the players share an IID box $\O_{AB|XY}^{\otimes n}$ and let $\zeta,\epsilon>0$ be the the parameters defined  as in Equation~\eqref{eq:test_definition}. For every $\mathcal{T}^{(A\rightarrow B,x,y,b)}$,
			\begin{enumerate}
				\item If $\mathrm{Sig}^{(A\rightarrow B, x, y, b)}\left( \O_{AB|XY}\right)\geq \zeta$ then 
				\begin{equation}\label{eq:reliable_p1}
					\mathrm{Pr}_{\textup{data}\sim\O_{AB|XY}^{\otimes n}} \left[ \passtest \right] > 1- \delta
				\end{equation}
				\item If $\mathrm{Sig}^{(A\rightarrow B, x, y, b)}\left( \O_{AB|XY}\right)= 0$ then 
				\begin{equation}\label{eq:reliable_p2}
					\mathrm{Pr}_{\textup{data}\sim\O_{AB|XY}^{\otimes n}} \left[ \lnot\passtest\right] > 1 - \delta
				\end{equation}
			\end{enumerate}
			where $\delta=\delta\left(\frac{n}{2},\epsilon\right)=\left(\frac{n}{2}+1\right)^{|\mathcal{A}|\cdot|\mathcal{Q}|-1}e^{-n\epsilon^2/4}$.
		\end{lem}
		
		\begin{proof}
			For the first part of the lemma assume that $\mathrm{Sig}^{(A\rightarrow B, x, y, b)}\left( \O_{AB|XY}\right) \geq \zeta$. Then
			\begin{align*}
				\mathrm{Pr}_{\textup{data}\sim\O_{AB|XY}^{\otimes n}} \left[ \lnot\passtest \right] &= 
				\mathrm{Pr}_{\textup{data}\sim\O_{AB|XY}^{\otimes n}} \left[  \mathrm{Sig}^{(A\rightarrow B, x, y, b)}\left( \O_{AB|XY}^{\textup{freq}(\textup{data}_2)} \right) < \zeta - 2\epsilon  \right] \\
				&\leq  \mathrm{Pr}_{\textup{data}\sim\O_{AB|XY}^{\otimes n}} \left[ |\O_{AB|XY}^{\textup{freq}(\textup{data}_2)} -  \O_{AB|XY}|_1 > \epsilon \right] \\
				&\leq \delta
			\end{align*}
			where the first inequality is due to Lemma \ref{lem:continuity_sig} and the second due to Lemma \ref{lem:sanovs_theorem}. This implies Equation~\eqref{eq:reliable_p1}. Equation \eqref{eq:reliable_p2} can be proven in an analogous way. 
		\end{proof}


		\begin{customlemma}{\ref{lem:de_finetti_prop}}
			Let $\tau_{\mr{A}\mr{B}\mr{X}\mr{Y}}=\Q_{XY}^{\otimes n} \tau_{\mr{A}\mr{B}|\mr{X}\mr{Y}}$, where $\tau_{\mr{A}\mr{B}|\mr{X}\mr{Y}}$ is a de Finetti box. 
			For every $\mathcal{T}^{(A\rightarrow B,x,y,b)}$
			\begin{enumerate}
				\item $\mathrm{Pr}_{\textup{data}\sim\tau_{\mr{A}\mr{B}\mr{X}\mr{Y}}} \left[ \lnot\inSigma \land \passtest  \right] \leq \delta $
				\item $\mathrm{Pr}_{\textup{data}\sim\tau_{\mr{A}\mr{B}\mr{X}\mr{Y}}} \left[  \insigma \land \lnot\passtest \right] \leq \delta $, 
			\end{enumerate}
			where $\delta=\delta\left(\frac{n}{2},\epsilon\right)=\left(\frac{n}{2}+1\right)^{|\mathcal{A}||\mathcal{B}||\mathcal{X}||\mathcal{Y}|-1}e^{-n\epsilon^2/4}$.
		\end{customlemma}

	\begin{proof}
		Since a de Finetti box is a convex combination of IID boxes, it is sufficient to prove this for IID boxes $\O_{AB|XY}^{\otimes n}$ and the lemma will follow. We start by proving the first part of the lemma.
		
		If $\mathrm{Pr}_{\textup{data}\sim\O_{AB|XY}^{\otimes n}} \left[ \passtest \right] \leq \delta$ then we are done. Consider therefore single-round boxes $\O_{AB|XY}$ such that 
		\[
			\mathrm{Pr}_{\textup{data}\sim\O_{AB|XY}^{\otimes n}} \left[ \passtest \right] > \delta \;.
		\]
		For such boxes
		\[
			\mathrm{Pr}_{\textup{data}\sim\O_{AB|XY}^{\otimes n}} \left[\lnot\inSigma \right] \leq \mathrm{Pr}_{\textup{data}\sim\O_{AB|XY}^{\otimes n}}\left[ |\O_{AB|XY}^{\textup{freq}(\textup{data}_1)} -  \O_{AB|XY}|_1 > \epsilon \right] \leq \delta
		\]
		where the first inequality follows from the definition of $\Sigma^{(A\rightarrow B,x,y,b)}$ and the second from Lemma \ref{lem:sanovs_theorem}. 
		All together we get $\mathrm{Pr}_{\textup{data}\sim\O_{AB|XY}^{\otimes n}} \left[  \lnot\inSigma \land \passtest  \right] \leq \delta$ as required for the first part of the lemma.  
		
		We now proceed to the second part of the lemma. 
		If $\mathrm{Pr}_{\textup{data}\sim\O_{AB|XY}^{\otimes n}} \left[ \insigma\right] \leq \delta$ then we are done. 
		Consider therefore boxes $\O_{AB|XY}$ such that 
		\[
			\mathrm{Pr}_{\textup{data}\sim\O_{AB|XY}^{\otimes n}} \left[ \insigma\right] > \delta\;.
		\]
		Using Lemma \ref{lem:sanovs_theorem} we know that there exists a state $\O_{AB|XY}^{\textup{freq}(\textup{data}_1)}\in \Sigma^{(A\rightarrow B,x,y,b)}$ such that $|\O_{AB|XY}^{\textup{freq}(\textup{data}_1)}-\O_{AB|XY}|_1 \leq \epsilon$ and according to the definition of $\Sigma^{(A\rightarrow B,x,y,b)}$ this implies that $\O_{AB|XY}$ is $\zeta$ signalling or more. Therefore, according to Lemma~\ref{lem:reliable_test}, $\mathrm{Pr}_{\textup{data}\sim\O_{AB|XY}^{\otimes n}} \left[ \lnot\passtest  \right] \leq \delta$. All together we get  
		\[
			\mathrm{Pr}_{\textup{data}\sim\O_{AB|XY}^{\otimes n}} \left[  \insigma \land \lnot\passtest   \right] \leq \delta \;. \qedhere
		\]
		
	\end{proof}


	\section{Sensitivity analysis}\label{appsec:pr_sen_analy}

		Linear programs (see, e.g., \cite{schrijver1998theory}) are a useful tool when considering non-signalling games, as the non-signalling constraints are linear. The following general results regarding the sensitivity of linear programs will be of use for us. 
		\sloppy
		\begin{lem}[Sensitivity analysis of linear programs, \cite{schrijver1998theory} Section 10.4]\label{lem:sensitivity1}
			Let $\max\{c^Tx | Ax \leq b\}$ be a primal linear program and $min\{b^Ty|A^Ty=c,y\geq 0\}$ its dual. Denote the optimal value of the programs by $w$ and the optimal dual solution by $y^{\star}$. Then the optimal value of the perturbed program $w_e = max\{c^Tx | Ax \leq b+e\}$ for some perturbation $e$ is bounded by $w_e \leq w + e^Ty^{\star}$. 
		\end{lem}
		
		\begin{lem}[Dual optimal solution bound, \cite{schrijver1998theory} Section 10.4]\label{lem:sensitivity2}
			Let $A$ be an $r_1\times r_2$-matrix and let $\Delta$ be such that for each non-singular sub-matrix $B$ of $A$ all entries of~$B^{-1}$ are at most $\Delta$ in absolute value. Let $c$ be a row vector of dimension $r_2$ and let~$y^{\star}$ be the optimal dual solution of $min\{b^Ty|A^Ty=c,y\geq 0\}$. Then 
			\[
				 \kappa = \sum_{j=1}^{r_1} |y^{\star}_j| \leq  r_2\Delta \sum_{j=1}^{r_2} |c_j| \;.
			\]
		\end{lem}

		We start with the following program from Section~\ref{sec:lin_prog_ns_wp}:
		\begin{subequations} \label{eq:app_linear_non_relaxed}
			\begin{align}
				\max \quad &\sum_{a,b,x,y} \Q_{XY}(xy) R(a,b,x,y) \O_{AB|XY}(ab|xy) \nonumber \\
				\text{s.t.} \quad &   
					\mathrm{Sig}^{(A\rightarrow B, x, y, b)} \left(\O_{AB|XY}(ab|xy)\right) = 0 &\forall x,y,b \label{eq:linear1_ns_a} \\
					& \mathrm{Sig}^{(B\rightarrow A, x, y, b)} \left(\O_{AB|XY}(ab|xy)\right) = 0 &\forall x,y,a \label{eq:linear1_ns_b} \\
				&\sum_{a,b} \O_{AB|XY}(ab|xy) = 1  &\forall x,y \nonumber \\
				& \O_{AB|XY}(ab|xy) \geq 0 &\forall a,b,x,y \nonumber
			\end{align} 
		\end{subequations}
			
			To apply Lemma~\ref{lem:sensitivity1} we first need to write the program in the form $\max\{c^Tx | Ax \leq b\}$. 
			For this purpose, one can relax the linear program \eqref{eq:app_linear_non_relaxed} to the following:
			\begin{subequations} \label{eq:linear_program}
			\begin{align}
				\max \quad &\sum_{a,b,x,y} \Q_{XY}(xy) R(a,b,x,y) \O_{AB|XY}(ab|xy)\nonumber \\
				\text{s.t.} \quad &   
				\mathrm{Sig}^{(A\rightarrow B, x, y, b)} \left(\O_{AB|XY}(ab|xy)\right) \leq 0 &\forall x,y,b  \label{eq:relaxed_a} \\
				& \mathrm{Sig}^{(B\rightarrow A, x, y, b)} \left(\O_{AB|XY}(ab|xy)\right) \leq 0 &\forall x,y,a \label{eq:relaxed_b}  \\
				&\sum_{a,b} \O_{AB|XY}(ab|xy) = 1  &\forall x,y  \nonumber\\
				& \O_{AB|XY}(ab|xy) \geq 0 &\forall a,b,x,y \nonumber
			\end{align}
			\end{subequations}
			
			To see that the relaxation of the non-signalling constraints~\eqref{eq:linear1_ns_a} and~\eqref{eq:linear1_ns_b} to the constraints \eqref{eq:relaxed_a} and~\eqref{eq:relaxed_b} does not change the program, i.e., does not change the value of the optimal solution, recall Equation~\eqref{eq:ns_cond_par_rep} and assume there exists $x,y,b$ for which
			\[
				\Q_{XY}(x,y)\left[ \O_{B|XY}(b | x,y) - \sum_{\tilde{x}} \Q_{X|Y}(\tilde{x}|y)\O_{B|XY}(b | \tilde{x},y)\right] < 0 \;.
			\]
			That is, $\O_{B|XY}(b | x,y)$ is smaller than the average $\sum_{\tilde{x}} \Q_{X|Y}(\tilde{x}|y)\O_{B|XY}(b | \tilde{x},y)$, and therefore there must be some $x'$ for which $\O_{B|XY}(b|x' , y)$ is larger than the average, meaning,  
			\[
				\Q_{XY}(x' , y)\left[ \O_{B|XY}(b|x' , y) - \sum_{\tilde{x}} \Q_{X|Y}(\tilde{x}|y)\O_{B|XY}(b | \tilde{x},y)\right] > 0 \;,
			\]
			but this contradicts the constraints~\eqref{eq:relaxed_a} and~\eqref{eq:relaxed_b}.
	
			The dual program of the primal \eqref{eq:linear_program} is given below. 
			\begin{subequations}\label{eq:dual}
			\begin{align}
				\min \quad & \sum_{x,y} z(x,y) \nonumber \\
				\text{s.t.} \quad & z(x,y) +  y_A(x,y,b)\Q_{XY}(x,y) +  y_B(x,y,a)\Q_{XY}(x,y) \nonumber \\
				& \qquad - \sum_{\tilde{x}} y_A(\tilde{x},y,b) \Q_{XY}(\tilde{x},y) \Q_{X|Y}(x|y) \nonumber \\
				& \qquad - \sum_{\tilde{y}} y_B(x,\tilde{y},a) \Q_{XY}(x,\tilde{y}) \Q_{Y|X}(y|x) \nonumber \\
				& \qquad \qquad \qquad \geq \Q_{XY}(x,y)R(a,b,x,y) & \forall a,b,x,y \label{eq:dual_constraint} \\
				& y_A(x,y,b) \geq 0 &\forall x,y,b \nonumber \\
				& y_B(x,y,a) \geq 0 &\forall x,y,a \nonumber
			\end{align}
			\end{subequations}

	\begin{lem}\label{lem:distance_transformation}
		Let $\kappa =  \sum_{j=1}^{d} |y^{\star}_j|$ where $d$ is the number of signalling tests and $y^{\star}$ is an optimal solution of the dual program \eqref{eq:dual}. 
		Let $\O_{AB|XY}$ be a strategy such that the following holds for all $a,b,x,y$
			\begin{equation}\label{eq:not_too_sig}
				\begin{split}
					&\mathrm{Sig}^{(A\rightarrow B, x, y, b)}\left(\O_{AB|XY}\right) \leq  \zeta + 2\epsilon \\
					&\mathrm{Sig}^{(B\rightarrow A, x, y, a)}\left(\O_{AB|XY}\right) \leq  \zeta + 2\epsilon \;.
				\end{split}
			\end{equation}
			Then $w\left(\O_{AB|XY}\right)\leq 1-\alpha + \left(\zeta+2\epsilon\right)d$.
	\end{lem}
	
	\begin{proof}
		The fact that $\O_{AB|XY}$  is not ``too signalling'' in any direction can be used to bound its winning probability in the game $\G$.
		
		The following linear program describes the optimal winning probability of a strategy $\O_{AB|XY}$ which fulfils Equation~\eqref{eq:not_too_sig}: 
		\begin{equation} \label{eq:app_perturbation}
		\begin{aligned}
			\max \quad &\sum_{a,b,x,y} \Q_{XY}(xy) R(a,b,x,y) \O_{AB|XY}(ab|xy) \\
			\text{s.t.} \quad &  
			 \mathrm{Sig}^{(A\rightarrow B, x, y, b)} \left(\O_{AB|XY}(ab|xy)\right)  \leq \zeta+2\epsilon &\forall x,y,b  \\
					& \mathrm{Sig}^{(B\rightarrow A, x, y, b)} \left(\O_{AB|XY}(ab|xy)\right)  \leq \zeta+2\epsilon &\forall x,y,a \\
			&\sum_{a} \O_{AB|XY}(ab|xy) = 1  &\forall x,y  \\
			& \O_{AB|XY}(ab|xy) \geq 0 &\forall a,b,x,y
		\end{aligned}
		\end{equation}
	
		Program \eqref{eq:app_perturbation} can be seen as a perturbation of the linear program~\eqref{eq:linear_program}, we can therefore bound its optimal value by using known tools for sensitivity analysis of linear programs, stated in Lemmas~\ref{lem:sensitivity1} and \ref{lem:sensitivity2}. 
		
		Denote by $y^{\star}$ an optimal solution of the dual program\footnote{We are only interested in the value of $y^{\star}$ as $z^{\star}$ will not affect the bound.} \eqref{eq:dual} and let $\kappa =  \sum_{j=1}^{d} |y^{\star}_j|$ where $d$ is the number of signalling tests. That is, $\kappa$ is the sum of all the dual variables which are associated to the non-signalling constraints. 
		
		According to Lemma \ref{lem:sensitivity1} the perturbed winning probability is then bounded by 
		\[
			w_e \leq 1-\alpha + \left(\zeta+2\epsilon\right)\kappa .
		\]
		In the case of a game with two players, using \cite[Section~4]{ito2010polynomial}, one can show that $\kappa \leq d$ where $d$ is the number of different signalling tests, i.e., $d=|\mathcal{X}||\mathcal{Y}|(|\mathcal{A}|+|\mathcal{B}|)$).
	\end{proof}

\chapter{Additional proofs: device-independent quantum cryptography}

	This appendix is devoted to presenting the technical proofs of the statements made in this thesis related to the device-independent cryptography. The proofs previously appeared in~\cite{arnon2016simple}.

	\section{Single-round statement}\label{appsec:crypto_single_proof}
	
		As mentioned in Section~\ref{sec:randomness_single_round}, Lemma~\ref{lem:single_round_secrecy} (restated below) follows, almost directly, from~\cite{pironio2010random,acin2012randomness}. We show here how the bound derived in these works can be manipulated in a simple way to get the bound used in this thesis. 
		
		\begin{customlemma}{\ref{lem:single_round_secrecy}}
			For any quantum single-round box $\P_{AB|XY}$ with winning probability $\omega\in\left[\frac{3}{4},\frac{2+\sqrt{2}}{4}\right]$ in the \textup{CHSH} game,
				\begin{equation*}
	 				H(A|XYE) \geq 1 - h\left( \frac{1}{2} + \frac{1}{2}\sqrt{16\omega \left(\omega-1\right) +3}  \right) \;,
				\end{equation*}
				where $E$ denotes the quantum side-information belonging to the adversary and $h(\cdot)$ is the binary entropy function.
		\end{customlemma}

		\begin{proof}
						
				Our starting point is the result of~\cite{pironio2009device}. There, a quantum single-round box $\P_{\hat{A}\hat{B}|XY}$ with symmetric  marginals on $\hat{A}$ and $\hat{B}$ was considered (i.e., $\hat{A}$ and $\hat{B}$ are uniformly distributed). 
				To derive a bound which holds for any $\P_{AB|XY}$ we do the following: 
				\begin{enumerate}
					\item Symmetrisation of $\P_{AB|XY}$ --  Alice chooses a bit $F$ uniformly at random and communicates it to Bob. They then symmetrise their marginals by setting $\hat{A} = A \oplus F$ and $\hat{B} = B \oplus F$. 
					\item Use~\cite{pironio2009device} to lower-bound $H\left( \hat{A}| X Y  F E \right)$.
					\item Derive a bound on $H\left( A| X Y   E \right)$ from $H\left( \hat{A}| X Y  F E \right)$.
				\end{enumerate}  
				
				Let us follow the above steps. After applying $\hat{A} = A \oplus F$ and $\hat{B} = B \oplus F$, with $F$ uniformly distributed, $A$ and $B$ are unbiased.   
				In our notation, \cite{pironio2009device} considered the following Holevo quantity
				\[
					\chi\left( \hat{A} : F E | X=0\right)=H\left( FE |  X=0 \right)-H\left( FE | \hat{A}, X=0 \right)\;.
				\]
				and showed that for states leading to a CHSH violation of $\beta\in[2,2\sqrt{2}]$, related to the winning probability via $\omega=1/2+\beta/8$, the following tight bound holds~\cite[Section 2.3]{pironio2009device}:
				\begin{equation*}
					 \chi\left( \hat{A} : F E | X=0\right) \leq h\left( \frac{1}{2} + \frac{1}{2}\sqrt{\frac{\beta^2}{4} -1} \right) \;.
				\end{equation*}
				Rewriting the bound in terms of the winning probability $\omega$ one gets that for all $\omega\in\left[\frac{3}{4},\frac{2+\sqrt{2}}{4}\right]$ (i.e. a winning probability in the quantum regime) 
				\begin{equation}\label{eq:chi-bound}
						 \chi\left( \hat{A} : F E | X=0 \right) \leq h\left( \frac{1}{2} + \frac{1}{2}\sqrt{16\omega \left(\omega-1\right) +3}  \right) \;.
				\end{equation}
				
				We now wish to related the above Holevo quantities to our von Neumann entropy. Using the definition of the conditional von Neumann entropy one can rewrite $H\left( A | F E , X=0  \right)(\sigma)$ as follows:
				\begin{align}
					H\left(  \hat{A} | F E , X=0\right) &= H\left( \hat{A} F E |  X=0 \right) - H\left( F E | X=0 \right) \notag \\
					&= H\left( \hat{A} | X=0\right) + H\left(F E | \hat{A}, X=0 \right) - H\left(F E | X=0\right)\notag  \\
					&= H\left( \hat{A} |  X=0 \right)  - \chi\left( \hat{A} : FE | X=0 \right) \notag\\
					&= 1  - \chi\left( \hat{A} : FE | X=0 \right) \;, \label{eq:h-chain-1}
				\end{align}
				where the last equality holds since $A$ is uniformly random due to the symmetrisation step (note that we do not condition on $F$ in $ H\left( A |  X=0 \right)$).
				Furthermore,  
				\begin{equation}\label{eq:entropy_input_split}
				\begin{split}
					H\left( \hat{A} | XY F E \right)_{\mathcal{M}_i(\sigma)} &= \Pr\left[ X_i=0\right] \cdot H\left( \hat{A} | Y F E, X_i=0 \right)_{\mathcal{M}_i(\sigma)} \\
					&+\Pr\left[ X_i=1\right] \cdot H\left( \hat{A} | Y F E, X_i=1 \right)_{\mathcal{M}_i(\sigma)} \;.
				\end{split}
				\end{equation}
				Combining Equations~\eqref{eq:chi-bound} and~\eqref{eq:h-chain-1} while noting that, due to the symmetry between the cases $X=0$ and $X=1$, the same relations can be written for $X=1$ and plugging the bounds in Equation~\eqref{eq:entropy_input_split} we get		
				\begin{equation}\label{eq:entropy_bound_single_round_symm}
					H\left( \hat{A} | X Y F E \right) \geq 1 - h\left( \frac{1}{2} + \frac{1}{2}\sqrt{16\omega \left(\omega-1\right) +3}  \right) \;.
				\end{equation}
				
				The only think left to do is to use the above to get our bound for the original box  $\P_{AB|XY}$. For this simply observe that
				\[
					H\left( \hat{A} | X Y F E \right) = H\left( A | X Y F E \right) = H\left( A | X Y E \right) \;,
				\]
				where the first equality holds since $\hat{A} = A \oplus F$ and the second follows since $F$ is independent of $A$, $X$, $Y$, and $E$. The lemma therefore follows. \qedhere	

		\end{proof}

	\section{An improved dependency on the test probability}\label{sec:better_rate}

		In this section we show how the EAT can be used in a slightly different way than what was done in the main text. This results in an entropy rate which has a better dependency on the probability of a test round $\gamma$, compared to the entropy rate given in Equation~\eqref{eq:eta_opt}. The improved entropy rate derived here is the one used for calculating the  key rates of the DIQKD protocol is Section~\ref{sec:qkd_curves}.
		
		\subsection{Modified entropy accumulation protocol}

			We use a different entropy accumulation protocol, given as Protocol~\ref{pro:randomness_generation_mod}.
			In this modified protocol instead of considering each round separately we consider blocks of rounds. A block is defined by a sequence of rounds: in each round a test is carried out with probability $\gamma$ (and otherwise the round is a generation round). The block ends when a test round is being performed and then the next block begins.  If for~$s_{\max}$ rounds there was no test, the block ends without performing a test and the next begins.  Thus, the blocks can be of different length, but they all consist at most~$s_{\max}$ rounds.
			
			In this setting, instead of fixing the number of rounds $n$ in the beginning of the protocol, we fix the number of blocks $m$. The expected length of  block is 
			\begin{align}
				\bar{s} = \sum_{s\in [s_{\max}]}\left[ s (1-\gamma)^{(s-1)}\gamma \right] +s_{\max}(1-\gamma)^{s_{\max}} &=  \frac{1- (1-\gamma)^{s_{\max}}}{\gamma} \nonumber \\
				&= \sum_{s\in [s_{\max}]}\left[ (1-\gamma)^{(s-1)} \right] \;.\label{eq:exp_size}
			\end{align}
			The expected number of rounds is denoted by $\bar{n} = m\cdot \bar{s}$.
			
			Compared to the main text, we now have a RV $\tilde{W}_j\in\{0,1,\perp\}$ for each block, instead of each round. Alice and Bob set $\tilde{W}_j$ to be $0$ or $1$ depending on the result of the game in the block's test round (i.e., the last round of the block), or $\tilde{W}_j=\perp$ if a test round was not carried out in the block. By the definition of the blocks we have $\Pr[\tilde{W}_j=\perp]=(1-\gamma)^{s_{\max}}$. 
			
			\begin{algorithm}
			\caption{Modified entropy accumulation protocol}
			\label{pro:randomness_generation_mod}
			\begin{algorithmic}[1]
				\STATEx \textbf{Arguments:} 
					\STATEx\hspace{\algorithmicindent} $G$ -- two-player non-local game
					\STATEx\hspace{\algorithmicindent} $\mathcal{X}_g,\mathcal{X}_t \subset \mathcal{X}$ -- generation and test inputs for Alice
					\STATEx\hspace{\algorithmicindent} $\mathcal{Y}_g,\mathcal{Y}_t \subset \mathcal{Y}$ -- generation and test inputs for Bob
					\STATEx\hspace{\algorithmicindent} $D$ -- untrusted device of (at least) two components that can play $G$ repeatedly
					\STATEx\hspace{\algorithmicindent} $m\in \mathbb{N}_+$ -- number of blocks
					\STATEx\hspace{\algorithmicindent} $s_{\max} \in \mathbb{N}_+$ --maximal length of a block
					\STATEx\hspace{\algorithmicindent} $\gamma \in (0,1]$ -- probability of a test round 
					
					\STATEx\hspace{\algorithmicindent} $\omega_{\mathrm{exp}}$ -- expected winning probability in $G$ for an honest (perhaps noisy) implementation    
					\STATEx\hspace{\algorithmicindent} $\delta_{\mathrm{est}} \in (0,1)$ -- width of the statistical confidence interval for the estimation test
					
				\STATEx
				
				\STATE For every block $j\in[m]$ do Steps~\ref{prostep:ini_block}-\ref{prostep:calculate_Ci_EA_mod}:
					\STATE\hspace{\algorithmicindent} Set $i=0$ and $W_j=\perp$.\label{prostep:ini_block}
					\STATE\hspace{\algorithmicindent} If $i \leq s_{\max}$:
					
					\STATE\hspace{\algorithmicindent}\hspace{\algorithmicindent} Set $i=i+1$.
					
					\STATE\hspace{\algorithmicindent}\hspace{\algorithmicindent}Alice and Bob choose $T_i\in\{0,1\}$ at random such that $\Pr(T_i=1)=\gamma$. 
					\STATE\hspace{\algorithmicindent}\hspace{\algorithmicindent}If $T_i=0$ Alice and Bob choose inputs $X_i\in\mathcal{X}_g$ and $Y_i\in \mathcal{Y}_g$ respectively. If $T_i=1$ they choose inputs $X_i\in\mathcal{X}_t$ and $Y_i\in \mathcal{Y}_t$. 
					\STATE\hspace{\algorithmicindent}\hspace{\algorithmicindent}Alice and Bob use $D$ with $X_i,Y_i$ and record their outputs as $A_i$ and $B_i$ respectively. \label{prostep:measurementb}
					\STATE\hspace{\algorithmicindent}\hspace{\algorithmicindent}If $T_i=0$ Bob updates $B_i$ to $B_i = \perp$.
					\STATE\hspace{\algorithmicindent}\hspace{\algorithmicindent}If $T_i=1$ they set $\tilde{W}_j =w\left(A_i,B_i,X_i,Y_i\right)$ and $i=s_{\max}+1$.\label{prostep:calculate_Ci_EA_mod}

				\STATE Alice and Bob abort if $\sum_{j\in[m]} \tilde{W}_j < \left[\omega_{\mathrm{exp}} \left(1-(1-\gamma)^{s_{\max}}\right) - \delta_{\mathrm{est}}\right] \cdot m$. \label{prostep:abort_general_EA}
			\end{algorithmic}
			\end{algorithm}

		\subsection{Modified min-tradeoff function}

			Below, we apply the EAT on blocks of outputs instead of single rounds directly. Let $\mathcal{M}_j$ denote the EAT channels defined by the actions of Steps~\ref{prostep:ini_block}-\ref{prostep:calculate_Ci_EA_mod} in Protocol~\ref{pro:randomness_generation_mod} together with the behaviour of the device.
			It is easy to verify that~$\mathcal{M}_j$ fulfil the necessary conditions given in Definition~\ref{def:eat_channels}. 
			
			We now construct a min-tradeoff function for $\mathcal{M}_j$. Let $\tilde{p}$ be a probability distribution over $\{0,1,\perp\}$. Our goal is to find $F_{\min}$ such that 
			\begin{equation}\label{eq:eat_f_min_bound_mod}
				\forall j\in[m]\qquad	F_{\min}(\tilde{p}) \leq \inf_{\sigma_{R_{j-1}R'}:\mathcal{M}_j(\sigma)_{\tilde{W}_j}=\tilde{p}} H\left( \vec{A_j} \vec{B_j} | \vec{X_j} \vec{Y_j} \vec{T_j}  R' \right)_{\mathcal{M}_j(\sigma)} \;,
			\end{equation}
			where $\vec{A_j}$ is a vector of varying length (but at most $s_{\max}$). We use $A_{j,i}$ to denote the $i$'th entry of $\vec{A_j}$ and $A_{j,1}^{j,i-1} = A_{j,1}\dotsc A_{j,i-1}$. Since we will only be interested in the entropy of $\vec{A_j}$ we can also describe it as a vector of length $s_{\max}$ which is initialised to be all $\perp$. For every actual round being performed in the block the value of $A_{j,i}$ is updated. Thus, the entries of  $\vec{A_j}$ which correspond to rounds which were not performed do not contribute to the entropy. We use similar notation for the other vectors of RVs.

			To lower-bound the right-hand side of Equation~\eqref{eq:eat_f_min_bound_mod} we first use the chain rule
			\begin{align}\label{eq:block_cr}
				H\left( \vec{A_j} \vec{B_j} | \vec{X_j} \vec{Y_j} \vec{T_j}  R' \right) = \sum_{i\in [s_{\max}]}  H(A_{j,i}B_{j,i}|\vec{X_j} \vec{Y_j} \vec{T_j}  R' A_{j,1}^{j,i-1}B_{j,1}^{j,i-1} )\;.
			\end{align}
			Next, for every $i\in[s_{\max}]$,
			\begin{align*}
				&H(A_{j,i}B_{j,i}|\vec{X_j} \vec{Y_j} \vec{T_j}  R' A_{j,1}^{j,i-1}B_{j,1}^{j,i-1} ) =\\
				&\qquad \Pr[T_{j,1}^{j,i-1} = \vec{0}] H(A_{j,i}B_{j,i}|\vec{X_j} \vec{Y_j}  R' A_{j,1}^{j,i-1}B_{j,1}^{j,i-1} T_{j,i}^{j,s_{\max}} T_{j,1}^{j,i-1} = \vec{0} ) \\
				&\qquad+ \Pr[T_{j,1}^{j,i-1} \neq \vec{0}] H(A_{j,i}B_{j,i}|\vec{X_j} \vec{Y_j}  R' A_{j,1}^{j,i-1}B_{j,1}^{j,i-1} T_{j,i}^{j,s_{\max}} T_{j,1}^{j,i-1} \neq \vec{0} ) \\
				=&\qquad (1-\gamma)^{(i-1)} H(A_{j,i}B_{j,i}|\vec{X_j} \vec{Y_j}  R' A_{j,1}^{j,i-1}B_{j,1}^{j,i-1} T_{j,i}^{j,s_{\max}} T_{j,1}^{j,i-1} = \vec{0} )
			\end{align*}
			since the entropy is not zero only if the $i$'th round is being performed in the block, i.e., if a test was not performed before that round. Plugging this into Eq.~\eqref{eq:block_cr} we get
			\begin{align*}
				&H\left( \vec{A_j} \vec{B_j} | \vec{X_j} \vec{Y_j} \vec{T_j}  R' \right) \\
				& \qquad = \sum_{i\in [s_{\max}]} (1-\gamma)^{(i-1)} H(A_{j,i}B_{j,i}|\vec{X_j} \vec{Y_j}  R' A_{j,1}^{j,i-1}B_{j,1}^{j,i-1} T_{j,i}^{j,s_{\max}} T_{j,1}^{j,i-1} = \vec{0} )\;.
			\end{align*}

			Each term in the sum can now be identified as the entropy of a single round. We can therefore use the bound derived in the main text, as given in Equation~\eqref{eq:entropy_bound_for_min_tradeoff}. For this we denote by $\omega_i$ the winning probability in the $i$'th round (given that a test was not performed before). Then it holds that
			\begin{equation}\label{eq:to_mini}
				H\left( \vec{A_j} \vec{B_j} | \vec{X_j} \vec{Y_j} \vec{T_j}  R' \right) \geq \sum_{i\in [s_{\max}]} (1-\gamma)^{(i-1)} \left[1 - h\left( \frac{1}{2} + \frac{1}{2}\sqrt{16\omega_i \left(\omega_i-1\right) +3}  \right) \right] \;,
			\end{equation}
			where, by the actions of the EAT channel $\mathcal{M}_j$, the $\omega_i$'s must fulfil the constraint
			\begin{equation}\label{eq:constraint}
				\tilde{p}(1) = \sum_{i\in[s_{\max}]} \gamma (1-\gamma)^{(i-1)}\omega_i \;.
			\end{equation}
			
			Note that, similarly to what was done in the main text,  we only need to consider~$\tilde{p}$ for which $\tilde{p}(1) + \tilde{p}(0) = 1-(1-\gamma)^{s_{\max}}$ (otherwise the condition on the min-tradeoff function is trivial, as the infimum is over an empty set). 
			
			To find the min-tradeoff function $F_{\min}(\tilde{p})$ we therefore need to minimise Equation~\eqref{eq:to_mini} under the constraint of Equation~\eqref{eq:constraint}. The following lemma shows that the minimum is achieved when all $\omega_i$ are equal.
			
			\begin{lem}
				The minimum of the function given on the righthand-side of Equation~\eqref{eq:to_mini} over $\omega_i$ constrained by Equation~\eqref{eq:constraint} is achieved for $\omega^*_i=\frac{\tilde{p}(1)}{1-(1-\gamma)^{s_{\max}}}$ for all $i\in[s_{\max}]$.
			\end{lem}
			\begin{proof}
				Let $\vec{\omega} = \omega_1,\dotsc,\omega_{s_{\max}}$ and 
				\begin{align*}
					&f(\vec{\omega}) \equiv \sum_{i\in [s_{\max}]} (1-\gamma)^{(i-1)} \left[1 - h\left( \frac{1}{2} + \frac{1}{2}\sqrt{16\omega_i \left(\omega_i-1\right) +3}  \right) \right] \; ; \\
					&g(\vec{\omega}) \equiv \sum_{i\in[s_{\max}]} \gamma (1-\gamma)^{(i-1)}\omega_i  - \tilde{p}(1) \;. 
				\end{align*}
				Using the method of Lagrange multipliers, we should look for $\vec{\omega}^{*}$ such that $g(\vec{\omega}^{*})=0$ and $\nabla f (\vec{\omega}^{*})= - \lambda \nabla g (\vec{\omega}^{*})$ for some constant $\lambda$. $\nabla f (\vec{\omega}^{*})= - \lambda \nabla g (\vec{\omega}^{*})$ implies that for any $i$, 
				\[
					(1-\gamma)^{(i-1)} \frac{\mathrm{d}}{\mathrm{d}\omega_i}\left[1 - h\left( \frac{1}{2} + \frac{1}{2}\sqrt{16\omega_i \left(\omega_i-1\right) +3}  \right) \right] \Big|_{\omega^*_i}= - \lambda \gamma (1-\gamma)^{(i-1)} 
				\]
				and therefore
				\[
					 \frac{\mathrm{d}}{\mathrm{d}\omega_i}\left[1 - h\left( \frac{1}{2} + \frac{1}{2}\sqrt{16\omega_i \left(\omega_i-1\right) +3}  \right) \right] \Big|_{\omega^*_i}= -\lambda \gamma \;.
				\]
				The function on the left-hand side of the above equation is strictly increasing. Hence, it must be that all $\omega^*_i$ are equal to some constant $\omega^*$.
				
				Lastly,  we must have  $g(\vec{\omega}^{*})=0$. Thus,
				\[
					\sum_{i\in[s_{\max}]} \gamma (1-\gamma)^{(i-1)}\omega^*  - \tilde{p}(1) =0 
				\]
				which means
				\[
					\omega^* = \frac{\tilde{p}(1)}{\sum_{i\in[s_{\max}]} \gamma (1-\gamma)^{(i-1)}} =\frac{\tilde{p}(1)}{1-(1-\gamma)^{s_{\max}}} \;. \qedhere
				\]
			\end{proof}
			
			Plugging the minimal values of $\omega_i$ into Equation~\eqref{eq:to_mini} we get that
			\begin{align*}
				&H\left( \vec{A_j} \vec{B_j} | \vec{X_j} \vec{Y_j} \vec{T_j}  R' \right)  \\
				&\quad \geq \sum_{i\in [s_{\max}]} (1-\gamma)^{(i-1)} \left[1 - h\left( \frac{1}{2} + \frac{1}{2}\sqrt{16\frac{\tilde{p}(1)}{1-(1-\gamma)^{s_{\max}}} \left(\frac{\tilde{p}(1)}{1-(1-\gamma)^{s_{\max}}}-1\right) +3}  \right) \right] \\
				&\quad = \bar{s} \left[1 - h\left( \frac{1}{2} + \frac{1}{2}\sqrt{16\frac{\tilde{p}(1)}{1-(1-\gamma)^{s_{\max}}} \left(\frac{\tilde{p}(1)}{1-(1-\gamma)^{s_{\max}}}-1\right) +3}  \right) \right] \;,
			\end{align*}
			where we used Equation~\eqref{eq:exp_size} to get the last equality. 
			
			From this point we can follow the same steps as in Section~\ref{sec:crypto_min_tradeoff} (cutting and gluing the function etc.\@). The resulting min-tradeoff function is given by
			
			\begin{align}
				&g(\tilde{p}) =  \\
				& \begin{cases}
						 \bar{s} \left[1 - h\left( \frac{1}{2} + \frac{1}{2}\sqrt{16\frac{\tilde{p}(1)}{1-(1-\gamma)^{s_{\max}}} \left(\frac{\tilde{p}(1)}{1-(1-\gamma)^{s_{\max}}}-1\right) +3}  \right) \right] &  \frac{\tilde{p}(1)}{1-(1-\gamma)^{s_{\max}}}\in\left[0,\frac{2+\sqrt{2}}{4}\right] \\
						\bar{s}& \frac{\tilde{p}(1)}{1-(1-\gamma)^{s_{\max}}}\in\left[\frac{2+\sqrt{2}}{4},1\right]\;,
						\end{cases}\notag\\
				&F_{\min}\left(\tilde{p},\tilde{p}_t\right) = \begin{cases}
				g\left(\tilde{p}\right)&  \tilde{p}(1) \leq \tilde{p}_t(1) \;  \\
				\frac{\mathrm{d}}{\mathrm{d}\tilde{p}(1)} g(\tilde{p})\big|_{\tilde{p}_t}  \cdot \tilde{p}(1)+ \Big( g(\tilde{p}_t) -	\frac{\mathrm{d}}{\mathrm{d}\tilde{p}(1)} g(\tilde{p})\big|_{\tilde{p}_t} \cdot \tilde{p}_t(1) \Big)& \tilde{p}(1)> \tilde{p}_t(1)\;.
				\end{cases} \nonumber
			\end{align}
			The min-tradeoff function given above is effectively identical to the one derived in the main text; although it gives us a bound on the von Neumann entropy in a block, instead of a single round, this bound is exactly the expected length of a block, $\bar{s}$, times the entropy in one round.  For $s_{\max}=1$ the min-tradeoff function constructed in the main text is retrieved. 

		\subsection{Modified entropy rate}

			Since we apply the EAT on the blocks, the entropy rate is now defined to be the entropy \emph{per block}. We therefore get
			\begin{align*}
				\mu(\tilde{p},\tilde{p}_t,\varepsilon_{\mathrm{s}},\varepsilon_{\text{e}}) = & F_{\min}\left(\tilde{p}, \tilde{p}_t\right) \\
				& - \frac{1}{\sqrt{m}}2\left( \log (1+2\cdot 2^{s_{\max}} 3^{s_{\max}}) + \| \frac{\mathrm{d}}{\mathrm{d}\tilde{p}(1)} g(\tilde{p})\|_{\infty}  \right)\sqrt{1-2 \log (\varepsilon_{\mathrm{s}} \cdot \varepsilon_{\text{e}})}\;, \nonumber\\
				\mu_{\mathrm{opt}}(\varepsilon_{\mathrm{s}}, \varepsilon_{\text{e}}) = & \max_{\frac{3}{4} < \tilde{p}_t(1) < \frac{2+\sqrt{2}}{4}} \; \mu(\omega_{\mathrm{exp}}\left[ 1- (1-\gamma)^{s_{\max}} \right] - \delta_{\mathrm{est}},\tilde{p}_t,\varepsilon_{\mathrm{s}},\varepsilon_{\text{e}}) \;,
			\end{align*}
			and the total amount of entropy is given by
			\begin{equation}\label{eq:full_entropy_mod}
					 H^{\varepsilon_{\mathrm{s}}}_{\min} \left( \mr{A}\mr{B} | \mr{X}\mr{Y}\mr{T} E \right)_{\rho_{|\Omega}} > m\cdot \mu_{\mathrm{opt}}(\varepsilon_{\mathrm{s}},\varepsilon_{\mathrm{EA}}) = \frac{\bar{n}}{\bar{s}}\cdot \mu_{\mathrm{opt}}(\varepsilon_{\mathrm{s}},\varepsilon_{\mathrm{EA}})  \;.
			\end{equation}
			
			By choosing  $s_{\max} = \lceil\frac{1}{\gamma}\rceil$ the scaling of the entropy rate with $\gamma$ is better than the rate derived in the main text.
			In particular, a short calculation reveals that the second order term scales, roughly, as $\sqrt{\bar{n}/\gamma}$ instead of $\sqrt{n}/\gamma$.

		\subsection{Modified key rate}

			To get the final key rate we need to repeat the same steps from the main text, but this time applied to random variables of varying length. 
			
			For this we first observe that, with high probability, the actual number of rounds, $n$, cannot be much larger than the expected number of rounds $\bar{n}$. Let $S_i$ be the RV describing the length of block $i$, for $i\in[m]$, and $N$ the RV describing the total number of rounds. Then $N=S_1+\dots +S_m$. Since all the $S_i$ are independent, identical, and have values in $\left[1,\frac{1}{\gamma}\right]$ we have
			\[
				\Pr[N\geq \bar{n}+t] \leq \exp\left[ - \frac{2t^2\gamma^2}{m(1-\gamma)^2}\right] \;.
			\]
			Let $\varepsilon_t=\exp\left[ - \frac{2t^2\gamma^2}{m(1-\gamma)^2}\right]$ then
			\[
				t=\sqrt{-\frac{m(1-\gamma)^2\log\varepsilon_t}{2\gamma^2}} \;.
			\]
			
			The first step in the derivation of the key rate which needs to be changed is the one given in Equation~\eqref{eq:bound_max_entropy}. The quantity that needs to be upper bounded is $H^{\frac{\varepsilon_{\mathrm{s}}}{4}}_{\max} \left( \mr{B} | \mr{ T}  E N \right)_{\rho_{|\hat{\Omega}}} $; $N$ can be included in the entropy since its value is fixed by $\mathbf{T}$. 
			By the definition of the smooth max-entropy we have
			\[
				H^{\frac{\varepsilon_{\mathrm{s}}}{4}}_{\max} \left( \mr{B} | \mr{ T}  E N \right) \leq H^{\frac{\varepsilon_{\mathrm{s}}}{4} - \sqrt{\varepsilon_t}}_{\max} \left( \mr{B} | \mr{ T}  E N ,N\leq \bar{n}+t \right) \;.
			\]
			Following the same steps as in the proof of Lemma~\ref{lem:smooth_bound_qkd} we have
			\begin{equation*}
				\begin{split}
					H^{\frac{\varepsilon_{\mathrm{s}}}{4} - \sqrt{\varepsilon_t}}_{\max} \left( \mr{B} | \mr{ T}  E N ,N\leq \bar{n}+t \right) _{\rho_{|\hat{\Omega}}} < \\
					\gamma (\bar{n} + t) + \sqrt{\bar{n} + t} 2\log7\sqrt{1-2\log \left(( \varepsilon_{\mathrm{s}}/4 -\sqrt{\varepsilon_t} )\cdot \left(\varepsilon_{\mathrm{EA}} + \varepsilon_{\mathrm{EC}}\right) \right)} \;.
				\end{split}
			\end{equation*}
			
			With this modification and the modified entropy rate given in Equation~\eqref{eq:full_entropy_mod} we get
			\begin{equation*}
				\begin{split}
				H^{\varepsilon_{\mathrm{s}}}_{\min} \left( \mr{A} | \mr{X}\mr{Y}\mr{T} O E \right)_{\tilde{\rho}_{|\tilde{\Omega}}} \geq &\frac{\bar{n}}{\bar{s}} \cdot \mu_{\mathrm{opt}}\left(\varepsilon_{\mathrm{s}}/4,\varepsilon_{\mathrm{EA}} + \varepsilon_{\mathrm{EC}}\right) - \mathrm{leak_{EC}} \\
				& -  3 \log\left(1-\sqrt{1-(\varepsilon_{\mathrm{s}}/4)^2}\right)  - \gamma (\bar{n} + t) \\
				& - \sqrt{\bar{n} + t} 2\log7\sqrt{1-2\log \left(( \varepsilon_{\mathrm{s}}/4 -\sqrt{\varepsilon_t})\cdot \left(\varepsilon_{\mathrm{EA}} + \varepsilon_{\mathrm{EC}}\right) \right)}  \;.\qedhere
			\end{split}
			\end{equation*}
			
			Similarly, the amount of leakage due to the error correction step $ \mathrm{leak_{EC}}$ should be modified as well. Following the steps in Section~\ref{sec:leakage_ec_calc}, the quantity to be upper bounded is $H_{\max}^{\frac{\varepsilon'_{\mathrm{EC}}}{2}}\left(\mr{A}|\tilde{\mr{B}}\mr{X}\mr{Y}\mr{T}N\right)$. Here as well we have
			\[
				H_{\max}^{\frac{\varepsilon'_{\mathrm{EC}}}{2}}\left(\mr{A}|\tilde{\mr{B}}\mr{X}\mr{Y}\mr{T}N\right) \leq H_{\max}^{\frac{\varepsilon'_{\mathrm{EC}}}{2}-\sqrt{\varepsilon_t}}\left(\mr{A}|\tilde{\mr{B}}\mr{X}\mr{Y}\mr{T}N,N\leq \bar{n}+t\right) \;.
			\]
			The asymptotic equipartition property can be used with the maximal length $\bar{n}+t$ to get
			\begin{equation*}
				\begin{split}
					&H_{\max}^{\frac{\varepsilon'_{\mathrm{EC}}}{2}-\sqrt{\varepsilon_t}}\left(\mr{A}|\tilde{\mr{B}}\mr{X}\mr{Y}\mr{T}N,N\leq \bar{n}+t\right) \\
					&\qquad \leq (\bar{n}+t) \cdot H(A_i|\tilde{B}_iX_i Y_i T_i) + \sqrt{\bar{n}+t}\; \delta (\varepsilon'_{\mathrm{EC}}-2\sqrt{\varepsilon_t}, \tau) \;,
				\end{split}
			\end{equation*}
			for 
			\begin{align*}
				&\tau =2 \sqrt{2^{H_{\max}(A_i|\tilde{B}_iX_i Y_i T_i)}}+1\\
				&\delta (\varepsilon'_{\mathrm{EC}}-2\sqrt{\varepsilon_t},\tau) = 4\log \tau \sqrt{2 \log \left(8/ (\varepsilon'_{\mathrm{EC}}-2\sqrt{\varepsilon_t})^2\right)} \;.
			\end{align*}
			Continuing exactly as in Section~\ref{sec:leakage_ec_calc} we get
			\begin{equation*}
				\begin{split}
					\mathrm{leak_{EC}} \leq (\bar{n}+t) \cdot  \left[\left( 1-\gamma \right)  h(Q) + \gamma h(\omega_{\mathrm{exp}}) \right] \\
					+ \sqrt{\bar{n}+t} \; 4\log \left(2\sqrt{2} +1\right) \sqrt{2 \log \left(8/ (\varepsilon'_{\mathrm{EC}}-2\sqrt{\varepsilon_t})^2\right)}\\
					+ \log \left( 8/\varepsilon'^2_{\mathrm{EC}} + 2/\left(2-\varepsilon'_{\mathrm{EC}}\right)\right) + \log\left(\frac{1}{\varepsilon_{\mathrm{EC}}}\right) \;.
				\end{split}
			\end{equation*}
			
			The parameter $\varepsilon_t$  should be chosen such that the key rate is optimised. The resulting key rates are shown in Figures~\ref{fig:qkd_rates_Q_mod} and~\ref{fig:qkd_rates_n_mod} in the main text.

	\section{Summary of parameters and variables}\label{appsec:qkd_param_summ}

		\begin{table}[H]
			\begin{center}
			\begin{tabular}{| l | l | } 
			\hline
				Symbol & Meaning  \\ \hhline{|=|=|}
				$n\in\mathbb{N}_+$ & Number of rounds   \\ \hline
				$\gamma\in(0,1]$ & Expected fraction of Bell violation estimation rounds  \\ \hline
				$\omega_{\mathrm{exp}}\in[0,1]$ & \specialcell{Expected winning probability in an honest  (perhaps  \\ noisy) implementation}  \\ \hline
				$\delta_{\mathrm{est}} \in (0,1)$ & \specialcell{Width of the statistical confidence interval \\ for the Bell violation estimation test}  \\ \hline
				$\varepsilon_{\text{s}}$ & Smoothing parameter  \\ \hline
				$\varepsilon^c_{EA} $ & Completeness error of the entropy accumulation protocol \\ \hline
				$\varepsilon_{\mathrm{EA}}$ & The error probability of the entropy accumulation protocol  \\ \hline
				$\mathrm{leak_{EC}}$ & The leakage of the error correction protocol  \\ \hline
				$\varepsilon_{\mathrm{EC}},\varepsilon'_{\mathrm{EC}}$ & Error probabilities of the error correction protocol  \\ \hline
				$\varepsilon_{\mathrm{EC}}^c$ & Completeness error of the error correction protocol \\ \hline
				$\varepsilon_{\mathrm{PE}}^c$ & Completeness error of the parameter estimation step \\ \hline
				$\varepsilon_{\mathrm{PA}}$ & Error probability of the privacy amplification protocol \\ \hline
				$\ell$ & Final key length in the DIQKD protocol \\ \hline
				$\varepsilon^c_{QKD}$ & Completeness error of the DIQKD protocol  \\ \hline
				$\varepsilon^s_{QKD}$ & Soundness error of the DIQKD protocol  \\ \hline
			 \end{tabular}
			\end{center}\caption{Parameters used in Chapter~\ref{ch:crypto_showcase}}\label{tb:parameters}
		\end{table}

		\begin{table}[H]
			\begin{center}
			\begin{tabular}{| l | l |} 
			\hline
				Random variables and systems & Meaning \\ \hhline{|=|=|}
				$X_i\in\{0,1\}$ & Alice's input in round $i\in[n]$ \\ \hline
				$Y_i\in\{0,1\}$ & Bob's input in round $i\in[n]$ \\ \hline
				$A_i\in\{0,1\}$ & Alice's output in round $i\in[n]$ \\ \hline
				$B_i\in\{0,1,\perp\}$, $\tilde{B}_i\in\{0,1\}$  & Bob's output in round $i\in[n]$ \\ \hline
				$T_i\in\{0,1\}$ & \specialcell{Indicator of the estimation test in round $i$: \\ 
							$T_i = \begin{cases}
							0 &\text{$i$'th round is not a test round} \\
							1 &\text{$i$'th round is a test round} 
							\end{cases}$} \\ \hline
				$W_i \in \{\perp,0,1\}$ & \specialcell{Indicator of the correlation in the test rounds: \\ 
							$W_i = \begin{cases}
							\perp &T_i =0 \\
							0 &\text{$T_i=1$ and the test fails} \\
							1 &\text{$T_i=1$ and the test succeeds} \;.
							\end{cases}$} \\ \hhline{|=|=|}
				$E$ & Register of Eve's quantum state \\ \hline
				$R_i$ & \specialcell{Register of the (unknown) quantum state $\rho_{Q_AQ_B}^i$ \\ of Alice and Bob's devices after step $i$ \\ of the protocol, for $i\in\{0\}\cup[n]$.} \\ \hline
			 \end{tabular}
			\end{center}\caption{Random variables and quantum systems used in Chapter~\ref{ch:crypto_showcase}}\label{tb:RV_table}
		\end{table}

\addcontentsline{toc}{chapter}{Bibliography}
\bibliographystyle{apalike}
\bibliography{thesis}

\begin{thebibliography}{}

\bibitem[Ac{\'\i}n et~al., 2007]{acin2007device}
Ac{\'\i}n, A., Brunner, N., Gisin, N., Massar, S., Pironio, S., and Scarani, V.
  (2007).
\newblock Device-independent security of quantum cryptography against
  collective attacks.
\newblock {\em Physical Review Letters}, 98(23):230501.

\bibitem[Ac{\'\i}n et~al., 2006a]{acin2006bell}
Ac{\'\i}n, A., Gisin, N., and Masanes, L. (2006a).
\newblock From {B}ell's theorem to secure quantum key distribution.
\newblock {\em Physical review letters}, 97(12):120405.

\bibitem[Ac{\'\i}n et~al., 2006b]{acin2006efficient}
Ac{\'\i}n, A., Massar, S., and Pironio, S. (2006b).
\newblock Efficient quantum key distribution secure against no-signalling
  eavesdroppers.
\newblock {\em New Journal of Physics}, 8(8):126.

\bibitem[Ac{\'\i}n et~al., 2012]{acin2012randomness}
Ac{\'\i}n, A., Massar, S., and Pironio, S. (2012).
\newblock Randomness versus nonlocality and entanglement.
\newblock {\em Physical review letters}, 108(10):100402.

\bibitem[Aharon et~al., 2015]{aharon2015device}
Aharon, N., Massar, S., Pironio, S., and Silman, J. (2015).
\newblock Device-independent bit commitment based on the {CHSH} inequality.
\newblock {\em arXiv preprint arXiv:1511.06283}.

\bibitem[Arnon-Friedman and Bancal, 2017]{arnon2017device}
Arnon-Friedman, R. and Bancal, J.-D. (2017).
\newblock Device-independent certification of one-shot distillable
  entanglement.
\newblock {\em arXiv preprint arXiv:1712.09369}.

\bibitem[Arnon-Friedman et~al., 2018]{arnon2018practical}
Arnon-Friedman, R., Dupuis, F., Fawzi, O., Renner, R., and Vidick, T. (2018).
\newblock Practical device-independent quantum cryptography via entropy
  accumulation.
\newblock {\em Nature communications}, 9(1):459.

\bibitem[Arnon-Friedman et~al., 2016a]{arnon2015quantum}
Arnon-Friedman, R., Portmann, C., and Scholz, V.~B. (2016a).
\newblock Quantum-proof multi-source randomness extractors in the markov model.
\newblock In {\em 11th Conference on the Theory of Quantum Computation,
  Communication and Cryptography}, page~1.

\bibitem[Arnon-Friedman and Renner, 2015]{arnon2013finetti}
Arnon-Friedman, R. and Renner, R. (2015).
\newblock de {F}inetti reductions for correlations.
\newblock {\em Journal of Mathematical Physics}, 56(5):052203.

\bibitem[Arnon-Friedman et~al., 2016b]{arnon2016non}
Arnon-Friedman, R., Renner, R., and Vidick, T. (2016b).
\newblock Non-signaling parallel repetition using de finetti reductions.
\newblock {\em IEEE Transactions on Information Theory}, 62(3):1440--1457.

\bibitem[Arnon-Friedman et~al., 2016c]{arnon2016simple}
Arnon-Friedman, R., Renner, R., and Vidick, T. (2016c).
\newblock Simple and tight device-independent security proofs.
\newblock {\em arXiv preprint arXiv:1607.01797}.

\bibitem[Arnon-Friedman and Ta-Shma, 2012]{arnon2012limits}
Arnon-Friedman, R. and Ta-Shma, A. (2012).
\newblock Limits of privacy amplification against nonsignaling memory attacks.
\newblock {\em Physical Review A}, 86(6):062333.

\bibitem[Arnon-Friedman and Yuen, 2018]{arnon2017noise}
Arnon-Friedman, R. and Yuen, H. (2018).
\newblock Noise-tolerant testing of high entanglement of formation.
\newblock {\em International Colloquium of Automata, Languages, and
  Programming}.

\bibitem[Bamps et~al., 2017]{bamps2017device}
Bamps, C., Massar, S., and Pironio, S. (2017).
\newblock Device-independent randomness generation with sublinear shared
  quantum resources.
\newblock {\em arXiv preprint arXiv:1704.02130}.

\bibitem[Bancal et~al., 2015]{bancal2015physical}
Bancal, J.-D., Navascu{\'e}s, M., Scarani, V., V{\'e}rtesi, T., and Yang, T.~H.
  (2015).
\newblock Physical characterization of quantum devices from nonlocal
  correlations.
\newblock {\em Physical Review A}, 91(2):022115.

\bibitem[Barrett, 2007]{barrett2007information}
Barrett, J. (2007).
\newblock Information processing in generalized probabilistic theories.
\newblock {\em Physical Review A}, 75(3):032304.

\bibitem[Barrett et~al., 2013]{barrett2013memory}
Barrett, J., Colbeck, R., and Kent, A. (2013).
\newblock Memory attacks on device-independent quantum cryptography.
\newblock {\em Physical review letters}, 110(1):010503.

\bibitem[Barrett et~al., 2002]{barrett2002quantum}
Barrett, J., Collins, D., Hardy, L., Kent, A., and Popescu, S. (2002).
\newblock Quantum nonlocality, bell inequalities, and the memory loophole.
\newblock {\em Physical Review A}, 66(4):042111.

\bibitem[Barrett et~al., 2005]{barrett2005no}
Barrett, J., Hardy, L., and Kent, A. (2005).
\newblock No signaling and quantum key distribution.
\newblock {\em Physical Review Letters}, 95(1):010503.

\bibitem[Barrett and Leifer, 2009]{barrett2009finetti}
Barrett, J. and Leifer, M. (2009).
\newblock The de {F}inetti theorem for test spaces.
\newblock {\em New Journal of Physics}, 11(3):033024.

\bibitem[Bavarian et~al., 2017a]{bavarian2015anchoring}
Bavarian, M., Vidick, T., and Yuen, H. (2017a).
\newblock Anchoring games for parallel repetition.
\newblock {\em Symposium on the Theory of Computing}.

\bibitem[Bavarian et~al., 2017b]{bavarian2016parallel}
Bavarian, M., Vidick, T., and Yuen, H. (2017b).
\newblock Parallel repetition via fortification: analytic view and the quantum
  case.
\newblock {\em Innovations in Theoretical Computer Science}.

\bibitem[Beaudry, 2015]{beaudry2015assumptions}
Beaudry, N.~J. (2015).
\newblock Assumptions in quantum cryptography.
\newblock {\em arXiv preprint arXiv:1505.02792}.

\bibitem[Bell, 1964]{bell1964einstein}
Bell, J.~S. (1964).
\newblock On the {E}instein-{P}odolsky-{R}osen paradox.
\newblock {\em Physics}, 1(3):195--200.

\bibitem[Ben-Or and Mayers, 2004]{ben2004general}
Ben-Or, M. and Mayers, D. (2004).
\newblock General security definition and composability for quantum \&
  classical protocols.
\newblock {\em arXiv preprint quant-ph/0409062}.

\bibitem[Bennett and Brassard, 1984]{bennett1984proceedings}
Bennett, C.~H. and Brassard, G. (1984).
\newblock Proceedings of the ieee international conference on computers,
  systems, and signal processing, bangalore, india, 1984.

\bibitem[Berta et~al., 2011]{berta2011reverse}
Berta, M., Christandl, M., and Renner, R. (2011).
\newblock The quantum reverse {S}hannon {T}heorem based on one-shot information
  theory.
\newblock {\em Communications in Mathematical Physics}, 306(3):579--615.

\bibitem[Brandao and Harrow, 2013]{brandao2013quantum}
Brandao, F.~G. and Harrow, A.~W. (2013).
\newblock Quantum de {F}inetti theorems under local measurements with
  applications.
\newblock In {\em Proceedings of the forty-fifth annual ACM symposium on Theory
  of computing}, pages 861--870. ACM.

\bibitem[Brand{\~a}o et~al., 2016]{brandao2016realistic}
Brand{\~a}o, F.~G., Ramanathan, R., Grudka, A., Horodecki, K., Horodecki, M.,
  Horodecki, P., Szarek, T., and Wojew{\'o}dka, H. (2016).
\newblock Realistic noise-tolerant randomness amplification using finite number
  of devices.
\newblock {\em Nature communications}, 7:11345.

\bibitem[Brassard and Salvail, 1993]{brassard1993secret}
Brassard, G. and Salvail, L. (1993).
\newblock Secret-key reconciliation by public discussion.
\newblock In {\em advances in Cryptology EUROCRYPT 93}, pages 410--423.
  Springer.

\bibitem[Brunner et~al., 2014]{brunner2014bell}
Brunner, N., Cavalcanti, D., Pironio, S., Scarani, V., and Wehner, S. (2014).
\newblock Bell nonlocality.
\newblock {\em Reviews of Modern Physics}, 86(2):419.

\bibitem[Buhrman et~al., 2013]{buhrman2013parallel}
Buhrman, H., Fehr, S., and Schaffner, C. (2013).
\newblock On the parallel repetition of multi-player games: The no-signaling
  case.
\newblock {\em arXiv preprint arXiv:1312.7455}.

\bibitem[Canetti, 2001]{canetti2001universally}
Canetti, R. (2001).
\newblock Universally composable security: A new paradigm for cryptographic
  protocols 2005.
\newblock {\em Revision 3 of ECCC Report}.

\bibitem[Caves et~al., 2002]{caves2002unknown}
Caves, C.~M., Fuchs, C.~A., and Schack, R. (2002).
\newblock Unknown quantum states: the quantum de-{F}inetti representation.
\newblock {\em Journal of Mathematical Physics}, 43:4537.

\bibitem[Chiribella et~al., 2010]{chiribella2010probabilistic}
Chiribella, G., D'Ariano, G.~M., and Perinotti, P. (2010).
\newblock Probabilistic theories with purification.
\newblock {\em Physical Review A}, 81(6):062348.

\bibitem[Christandl et~al., 2007]{christandl2007one}
Christandl, M., K{\"o}nig, R., Mitchison, G., and Renner, R. (2007).
\newblock One-and-a-half quantum de {F}inetti theorems.
\newblock {\em Communications in Mathematical Physics}, 273(2):473--498.

\bibitem[Christandl et~al., 2009]{christandl2009postselection}
Christandl, M., K{\"o}nig, R., and Renner, R. (2009).
\newblock Postselection technique for quantum channels with applications to
  quantum cryptography.
\newblock {\em Physical Review Letters}, 102(2):020504.

\bibitem[Christandl and Renner, 2012]{christandl2012reliable}
Christandl, M. and Renner, R. (2012).
\newblock Reliable quantum state tomography.
\newblock {\em Physical Review Letters}, 109(12):120403.

\bibitem[Christandl and Toner, 2009]{christandl2009finite}
Christandl, M. and Toner, B. (2009).
\newblock Finite de {F}inetti theorem for conditional probability distributions
  describing physical theories.
\newblock {\em Journal of Mathematical Physics}, 50:042104.

\bibitem[Chung et~al., 2014]{chung2014physical}
Chung, K.-M., Shi, Y., and Wu, X. (2014).
\newblock Physical randomness extractors: Generating random numbers with
  minimal assumptions.
\newblock {\em arXiv preprint arXiv:1402.4797}.

\bibitem[Chung et~al., 2015]{chung2015parallel}
Chung, K.-M., Wu, X., and Yuen, H. (2015).
\newblock Parallel repetition for entangled k-player games via fast quantum
  search.
\newblock In {\em Proceedings of the 30th Conference on Computational
  Complexity}, pages 512--536. Schloss Dagstuhl--Leibniz-Zentrum fuer
  Informatik.

\bibitem[Clauser et~al., 1969]{clauser1969proposed}
Clauser, J.~F., Horne, M.~A., Shimony, A., and Holt, R.~A. (1969).
\newblock Proposed experiment to test local hidden-variable theories.
\newblock {\em Physical review letters}, 23(15):880.

\bibitem[Cleve et~al., 2008]{cleve2008perfect}
Cleve, R., Slofstra, W., Unger, F., and Upadhyay, S. (2008).
\newblock Perfect parallel repetition theorem for quantum xor proof systems.
\newblock {\em Computational Complexity}, 17(2):282--299.

\bibitem[Coladangelo et~al., 2017]{coladangelo2017verifier}
Coladangelo, A., Grilo, A., Jeffery, S., and Vidick, T. (2017).
\newblock Verifier-on-a-leash: new schemes for verifiable delegated quantum
  computation, with quasilinear resources.
\newblock {\em arXiv preprint arXiv:1708.07359}.

\bibitem[Colbeck, 2006]{Colbeck09}
Colbeck, R. (2006).
\newblock {\em Quantum And Relativistic Protocols For Secure Multi-Party
  Computation}.
\newblock PhD thesis, Trinity College, University of Cambridge.

\bibitem[Colbeck and Renner, 2012]{colbeck2012free}
Colbeck, R. and Renner, R. (2012).
\newblock Free randomness can be amplified.
\newblock {\em Nature Physics}, 8(6):450--453.

\bibitem[Coudron and Yuen, 2013]{coudron2013infinite}
Coudron, M. and Yuen, H. (2013).
\newblock Infinite randomness expansion and amplification with a constant
  number of devices.
\newblock {\em arXiv preprint arXiv:1310.6755}.

\bibitem[Cover and Thomas, 2012]{cover2012elements}
Cover, T.~M. and Thomas, J.~A. (2012).
\newblock {\em Elements of information theory}.
\newblock John Wiley \& Sons.

\bibitem[De et~al., 2012]{de2012trevisan}
De, A., Portmann, C., Vidick, T., and Renner, R. (2012).
\newblock Trevisan's extractor in the presence of quantum side information.
\newblock {\em SIAM Journal on Computing}, 41(4):915--940.

\bibitem[de~Finetti, 1969]{deFinetti69}
de~Finetti, B. (1969).
\newblock Sulla proseguibilit\`a di processi aleatori scambiabili.
\newblock {\em Rend. Matem. Trieste}, pages 53--67.

\bibitem[Devetak and Winter, 2005]{devetak2005distillation}
Devetak, I. and Winter, A. (2005).
\newblock Distillation of secret key and entanglement from quantum states.
\newblock In {\em Proceedings of the Royal Society of London A: Mathematical,
  Physical and Engineering Sciences}, volume 461, pages 207--235. The Royal
  Society.

\bibitem[Diaconis and Freedman, 1980a]{DiaconisF80finite}
Diaconis, P. and Freedman, D. (1980a).
\newblock Finite exchangeable sequences.
\newblock {\em The Annals of Probability}, pages 745--764.

\bibitem[Diaconis and Freedman, 1980b]{diaconis1980finite}
Diaconis, P. and Freedman, D. (1980b).
\newblock Finite exchangeable sequences.
\newblock {\em The Annals of Probability}, pages 745--764.

\bibitem[Dinur et~al., 2015]{dinur2015parallel}
Dinur, I., Steurer, D., and Vidick, T. (2015).
\newblock A parallel repetition theorem for entangled projection games.
\newblock {\em computational complexity}, 24(2):201--254.

\bibitem[Dupuis and Fawzi, 2018]{dupuis2018entropy}
Dupuis, F. and Fawzi, O. (2018).
\newblock Entropy accumulation with improved second-order.
\newblock {\em arXiv preprint arXiv:1805.11652}.

\bibitem[Dupuis et~al., 2016]{dupuis2016entropy}
Dupuis, F., Fawzi, O., and Renner, R. (2016).
\newblock Entropy accumulation.
\newblock {\em arXiv preprint arXiv:1607.01796}.

\bibitem[Einstein et~al., 1935]{einstein1935can}
Einstein, A., Podolsky, B., and Rosen, N. (1935).
\newblock Can quantum-mechanical description of physical reality be considered
  complete?
\newblock {\em Physical review}, 47(10):777.

\bibitem[Ekert and Renner, 2014]{ekert2014ultimate}
Ekert, A. and Renner, R. (2014).
\newblock The ultimate physical limits of privacy.
\newblock {\em Nature}, 507(7493):443--447.

\bibitem[Ekert, 1991]{ekert1991quantum}
Ekert, A.~K. (1991).
\newblock Quantum cryptography based on {B}ell's theorem.
\newblock {\em Physical review letters}, 67(6):661.

\bibitem[Fehr and Schaffner, 2008]{fehr2008randomness}
Fehr, S. and Schaffner, C. (2008).
\newblock Randomness extraction via $\delta$-biased masking in the presence of
  a quantum attacker.
\newblock In {\em Theory of Cryptography Conference}, pages 465--481. Springer.

\bibitem[Feige, 1991]{feige1991success}
Feige, U. (1991).
\newblock On the success probability of the two provers in one-round proof
  systems.
\newblock In {\em 1991 Proceedings of the Sixth Annual Structure in Complexity
  Theory Conference}, pages 116--123. IEEE.

\bibitem[Fu and Miller, 2018]{fu2018local}
Fu, H. and Miller, C.~A. (2018).
\newblock Local randomness: Examples and application.
\newblock {\em Physical Review A}, 97(3):032324.

\bibitem[Fung et~al., 2007]{fung2007phase}
Fung, C.-H.~F., Qi, B., Tamaki, K., and Lo, H.-K. (2007).
\newblock Phase-remapping attack in practical quantum-key-distribution systems.
\newblock {\em Phys. Rev. A}, 75(3):032314.

\bibitem[Gallego et~al., 2013a]{gallego2013full}
Gallego, R., Masanes, L., De~La~Torre, G., Dhara, C., Aolita, L., and
  Ac{\'\i}n, A. (2013a).
\newblock Full randomness from arbitrarily deterministic events.
\newblock {\em Nature communications}, 4:2654.

\bibitem[Gallego et~al., 2013b]{gallego6514full}
Gallego, R., Masanes, L., De~La~Torre, G., Dhara, C., Aolita, L., and
  Ac{\'\i}n, A. (2013b).
\newblock Full randomness from arbitrarily deterministic events.
\newblock {\em Nature communications}, 4.

\bibitem[Gerhardt et~al., 2011]{gerhardt2011full}
Gerhardt, I., Liu, Q., Lamas-Linares, A., Skaar, J., Kurtsiefer, C., and
  Makarov, V. (2011).
\newblock Full-field implementation of a perfect eavesdropper on a quantum
  cryptography system.
\newblock {\em Nature communications}, 2:349.

\bibitem[Gheorghiu et~al., 2015]{gheorghiu2015robustness}
Gheorghiu, A., Kashefi, E., and Wallden, P. (2015).
\newblock Robustness and device independence of verifiable blind quantum
  computing.
\newblock {\em New Journal of Physics}, 17(8):083040.

\bibitem[Giustina et~al., 2015]{giustina2015significant}
Giustina, M., Versteegh, M.~A., Wengerowsky, S., Handsteiner, J., Hochrainer,
  A., Phelan, K., Steinlechner, F., Kofler, J., Larsson, J.-{\AA}.,
  Abell{\'a}n, C., et~al. (2015).
\newblock Significant-loophole-free test of {B}ell's theorem with entangled
  photons.
\newblock {\em Physical review letters}, 115(25):250401.

\bibitem[Gottesman, 2010]{gottesman2010introduction}
Gottesman, D. (2010).
\newblock An introduction to quantum error correction and fault-tolerant
  quantum computation.
\newblock In {\em Quantum information science and its contributions to
  mathematics, Proceedings of Symposia in Applied Mathematics}, volume~68,
  pages 13--58.

\bibitem[Hajdu{\v{s}}ek et~al., 2015]{hajduvsek2015device}
Hajdu{\v{s}}ek, M., P{\'e}rez-Delgado, C.~A., and Fitzsimons, J.~F. (2015).
\newblock Device-independent verifiable blind quantum computation.
\newblock {\em arXiv preprint arXiv:1502.02563}.

\bibitem[H{\"a}nggi, 2010]{hanggi2010device_thesis}
H{\"a}nggi, E. (2010).
\newblock {\em Device-independent quantum key distribution}.
\newblock PhD thesis.

\bibitem[H{\"a}nggi and Renner, 2010]{hanggi2010device}
H{\"a}nggi, E. and Renner, R. (2010).
\newblock Device-independent quantum key distribution with commuting
  measurements.
\newblock {\em arXiv preprint arXiv:1009.1833}.

\bibitem[H{\"a}nggi et~al., 2009]{hanggi2009ns}
H{\"a}nggi, E., Renner, R., and Wolf, S. (2009).
\newblock Quantum cryptography based solely on bell's theorem.
\newblock {\em arXiv preprint arXiv:0911.4171}.

\bibitem[H{\"a}nggi et~al., 2010a]{hanggi2010efficient}
H{\"a}nggi, E., Renner, R., and Wolf, S. (2010a).
\newblock Efficient device-independent quantum key distribution.
\newblock In {\em Advances in Cryptology--EUROCRYPT 2010}, pages 216--234.
  Springer.

\bibitem[H{\"a}nggi et~al., 2010b]{hanggi2009quantum}
H{\"a}nggi, E., Renner, R., and Wolf, S. (2010b).
\newblock Efficient device-independent quantum key distribution.
\newblock In {\em Advances in Cryptology--EUROCRYPT 2010}, pages 216--234.
  Springer.

\bibitem[H{\aa}stad, 2001]{haastad2001some}
H{\aa}stad, J. (2001).
\newblock Some optimal inapproximability results.
\newblock {\em Journal of the ACM (JACM)}, 48(4):798--859.

\bibitem[Hayden et~al., 2004]{hayden2004structure}
Hayden, P., Jozsa, R., Petz, D., and Winter, A. (2004).
\newblock Structure of states which satisfy strong subadditivity of quantum
  entropy with equality.
\newblock {\em Communications in mathematical physics}, 246(2):359--374.

\bibitem[Hensen et~al., 2015]{hensen2015loophole}
Hensen, B., Bernien, H., Dr{\'e}au, A., Reiserer, A., Kalb, N., Blok, M.,
  Ruitenberg, J., Vermeulen, R., Schouten, R., Abell{\'a}n, C., et~al. (2015).
\newblock Loophole-free {B}ell inequality violation using electron spins
  separated by 1.3 kilometres.
\newblock {\em Nature}, 526(7575):682--686.

\bibitem[Holenstein, 2007]{holenstein2007parallel}
Holenstein, T. (2007).
\newblock Parallel repetition: simplifications and the no-signaling case.
\newblock In {\em Proceedings of the thirty-ninth annual ACM symposium on
  Theory of computing}, pages 411--419. ACM.

\bibitem[Holenstein and Renner, 2011]{holenstein2011randomness}
Holenstein, T. and Renner, R. (2011).
\newblock On the randomness of independent experiments.
\newblock {\em IEEE transactions on information theory}, 57(4):1865--1871.

\bibitem[Holmgren and Yang, 2017]{holmgren2017counterexample}
Holmgren, J. and Yang, L. (2017).
\newblock (a counterexample to) parallel repetition for non-signaling
  multi-player games.
\newblock In {\em Electronic Colloquium on Computational Complexity (ECCC)},
  volume~24, page 178.

\bibitem[Horodecki and Ramanathan, 2016]{horodecki2016relativistic}
Horodecki, P. and Ramanathan, R. (2016).
\newblock Relativistic causality vs. no-signaling as the limiting paradigm for
  correlations in physical theories.
\newblock {\em arXiv preprint arXiv:1611.06781}.

\bibitem[Ito, 2010]{ito2010polynomial}
Ito, T. (2010).
\newblock Polynomial-space approximation of no-signaling provers.
\newblock In {\em Automata, Languages and Programming}, pages 140--151.
  Springer.

\bibitem[Jain et~al., 2017]{jain2017parallel}
Jain, R., Miller, C.~A., and Shi, Y. (2017).
\newblock Parallel device-independent quantum key distribution.
\newblock {\em arXiv preprint arXiv:1703.05426}.

\bibitem[Kaniewski and Wehner, 2016]{kaniewski2016device}
Kaniewski, J. and Wehner, S. (2016).
\newblock Device-independent two-party cryptography secure against sequential
  attacks.
\newblock {\em New Journal of Physics}, 18(5):055004.

\bibitem[Kempe and Regev, 2010]{kempe2010no}
Kempe, J. and Regev, O. (2010).
\newblock No strong parallel repetition with entangled and non-signaling
  provers.
\newblock In {\em Computational Complexity (CCC), 2010 IEEE 25th Annual
  Conference on}, pages 7--15. IEEE.

\bibitem[Kempe et~al., 2010]{kempe2010unique}
Kempe, J., Regev, O., and Toner, B. (2010).
\newblock Unique games with entangled provers are easy.
\newblock {\em SIAM Journal on Computing}, 39(7):3207--3229.

\bibitem[Kempe and Vidick, 2011]{kempe2011parallel}
Kempe, J. and Vidick, T. (2011).
\newblock Parallel repetition of entangled games.
\newblock In {\em Proceedings of the forty-third annual ACM symposium on Theory
  of computing}, pages 353--362. ACM.

\bibitem[Kessler and Arnon-Friedman, 2017]{kessler2017device}
Kessler, M. and Arnon-Friedman, R. (2017).
\newblock Device-independent randomness amplification and privatization.
\newblock {\em arXiv preprint arXiv:1705.04148}.

\bibitem[Kitaev, 1997]{kitaev1997quantum}
Kitaev, A.~Y. (1997).
\newblock Quantum computations: algorithms and error correction.
\newblock {\em Russian Mathematical Surveys}, 52(6):1191--1249.

\bibitem[K{\"o}nig and Renner, 2005]{konig2005finetti}
K{\"o}nig, R. and Renner, R. (2005).
\newblock A de finetti representation for finite symmetric quantum states.
\newblock {\em Journal of Mathematical physics}, 46(12):122108.

\bibitem[Konig et~al., 2009]{konig2009operational}
Konig, R., Renner, R., and Schaffner, C. (2009).
\newblock The operational meaning of min-and max-entropy.
\newblock {\em IEEE Transactions on Information theory}, 55(9):4337--4347.

\bibitem[Konig and Terhal, 2008]{konig2008bounded}
Konig, R.~T. and Terhal, B.~M. (2008).
\newblock The bounded-storage model in the presence of a quantum adversary.
\newblock {\em IEEE Transactions on Information Theory}, 54(2):749--762.

\bibitem[Lancien and Winter, 2016]{lancien2016parallel}
Lancien, C. and Winter, A. (2016).
\newblock Parallel repetition and concentration for (sub-)no-signalling games
  via a flexible constrained de finetti reduction.
\newblock {\em Chicago Journal of Theoretical Computer Science}, (11).

\bibitem[Lancien and Winter, 2017]{lancien2017flexible}
Lancien, C. and Winter, A. (2017).
\newblock Flexible constrained de finetti reductions and applications.
\newblock {\em Journal of Mathematical Physics}, 58(9):092203.

\bibitem[Leverrier, 2014]{leverrier2014composable}
Leverrier, A. (2014).
\newblock Composable security proof for continuous-variable quantum key
  distribution with coherent states.
\newblock {\em arXiv preprint arXiv:1408.5689}.

\bibitem[Lin et~al., 2018]{lin2018device}
Lin, P.-S., Rosset, D., Zhang, Y., Bancal, J.-D., and Liang, Y.-C. (2018).
\newblock Device-independent point estimation from finite data and its
  application to device-independent property estimation.
\newblock {\em Physical Review A}, 97(3):032309.

\bibitem[Liu et~al., 2017]{liu2017high}
Liu, Y., Yuan, X., Li, M.-H., Zhang, W., Zhao, Q., Zhong, J., Cao, Y., Li,
  Y.-H., Chen, L.-K., Li, H., et~al. (2017).
\newblock High speed self-testing quantum random number generation without
  detection loophole.
\newblock In {\em Frontiers in Optics}, pages FTh2E--1. Optical Society of
  America.

\bibitem[Lydersen et~al., 2010]{lydersen2010hacking}
Lydersen, L., Wiechers, C., Wittmann, C., Elser, D., Skaar, J., and Makarov, V.
  (2010).
\newblock Hacking commercial quantum cryptography systems by tailored bright
  illumination.
\newblock {\em Nat. photonics}, 4(10):686--689.

\bibitem[Masanes, 2009]{masanes2009universally}
Masanes, L. (2009).
\newblock Universally composable privacy amplification from causality
  constraints.
\newblock {\em Physical review letters}, 102(14):140501.

\bibitem[Masanes et~al., 2011]{masanes2011secure}
Masanes, L., Pironio, S., and Ac{\'\i}n, A. (2011).
\newblock Secure device-independent quantum key distribution with causally
  independent measurement devices.
\newblock {\em Nature communications}, 2:238.

\bibitem[Masanes et~al., 2014]{masanes2014full}
Masanes, L., Renner, R., Christandl, M., Winter, A., and Barrett, J. (2014).
\newblock Full security of quantum key distribution from no-signaling
  constraints.
\newblock {\em Information Theory, IEEE Transactions on}, 60(8):4973--4986.

\bibitem[Mayers and Yao, 1998]{mayers1998quantum}
Mayers, D. and Yao, A. (1998).
\newblock Quantum cryptography with imperfect apparatus.
\newblock In {\em Foundations of Computer Science, 1998. Proceedings. 39th
  Annual Symposium on}, pages 503--509. IEEE.

\bibitem[Mayers and Yao, 2003]{mayers2003self}
Mayers, D. and Yao, A. (2003).
\newblock Self testing quantum apparatus.
\newblock {\em arXiv preprint quant-ph/0307205}.

\bibitem[McKague et~al., 2012]{mckague2012robust}
McKague, M., Yang, T.~H., and Scarani, V. (2012).
\newblock Robust self-testing of the singlet.
\newblock {\em Journal of Physics A: Mathematical and Theoretical},
  45(45):455304.

\bibitem[Miller and Shi, 2014]{miller2014robust}
Miller, C.~A. and Shi, Y. (2014).
\newblock Robust protocols for securely expanding randomness and distributing
  keys using untrusted quantum devices.
\newblock In {\em Proceedings of the 46th Annual ACM Symposium on Theory of
  Computing}, pages 417--426. ACM.

\bibitem[Natarajan and Vidick, 2017]{natarajan2017quantum}
Natarajan, A. and Vidick, T. (2017).
\newblock A quantum linearity test for robustly verifying entanglement.
\newblock In {\em Proceedings of the 49th Annual ACM SIGACT Symposium on Theory
  of Computing}, pages 1003--1015. ACM.

\bibitem[Navascu{\'e}s et~al., 2008]{navascues2008convergent}
Navascu{\'e}s, M., Pironio, S., and Ac{\'\i}n, A. (2008).
\newblock A convergent hierarchy of semidefinite programs characterizing the
  set of quantum correlations.
\newblock {\em New Journal of Physics}, 10(7):073013.

\bibitem[Nielsen and Chuang, 2002]{nielsen2002quantum}
Nielsen, M.~A. and Chuang, I. (2002).
\newblock Quantum computation and quantum information.

\bibitem[Nieto-Silleras et~al., 2016]{nieto2016device}
Nieto-Silleras, O., Bamps, C., Silman, J., and Pironio, S. (2016).
\newblock Device-independent randomness generation from several bell
  estimators.
\newblock {\em arXiv preprint arXiv:1611.00352}.

\bibitem[Nieto-Silleras et~al., 2018]{nieto2018device}
Nieto-Silleras, O., Bamps, C., Silman, J., and Pironio, S. (2018).
\newblock Device-independent randomness generation from several bell
  estimators.
\newblock {\em New Journal of Physics}, 20(2):023049.

\bibitem[Paris and {\v{R}}eh{\'a}{\v{c}}ek, 2004]{tomography}
Paris, M.~G. and {\v{R}}eh{\'a}{\v{c}}ek, J., editors (2004).
\newblock {\em Quantum State Estimation}.
\newblock Springer.

\bibitem[Pironio et~al., 2009]{pironio2009device}
Pironio, S., Ac{\'\i}n, A., Brunner, N., Gisin, N., Massar, S., and Scarani, V.
  (2009).
\newblock Device-independent quantum key distribution secure against collective
  attacks.
\newblock {\em New Journal of Physics}, 11(4):045021.

\bibitem[Pironio et~al., 2010]{pironio2010random}
Pironio, S., Ac{\'\i}n, A., Massar, S., de~La~Giroday, A.~B., Matsukevich,
  D.~N., Maunz, P., Olmschenk, S., Hayes, D., Luo, L., Manning, T.~A., et~al.
  (2010).
\newblock Random numbers certified by {B}ell's theorem.
\newblock {\em Nature}, 464(7291):1021--1024.

\bibitem[Popescu and Rohrlich, 1992]{popescu1992states}
Popescu, S. and Rohrlich, D. (1992).
\newblock Which states violate bell's inequality maximally?
\newblock {\em Physics Letters A}, 169(6):411--414.

\bibitem[Portmann and Renner, 2014]{portmann2014cryptographic}
Portmann, C. and Renner, R. (2014).
\newblock Cryptographic security of quantum key distribution.
\newblock {\em arXiv preprint arXiv:1409.3525}.

\bibitem[Raggio and Werner, 1989]{raggio1989quantum}
Raggio, G. and Werner, R. (1989).
\newblock Quantum statistical mechanics of general mean field systems.
\newblock {\em Helvetica Physica Acta}, 62(8):980--1003.

\bibitem[Rao, 2011]{rao2011parallel}
Rao, A. (2011).
\newblock Parallel repetition in projection games and a concentration bound.
\newblock {\em SIAM Journal on Computing}, 40(6):1871--1891.

\bibitem[Raz, 1998]{raz1998parallel}
Raz, R. (1998).
\newblock A parallel repetition theorem.
\newblock {\em SIAM Journal on Computing}, 27(3):763--803.

\bibitem[Raz, 2011]{raz2011counterexample}
Raz, R. (2011).
\newblock A counterexample to strong parallel repetition.
\newblock {\em SIAM Journal on Computing}, 40(3):771--777.

\bibitem[Reichardt et~al., 2013]{reichardt2013classical}
Reichardt, B.~W., Unger, F., and Vazirani, U. (2013).
\newblock Classical command of quantum systems.
\newblock {\em Nature}, 496(7446):456--460.

\bibitem[Renner, 2007]{renner2007symmetry}
Renner, R. (2007).
\newblock Symmetry of large physical systems implies independence of
  subsystems.
\newblock {\em Nature Physics}, 3(9):645--649.

\bibitem[Renner, 2008]{renner2008security}
Renner, R. (2008).
\newblock Security of quantum key distribution.
\newblock {\em International Journal of Quantum Information}, 6(01):1--127.

\bibitem[Renner, 2010]{renner2010simplifying}
Renner, R. (2010).
\newblock Simplifying information-theoretic arguments by post-selection.
\newblock In {\em NATO Advanced Research Workshop Quantum Cryptography and
  Computing: Theory and Implementation}, volume~26, pages 66--75. IOS Press.

\bibitem[Renner and K{\"o}nig, 2005]{renner2005universally}
Renner, R. and K{\"o}nig, R. (2005).
\newblock Universally composable privacy amplification against quantum
  adversaries.
\newblock In {\em Theory of Cryptography}, pages 407--425. Springer.

\bibitem[Renner and Wolf, 2004]{renner2004smooth}
Renner, R. and Wolf, S. (2004).
\newblock Smooth r{\'e}nyi entropy and applications.
\newblock In {\em Information Theory, 2004. ISIT 2004. Proceedings.
  International Symposium on}, page 233. IEEE.

\bibitem[Renner and Wolf, 2005]{renner2005simple}
Renner, R. and Wolf, S. (2005).
\newblock Simple and tight bounds for information reconciliation and privacy
  amplification.
\newblock In {\em Advances in cryptology-ASIACRYPT 2005}, pages 199--216.
  Springer.

\bibitem[Ribeiro et~al., 2018]{ribeiro2018device}
Ribeiro, J., Kaniewski, J., Helsen, J., Wehner, S., et~al. (2018).
\newblock Device independence for two-party cryptography and position
  verification with memoryless devices.
\newblock {\em Physical Review A}, 97(6):062307.

\bibitem[Ribeiro et~al., 2017]{ribeiro2017fully}
Ribeiro, J., Murta, G., and Wehner, S. (2017).
\newblock Fully device independent conference key agreement.
\newblock {\em arXiv preprint arXiv:1708.00798}.

\bibitem[Scarani, 2013]{scarani2013device}
Scarani, V. (2013).
\newblock The device-independent outlook on quantum physics (lecture notes on
  the power of {B}ell's theorem).
\newblock {\em arXiv preprint arXiv:1303.3081}.

\bibitem[Scarani et~al., 2006]{scarani2006secrecy}
Scarani, V., Gisin, N., Brunner, N., Masanes, L., Pino, S., and Ac{\'\i}n, A.
  (2006).
\newblock Secrecy extraction from no-signaling correlations.
\newblock {\em Physical Review A}, 74(4):042339.

\bibitem[Scarani and Renner, 2008a]{scarani2008quantum}
Scarani, V. and Renner, R. (2008a).
\newblock Quantum cryptography with finite resources: Unconditional security
  bound for discrete-variable protocols with one-way postprocessing.
\newblock {\em Physical review letters}, 100(20):200501.

\bibitem[Scarani and Renner, 2008b]{scarani2008security}
Scarani, V. and Renner, R. (2008b).
\newblock Security bounds for quantum cryptography with finite resources.
\newblock In {\em Theory of Quantum Computation, Communication, and
  Cryptography}, pages 83--95. Springer.

\bibitem[Schrijver, 1998]{schrijver1998theory}
Schrijver, A. (1998).
\newblock {\em Theory of linear and integer programming}.
\newblock John Wiley \& Sons.

\bibitem[Schwemmer et~al., 2015]{schwemmer2015systematic}
Schwemmer, C., Knips, L., Richart, D., Weinfurter, H., Moroder, T., Kleinmann,
  M., and G{\"u}hne, O. (2015).
\newblock Systematic errors in current quantum state tomography tools.
\newblock {\em Physical review letters}, 114(8):080403.

\bibitem[Shalm et~al., 2015]{shalm2015strong}
Shalm, L.~K., Meyer-Scott, E., Christensen, B.~G., Bierhorst, P., Wayne, M.~A.,
  Stevens, M.~J., Gerrits, T., Glancy, S., Hamel, D.~R., Allman, M.~S., et~al.
  (2015).
\newblock Strong loophole-free test of local realism.
\newblock {\em Physical review letters}, 115(25):250402.

\bibitem[Shannon, 1948]{shannon2001mathematical}
Shannon, C.~E. (1948).
\newblock A mathematical theory of communication.
\newblock {\em The Bell System Technical Journal}, 27.

\bibitem[Slofstra, 2017]{slofstra2017set}
Slofstra, W. (2017).
\newblock The set of quantum correlations is not closed.
\newblock {\em arXiv preprint arXiv:1703.08618}.

\bibitem[Tomamichel, 2012]{tomamichel2012framework}
Tomamichel, M. (2012).
\newblock A framework for non-asymptotic quantum information theory.
\newblock {\em arXiv preprint arXiv:1203.2142}.

\bibitem[Tomamichel, 2015]{tomamichel2015quantum}
Tomamichel, M. (2015).
\newblock {\em Quantum Information Processing with Finite Resources:
  Mathematical Foundations}, volume~5.
\newblock Springer.

\bibitem[Tomamichel et~al., 2009]{tomamichel2009fully}
Tomamichel, M., Colbeck, R., and Renner, R. (2009).
\newblock A fully quantum asymptotic equipartition property.
\newblock {\em Information Theory, IEEE Transactions on}, 55(12):5840--5847.

\bibitem[Tomamichel et~al., 2010a]{tomamichel2010entropyduality}
Tomamichel, M., Colbeck, R., and Renner, R. (2010a).
\newblock Duality between smooth min- and max-entropies.
\newblock {\em IEEE Transactions on Information Theory}, 56(9):4674--4681.

\bibitem[Tomamichel and Hayashi, 2013]{tomamichel2013hierarchy}
Tomamichel, M. and Hayashi, M. (2013).
\newblock A hierarchy of information quantities for finite block length
  analysis of quantum tasks.
\newblock {\em IEEE Transactions on Information Theory}, 59(11):7693--7710.

\bibitem[Tomamichel and Leverrier, 2017]{tomamichel2017largely}
Tomamichel, M. and Leverrier, A. (2017).
\newblock A largely self-contained and complete security proof for quantum key
  distribution.
\newblock {\em Quantum}, 1:14.

\bibitem[Tomamichel et~al., 2010b]{tomamichel2010leftover}
Tomamichel, M., Renner, R., Schaffner, C., and Smith, A. (2010b).
\newblock Leftover hashing against quantum side information.
\newblock In {\em Information Theory Proceedings (ISIT), 2010 IEEE
  International Symposium on}, pages 2703--2707. IEEE.

\bibitem[Tomamichel et~al., 2011]{tomamichel2011leftover}
Tomamichel, M., Schaffner, C., Smith, A., and Renner, R. (2011).
\newblock Leftover hashing against quantum side information.
\newblock {\em Information Theory, IEEE Transactions on}, 57(8):5524--5535.

\bibitem[Vazirani and Vidick, 2012]{vazirani2012certifiable}
Vazirani, U. and Vidick, T. (2012).
\newblock Certifiable quantum dice: or, true random number generation secure
  against quantum adversaries.
\newblock In {\em Proceedings of the forty-fourth annual ACM symposium on
  Theory of computing}, pages 61--76. ACM.

\bibitem[Vazirani and Vidick, 2014]{vazirani2014fully}
Vazirani, U. and Vidick, T. (2014).
\newblock Fully device-independent quantum key distribution.
\newblock {\em Physical review letters}, 113(14):140501.

\bibitem[Weier et~al., 2011]{weier2011quantum}
Weier, H., Krauss, H., Rau, M., F{\"u}rst, M., Nauerth, S., and Weinfurter, H.
  (2011).
\newblock Quantum eavesdropping without interception: an attack exploiting the
  dead time of single-photon detectors.
\newblock {\em New J. Phys.}, 13(7):073024.

\bibitem[Winick et~al., 2017]{winick2017reliable}
Winick, A., L{\"u}tkenhaus, N., and Coles, P.~J. (2017).
\newblock Reliable numerical key rates for quantum key distribution.
\newblock {\em arXiv preprint arXiv:1710.05511}.

\bibitem[Yuen, 2016]{yuen2016parallel}
Yuen, H. (2016).
\newblock A parallel repetition theorem for all entangled games.
\newblock {\em International Colloquium of Automata, Languages, and
  Programming}.

\end{thebibliography}

\end{document}